\documentclass[timesnewroman,12pt, twoside]{report}
\usepackage[titletoc]{appendix}
\usepackage[a4paper]{geometry}
\usepackage{authblk, lineno, color, lscape, url, sectsty}
\usepackage[autostyle]{csquotes}
\usepackage[usenames,dvipsnames,table]{xcolor}
\usepackage[labelfont=bf]{caption}
\usepackage{emptypage}
\usepackage{adjustbox}
\usepackage{relsize}
\usepackage[11pt]{moresize}
\usepackage[normalem]{ulem}
\usepackage[verbose]{placeins}
\usepackage{makecell}
\usepackage{array, float, multirow, graphicx, setspace, cite, url, titlesec, tocloft, etoolbox}
\usepackage{amsmath, amssymb, amsfonts, mathtools, notoccite}
\usepackage{chngcntr, rotating, pdflscape, enumitem, algpseudocode, algorithm, subcaption}
\usepackage[square,sort,comma,numbers,sort&compress]{natbib}
\usepackage[colorlinks=True,citecolor=blue,linkcolor=blue,anchorcolor=blue,filecolor=blue,urlcolor=blue]{hyperref}
\allowdisplaybreaks
\usepackage{tensor}
\usepackage{braket}
\usepackage{booktabs}
\usepackage{lipsum}
\usepackage[Sonny]{fncychap}
\usepackage{tikz, tikz-cd}

\usepackage{latexsym}
\usepackage{epsfig}  % Include figure files
\usepackage{epsf}    % Include figure files
\usepackage{bm}% bold math
\usepackage{dcolumn}% Align table columns on decimal point
\usepackage{textcomp}% Align table columns on decimal point
\usepackage{hypcap}
\usepackage{pifont}
\usepackage{appendix}
\usepackage{multirow,bigdelim}
\usepackage{multicol}  
\usepackage{diagbox}
\usepackage{afterpage}
\usepackage{capt-of}
\usepackage{scalerel}
\usepackage{wasysym}
\usepackage{mathrsfs}
\usepackage{tablefootnote}
\usepackage{fontawesome5}
\usepackage{arydshln}

\def\blankpage{%
	\clearpage%
	\thispagestyle{empty}%
	\addtocounter{page}{-1}%
	\null%
	\clearpage}
\newcommand{\mc}[3]{\multicolumn{#1}{#2}{#3}}
\newcommand{\mr}[3]{\multirow{#1}{#2}{#3}}

\newcommand{\capdef}{}
\newcommand{\mycaption}[2][\capdef]{\renewcommand{\capdef}{#2}
	\caption[#1]{{\footnotesize #2}}}
\makeatletter
\renewcommand{\fnum@table}{\textbf{\tablename~\thetable}}
\renewcommand{\fnum@figure}{\textbf{\figurename~\thefigure}}

\newcommand{\ldm}{\ensuremath{\Delta m_{31}^2}}
\newcommand{\sdm}{\ensuremath{\Delta m_{21}^2}}

\newcommand{\tym}{\ensuremath{\theta_{13}}}
\newcommand{\tzm}{\ensuremath{\theta_{23}}}
\newcommand{\aem}{\ensuremath{a_{e\mu}}}

\newcommand{\aet}{\ensuremath{a_{e\tau}}}
\newcommand{\amt}{\ensuremath{a_{\mu\tau}}}

\newcommand{\cem}{\ensuremath{c_{e\mu}}}

\newcommand{\cet}{\ensuremath{c_{e\tau}}}
\newcommand{\cmt}{\ensuremath{c_{\mu\tau}}}

\newcommand{\ie}{\textit{i.e.}}
\newcommand{\eg}{{\it e.g.}}
\newcommand{\nue}{\mbox{$\nu_e$}}

\newcommand{\numu}{\mbox{$\nu_{\mu}$}}

\newcommand{\fig}{fig.}

\newcommand{\Refe}{Ref.}
\newcommand{\Refes}{Refs.}

\newcommand{\equ}[1]{eq.~(\ref{equ:#1})}
\newcommand{\figu}[1]{\fig~\ref{fig:#1}}

\newcommand{\tx}{{\theta_{12}}}
\newcommand{\ty}{{\theta_{13}}}
\newcommand{\tz}{{\theta_{23}}}

\newcommand{\da}{\delta_{13}}
\newcommand{\db}{\delta_{14}}

\newcommand{\tmet}{\theta^{3\nu}_{\mu e}}
\newcommand{\tmef}{\theta^{4\nu}_{\mu e}}
\newcommand{\stfr}{\Delta{m}^{2}_{41}}
\newcommand{\pab}{P(\nu_{\alpha} \rightarrow \nu_{\beta})}

\newcommand{\barparen}[1]{\ensuremath{\rlap{\kern-2.5pt\ensuremath{\overset{\scriptscriptstyle(-)}{\phantom{\nu_{#1}}}}}{\ensuremath{{\nu}_{#1}}}}}

\makeatother

\newenvironment{conditions*}
{\par\vspace{\abovedisplayskip}\noindent
	\tabularx{\columnwidth}{>{$}l<{$} @{${}={}$} >{\raggedright\arraybackslash}X}}
{\endtabularx\par\vspace{\belowdisplayskip}}
\begin{document}
%%%%%%%%%%%%%%%%%%%  Title page %%%%%%%%%%%%%%%%%%%%%%%%%%%%%%%%%%%
\pagenumbering{roman}
\thispagestyle{empty}
%%%%%%%%%%%%%%%%%%%%%%%%%%%%%%%%%%%%%%%
\thispagestyle{empty}
\graphicspath{{Figures/PNG/}{Figures/}}
%%%%%%%%%%%%%%%%%%%%%%%%%%%%%%%%%%%%%%%
\begin{center}
{\bf {\Large Testing beyond the Standard Model scenarios in next-generation long-baseline neutrino oscillation experiments} }
\end{center}
%%%%%%%%%%%%
\vspace{0.3 cm}
\begin{center}
    {\it \large By}
\end{center}
%%%%%%%%%%%%%%
\begin{center}
    {\bf {\Large Pragyanprasu Swain } \\ PHYS07201804014}
\end{center}
%%%%%%%%%%%%%%%
\begin{center}
\bf {{\large Institute of Physics, Bhubaneswar, India }}
\end{center}
%%%%%%%%%%%%%%
\vskip 2.0 cm
\begin{center}
%%%%%%%%%%%%%%
\large{
{ A thesis submitted to the }  \\
 {Board of Studies in Physical Sciences }\\

In partial fulfillment of requirements \\
For the Degree of } \\
{\bf  DOCTOR OF PHILOSOPHY} \\
\emph{of} \\
{\bf HOMI BHABHA NATIONAL INSTITUTE}
%%%%%%%%%%%%%%%%%
\vskip 1.0 cm
%%%%%%%%%%%%%%%%%
\begin{figure}[H]
	\begin{center}
    \includegraphics[width=4.0cm, height= 4.0cm]{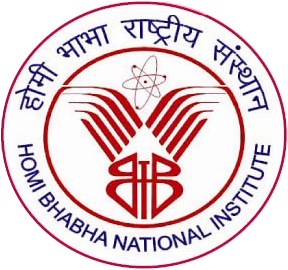}
	\end{center}
\end{figure}
%%%%%%%%%%%%%%%%%
\vskip 0.2 cm
{\bf August 12, 2025}
\end{center}
\thispagestyle{empty}
%%%%%%%%%%%%%%%%%%%%%  
\newpage
%\blankpage     
%%%%%%%%%%%%%%%%%%%% Recommendation page %%%%%%%%%%%%%%%%%%%%%%%%%%%
\begin{center}
{\Large\textbf{Homi Bhabha National Institute \\}}
{\large\textbf{\ \\Recommendations of the Viva Voce Committee}}
\end{center}
As members of the Viva Voce Committee, we certify that we have read the
dissertation prepared by \textbf{Pragyanprasu Swain} entitled ``\textbf{Testing beyond the Standard Model scenarios in next-generation long-baseline neutrino oscillation experiments}", and recommend that it may be accepted as fulfilling the thesis requirement for the award of Degree of Doctor of Philosophy.
\begin{center}
\begin{tabular}{p{0.74\linewidth}p{0.19\linewidth}}
\hline
Chairman - Prof. Suresh Kumar Patra & \textbf{Date:} \\
 & \\ \hline
Guide / Convener - Prof. Sanjib Kumar Agarwalla & \textbf{Date:}\\
 & \\ \hline
Co-Guide (if any)  & \textbf{Date:} \\
 & \\ \hline
Examiner -  Prof. S. Uma Sankar & \textbf{Date:} \\
 & \\ \hline
Member 1 - Prof. Aruna Kumar Nayak & \textbf{Date:}\\
 & \\ \hline
Member 2 - Prof. Kirtiman Ghosh  & \textbf{Date:}\\
 & \\ \hline
 Member 2 - Prof. Sanjay Kumar Swain & \textbf{Date:}
\end{tabular}
\end{center}
\rule{\linewidth}{1.0pt}

Final approval and acceptance of this thesis is contingent upon the candidate's
submission of the final copies of the thesis to HBNI.

I/We hereby certify that I/We have read this thesis prepared under my direction and recommend that it may be accepted as fulfilling the thesis requirement.\\
\begin{center}
\begin{tabular}{p{0.185\linewidth}p{0.36\linewidth}>{\centering\arraybackslash}p{0.36\linewidth}}
\textbf{Date:}& & \\[3 pt]
\textbf{Place:} & & \\[3 pt]
&  \textbf{Co-Guide (if applicable)}& \textbf{Guide} \hfill
\end{tabular}
\end{center}
%%%%%%%%%%%%%%%%%%%%%%%%%%%%%%%%%%%%%%
\newpage
%\blankpage
%%%%%%%%%%%%%%%%%%% Statement page %%%%%%%%%%%%%%%%%%%%%%%%%%%%%%%%
%%%%%%%%%%%%%%
%%%%%%%%%%%%%%%%%%%%
\begin{center}
\large \bf
STATEMENT BY AUTHOR
\end{center}
%%%%%%%%%%%%%%%%%%%%
This dissertation has been submitted in partial fulfillment of the requirements
for an advanced degree at Homi Bhabha National Institute (HBNI) and deposited
in the Library to be made available to borrowers under the rules of the HBNI.
\paragraph{}
Brief quotations from this dissertation are allowable without special
permission, provided that accurate acknowledgment of the source is made.
Requests for permission for extended  quotation from or reproduction
of this manuscript in whole or in part may be granted by the competent
Authority of HBNI when in his or her judgment the proposed use of the
material is in the interests of scholarship. In all other instances,
however, permission must be obtained from the author.
%%%%%%%%%%%%%%%%%%%%%%%%%%%%%%%%%
\vspace{1.2cm}
\begin{flushright}
	\textbf{(Pragyanprasu Swain)}
\end{flushright}
%%%%%%%%%%%%%%%%%%%%%%%%%%%%%%%%%
\newpage 
%\blankpage   
%%%%%%%%%%%%%%%%%%% Declaration page %%%%%%%%%%%%%%%%%%%%%%%%%%%%%%
%%%%%%%%%%%%%%%
\begin{center}
\large \bf
DECLARATION
\end{center}
%%%%%%%%%%%%%%%
I, {\textbf {Pragyanprasu Swain}}, hereby declare that the investigations presented in the thesis have been carried out by
me. The work is original and has not been submitted earlier as a whole or in part for a
degree/diploma at this or any other Institution/University.
%%%%%%%%%%%%%%%%%%%%%%%%%%%%%%%%%
\vspace{1.2cm}
\begin{flushright}
	\textbf{(Pragyanprasu Swain)}
\end{flushright}
%%%%%%%%%%%%%%%%%%%%%%%%%%%%%%%%%
 
\newpage    
%\blankpage
%%%%%%%%%%%%%%%%%%% List of publications %%%%%%%%%%%%%%%%%%%%%%%%%%
%%%%%%%%%%%%%%%%%%%%%%%%%%%%%%%%%%%%%%%%%%%%%%%%%%
\begin{center}
\textbf{\large \underline{List of publications arising from this thesis}}
\end{center}
%%%%%%%%%%%%%%%%%%%%%%%%%%%%%%%%%%%%%%%%%%%%%%%%%%
\begin{itemize}
	\item{\textbf{Papers published in peer-reviewed journals}}
\begin{enumerate}
%%%%%%%%%%%%%%
\item \textit{Constraining Lorentz invariance violation with next-generation long-baseline experiments} \\
Sanjib Kumar Agarwalla, Sudipta Das, Sadashiv Sahoo, \textbf{Pragyanprasu Swain},\\
\href{https://link.springer.com/article/10.1007/JHEP07(2023)216}{\textcolor{blue}{\it JHEP 07 (2023) 216}}
%%%%%%%%%%%%%%
\item \textit{A plethora of long-range neutrino interactions probed by DUNE and T2HK}\\
Sanjib Kumar Agarwalla, Mauricio Bustamante, Masoom Singh,\\ \textbf{Pragyanprasu Swain},\\ \href{https://link.springer.com/article/10.1007/JHEP09(2024)055}{\textcolor{blue}{\it JHEP 09 (2024) 055}}
%%%%%%%%%%%%%%%
\end{enumerate}
%%%%%%%%%%%%%%%%%%%%%%%%%%%%%%%%%%%%%%%%%%%%%%%%%%
\item \textbf{Papers to be submitted in peer-reviewed journals}
%%%%%%%%%%%%%%%%%%%%%%%%%%%%%%%%%%%%%%%%%%%%%%%%%%
\begin{enumerate}
%%%%%%%%%%%%%%%
\item  \textit{Active-sterile neutrino oscillations in long-baseline experiments over a wide mass-squared range}\\
Sanjib Kumar Agarwalla, Suprabh Prakash, Samiran Roy, \textbf{Pragyanprasu Swain},\\
{\textcolor{blue} {Preprint: IOP/BBSR/2024-07}}
%%%%%%%%%%%%%%%
\end{enumerate}
%%%%%%%%%%%%%%%%%%%%%%%%%%%%%%%%%%%%%%%%%%%%%%%%%
\item \textbf{{Conference Proceedings}}
%%%%%%%%%%%%%%%%%%%%%%%%%%%%%%%%%%%%%%%%%%%%%%%%%
\begin{enumerate}
\item \textit{A plethora of long-range neutrino interactions probed by DUNE and T2HK}\\
Sanjib Kumar Agarwalla, Mauricio Bustamante, Masoom Singh,\\ \textbf{Pragyanprasu Swain},\\
\href{https://arxiv.org/abs/2501.14835}{\textcolor{blue}{\it arXiv: 2501.14835}}
\end{enumerate}
\end{itemize}
%%%%%%%%%%%%%%%%%%%%%%%%%%%%%%%%%%%%%%%%%%%%%%%%%%
\begin{center}
{\textbf{\large\underline{Talks/Posters}}}
\end{center}
%%%%%%%%%%%%%%%%%%%%%%%%%%%%%%%%%%%%%%%%%%%%%%%%%%
%%%%%%%%%%%%%%%
\begin{enumerate}
	
\item {\bf Talk:} \textcolor{blue}{\textit{A plethora of long-range neutrino interactions probed by DUNE and T2HK}}\\
{ NuFact 2024 - The 25th International Workshop on Neutrinos from Accelerators,  16–21 Sep 2024, Argonne National Laboratory, USA}

\item {\bf Talk:} \textcolor{blue}{\textit{Probing Lorentz Invariance Violation with Next-Generation Long-Baseline Neutrino Oscillation Experiments}}\\
{ National Conference on Recent Advances in Material and Particle Physics \\(NCRAMPP – 2023), Berhampur University, Odisha}
	
\item {\bf Poster:} \textcolor{blue}{\it Constraining Lorentz Invariance Violation with Future Long-Baseline Experiments}\\
{24th International Workshop on Neutrinos from Accelerators (NuFact 2023)}

\item {\bf Talk:} \textcolor{blue}{\textit{Probing active-sterile neutrino oscillations in long-baseline experiments for a wide
		mass-squared range}}\\
{Neutrino Workshop at IFIRSE, ICISE, Quy Nhon, Vietnam on July 19, 2023}
	
\item {\bf Poster:} \textcolor{blue}{\textit{Active-sterile neutrino oscillation in long-baseline experiments for a wide $\Delta m^2$ range}}\\
{XXX International conference on Neutrino Physics and Astrophysics (Neutrino 2022)}
\end{enumerate}
%%%%%%%%%%%%%%%%%%%%%%%%%%%%%%%%
\vspace{1cm}
\begin{flushright}
	\textbf{(Pragyanprasu Swain)}
\end{flushright}
%%%%%%%%%%%%%%%%%%%%%%%%%%%%%%%%%
\newpage
%\blankpage
%%%%%%%%%%%%%%%%%%% Dedication page %%%%%%%%%%%%%%%%%%%%%%%%%%%%%%%
%%%%%%%%%%%%%%%%%%%%%%%%%%%%%
\begin{center}
	{\large\textbf{DEDICATIONS}}
\end{center}
%%%%%%%%%%%%%%%%%%%%%%%%%%%%
\vspace*{3.0in}
\hspace*{2.0in}
\begin{center}
        {\Large \emph{Dedicated To\\ My\\ Father Late Mr. Surendra Narayan Swain \\and \\Mother Mrs. Pratima Swain}\\}
        \vspace{0.1cm}
%\hrule
\end{center}
    
\newpage
%\blankpage  
%%%%%%%%%%%%%%%%%%% Acknowledgement page %%%%%%%%%%%%%%%%%%%%%%%%%%
%%%%%%%%%%%%%%%%
\begin{center}
\large \bf
ACKNOWLEDGMENTS
\end{center}
%%%%%%%%%%%%%%%%
\vspace{1.5cm}
%%%%%%%%%%%%%%%%
Completing this thesis has been a significant milestone in my life. I am fortunate enough to have received support from many remarkable individuals during this journey. I want to take the opportunity to thank them all.

First and foremost, I thank my supervisor, Prof. Sanjib Kumar Agarwalla, for constantly guiding me throughout this Ph.D. tenure. It has been great working under his supervision. He has been very supportive and encouraging during the difficult times. His insightful comments and suggestions have immensely helped me to complete this thesis.

I want to express my heartfelt gratitude to my collaborator, Prof. Mauricio Bustamante, who has been incredible in providing many to-the-point, in-depth suggestions and advice in my Ph.D. work. His critical perspective has been a constant source of inspiration during our last work, pushing me to refine my work. I also acknowledge the support from my collaborators, Dr. Mehedi Masud, Dr. Samiran Roy, and Dr. Suprabh Prakash. I thank my friends-cum-collaborators, Dr. Sudipta Das, Sadashiv Sahoo, and Dr. Masoom Singh, for several physics and non-physics discussions, all in a friendly and empathetic atmosphere. I thank my senior, Dr. Ashish Narang, who has been a very good friend for the last three years. I would also like to acknowledge my other group members, Dr. Anil Kumar, Ritam, Anuj, Krishna, Sharmishta, Gopal, and Dr. Thiru Senthil, for creating a family-like atmosphere. I thank all of you for your input in my Ph.D. journey.  

I thank my doctoral committee members, Prof. Suresh Kumar Patra, Prof. Aruna Kumar Nayak, Prof. Kirtiman Ghosh, and Prof. Sanjay Kumar Swain, for their valuable comments and suggestions at several stages of Ph.D., especially during the annual review talks in IOP. 

%I thank Prof. Shamik Banerjee, Prof. Debottam Das, and Prof. Manimala Mitra for their helpful lectures, respectively, on Advanced Quantum Mechanics, QFT-I, and QFT-II during our Predoctoral coursework, which helped me grow as a High-Energy Physics student.  

I want to express my sincere gratitude to Prof. 
Amitava Raychaudhuri for his generosity in providing invaluable comments on the papers included in this thesis.

I greatly acknowledge the IOP cluster facility SAMKHYA and the associated members for extending their support when necessary. 

I want to thank my batchmates, Abhishek, Arpan, Arnob, Ankit, Aisha, Chitrak, Harish, Mousam, Pritam, Ritam, Sameer, Sandhya, Siddharth, Sudipta, Juniors -- Moonsun, Subhadip, and Rameswar, and Seniors -- Saiyad Da and Rupam Da, for making my life at IOP full of friends and lively moments. Your camaraderie, laughter, and support have made this journey unforgettable.

My deepest gratitude goes to one of the great teachers I have had the privilege of learning from, Dr. Suchismita Mohanty. 
Her exceptional teaching style and a great way of inspiring students have encouraged me and many others to pursue physics.

My seniors, Dr. Krushna Chandra Barik and Dr. Durga Prasad Khatua, have always inspired me. From my college days, they have supported me with several prospects in life, whether academic, emotional, social, or financial.

I feel so fortunate to have friends like Tina, Bamadev, Sangram, Manas, and Anil, who have been there to share the ups and downs of my life. 

I extend my heartiest gratitude to my family -- Jitu Bhai, Situ Bhai, Litan, Sonali Bhauja, my uncle, Mr. Ramesh Chandra Swain, my Aunty, Mrs. Pramila Swain, and my mother, Mrs. Pratima Swain -- for their unwavering love and care, which have been a constant source of strength in my life. I am profoundly grateful to my father, late Mr. Surendra Narayan Swain whose studious nature, self-made journey, and hardworking personality have been a guiding light in my life. My mother's love for mathematics and my father's passion for study have both greatly shaped my academic path. My heartfelt gratitude goes to my Piusa, Mr. Nilamani Swain, and Piusi, Mrs. Sanjulata Swain, for their unwavering support during one of the most vulnerable times in my academic journey. I am thankful to my cousins, Bubu and Babi, for believing in my academic abilities. Solving their doubts and explaining concepts deepened my understanding and boosted my confidence during my college days.  Finally, I would like to thank my wife, Arati, for all her love and support, especially in times of self-doubt, disappointment, and countless failures. Thank you, Arati, for never giving up on me. 

%%%%%%%%%%%%%%%%%%%%%%%%%%%%%%%%%
\vspace{1.2cm}
\begin{flushright}
	\textbf{(Pragyanprasu Swain)}
\end{flushright}
%%%%%%%%%%%%%%%%%%%%%%%%%%%%%%%%%

%\blankpage
%%%%%%%%%%%%%%%%% Table of contents %%%%%%%%%%%%%%%%%%%%%%%%%%%%%
\hypersetup{linkcolor=blue}
\tableofcontents
\blankpage 
%%%%%%%%%%%%%%%%%%%%%%%%%%%%%%%%
\numberwithin{equation}{chapter}
\numberwithin{figure}{chapter}
\numberwithin{table}{chapter}
%%%%%%%%%%%%%%%%%%%%%%%%%%%%%%

%%%%%%%%%%%%%%%%%%% Table of figures %%%%%%%%%%%%%%%%%%%%%%%%%%%%%%
\addcontentsline{toc}{chapter}{List of Figures}
\listoffigures
\blankpage 
%%%%%%%%%%%%%%%%%%% Table of tables %%%%%%%%%%%%%%%%%%%%%%%%%%%%%%%
\addcontentsline{toc}{chapter}{List of Tables}
\listoftables
\blankpage 
%%%%%%%%%%%%%%%%%%%%%%%%%%%%%%%%%%%%%%%%%%%%%%%%%%%%%%%%%%%%%%%%%%%
%%%%%%%%%%%%%%%%%%%% Synopsis %%%%%%%%%%%%%%%%%%%%%%%%%%%%%%%%%%%%%%
%%%%%%%%%%%%%%%%%%%%%%%%%%%%%%%%%%%%%%%%
\addcontentsline{toc}{chapter}{Summary}
%%%%%%%%%%%%%%%%%%%%%%%%%%%%%%%%%%%%%%%%
\begin{center}
	{\large\textbf{Summary}}
\end{center}
%%%%%%%%%%%%%%%%%%%%%%%%%%%%%
The three-flavor neutrino oscillation framework involves three mixing angles, $\theta_{12}$, $\theta_{13}$, and $\theta_{23}$, one Dirac CP violating phase $\delta_{\rm CP}$, and two mass-squared splittings, $\Delta m^2_{31}$ and  $\Delta m^2_{21}$. They have been measured with unprecedented precision using the data from various neutrino experiments using solar, reactor, atmospheric, and accelerator neutrinos. There are still some open issues to be resolved in this framework, which are (i) the octant issue of  $\theta_{23}$, (ii) the absolute value of $\delta_{\rm CP}$, and (iii) if the neutrino mass ordering is normal ($\Delta m^2_{31}>0$) or inverted ($\Delta m^2_{31} < 0$). The next generation long-baseline neutrino oscillation experiments, the Deep Underground Neutrino Experiment (DUNE), Tokai-to-Hyper-Kamiokande (T2HK), and Tokai-to-Hyper-Kamiokande with a second detector in Korea (T2HKK), hold promise to resolve these issues. Apart from solving the above issues, these experiments offer an excellent platform to look for any sub-leading Beyond the Standard Model (BSM) physics scenarios which can modify neutrino oscillations.  In this thesis, we probe three popular BSM scenarios, namely, (i) long-range neutrino-matter interaction, (ii) Lorentz invariance violation, and (iii) active-sterile neutrino oscillations. 

In the first project, we probe flavor-dependent long-range neutrino interactions with matter mediated by an ultra-light vector boson with mass between $10^{-35}-10^{-10}$ eV, in the context of DUNE and T2HK. We introduce the new neutrino-matter interactions by gauging fourteen different anomaly-free $U(1)^\prime$ symmetries that are combinations of baryon and lepton numbers. We find in this study that DUNE and T2HK both can constrain the new interactions to the same order as the standard matter potential, discover them in some favorable cases, and distinguish between the alternate symmetries.

In the second project, we test the possibility of Lorentz invariance violation, a Planck-scale ($10^{19}$ GeV) phenomenon, in the above-mentioned experiments, DUNE and T2HK. We probe both the CPT-conserving and CPT-violating Lorentz invariance violating (LIV) parameters. We find through analytical derivation of the oscillation probabilities that the strength of the CPT-violating LIV parameters is proportional to the baseline and that of the CPT-conserving ones is proportional to the product of energy and baseline.  DUNE, with its longer baseline (1285 km) and access to high-energy neutrinos (GeV), has an upper hand in constraining both the CPT-conserving and CPT-violating LIV parameters. T2HK, due to its relatively shorter baseline (295 km) and access to low energy neutrinos (sub-GeV) remains almost blind to the CPT-conserving LIV parameters.  

In the third project, we probe the active-sterile neutrino oscillations in DUNE, T2HK (Japanese detector, JD), and T2HKK (JD with another detector in Korea, JD+KD), over a wide range of sterile mass-squared splitting $\Delta m^2_{41}$ ($10^{-5}- 10^2$ eV$^2$). In the 3+1 scheme of neutrino active-sterile oscillation, we have extra CP phases that induce additional CP violation. We find that the CP violation discovery potential and the CP phase reconstruction capabilities of the above experiments are maximum in the case of $\Delta m^2_{41}$ = $\Delta m^2_{31}$, compared to the cases where  $\Delta m^2_{41}$ = $\Delta m^2_{21}$ and  $\Delta m^2_{41}$ = 1 eV$^2$. We place exclusion limits in the plane of  $\Delta m^2_{41}-\sin^2\theta_{14}$, $\Delta m^2_{41}-\sin^2\theta_{24}$, and $\Delta m^2_{41}-\sin^22\theta_{\mu e}$ ($\sin^22\theta_{\mu e}\equiv  4 \sin^2\theta_{14}\cos^2 \theta_{14} \sin^2
\theta_{24}$). In deriving these exclusion limits, we consider the presence of DUNE Near Detector (ND) along with the Far Detector (FD) and one Intermediate Water Cherenkov Detector (IWCD) for the T2HK/T2HKK setups. We find that the inclusion of these near detectors enhances the sensitivity at the higher $\Delta m^2_{41}$  (0.1 to 10 eV$^2$) region. At 90\% C.L., DUNE (FD+ND) and JD+KD+IWCD exclude $\theta_{14} \geq 1.9^\circ$ for $\Delta m^2_{41}\sim$10 eV$^2$, and  $\theta_{14}\geq 3^\circ$ for $\Delta m^2_{41} \sim 1~{\rm eV}^2$, respectively. Similarly, for $\theta_{24}$, at 90\% C.L.,  DUNE (FD+ND) and JD+KD+IWCD exclude $\theta_{24} \geq 0.6^\circ$ for $\Delta m^2_{41}\sim 5-10$ eV$^2$, and  $\theta_{24}\geq 1.2^\circ$ for $\Delta m^2_{41}\sim 1~{\rm eV}^2$, respectively. We find that in  $\Delta m^2_{41}-\sin^22\theta_{\mu e}$ plane, our setups, DUNE (FD+ND) and JD+KD+IWCD, can probe the anomalous allowed regions from the LSND and MiniBooNE experiments. 

\blankpage 
\newpage
\setcounter{page}{1}
\pagenumbering{arabic}
%%%%%%%%%%%%%%%%%%%%%%%%% Chapter-1 %%%%%%%%%%%%%%%%%%%%%%%%%%%%%%%
%\blankpage
%%%%%%%%%%%%%%%%%%%%%%% CHAPTER - 1 %%%%%%%%%%%%%%%%%%%%
\chapter{Introduction}
\label{C1} 
%%%%%%%%%%%%%%%%%%%%%%%%%%%%
%\graphicspath{{Figures/Chapter-1figs/PDF/}{Figures/Chapter-1figs/}}
%%%%%%%%%%%%%%%%%%%%%%%%%%%%
Neutrinos are the second most abundant particles after photon, but still the most elusive ones. Within the Standard Model (SM) of particle physics, neutrinos are massless and electrically neutral, interacting only via weak interactions. The extremely feeble interaction of neutrinos with matter makes their detection very challenging. That is why neutrinos draw the attention of most of the scientific community. Adding to their intrigue, neutrinos are known to oscillate between different flavors while propagating in space and time. Thanks to the relentless efforts of our scientists and engineers, today, we have advanced detector techniques to catch these mysterious particles. Experimental results from the solar, atmospheric, reactor, and accelerator-based neutrino experiments have firmly established the phenomenon of neutrino oscillation, which necessitates non-zero neutrino masses and mixing among their flavors. Their non-zero masses, inferred from neutrino oscillation, compel the physicists to go beyond the Standard Model (BSM). The BSM theories developed to accommodate non-zero neutrino masses and mixing often lead to new neutrino species as well as new neutrino interactions. Additionally, in the context of unifying the four fundamental forces of Nature to a single theoretical framework, the canonical theories like quantum loop gravity and string theory suggest violation of fundamental symmetries like Lorentz invariance, which necessarily affects neutrino propagation.

Despite remarkable experimental progress, several key questions still need to be answered within the standard oscillation framework. Among these, the determination of CP violation in the lepton sector and the neutrino mass ordering--normal or inverted--remain unresolved. Next-generation long-baseline neutrino oscillation experiments like the Deep Underground Neutrino Experiment (DUNE) and Tokai-to-Hyper-Kamiokande (T2HK) hold promise to resolve these issues with high significance.

This thesis focuses on exploring some interesting BSM scenarios in the context of neutrino oscillations, leveraging the precision capabilities of next-generation long-baseline experiments such as DUNE and T2HK. Specifically, this work investigates how new physics scenarios, including long-range neutrino interactions, Lorentz invariance violation, and the presence of sterile neutrinos, modify neutrino oscillation probabilities, leaving their signatures in oscillation experiments.

We start this thesis with the basic properties of neutrinos within the Standard Model of particle physics.
%%%%%%%%%%%%%%%%%%%%%%%%%%%%
%%%%%%%%%%%%%%%%%%%%%%%%%%%%
\section{The Standard Model and Neutrinos}
%%%%%%%%%%%%%%%%%%%%%%%%%%%%
The Standard Model (SM) has been proven to be the best theory for understanding almost all the fundamental particles and their interactions. The SM unifies three of the four fundamental forces -- the strong nuclear force, the weak nuclear force, and the electromagnetic force. Gauge invariance is the key to the SM. It is based on the gauge group $SU(3)_{C} \times SU(2)_{L} \times U(1)_{Y}$, where $C$ stands for color, $L$ for weak isospin, and $Y$ for hypercharge. The SM matter particles are leptons and quarks and are fermionic in nature; however, the force carriers are bosons. The Higgs boson discovery in 2012~\cite{CMS:2013btf, ATLAS:2012yve} completed the SM particle spectrum. The so-called Brout-Englert-Higgs mechanism~\cite{Englert:1964et, Higgs:1964pj} (commonly referred to as the Higgs Mechanism) provides the mass generation mechanism for the weak gauge bosons, all the SM fermions but neutrinos. The neutrinos in the SM are colorless, chargeless, and massless. In other words, neutrinos do not take part in the strong and  electromagnetic interactions neither do they have gravitational interaction. They interact only via weak interaction. Below, we discuss the properties of neutrinos in the SM. 

\vspace{0.1cm}
\noindent\textbf{(i) Number of neutrino species in the SM:}

The number of light neutrino species (typically dubbed as active neutrinos) is known to be three from the measurement of the invisible decay width of $Z^0$ in the LEP experiment~\cite{ALEPH:2005ab}. Anything beyond these three flavors must not have the SM interactions and, hence, should be sterile with respect to the SM gauge group. Considering the presence of one light sterile neutrino, we discuss the possibility of active-sterile neutrino oscillation in chapter~\ref{C5}.

\vspace{0.1cm}
\noindent\textbf{(ii) Lepton flavor and lepton number violation:}

The Standard Model respects the lepton flavor conservation, as a result, neutrino flavor remains associated with its corresponding charged lepton. For an instance, an electron neutrino will produce an electron via the charged-current interaction. The Standard Model also respects lepton number conservation, where the total number of leptons remains constant in a given interaction. However, the discovery of neutrino oscillations indicates that the lepton flavor is not strictly conserved in propagation, as neutrinos can change flavor.

In theories that extend the Standard Model, such as those involving Majorana neutrinos, lepton number conservation can be violated. The neutrinoless double beta decay is one such example that would provide a direct probe of the nature of neutrinos. 

\vspace{0.1cm}
\noindent\textbf{(iii) Neutrino interactions in the SM:}\\
The charged-current (CC) and neutral current (NC) neutrino interactions with the SM leptons and quarks mediated by the weak gauge bosons $W^{\pm}$ and $Z$ are described by the following CC and NC Lagrangians, 
\begin{eqnarray}
	\label{equ:lagrangian_CC_sm}
	\mathcal{L}_{\text{CC}}
	&=&
	\frac{e}{\sqrt{2} \sin \theta_{W}} 
	\left[W^{+}_\mu\{{\bar{\nu}_{\alpha}\gamma^\mu P_{L} l_{\alpha}+\bar{u}\gamma^\mu P_{L} d}\}+ \textrm{h.c.}\right] \,,
	\\
\label{equ:lagrangian_NC_sm}
\mathcal{L}_{\text{NC}}
&=&
\frac{e}{\sin \theta_{W} \cos \theta_{W}}Z_\mu 
\left[\bar{\nu}_{\alpha}\gamma^\mu P_{L} \nu_{\alpha}+\sum_{f = l,q}\bar f\gamma^\mu (c^f_L P_{L} + c^f_R P_{R} )f
\right] \,,
\end{eqnarray}
where,  
\begin{equation}
c^f_L = T^3-Q_f\sin^2\theta_W\,\, \text{and}\,\, c^f_R =-Q_f\sin^2\theta_W\,.
\end{equation}
Here, $l$ and $q$ are the charged leptons and the quarks, respectively. $e$ is the unit electric charge, $T^3$ is the isospin third component, $Q_f$ is the electric charge of the fermion $f$, $\theta_W$ is the Weinberg angle, and $P_L$ and $P_R$ are, respectively, the left-handed and right-handed projection operators.

\noindent\textbf{Neutrino interactions across various energies:}

\textbf{Elastic scattering:} In the elastic scattering of neutrinos with matter, the initial and final state particles remain the same. These processes are thresholdless and can occur at any energy scale. These processes can be both NC and CC types.
\begin{equation}
\nu_\alpha+l_\alpha \to \nu_\alpha+l_\alpha~(\rm CC~and~NC)\,,
\end{equation}
\begin{equation}
\nu_\alpha+ N \to \nu_\alpha+N~(\rm NC)\,,
\end{equation}
where, N stands for nucleon, which can be proton ($p$) or neutron ($n$).

\textbf{Quasi-Elastic scattering (QE):} In these processes, the neutrinos scatter off the nucleon to produce a lepton in the final state along with another nucleon. These processes, for $\nu_e$ and $\nu_\mu$, mostly happen in sub-GeV energies, dominate in the range of 0.1 to 1 GeV. However, due to heavier mass of tau, QE processes become possible at energies around 4 GeV in case of $\nu_\tau$. These processes are described in the following fashion:  
\begin{equation}
\nu_\alpha~({\bar\nu_{\alpha}})+ N ~(p~or~n) \to l_\alpha~(\bar l_\alpha)+N^\prime~(n~or~p)\,.
\end{equation}

\textbf{Inelastic Resonance scattering (RES):} In such processes, neutrinos interact with the nucleons and excite them to resonant states, for example, Delta resonance ($\Delta$), having a mass of 1232 MeV,
\begin{equation}
\nu_\alpha+N \to \nu_\alpha+\Delta\,,
\end{equation}
where, $\Delta$ can further decay to pions,
\begin{equation}
\Delta \to N+\pi.
\end{equation}

\textbf{Deep Inelastic scattering (DIS):} In these processes, the neutrinos directly interact with the quarks inside the nucleon, breaking apart the nucleon and resulting in multiple particles (leptons and hadrons) in the final state. These are high-energy processes that dominate above 2 GeV. 
\begin{equation}
\nu_\alpha+N \to \l_\alpha+X\,,
\end{equation}
where, X is the shower of hadrons.

\begin{figure}[h]
	\centering
	\includegraphics[width=0.5\textwidth]{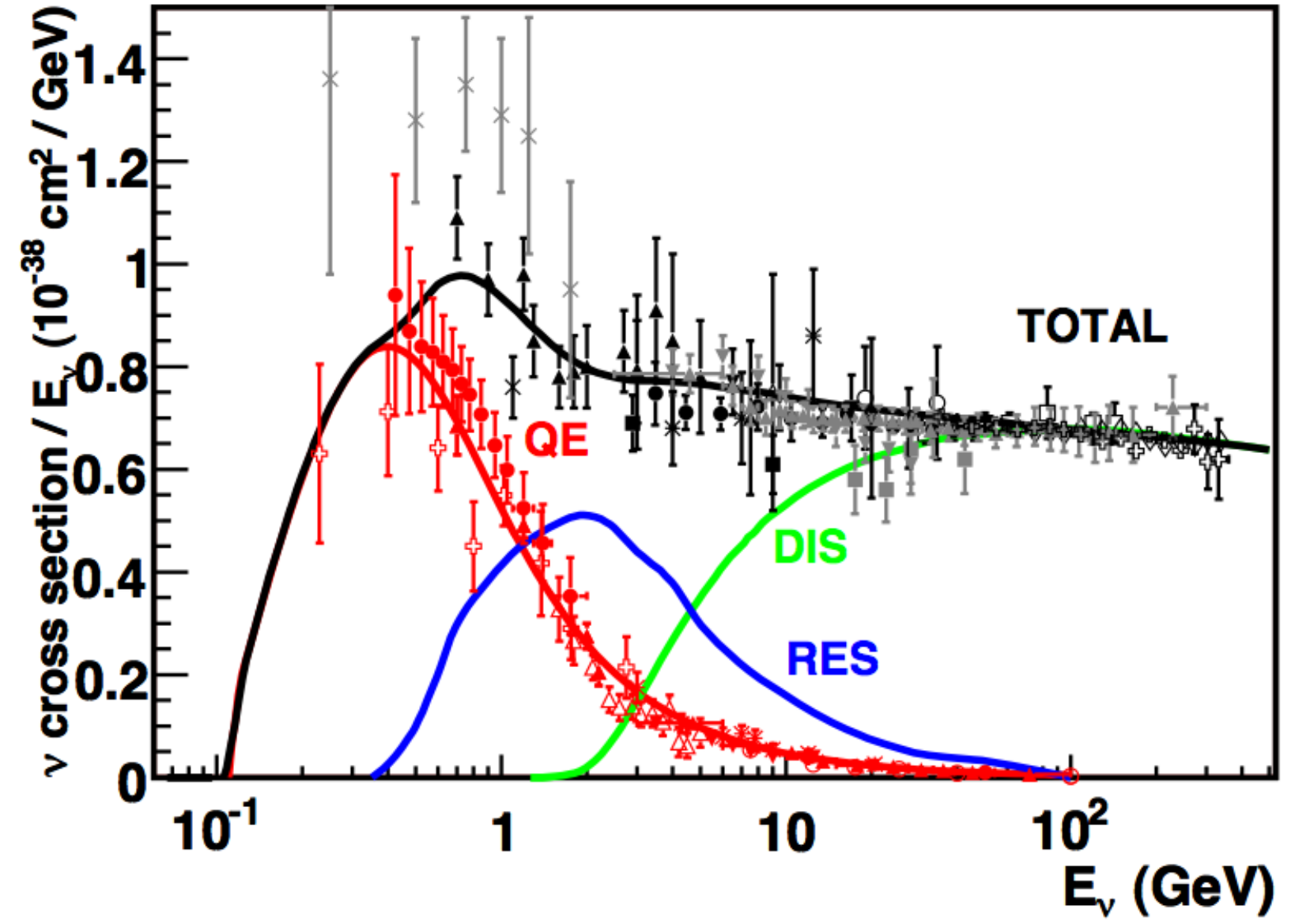}\hfill
	\includegraphics[width=0.5\textwidth]{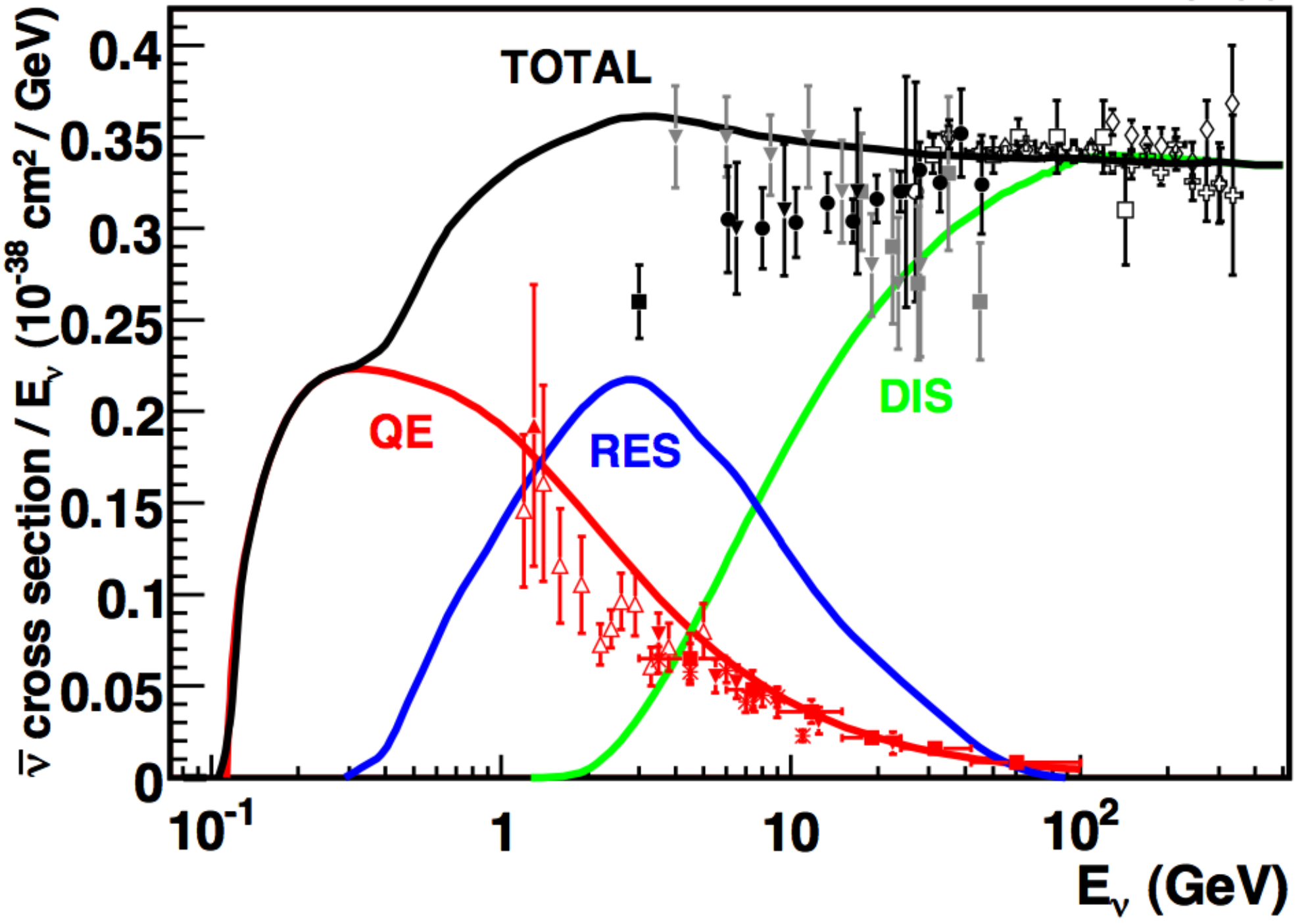}
	\caption{Neutrino (left panel) and antineutrino (right panel) charged-current cross sections across different energies. These figures are taken from~\Refe~\cite{nuSTORM:2012jbd}.}
	\label{fig:neutrino-interactions}
\end{figure}
Figure~\ref{fig:neutrino-interactions} summarizes the neutrino (antineutrino) CC cross sections per nucleon per unit energy of the incoming neutrino (antineutrino) versus the neutrino (antineutrino) energy. As seen from this figure, the error bars in the cross sections are still large, contributing to larger systematics in the present-day experiments. Since the number of events in the neutrino oscillation experiments is proportional to the neutrino cross sections, its measurement with higher accuracy is inevitable in the precision era of neutrino physics. 

\noindent\textbf{(iv) Neutrino mass in the SM:}

The Dirac mass term for any fermion $\psi$ is given by,
$m\bar\psi\psi$. Decomposing $\psi$ as $\psi_{L}+\psi_{R}$ and using $\overline{\psi_{L}}\,\psi_{L}= \overline{\psi_{R}}\,\psi_{R}=0$, we have,
\begin{equation}
\mathcal{L}_{\rm Dirac}=\bar\psi\psi = m\overline{\psi_{L}}\,\psi_{R}+m\overline{\psi_{R}}\,\psi_{L}\,.
\end{equation}
Hence, the Dirac mass term requires the existence of both the left and right chiral fields in the theory. Note that these mass terms are not gauge invariant and, hence, can not be directly added to the SM Lagrangian. So, it seems all the SM fermions are massless. However, due to the existence of the SM Higgs, we can have the Yukawa couplings between the two chiral partners via the Higgs doublet ($\Phi$) as,  
\begin{equation}\label{equ:yukawa}
\mathcal{L}_{\rm Yukawa}=-Y_{l}\overline{ L_{L}}\Phi l_{R}-Y_{u} \overline{ Q_{L}}\tilde\Phi u_{R} -Y_{d} \overline{ Q_{L}}\Phi d_{R} + h.c.\,,
\end{equation}
where, $L_{L}$ are the lepton doublets, $l_{R}$ are the right-handed leptons, $Q_{L}$ are the quark doublets, and $q_{R}$ are the right-handed quarks. The conjugate Higgs doublet for the up-type quark coupling is, $\tilde\Phi=i\sigma_2\Phi^*$, $\sigma_2$ being the Pauli matrix. $Y_l$, $Y_u$, and $Y_d$ are the Yukawa couplings of leptons and up-type and down-type quarks with the SM Higgs.

All the charged fermions have the Dirac masses through the Higgs mechanism once the Higgs field gets the vacuum expectation value ($\it vev$, $v$) after the electroweak symmetry breaking, \ie,~$SU(3)_C \times SU(2)_L \times U(1)_Y \to SU(3)_C \times U(1)_{em}$. So, as $\Phi$ acquires $vev$,
\begin{equation}
\Phi = \begin{pmatrix}
0 \\ v/\sqrt{2}
\end{pmatrix}\\,
\end{equation}
the mass matrices of the SM charged leptons and quarks are, respectively, 
\begin{equation}
m_l = Y_l \frac{v}{\sqrt{2}}\,\,\,{\rm and}\,\,\, m_Q = Y_Q \frac{v}{\sqrt{2}} \,.  
\end{equation}
However, due to the absence of right-handed neutrinos in the SM, neutrinos are massless in the SM of particle physics. 

In contrast, neutrino oscillation confirms neutrinos to be massive. For generating the neutrino masses, one of the simplest possible ways could be assuming the presence of right-handed sterile neutrinos, which have no SM interactions. Their presence allows the Dirac neutrino mass of the form, $m_\nu=Y_\nu \frac{v}{\sqrt{2}}$. However, to get the tiny neutrino mass~$\sim0.1~{\rm eV}$ as evident from neutrino oscillation, one needs $Y_\nu\sim10^{-12}$, which is quite small, raising questions about the naturalness of such scenario.
%, and to the theoretical physicists, it feels {\it philosophically unpleasant}.

\noindent Moreover, the Majorana mass term for neutrinos with only the SM left-handed neutrinos is given by,
\begin{equation}\label{equ:majorana}
\mathcal{L}_{\rm Majorana}=-\frac{1}{2}m_L \overline{\nu^c_{L}}\nu_L+h.c.\,,
\end{equation}
where, $\nu^c_L=C\overline\nu^T_L$, $C$ being the charge conjugation operator. This term violates the SM gauge invariance, and the lepton number is also violated by two units. For this term, the isospin third component, $T_3=1$, so the Yukawa term needs a triplet scalar for neutrino mass generation after symmetry breaking. However, the SM is only available with a doublet scalar.  
Hence, we need new mass generation mechanisms. While preserving the SM gauge invariance, the minimal SM extensions describing the neutrino mass include additional triplet scalar, singly charged and doubly charged singlet scalar, or fermions in singlet and doublet representations~\cite{Cheng:1980qt, King:2003jb, Cai:2017jrq}. Below, we list and briefly describe the popular neutrino mass models that are based on the above-said extensions.   

%%%%%%%%%%%%%%%%%%%%%%%%%%%%%%%%%%%%%%%%%%%
\section{See-saw mechanism for neutrino mass generation} 
%%%%%%%%%%%%%%%%%%%%%%%%%%%%%%%%%%%%%%%%%%%
As we discussed previously, with the addition of a right-handed neutrino, a Dirac-type neutrino mass is unnatural in the sense that the Yukawa coupling is unnaturally small. However, a Majorana-type neutrino mass can be achieved by introducing new particles to the SM, basically at a very high-energy scale. After integrating out the heavy new particles, we get the Majorana neutrino masses. Such high-energy operator at the lowest order is the most famous dimension-five Weinberg operator~\cite{Weinberg:1979sa},
\begin{equation}
\mathcal L (d=5) = \frac{c^{ d=5}}{\Lambda}(\overline{L^c}  \tilde \Phi^*)(  \tilde\Phi^\dagger L)\,.
\end{equation}  
Here, $L$ and $\Phi$ are, respectively, the lepton doublet and scalar doublet of the SM. This operator violates the lepton number by two units; $\Lambda$ is the lepton number violating scale. $c^{ d=5}$ is a coefficient which depends on the high energy theory that leads to the effective operator at low energies. After electroweak symmetry breaking (EWSB), such an operator gives rise to the Majorana mass as in \equ{majorana}, where the light neutrino mass is $m_\nu \sim \frac{v^2}{\Lambda}$. The small neutrino mass is a consequence of high scale $\Lambda$. 
Such masses can be realized at the tree level via the see-saw mechanisms, namely, type-I~\cite{Minkowski:1977sc, Mohapatra:1979ia, Gell-Mann:1979vob, Yanagida:1979as}, type-II~\cite{Magg:1980ut, Cheng:1980qt, Lazarides:1980nt}, and type-III~\cite{Foot:1988aq} see-saw. We expand on them below.

\noindent{\textbf {The type-I see-saw mechanism:}} In the type-I see-saw mechanism, heavy right-handed neutrinos are added to the SM framework in order to get neutrino mass. Here, the right-handed neutrino gets the Majorana mass, and the left-handed neutrino mass is an admixture of Dirac and Majorana masses and naturally results in a tiny neutrino mass without invoking the problem of unnaturally small Yukawa coupling.    

\noindent The corresponding Lagrangian generating Dirac-Majorana neutrino mass is,
\begin{equation}\label{equ:typeI-lagrangian}
\mathcal L _{\rm type{\text{-}}I} = -Y_{\nu}\overline{ L_L} \tilde \Phi N_R-\frac{1}{2}M_R\overline {N^c_R} N_R+h.c.\,,
\end{equation} 
where, $L_L$ is the SM lepton doublet, $\Phi$ is the SM Higgs doublet with $\tilde\Phi=i\sigma_2\Phi^*$. Here, $Y_\nu$ is the neutrino Yukawa matrix and $M_R$ is the Majorana mass matrix for the right-handed neutrinos ($N_R$). After the EWSB, the Dirac mass of the neutrinos in terms of Higgs $vev~(v)$ is $m_D=Y_\nu v/\sqrt 2$. Hence, \equ{typeI-lagrangian} gives,
\begin{equation}\label{equ:typeI-lagrangian-mass}
\mathcal L _{{\rm mass},~\rm type{\text{-}}I} = -\frac{1}{2}
\begin{pmatrix}
\overline{\nu_L} & \overline{N^c_{R}}
\end{pmatrix}
\begin{pmatrix}
0 & m^T_D\\
m_D & M_R
\end{pmatrix}
\begin{pmatrix}
{\nu^c_L} \\ {N_{R}}
\end{pmatrix}+h.c.\,.
\end{equation} 
Upon diagonalising the above mass matrix, and taking $M_R>>m_D$, we get the light neutrino mass as,
\begin{equation}\label{equ:light-nu-mass}
m_{\nu,~{\rm light}}=-m^T_D M^{-1}_{R}m_D\,,
\end{equation}
and the heavy neutrino mass,
\begin{equation}
m_{\nu,~{\rm heavy}}=M_{R}\,.
\end{equation}
As evident from \equ{light-nu-mass}, the heaviness of $M_R$ results in the smallness of the neutrino mass.

The mixing angle between the light and heavy states is given by,
\begin{equation}
\theta\sim m_D/M_{R}\,.
\end{equation}

\noindent{\textbf {Type-II see-saw mechanism:}}
Here, a $SU(2)_L$ triplet scalar, $S_t$, is introduced in order to preserve the gauge invariance of the mass term in \equ{majorana}. The triplet scalar is written in the following way, 
\begin{equation}
S_t=\begin{pmatrix}
S_t^+/{\sqrt{2}} & S_t^{++}\\
S_t^0 & -S_t^+/\sqrt{2}
\end{pmatrix}\,.
\end{equation}
The gauge invariant Lagrangian for generating Majorana neutrino mass is,
\begin{equation}
\mathcal L_{\rm type{\text{-}}II} = -Y_\nu \overline {L^c_L}i\sigma_2 S_t L_L+\mu \Phi^\dagger i\sigma_2 S^\dagger_t \Phi + h.c.\,.
\end{equation}
After EWSB, the doublet ($\Phi$) and triplet ($S_t$) $vevs$ are, respectively, $v$ and $v_S$ that are related via the equation,
\begin{equation}
v_S=\frac{\mu v^2}{\sqrt{2} M^2_{S}}\,,
\end{equation} 
where, $v^2_S+v^2=(246~ {\rm GeV})^2$; $\mu$ being the lepton number violating coupling. $M_{S}$ is the mass of the triplet scalar.
The neutrino mass is given by,
\begin{equation}
m_\nu = \sqrt{2}Y_\nu v_S\,.
\end{equation} 
The smallness of the neutrino mass ($m_\nu$) is implied by the largeness of the triplet scalar mass ($M_S$).

\noindent{\textbf {Type-III see-saw mechanism:}}
In type-III see-saw model, a $SU(2)_L$ triplet fermion is added to the SM particle content, which is not charged under $U(1)_Y$,
\begin{equation}
F_t=\begin{pmatrix}
F_t^0/{\sqrt{2}} & F_t^{+}\\
F_t^- & -F_t^0/{\sqrt{2}}
\end{pmatrix}\,.
\end{equation}
The gauge invariant Lagrangian for generating the neutrino mass is,
\begin{equation}
\mathcal L_{\rm type{\text{-}}III} = -\frac{1}{2}M_{F}{\rm Tr}(\overline {{F^c_t}}F_t)-Y_F \overline {L_L} \,\overline {F^c_t}\, i\sigma_2 \Phi^* + h.c.\,.
\end{equation}
Here, $M_{F}$ is the Majorana mass matrix and $Y_F$ is the Yukawa matrix. After EWSB, the see-saw formula for light neutrino mass is,
\begin{equation}
m_\nu=-\frac{v^2}{2}Y^T_F M^{-1}_FY_F\,.
\end{equation}
Note that the neutrino mass is suppressed by $M_F$ ($\gtrsim$ 880 GeV at 95\% confidence level as reported by CMS collaboration~\cite{CMS:2019lwf}), making it naturally small. 
%%%%%%%%%%%%%%%%%%%%%%%%%%%%%%%%%%%%%%%%%%%%%%%%%%%%%%%%%%
\section{Neutrino mass generation at loop level}
%%%%%%%%%%%%%%%%%%%%%%%%%%%%%%%%%%%%%%%%%%%%%%%%%%%%%%%%%%
Apart from the above-mentioned mass generation mechanisms at the tree level, the neutrino mass can also come from loop corrections that we commonly call as the radiative neutrino mass. Additional symmetries are imposed in this case to prevent the tree-level contributions to the neutrino mass via the Yukawa term in \equ{yukawa}. The popular radiative neutrino mass models are the Zee model~\cite{Zee:1980ai, Wolfenstein:1980sy} and Zee-Babu model~\cite{Babu:2002uu}.
%%%%%%%%%%%%%%%%%%%%%%%%%%%%%%%%%%%%%%%%%%%%%%%%%%%%%%%%%%
\section{Beyond the Standard Model scenarios with neutrinos}
%%%%%%%%%%%%%%%%%%%%%%%%%%%%%%%%%%%%%%%%%%%%%%%%%%%%%%%%%%
As discussed above, the SM fails to explain the non-zero neutrino mass and therefore, it is not a complete theory of the fundamental particles and their interactions. This invites beyond the Standard Model (BSM) theories to incorporate the neutrino mass and mixing. As a consequence of BSM theories, new neutrino states and interactions may naturally arise. Below, we discuss briefly some of the interesting BSM scenarios that can be probed in neutrino experiments. 

\noindent\textbf{Non-unitarity neutrino mixing:} We just discussed that to incorporate the neutrino mass, we must introduce right-handed neutrinos in the theory. This enhances the $3\times3$ oscillation picture to $(3+N_{\rm s})\times (3+N_{\rm s})$, where $N_{\rm s}$ is the number of SM gauge singlet right-handed neutrino species (typically known as ``sterile neutrino''). The number and mass scale of these sterile neutrinos can, in principle, be anything. When the mass is very high ($\sim$ TeV scale), it explains the tiny active neutrino mass. At the same time, being very heavy, they are challenging to detect in the experiments. Even though the heavy right-handed neutrinos are kinematically inaccessible in many present-day experiments and remain decoupled from active neutrino oscillation, their presence can still be realized indirectly by investigating the non-unitary $3\times3$ neutrino mixing matrix~\cite{Escrihuela:2015wra, Agarwalla:2021owd, Sahoo:2023mpj}. 

\noindent\textbf{eV-scale sterile neutrino:} When the mass of the right-handed neutrino is sufficiently low ($\sim$ eV-scale), it mixes significantly with the light active neutrinos. Some of the BSM theories that naturally lead to light sterile neutrino include the split see-saw models~\cite{Kusenko:2010ik, Adulpravitchai:2011rq}, Froggatt-
Nielsen mechanism~\cite{Froggatt:1978nt, Barry:2011wb, Merle:2011yv, Barry:2011fp}, and flavor symmetries~\cite{Mohapatra:2001ns, Babu:2003is, Babu:2004mj, Shaposhnikov:2006nn} for keV-scale sterile neutrino, and the minimal extended seesaw~\cite{Zhang:2011vh} for eV-scale sterile neutrino. The $A_{4}$ flavor symmetry discussed in \Refe~\cite{Barry:2011wb} also predicts eV-scale sterile neutrino. The presence of a light sterile neutrino with eV-scale mass can explain, via active-sterile neutrino oscillation, the so-called {\it short-baseline (SBL) anomalies}, \ie,~ the LSND~\cite{LSND:2001aii}, Gallium~\cite{Gariazzo:2015rra}, Reactor antineutrino~\cite{Mention:2011rk}, and MiniBooNE~\cite{MiniBooNE:2018esg} anomalies which otherwise can not be explained via the standard atmospheric or solar mass splittings. 

\noindent\textbf{Non-standard Interaction:} The radiative neutrino mass models such as the Zee-Babu model~\cite{Zee:1985id, Babu:1988ki, Babu:2002uu, Babu:2019mfe} can naturally lead to lepton flavor violating new neutrino interactions called the non-standard interactions (NSIs). They can affect neutrino production, detection, and also neutrino propagation. Production and detection are affected by the presence of charged-current NSIs (CC-NSIs)~\cite{Ohlsson:2008gx, Meloni:2009cg, Ohlsson:2012kf, Giarnetti:2020bmf}, while neutrino propagation is altered in the presence of neutral-current NSIs (NC-NSIs)~\cite{Ohlsson:2012kf, Bhupal:2019qno}. The corresponding Lagrangian in an effective field theory framework is given in terms of the dimension six four fermi operator as follows,
%=======================
\begin{align}
&\mathcal{L}_{\mathrm{CC-NSI}} = -2\sqrt{2} G_{F} \sum_{\alpha, \beta, f^{\prime}, f, C} \varepsilon^{ff^{\prime}C}_{\alpha\beta} ({\Bar\nu_{\alpha}}\gamma^{\mu}P_{L}l_{\beta}) (\Bar{f}^{\prime}\gamma_{\mu}P_Cf), 
\label{equ:Lnsi_CC}\\
&\mathcal{L}_{\mathrm{NC-NSI}} = -2\sqrt{2} G_{F} \sum_{\alpha, \beta, f, C} \varepsilon^{fC}_{\alpha\beta} ({\Bar\nu_{\alpha}}\gamma^{\mu}P_{L}\nu_{\beta}) (\Bar{f}\gamma_{\mu}P_Cf), 
\label{equ:Lnsi} 
\end{align}
where, $P_{C}=P_{L} ~{\rm or} ~P_{R}$, is the chiral projection
operator. The CC-NSI and NC-NSI strengths are, respectively, given by the dimensionless parameters
$\varepsilon^{ff^{\prime}C}_{\alpha\beta}$ and $\varepsilon^{fC}_{\alpha\beta}$. $\alpha,\beta$ are the flavor indices. $f$ and $f^\prime$ denote the first generation charged fermions, electron, up quark, and down quark.

%======================= 

\noindent\textbf{Long-range interactions:} The Standard Model extended with an extra $U(1)^\prime$ symmetry, results in additional neutrino-matter interactions different from those due to SM NC and CC interactions. If the gauge boson associated with the new symmetry is extremely light, say less than $10^{-10}$ eV, the range of the interactions becomes large -- the neutrino-matter interaction, in this case, is called the long-range interaction (LRI). The terrestrial neutrinos can feel the presence of matter all over the Universe via LRI if it is mediated by an extremely light gauge boson~\cite{Bustamante:2018mzu, Singh:2023nek, Agarwalla:2023sng}. 

\noindent\textbf{Neutrino decay:} As inspired by various neutrino mass models, the existence of BSM scalars in the theory may lead to neutrino decaying to light neutrino states~\cite{deGouvea:2019goq}.

For instance in case of Dirac neutrinos where lepton number conservation is mandatory,
dimension-5 operators of the kind,
\begin{equation}
\frac{f_{ij}}{\Lambda}(\overline{{L_i}}  \tilde H)(\nu_{jR} \phi_0)+h.c.\,,
\end{equation}
after electroweak symmetry breaking leads to neutrino interaction Lagrangian,
\begin{equation}\label{equ:decay-dim-5}
\frac{f_{ij}v}{\Lambda} \overline{\nu_{iL}}  \nu_{jR} \phi_0+h.c.\,,
\end{equation}
where, $L$ is the lepton doublet, $H$ is the SM Higgs doublet, and $\phi_0$ is the BSM scalar (singlet under SM) with vanishing lepton number. $f_{ij}$ is a dimensionless effective coupling, $\Lambda$ is the new physics scale, and $v$ is the vacuum expectation value of the SM Higgs (H). The operator in \equ{decay-dim-5} leads to neutrino decay via the BSM scalar ($\nu_i\to\nu_j+\phi_0$)~\cite{Chikashige:1980ui, Gelmini:1980re, Gelmini:1983ea}.

Neutrino decay can also be possible with the dimension-6 operator,
\begin{equation}\label{equ:decay-dim-6}
	\frac{g_{ij}}{\Lambda^2} (\overline{L_i^c}  \tilde H^*)(  \tilde H^\dagger L_j) \phi_2+h.c.\,,
\end{equation}
where, $g_{ij}$ is a dimensionless effective coupling and $\phi_2$ is the BSM scalar (singlet under SM) with lepton number -2~\cite{Berryman:2018ogk}. After symmetry breaking, the Lagrangian facilitating the neutrino decay is then,
\begin{equation}
\frac{g_{ij}v^2}{\Lambda^2} \bar{\nu^c_i} \nu_j\phi_2+h.c.\,.
\end{equation}
In the case of Majorana neutrinos, since the lepton number is violated by two units, there is no need to introduce additional light right-handed degrees of freedom. A dimension-6 operator with a scalar field $\phi_0$ (gauge singlet with zero lepton number) results in neutrino decay as follows,
\begin{equation}
	\frac{h_{ij}}{2\Lambda^2}(\overline{L_i^c}  \tilde H^*)(  \tilde H^\dagger L_j) \phi_0+h.c.~\xrightarrow{\rm EWSB}~\frac{h_{ij}v^2}{\Lambda^2} \bar{\nu^c_i} {\nu_j}\phi_0+h.c. \,.
\end{equation}

\noindent\textbf{Lorentz invariance violation:} The Standard Model (SM) is undergoing rigorous verification across various particle physics experiments. To ensure its completeness, it is equally important to test the consistency of any of its underlying symmetry. Among these, Lorentz invariance is a fundamental symmetry of Nature, which stands as a cornerstone of both the SM and Einstein's theory of relativity. Interestingly, some unified theories like loop quantum gravity and string theory suggest the violation of Lorentz invariance at a very high-energy scale (Planck scale, $M_P\sim10^{19}$ GeV). The Lorentz invariance violation (LIV) manifests itself in low-energy effective operators~\cite{Colladay:1996iz, Colladay:1998fq} and gives rise to new Lorentz-violating neutrino interactions, making them detectable in neutrino experiments. Testing LIV not only provides a stringent consistency check of the SM but also offers a window in exploring physics beyond it, potentially unveiling connections to quantum gravity and other high-energy phenomena.
 
Neutrino oscillation experiments offer a unique platform to study these BSM effects. Because of the tiny mass-squared differences involved in the neutrino oscillations, any sub-leading BSM effect can significantly modify the interference pattern among the neutrino mass eigenstates and, hence, the oscillation pattern, thus revealing its presence. This thesis explores three of the above-discussed BSM scenarios: long-range neutrino interaction (chapter~\ref{C3}), Lorentz invariance violation (chapter~\ref{C4}), and sterile neutrino (chapter~\ref{C5}) in the context of next-generation long-baseline neutrino oscillation experiments. 

\section{Layout of the thesis}  
The thesis is organized in the following fashion. In chapter~\ref{C2}, we describe in detail the theoretical formalism of neutrino oscillation in two flavor and three flavor scenarios, both in vacuum and matter. Then, we discuss the sources of neutrinos and the experiments designed to detect them. Also, we discuss the contributions from the already completed, ongoing, and future experiments in determining the oscillation parameters with great precision. Then, we discuss the present status of the oscillation parameters and briefly mention the remaining unsolved issues in three neutrino oscillation framework. In the following chapters, we mention the sensitivity of the next-generation long-baseline experiments, DUNE and T2HK, to various new physics scenarios.
In chapter~\ref{C3}, we explore the long-range interaction (LRI) of neutrinos from various gauged $U(1)^\prime$ symmetries. We derive constraints on the LRI potential arising from each $U(1)^\prime$ symmetry using DUNE and T2HK in isolation and combination. We also explore their discovery prospects and the possibilities of distinguishing these symmetries upon discovery.   
Next, in chapter~\ref{C4}, we place constraints on the CPT-conserving and CPT-violating Lorentz invariance violating parameters with the standalone DUNE and T2HK and in combination. Along with that, we also discuss in detail their possible correlations with the uncertain standard oscillation parameters. In chapter~\ref{C5}, we explore the possibility of going beyond the three neutrino oscillation scenario, in the presence of a light-sterile neutrino. We consider the presence of one extra neutrino species as predicted by the short-baseline anomalies. We derive the sensitivities of the future experiments DUNE, JD (T2HK), and JD+KD (T2HKK) in probing the active-sterile neutrino oscillation parameters. Here, we discuss the strength of these experiments in constraining and discovering the CP phases involved in the 3+1 neutrino oscillation scheme. Also, we derive exclusion limits on the active-sterile mixing angle over a wide range of sterile mass-squared difference. Finally, in chapter~\ref{C6}, we summarize the findings of this thesis and give the concluding remarks. 
%%%%%%%%%%%%%%%%%%%%%%%%%%%%
%\noindent \textbf{Arrangement of the Dissertation:}
%%%%%%%%%%%%%%%%%%%%%%%%%%%%
\blankpage
%%%%%%%%%%%%%%%%%%%%%%%%%% Chapter-2 %%%%%%%%%%%%%%%%%%%%%%%%%%%%%%
%%%%%%%%%%%%%%%%%%%%%%% CHAPTER - 2 %%%%%%%%%%%%%%%%%%%%%%%%%%
\chapter{Neutrino oscillation}
\label{C2} 
%%%%%%%%%%%%%%%%%%%%%%%%%%%%%%%%%%%%%%%%%%%%%%%%%%%%%%%%%%%%%%
%%%%%%%%%%%%%%%%%%%%%%%%%%%%
Inspired by the discovery of parity violation in weak interactions, Bruno Pontecorvo, in 1957, in light of $K^\circ\!\mathrel{\substack{\xleftarrow{\hspace{0.5em}} \\[-0.65ex] \xrightarrow{\hspace{0.5em}}}}\!\bar{K^\circ}$
oscillations~\cite{Gell-Mann:1955ipe}, proposed the possibility of neutrino-antineutrino oscillation for massive neutrinos. Later, with the discovery of distinct flavors of neutrinos, the above idea was extended to the mixing between various flavors of neutrinos, leading to the so-called neutrino oscillation as a consequence of non-zero neutrino mass. This chapter discusses the theoretical formalism of neutrino oscillation in vacuum and matter. We discuss in detail the two-flavor and three-flavor neutrino oscillation frameworks. Also, we include the discussion on various neutrino sources, the already completed, ongoing, and upcoming experiments for detecting neutrinos from various sources and their implications in neutrino oscillation studies. 
%%%%%%%%%%%%%%%%%%%%%%%%%%%%%%%%%%%%%%%%%%%%%%%%%%%%%%%%%%%%%%
\section{General formalism of neutrino oscillation in vacuum}
%%%%%%%%%%%%%%%%%%%%%%%%%%%%%%%%%%%%%%%%%%%%%%%%%%%%%%%%%%%%%%
Unlike charged leptons, the flavor and mass eigenstates of neutrinos are distinct from one another. Neutrinos are produced and detected via weak interaction as flavor eigenstates (equivalently called as weak or gauge eigenstates) and propagate as mass eigenstates (equivalently called physical eigenstates). The two eigen bases, flavor ($\ket{\nu_\alpha}$) and mass ($\ket{\nu_i} $), are related via a unitary matrix, known as the Pontecorvo-Maki-Nakagawa-Sakata (PMNS) matrix~\cite{Pontecorvo:1957qd, Maki:1962mu, Pontecorvo:1967fh}, $U$, as given below,
\begin{equation}
\ket{\nu_\alpha} = \sum_{i=1}^{n} U^{*}_{\alpha i} \ket{\nu_i}\,.
\end{equation}
The above equation is valid for $n$ flavors of neutrinos. The mass eigenstates evolve as plane waves in space-time,
\begin{equation}
\ket{\nu_i(t)} = e^{-i E_i t} \ket{\nu_i(0)}\,.
\end{equation}
Where, 
\begin{equation}
E_i = \sqrt {p^2+m^2_{i}}\,,
\end{equation}
$p$ is the neutrino momentum and $m_i$ is the mass of the $i$-th eigenstate.  Under ultra relativistic limit, where, $E_{i}\approx p\gg m_i$\,,
\begin{equation}\label{equ:energy}
E_i \simeq p + \frac{m^2_{i}}{2E}\,.
\end{equation}
In the above, we assume that all the mass eigenstates are propagating with equal momentum, ensuring the flavor eigenstates are coherent superposition of the states with definite mass.\\
The flavor state at a later time $t$ is then given by,
\begin{align}
\ket{\nu_\alpha (t)} \nonumber&= \sum_{i=1}^{n} U^{*}_{\alpha i} \ket{\nu_i(t)}\\
&= \sum_{i=1}^{n} U^{*}_{\alpha i} e^{-i E_i t} \ket{\nu_i(0)}\,.
\end{align}
Now, the flavor transition amplitude that a neutrino of flavor $\alpha$, after time $t$, has been converted to flavor $\beta$, is given by,
\begin{equation}
\mathcal{{A_{\alpha\beta}}}(t)=\braket{\nu_{\beta}|{\nu_{\alpha}}(t)}=\sum_{i,j=1}^{n} U^{*}_{\alpha i} U_{\beta j} e^{-i E_i t} \braket{\nu_j(0)|\nu_i(0)}\,.
\end{equation}
Since the mass eigenstates are orthogonal to each other, \ie~ $\braket{\nu_j|\nu_i}=\delta_{ij}$, the above equation becomes,
\begin{equation}
\mathcal{{A_{\alpha\beta}}}(t)=\sum_{j=1}^{n} U^{*}_{\alpha j} U_{\beta j} e^{-i E_j t}.
\end{equation}  
The probability of oscillation from $\nu_\alpha$ flavor to $\nu_\beta$ flavor is hence, 
\begin{align}\label{equ:prob-to-simplify}
P_{\nu_{\alpha}\to\nu_\beta}=|\mathcal{{A_{\alpha\beta}}}(t)|^2=\Big|\sum_{j=1}^{n} U^{*}_{\alpha j} U_{\beta j} e^{-i E_j t}\Big|^2=\sum_{j,k}U_{\alpha{j}}^*U_{\beta{j}}U_{\alpha{k}}U_{\beta{k}}^{*} e^{-i (E_j-E_k) t}\,.
\end{align}
Using \equ{energy}, the energy difference,
\begin{equation}
E_j-E_k= \frac{m^2_j-m^2_k}{2E}=\frac{\Delta{m^2_{jk}}}{2E}\,,
\end{equation}
where, $\Delta{m^2_{jk}}=m^2_j-m^2_k$ is known as the mass-squared difference or the mass-squared splitting between the two mass eigenstates $\nu_j$ and $\nu_k$.

Due to the relativistic nature of neutrinos, one can replace $t$ by the neutrino propagation length ($L$). Then \equ{prob-to-simplify} becomes,
\begin{align}\label{equ:prob-proceed}
P_{\nu_{\alpha}\to\nu_\beta}&\nonumber=\sum_{j,k}U_{\alpha{j}}^*U_{\beta{j}}U_{\alpha{k}}U_{\beta{k}}^{*} e^{-i \big(\frac{\Delta{m^2_{jk}}L}{2E}\big)}\\\nonumber
&=\sum_{k}|U_{\alpha{k}}|^2|U_{\beta{k}}|^2+\sum_{j>k}U_{\alpha{j}}^*U_{\beta{j}}U_{\alpha{k}}U_{\beta{k}}^{*} e^{-i \big(\frac{\Delta{m^2_{jk}}L}{2E}\big)}+\sum_{j<k}U_{\alpha{j}}^*U_{\beta{j}}U_{\alpha{k}}U_{\beta{k}}^{*} e^{-i \big(\frac{\Delta{m^2_{jk}}L}{2E}\big)}\\\nonumber
&=\sum_{k}|U_{\alpha{k}}|^2|U_{\beta{k}}|^2+\sum_{j>k}U_{\alpha{j}}^*U_{\beta{j}}U_{\alpha{k}}U_{\beta{k}}^{*} e^{-i \big(\frac{\Delta{m^2_{jk}}L}{2E}\big)}+\sum_{j<k}U_{\alpha{j}}^*U_{\beta{j}}U_{\alpha{k}}U_{\beta{k}}^{*} e^{i \big(\frac{\Delta{m^2_{kj}}L}{2E}\big)}\\\nonumber
&=\sum_{k}|U_{\alpha{k}}|^2|U_{\beta{k}}|^2+\sum_{j>k}U_{\alpha{j}}^*U_{\beta{j}}U_{\alpha{k}}U_{\beta{k}}^{*} e^{-i \big(\frac{\Delta{m^2_{jk}}L}{2E}\big)}+\sum_{k<j}U_{\alpha{k}}^*U_{\beta{k}}U_{\alpha{j}}U_{\beta{j}}^{*} e^{i \big(\frac{\Delta{m^2_{jk}}L}{2E}\big)}\\
&=\sum_{k}|U_{\alpha{k}}|^2|U_{\beta{k}}|^2+2\sum_{j>k}\mathcal{R}e\Big[U_{\alpha{j}}^*U_{\beta{j}}U_{\alpha{k}}U_{\beta{k}}^{*} e^{-i \big(\frac{\Delta{m^2_{jk}}L}{2E}\big)}\Big]\,.
\end{align}
We know for two complex variables $z_1$ and $z_2$,
\begin{equation}
\mathcal{R}e(z_1z_2)=\mathcal{R}e(z_1)\mathcal{R}e(z_2)-\mathcal{I}m(z_1)\mathcal{I}m(z_2)\,.
\end{equation}
Using the above equation in \equ{prob-proceed} we have,
\begin{align}\label{equ:prob-proceed2}
P_{\nu_{\alpha}\to\nu_\beta}=\sum_{k}|U_{\alpha{k}}|^2|U_{\beta{k}}|^2&\nonumber+2\sum_{j>k}\mathcal{R}e(U_{\alpha{j}}^*U_{\beta{j}}U_{\alpha{k}}U_{\beta{k}}^{*}) \mathcal{R}e\Big[e^{-i \big(\frac{\Delta{m^2_{jk}}L}{2E}\big)}\Big]
\\\nonumber&-2\sum_{j>k}\mathcal{I}m(U_{\alpha{j}}^*U_{\beta{j}}U_{\alpha{k}}U_{\beta{k}}^{*}) \mathcal{I}m\Big[e^{-i \big(\frac{\Delta{m^2_{jk}}L}{2E}\big)}\Big]\\\nonumber
=\sum_{k}|U_{\alpha{k}}|^2|U_{\beta{k}}|^2&+2\sum_{j>k}\mathcal{R}e(U_{\alpha{j}}^*U_{\beta{j}}U_{\alpha{k}}U_{\beta{k}}^{*})\cos\Big( \frac{\Delta{m^2_{jk}}L}{2E}\Big)
\\\nonumber&+2\sum_{j>k}\mathcal{I}m(U_{\alpha{j}}^*U_{\beta{j}}U_{\alpha{k}}U_{\beta{k}}^{*})\sin\Big(\frac{\Delta{m^2_{jk}}L}{2E}\Big)\\\nonumber
=\sum_{k}|U_{\alpha{k}}|^2|U_{\beta{k}}|^2&+2\sum_{j>k}\mathcal{R}e(U_{\alpha{j}}^*U_{\beta{j}}U_{\alpha{k}}U_{\beta{k}}^{*})-4\sum_{j>k}\mathcal{R}e(U_{\alpha{j}}^*U_{\beta{j}}U_{\alpha{k}}U_{\beta{k}}^{*})\sin^2\Big( \frac{\Delta{m^2_{jk}}L}{4E}\Big)
\\&+2\sum_{j>k}\mathcal{I}m(U_{\alpha{j}}^*U_{\beta{j}}U_{\alpha{k}}U_{\beta{k}}^{*})\sin\Big(\frac{\Delta{m^2_{jk}}L}{2E}\Big)\,.
\end{align}
%Further expanding \equ{prob-to-simplify} and making the use of the unitarity of the PMNS matrix, $U$, \ie~$\sum_{k}{U_{\alpha k}U_{\beta k}}=\delta_{\alpha\beta}$, we get the simplified version of the oscillation probability,
Now using the relation, 
\begin{equation}
\sum_{k}|U_{\alpha{k}}|^2|U_{\beta{k}}|^2=\delta_{\alpha\beta}-2\sum_{j>k}\mathcal{R}e(U_{\alpha{j}}^*U_{\beta{j}}U_{\alpha{k}}U_{\beta{k}}^{*})\\,
\end{equation}
we get the final expression for the oscillation probability as,
\begin{align}\label{equ:general-prob}
P_{\nu_\alpha{\rightarrow}\nu_\beta}
=\delta_{\alpha\beta}&-4\sum_{j > k}\mathcal{R}e(U_{\alpha{j}}^*U_{\beta{j}}U_{\alpha{k}}U_{\beta{k}}^*) \sin^2\Big(\frac{{\Delta{m^2_{jk}}L}}{4E}\Big)\nonumber
\\&+2\sum_{j > k}\mathcal{I}m(U_{\alpha{j}}^*U_{\beta{j}}U_{\alpha{k}}U_{\beta{k}}^*) \sin\Big(\frac{{\Delta{m^2_{jk}}L}}{2E}\Big)\,.
\end{align}
When $\alpha\neq\beta$, a new neutrino of flavor $\beta$ appears into the picture from the neutrino of flavor $\alpha$, and the corresponding probability is termed as the transition or appearance probability. However, when $\alpha=\beta$, there is no flavor change, and the corresponding probability is known as the survival or disappearance probability. 
%%%%%%%%%%%%%%%%%%%%%%%%%%%%%%%%%%%%%%%%%%%%%%%%%%%%%%%%%
\subsection{Two-flavor scenario in vacuum}
%%%%%%%%%%%%%%%%%%%%%%%%%%%%%%%%%%%%%%%%%%%%%%%%%%%%%%%%%
When two neutrino flavors are only involved in the oscillation framework, the flavor eigenstates ($\nu_\alpha,\nu_\beta$) and mass eigenstates ($\nu_1,\nu_2$) are related by a $2\times2$ unitary rotation matrix parameterized in terms of only one rotation angle ($\theta$).
\begin{equation}
\begin{pmatrix}
\nu_\alpha\\
\nu_\beta
\end{pmatrix}
=\begin{pmatrix}	
\cos\theta & \sin\theta\\
-\sin\theta & \cos\theta \\
\end{pmatrix}
\begin{pmatrix}
\nu_1\\
\nu_2
\end{pmatrix}\,.
\end{equation} 
From \equ{general-prob} with $\alpha\neq\beta$, we get the two-flavor appearance probability,
\begin{align}
P_{\nu_\alpha\rightarrow\nu_\beta}=\sin^2{2\theta}\sin^2\Big(\frac{\Delta m^2_{21}L}{4E}\Big)\,.
\end{align}
Here, $\Delta m^2_{21}=m^2_2-m^2_1$ is the mass-squared difference between $\nu_2$ and $\nu_1$. If we intend for all practical purposes to keep the baseline length ($L$) in km, the neutrino energy ($E$) in GeV, and the mass-squared difference in eV$^2$, the appearance probability simplifies to 
\begin{align}
	P_{\nu_\alpha\rightarrow\nu_\beta}=\sin^2{2\theta}\sin^2\Big[1.27\times\frac{\Delta m^2_{21}({\rm eV}^2)L(\rm km)}{E(\rm GeV)}\Big]\,.
\end{align}
The survival probability thus is given by,
\begin{align}
P_{\nu_\alpha\rightarrow\nu_\alpha}=1-P_{\nu_\alpha\rightarrow\nu_\beta}=1-\sin^2{2\theta}\sin^2\Big[1.27\times\frac{\Delta m^2_{21}({\rm eV}^2)L(\rm km)}{E(\rm GeV)}\Big]\,.
\end{align}
%%%%%%%%%%%%%%%%%%%%%%%%%%%%%%%%%%%%%%%%%%%%%%%%%%%%%%%%%
\subsection{Three-flavor scenario in vacuum}
%%%%%%%%%%%%%%%%%%%%%%%%%%%%%%%%%%%%%%%%%%%%%%%%%%%%%%%%%
The three neutrino framework of neutrino oscillation involves three active neutrino flavors, $\nu_e\,,\nu_\mu\,, {\rm and}\,\nu_\tau$. The $3\times3$ unitary rotation matrix relating the flavor and mass eigenstates is a product of three rotation matrices, which are parameterized in terms of three mixing angles, $\theta_{12}\,,\theta_{13}\,,{\rm and}\,\theta_{23}$, one Dirac CP phase ($\delta_{\rm CP}$).
\begin{align}
U&\nonumber=R_{23}(\theta_{23})R_{13}(\theta_{13},{\delta_{\rm CP}})R_{12}(\theta_{12})\\\nonumber
&=\begin{pmatrix}	
	1 & 0 & 0\\
	0 & c_{23} & s_{23}\\
	0 & -s_{23} & c_{23}
\end{pmatrix}
\begin{pmatrix}	
	c_{13} & 0 & {s_{13}}{e^{-i\delta_{\rm CP}}}\\
	0 & 1 & 0\\
	-s_{13}e^{i\delta_{\rm CP}} & 0 & c_{13}
\end{pmatrix}
\begin{pmatrix}	
	c_{12} &  {s_{12}} & 0\\
	-s_{12} & c_{12}& 0\\
	0 & 0 & 1
\end{pmatrix}\\
&=
\begin{pmatrix}	
	c_{13}c_{12} & c_{13}s_{12} & {s_{13}}{e^{-i\delta_{\rm CP}}}\\
	{-c_{23}s_{12}}-{{s_{23}s_{12}c_{12}}e^{i\delta_{\rm CP}}} & {c_{23}c_{12}}-{s_{12}s_{13}s_{23}}e^{i\delta_{\rm CP}}
	& s_{23}c_{13}\\
	{s_{23}s_{12}}-{{c_{23}c_{12}s_{13}}e^{i\delta_{\rm CP}}} & {-c_{12}s_{23}}-{s_{12}s_{13}c_{23}}e^{i\delta_{\rm CP}}
	& c_{13}c_{23}	
\end{pmatrix}\,.
\end{align}
Here, we use shorthand notations, $c_{ij}=\cos{\theta_{ij}}$ and $s_{ij}=\sin{\theta_{ij}}$. The above parameterization is the most famous PMNS parameterization which follows the conventional Cabibbo-Kobayashi-Maskawa (CKM) parameterization in the quark sector. This choice allows us to write the mixing angles in terms of the elements of the PMNS matrix $U$, as follows,

\begin{equation}
\frac{|U_{e2}|^2}{|U_{e1}|^2}=\tan^2{\theta_{12}}\,,
\end{equation}
\begin{equation}
	\frac{|U_{\mu 3}|^2}{|U_{\tau 3}|^2}=\tan^2{\theta_{23}}\,,
\end{equation}
\begin{equation}
|U_{e 3}|^2=\sin^2{\theta_{13}}\,.
\end{equation}
These simple ratios allow the clean and accurate determination of the mixing angles in specific experiments without interference from the other mixing angles. 
\begin{itemize}
\item $\theta_{12}$ is called the solar mixing angle and is determined by the experiments that observe solar neutrino oscillations (Homestake, SAGE, and Gallex) and long-baseline reactor neutrino oscillations (KamLAND). 
\item $\theta_{23}$ is known as the atmospheric mixing angle and is measured by the atmospheric neutrino experiments (Super-Kamiokande and IceCube DeepCore) that probe the oscillation between the $\nu_\mu$ and $\nu_\tau$ and the long-baseline experiments (K2K and MINOS).
\item $\theta_{13}$ is termed as the reactor mixing angle and is measured by both the reactor (Daya Bay, RENO, and Double CHOOZ) and long-baseline accelerator experiments (MINOS, T2K, and NO$\nu$A), which probe the mixing between the electron neutrino and the third mass eigenstate.
\end{itemize}
Note that the three-neutrino oscillation framework involves two independent mass-squared differences, $\Delta m^2_{31}=m^2_{3}-m^2_1$, known as the atmospheric mass-squared difference and $\Delta m^2_{21}=m^2_{2}-m^2_1$, called the solar mass-squared difference.   
Let us now define a variable,\\
\begin{equation}
\Delta_{ij}=\frac{{\Delta{m^2_{ij}}}}{2E}L=2.54\Big(\frac{{\Delta{m^2_{ij}}}}{{\rm eV}^2}\Big)\Big(\frac{{\rm GeV}}{E}\Big)\Big(\frac{L}{\rm km}\Big)\,.
\end{equation}
For $\alpha$=$\beta$, from \equ{general-prob}, we get the disappearance probability,
\begin{align}
P_{\nu_\alpha{\rightarrow}\nu_\alpha}
=&1-4|U_{\alpha{2}}|^2(1-|U_{\alpha{2}}|^2)\sin^2\frac{\Delta_{21}}{2}-4|U_{\alpha{3}}|^2(1-|U_{\alpha{3}}|^2)\sin^2\frac{\Delta_{31}}{2}\nonumber\\&+ 2|U_{\alpha{3}}|^2|U_{\alpha{2}}|^2\Big(4\sin^2\frac{\Delta_{21}}{2}\sin^2\frac{\Delta_{31}}{2}+\sin{\Delta_{21}}\sin{\Delta_{31}}\Big)\,.
\end{align}
For  $\alpha\neq\beta$, from \equ{general-prob} we get the appearance probability,
\begin{align}\label{equ:general-prob-3nu}
P_{\nu_\alpha{\rightarrow}\nu_\beta}=&4|U_{\alpha{2}}|^2|U_{\beta{2}}|^2\sin^2\frac{\Delta_{21}}{2}+4|U_{\alpha{3}}|^2|U_{\beta{3}}|^2\sin^2\frac{\Delta_{31}}{2}\nonumber
\\&+2\mathcal {R}e(U_{\alpha{3}}^*U_{\beta{3}}U_{\alpha{2}}U_{\beta{2}}^*)\Big(4\sin^2\frac{\Delta_{21}}{2}\sin^2\frac{\Delta_{31}}{2}+\sin{\Delta_{21}}\sin{\Delta_{31}}\Big)\nonumber
\\&+4J_{\alpha\beta}\Big(\sin^2\frac{\Delta_{21}}{2}\sin\Delta_{31}-\sin^2\frac{\Delta_{31}}{2}\sin{\Delta_{21}}\Big)\,,
\end{align}
where, $J_{\alpha\beta}$ is the Jarlskog invariant~\cite{Jarlskog:1985ht} with the following property,
\begin{align}
J_{\alpha\beta}&=\mathcal {I}m(U_{\alpha{1}}^*U_{\beta{1}}U_{\alpha{2}}U_{\beta{2}}^*)=\mathcal {I}m(U_{\alpha{2}}^*U_{\beta{2}}U_{\alpha{3}}U_{\beta{3}}^*)=\mathcal {I}m(U_{\alpha{3}}^*U_{\beta{3}}U_{\alpha{1}}U_{\beta{1}}^*)\nonumber
\\&=-\mathcal {I}m(U_{\alpha{2}}^*U_{\beta{2}}U_{\alpha{1}}U_{\beta{1}}^*)=-\mathcal {I}m(U_{\alpha{3}}^*U_{\beta{3}}U_{\alpha{2}}U_{\beta{2}}^*)=-\mathcal {I}m(U_{\alpha{1}}^*U_{\beta{1}}U_{\alpha{3}}U_{\beta{3}}^*)=-J_{\beta\alpha}\,.
\end{align}
We define a few more variables, $\alpha=\frac{\Delta m^2_{21}}{\Delta m^2_{31}}$ and $\Delta=\frac{\Delta_{31}}{2}$. Then we have, $\frac{\Delta_{21}}{2}=\alpha\Delta$. With the current knowledge of the oscillation parameters~\cite{NuFIT, deSalas:2020pgw, Capozzi:2021fjo}, one can see that $\alpha\simeq0.033$ and $s_{13}\simeq0.15$ (hence, $c_{13}\sim1$). Expanding \equ{general-prob-3nu} and keeping the terms up to second order in small parameters $\alpha$ and $s_{13}$, the $\nu_\mu\to\nu_e$ appearance probability takes the form, 
\begin{align}\label{equ:app_vac}
P_{\nu_\mu{\rightarrow}\nu_e}
\simeq~&\sin^2{\theta_{23}}\sin^2{2\theta_{13}}\sin^2{\Delta}\nonumber
\\&-(\alpha\Delta)\sin{2\theta_{13}}\sin{2\theta_{12}}\sin{2\theta_{23}}\sin\delta\sin\Delta\sin\Delta\nonumber
\\&+(\alpha\Delta)\sin{2\theta_{13}}\sin{2\theta_{12}}\sin{2\theta_{23}}\cos\delta\cos\Delta\sin\Delta\nonumber
\\&+ (\alpha\Delta)^2\sin^2{2\theta_{12}}\cos^2{\theta_{23}}\,,
\end{align}
and the $\nu_\mu\to\nu_\mu$ disappearance probability is given by,
\begin{align}\label{equ:disapp_vac}
P_{\nu_\mu{\rightarrow}\nu_\mu}\nonumber&\simeq 1-\sin^2{2\theta_{23}\sin^2\Delta}+4\sin^2{\theta_{13}}\cos{2\theta_{23}}\sin^2{\theta_{23}}\sin^2{\Delta}\\\nonumber
&+(\alpha\Delta)\sin 2\Delta[\cos^2\theta_{12}\sin^2{2\theta_{23}}-2\sin{\theta_{13}\sin2\theta_{12}\sin2\theta_{23}\sin^2\theta_{23}}\cos{\delta_{\rm CP}}]\\
&-(\alpha\Delta)^2[\sin^2{2\theta_{12}}\cos^2\theta_{23}+\cos^2\theta_{12}\sin^22\theta_{23}(\cos 2\Delta-\sin^2\theta_{12})]\,.
\end{align}
%%%%%%%%%%%%%%%%%%%%%%%%%%%%%%%%%%%%%%%%%%%%%%%
\section{Neutrino oscillation in matter}
%%%%%%%%%%%%%%%%%%%%%%%%%%%%%%%%%%%%%%%%%%%%%%%
\begin{figure}[h]
	\centering
	\includegraphics[width=\textwidth]{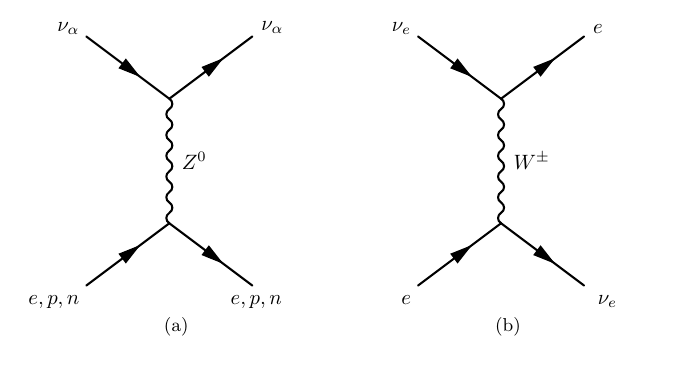}
	\caption{\textit{Feynman diagrams of the neutrino interactions with the ordinary matter.} Diagram (a) represents the NC interaction mediated by the neutral $Z^0$ vector boson.   Diagram (b) represents the CC interaction mediated by the $W^\pm$ vector boson.}
	\label{fig:feynman_diagram-mat}
\end{figure}
Neutrino oscillation is significantly affected when neutrinos travel through matter. While propagating through matter, neutrinos undergo {\it coherent forward elastic scattering} off the fermions present inside the ambient matter. 

Figure~\ref{fig:feynman_diagram-mat} represents the neutrino interactions with ordinary matter. As depicted in the left panel of \figu{feynman_diagram-mat}, all flavors of neutrinos ($\nu_e\,,\nu_\mu\,,\nu_\tau$) can have NC interactions with electrons ($e$), protons ($p$), and neutrons ($n$), and the resulting potential is given by~\cite{Linder:2005fc, Honda:2006gv},
\begin{equation}\label{equ:mat-potential-NC}
V_{\rm NC}=-\frac{1}{2}\sqrt{2}G_FN_n\,,
\end{equation}
where, $G_F$ is the Fermi coupling constant and $N_n$ is the neutron number density inside the matter. Due to the flavor-independent nature of the NC processes, the potential in \equ{mat-potential-NC} has no impact on neutrino oscillation.  

However, as shown in the right panel of \figu{feynman_diagram-mat}, only $\nu_e$ has the CC interaction with the ambient electrons. Due to the absence of muons and tau leptons, the other two flavors of neutrinos do not undergo the CC interaction. The resulting CC potential is given by, 
\begin{equation}\label{equ:mat-potential-CC}
V_{\rm CC}=\sqrt{2}G_FN_e\,.
\end{equation}
Here, $N_e$ is the electron number density. While going from neutrinos to antineutrinos, $V_{\rm NC}\to -V_{\rm NC}$ and $V_{\rm CC}\to -V_{\rm CC}$. Further simplifying, $V_{\rm CC}$ can be expressed as, 
\begin{equation}
V_{CC}\simeq 7.6\cdot10^{-14}\cdot{ Y}_e\cdot \Big(\frac{\rho_{\rm avg}}{\rm g \cdot cm^{-3}}\Big)~\rm eV\,.
\end{equation}
Here, $\rho_{\rm avg}$ is the average matter density along the neutrino path. $Y_e=\frac{N_e}{N_p+N_n}$ is the relative number density of electrons with respect to the protons and neutrons. For Earth matter, which is neutral ($N_e=N_p$) and almost isoscalar ($N_p \simeq N_n$), $Y_e$ is almost equal to 0.5.
%%%%%%%%%%%%%%%%%%%%%%%%%%%%%%%%%%%%%%%%%%%%%%%
\subsection{Two flavour oscillation in matter}
%%%%%%%%%%%%%%%%%%%%%%%%%%%%%%%%%%%%%%%%%%%%%%%
The propagation Hamiltonian in matter in flavor basis expressed in terms of the vacuum oscillation parameters is given by,
\begin{align}\label{equ:H_vac_param}
H&=U\begin{pmatrix}
0 & 0\\
0 & \frac{{\Delta m^2_{21}}}{2E}
\end{pmatrix}U^\dagger+\begin{pmatrix}
V_{\rm CC} & 0\\
0 & 0 
\end{pmatrix}\nonumber
\\&=\begin{pmatrix}
\cos\theta & \sin\theta\\
-\sin\theta & \cos\theta
\end{pmatrix}\begin{pmatrix}
0 & 0\\
0 & \frac{{\Delta m^2_{21}}}{2E}
\end{pmatrix}
\begin{pmatrix}
\cos\theta & -\sin\theta\\
\sin\theta & \cos\theta
\end{pmatrix}+\begin{pmatrix}
V_{\rm CC} & 0\\
0 & 0 
\end{pmatrix}\nonumber
\\&=\frac{{\Delta m^2_{21}}}{4E} 
\begin{pmatrix}
1-\cos2\theta+\frac{4EV_{\rm CC}}{\Delta m^2_{21}} & \sin2\theta\\
\sin2\theta & 1+\cos2\theta
\end{pmatrix}\,.
\end{align}
Diagonalizing the above Hamiltonian, we get the energy eigenvalues in the following form, 
\begin{equation}
\tilde E_{1}=\frac{V_{\rm CC}}{2}+\frac{{\Delta m^2_{21}}}{4E}\Bigg(1-\sqrt{\sin^22\theta+\Big(\cos2\theta-\frac{2EV_{\rm CC}}{\Delta m^2_{21}}\Big)^2}\Bigg)\,,
\end{equation}
\begin{equation}
\tilde E_{2}=\frac{V_{\rm CC}}{2}+\frac{{\Delta m^2_{21}}}{4E}\Bigg(1+\sqrt{\sin^22\theta+\Big(\cos2\theta-\frac{2EV_{\rm CC}}{\Delta m^2_{21}}\Big)^2}\Bigg)\,.
\end{equation}
Hence, 
\begin{equation}\label{equ:eigen_diff}
\tilde E_{2}-\tilde E_{1}=\frac{{\Delta m^2_{21}}}{2E}\sqrt{\sin^22\theta+\Big(\cos2\theta-\frac{2EV_{\rm CC}}{\Delta m^2_{21}}\Big)^2}\,.
\end{equation}
Let $\tilde U$ be the matter modified PMNS matrix that diagonalizes the effective Hamiltonian in matter. Then, $\tilde U$ can be parameterized in terms of the matter modifies mixing angle ($\tilde\theta$),
\begin{align}
\tilde U=\begin{pmatrix}
\cos\tilde\theta & \sin\tilde\theta\\
-\sin\tilde\theta & \cos\tilde\theta
\end{pmatrix}\,.
\end{align}
The propagation Hamiltonian in matter in tilde basis becomes,
\begin{align}\label{equ:H_mat_param}
\tilde{H}\nonumber&=\tilde{U}\,\begin{pmatrix}
0 & 0\\
0 & \frac{{\Delta \tilde m^2_{21}}}{2E}
\end{pmatrix}\,\tilde{U}^\dagger
=\tilde{U}\begin{pmatrix}
0 & 0\\
0 & \tilde{E_2}-\tilde{E1}
\end{pmatrix}\tilde{U}^\dagger
\\&= 
\begin{pmatrix}
\sin^2\tilde\theta(\tilde{E_2}-\tilde{E_1}) & \frac{1}{2}\sin 2\tilde\theta (\tilde{E_2}-\tilde{E_1})\\
\frac{1}{2}\sin 2\tilde\theta (\tilde{E_2}-\tilde{E_1}) & \cos^2\tilde\theta(\tilde{E_2}-\tilde{E_1})
\end{pmatrix}\,.
\end{align} 
Equating \equ{H_vac_param} and \equ{H_mat_param}, we get the following relation,
\begin{equation}\label{equ:mixing_angle_relation}
\frac{{\Delta m^2_{21}}}{4E}\sin 2\theta=\frac{1}{2}\sin{2\tilde\theta}(\tilde{E_2}-\tilde{E_1})\,.
\end{equation}
Using \equ{eigen_diff} in \equ{mixing_angle_relation}, we get the relation between the mixing angle in matter and the corresponding angle in vacuum as,
\begin{equation}
\sin{2\tilde\theta}=\frac{\sin{2\theta}}{\sqrt{\sin^22\theta+(\cos2\theta-\frac{2EV_{\rm CC}}{\Delta m^2_{21}})^2}}\,.
\end{equation}
The oscillation probability in matter is given by,
\begin{align}
P_{\nu_\alpha\rightarrow\nu_{\beta}}\nonumber&=\sin^2{2\tilde\theta}\sin^2\Big(\frac{\Delta \tilde m^2_{21}L}{4E}\Big)
\\&=(\sin^2{2\tilde\theta})\sin^2\Bigg(\frac{\Delta m_{21}^2L}{4E}\sqrt{\sin^2{2\theta}+\Big(\cos{2\theta}-\frac{2EV_{\rm CC}}{\Delta m_{21}^2}\Big)^2}\Bigg)\,.
\end{align}
Interestingly, when $\cos2\theta = 2EV_{\rm CC}/{\Delta m_{21}^2}$, the oscillation probability becomes maximum because $\sin2\tilde\theta = 1$. This resonance effect is commonly known as the Mikheyev-Smirnov-Wolfenstein (MSW) effect~\cite{Wolfenstein:1977ue,Barger:1980tf}. The resonance energy is given by,
\begin{equation}
E_{\rm res}=\frac{\Delta m^2_{21}\cos2\theta}{2V_{\rm CC}}\,.
\end{equation}
\subsection{Three-flavor oscillation in matter }
In the three-neutrino framework of neutrino oscillation, the effective Hamiltonian in flavor basis in matter is given by,
\begin{equation}
H=\frac{1}{2E}
U
\begin{pmatrix}
0 & 0 & 0\\
0 & {\Delta m^2_{21}} & 0\\
0 & 0 & {\Delta m^2_{31}}
\end{pmatrix}U^\dagger+\begin{pmatrix}
V_{\rm CC} & 0 & 0\\
0 & 0 & 0\\
0 & 0 & 0 
\end{pmatrix}\,.
\end{equation}
The first term denotes the neutrino oscillation in vacuum and the second term accounts for the matter effects.\\
For completeness, we give here the approximate expressions for the $\nu_{\mu}\to{\nu_{e}}$ appearance probability upto second order in the small parameters $\sin\theta_{13}$ and $\alpha~(=\Delta m_{21}^2/\Delta m_{31}^2)$~\cite{Akhmedov:2004ny},
\begin{align}\label{equ:app_mat}
P_{\nu_\mu{\rightarrow}\nu_e}
\simeq~\nonumber&\sin^2{\theta_{23}}\sin^2{2\theta_{13}}\frac{\sin^2[(1-\hat A)\Delta]}{(1-\hat A)^2}\nonumber\\&
-\alpha\sin{2\theta_{13}}\sin2\theta_{12}\sin2\theta_{23}\sin\delta_{\rm CP}\sin\Delta\frac{\sin(\hat A\Delta)}{\Delta}\frac{\sin[(1-\hat A)\Delta]}{(1-\hat A)}\nonumber
\\&+\alpha\sin{2\theta_{13}}\sin2\theta_{12}\sin2\theta_{23}\cos\delta_{\rm CP}\sin\Delta\frac{\sin(\hat A\Delta)}{\Delta}\frac{\sin[(1-\hat A)\Delta]}{(1-\hat A)}\nonumber
\\&+\alpha^2\cos^2{\theta_{23}}\sin^2{2\theta_{12}}\frac{\sin^2(\hat A\Delta)}{\hat A^2}\,,
\end{align}
and the $\nu_{\mu}\to{\nu_{\mu}}$ disappearance probability upto second order in $\sin\theta_{13}$ and $\alpha$ is given by~\cite{Akhmedov:2004ny},
\begin{align}\label{equ:disapp_mat}
P_{\nu_\mu{\rightarrow}\nu_\mu}
\simeq~\nonumber&1-\sin^2{2\theta_{23}}\sin^2{\Delta}-4\sin^2\theta_{13}\sin^2{\theta_{23}}\frac{\sin^2{{(1-\hat A)\Delta}}}{(1-\hat A)^2}\\\nonumber
&+\frac{2}{1-\hat A}\sin^2\theta_{13}\sin^2 2\theta_{23}\Big[\sin\Delta\cos\hat A\Delta\frac{\sin{{(1-\hat A)\Delta}}}{(1-\hat A)}-\frac{\hat A\Delta}{2}\sin2\Delta\Big]\\\nonumber
&-\frac{1}{1-\hat A}\alpha\sin\theta_{13}\sin 2\theta_{12}\sin 4\theta_{23}\cos\delta_{\rm CP}\sin\Delta\Big[\hat A\sin\Delta-\frac{\sin{\hat A\Delta}}{A}\cos(1-\hat A)\Delta]\\\nonumber
&-2\alpha\sin\theta_{13}\sin 2\theta_{12}\sin 2\theta_{23}\cos\delta_{\rm CP}\cos\Delta\frac{\sin\hat A\Delta}{\hat A}\frac{\sin{{(1-\hat A)\Delta}}}{(1-\hat A)}\\\nonumber
&+(\alpha\Delta)\cos^2\theta_{12}\sin^22\theta_{23}\sin 2\Delta
-\alpha^2\sin^22\theta_{12}\cos^2\theta_{23}\frac{\sin^2\hat A \Delta}{\hat A^2}\\\nonumber
&-(\alpha\Delta)^2\cos^4\theta_{12}\sin^22\theta_{23}\cos 2\Delta\\
&+\frac{1}{2\hat A}\alpha^2\sin^22\theta_{12}\sin^22\theta_{23}\Big(\sin\Delta\frac{\sin\hat A\Delta}{\hat A}\cos(1-\hat A)\Delta-\frac{\Delta}{2}\sin 2\Delta\Big)\,,
\end{align}
where, the matter effect term, $\hat A=2EV_{\rm CC}/\Delta{m^2_{31}}$ and $\Delta=\Delta m^2_{31}L/4E\,.$

%==========================================================================
\section{Various neutrino sources and detection experiments contributing to oscillation study}
\label{sec:global-osc-expt}
%==========================================================================
Neutrinos come from a variety of natural as well as artificial sources. Their energy can vary from sub-eV to TeV or even EeV, depending on the mechanism of neutrino production at the source. Below, we list some of the natural and artificial sources of neutrinos, experiments detecting them, and discuss how they contribute to the understanding of neutrino oscillation.

\noindent{\textbf {(i) Relic neutrinos:}} They form the cosmic neutrino background (C$\nu$B). Shortly after the Big Bang, when the Universe was around $10^{-6}$ seconds old, the neutrinos were in thermal equilibrium with the other matter particles (electron, positron, and nucleons) and photons via weak interaction, forming a thermal bath. As the Universe expanded and cooled down to below 1 MeV, the interaction rate got reduced significantly. After this point, when the expansion rate exceeds the interaction rate, neutrinos stop interacting with the other matter and radiation and decouple from the thermal bath. From then on, these relic neutrinos stream freely all over space. The present-day temperature of relic neutrinos is around 1.95 K $\sim 10^{-4}$ eV. The density of relic neutrinos is 336 (including neutrino and antineutrinos of all flavors) per centimeter cube, making them the second most abundant particle after the cosmic microwave background (CMB) photons. The extremely low energy of the relic neutrinos makes it challenging to detect them in the present-day experiments. However, Karlsruhe tritium neutrino experiment (KATRIN), recently looked for the relic neutrinos and put constraints on the local over-density of relic neutrinos~\cite{KATRIN:2022kkv}. The PTOLEMY experiment has a future plan for the direct detection of the relic neutrinos~\cite{PTOLEMY:2018jst, PTOLEMY:2024boh, PTOLEMY:2024lzs}.  

\noindent\textbf{(ii) Solar neutrinos:} Sun produces neutrinos via two nuclear processes called the $pp$ chain and CNO cycle; the net effect of them is the following fusion reaction,
\begin{equation}
4 p \to ^4{\rm He} + 2 e^+ + 2\nu_e+\gamma\,,
\end{equation}
The range of of solar neutrino energy is form 0.1 MeV to 15 MeV. 
Solar neutrinos are detected via inverse beta decay process,
\begin{equation}
\nu_e + ^{A}_N\!{\rm Z}~\to~ ^{A}_{N-1}{\rm (Z+1)}+ e^-\,,
\end{equation} 
where, $^{A}_{N-1}{\rm (Z+1)}$ is unstable and decays further to lighter nuclei.

\underline{\textbf{Homestake}} \\
This experiment~\cite{Cleveland:1998nv} was one of the earliest solar neutrino experiments, started in 1965. This experiment was situated at the Homestake mine in South Dakota that used a large chlorine tank to capture electron neutrino,
\begin{equation}
\nu_e + ^{37}\!{\rm Cl}~\to~ ^{37}{\rm Ar}+ e^-\,,
\end{equation}
with an energy threshold of 0.814 MeV. 

\noindent The $^{37}\!{\rm Ar}$ isotopes are radioactive and decay by electron capture with a half life of 35 days into $^{37}{\rm Cl}^*$,
\begin{equation}
 ^{37}\!{\rm Ar}+ e^-~\to~ ^{37}{\rm Cl}^* + \nu_e\,.
\end{equation}
The excited $^{37}{\rm Cl}^*$ emits characteristic X-rays of 2.82 keV (Augur electrons) to return to its ground state. 

The solar neutrino flux measured by this experiment (refer to~\cite{Cleveland:1998nv}) was about 30\% of the total flux predicted by the Standard Solar Model~\cite{Bahcall:2004fg}. This anomaly in the solar neutrino flux is known as the {\it solar neutrino anomaly}.
 
\underline{\textbf{Gallex and SAGE}}\\
%Gallex~\cite{GALLEX:1992gcp, GALLEX:1998kcz, Kaether:2010ag} and SAGE~\cite{SAGE:1999nng, SAGE:2009eeu} were two pioneering experiments to detect low energy solar neutrinos. 
%of the short-baseline (SBL) Gallium experiments designed to study the solar neutrino oscillation. To produce electron neutrinos, they used radiochemical processes that involve electron capture by $^{51}{\rm Cr}$ and $^{37}{\rm Ar}$,
Gallex~\cite{GALLEX:1992gcp} and SAGE~\cite{SAGE:1999nng} were two pioneering Gallium based experiments to detect low-energy solar neutrinos via the following inverse beta decay process, 
\begin{equation}
\nu_e+^{71}{\rm Ga}+  ~\to~ ^{71}{\rm Ge}+e^-\,,
\end{equation}
with a detection threshold of 0.233 MeV. The measured $\nu_e$ capture rate at Gallex was $77\pm8$ SNU~\cite{GALLEX:1998kcz} and at SAGE it was $70.8\pm5$ SNU~\cite{SAGE:2009eeu}. However, the predicted rate is $129\pm9$ SNU~\cite{Bahcall:1998wm} from the Standard Solar Model. This leads to a deficit of around $40\%$. These data also confirm the {\it solar neutrino anomaly}. 

Four calibration experiments, namely, Gallex Cr-1~\cite{GALLEX:1997lja, Kaether:2010ag}, Gallex Cr-2~\cite{GALLEX:1997lja, Kaether:2010ag}, SAGE-Ar~\cite{Abdurashitov:2005tb}, SAGE-Cr~\cite{SAGE:1998fvr}, with $^{51}{\rm Cr}$ and $^{37}{\rm Ar}$ as sources of $\nu_e$, were conducted to recheck the deficit. They reported an average ratio between the observed to the predicted flux to be $0.84\pm0.05$~\cite{Gariazzo:2015rra} which corresponds to a deficit in $\nu_e$ flux with a significance of about $2.9\sigma$.  
This is known as the {\it Gallium Anomaly}. A widely accepted solution towards resolving this anomaly is by considering active-sterile neutrino oscillation with a new mass-squared difference, $\Delta m^2\sim1~{\rm eV}^2$. 

\underline{\textbf{Super-Kamiokande (Super-K)}}\\
Super-K~\cite{Super-Kamiokande:2017yvm} uses water Cherenkov technique to detect neutrinos. It consists of a 50 kt water Cherenkov detector with a fiducial volume of 22.5 kt. The electron neutrinos from the Sun interact with the ambient electrons and nuclei inside the detector to produce secondary particles moving faster than light inside the medium, hence producing Cherenkov radiation. The threshold energy of the detector is around 5 MeV, because neutrino with energy below this will not produce charge particles with sufficient momentum to produce Cherenkov radiation. About 11147 Photomultiplier tubes (PMTs) are fixed on the detector walls to catch the emitted Cherenkov light. Super-K reported a capture rate of $0.45\pm0.02$ SNU, which is about half of the Standard Solar Model prediction.  

\underline{\textbf{Sudbury Neutrino Observatory (SNO)}}\\
SNO~\cite{SNO:1999crp} has a large Water Cherenkov detecor consisting of 1000 tonne of heavy water ($D_2O$). The use of heavy water target brings this experiment with the advantage of measuring both neutral current and charge current neutrino interactions as follows,
\begin{equation}
{\rm CC:}~ \nu_e + d ~\to~ e^-+p+p\,,
\end{equation}
\begin{equation}
{\rm NC:}~ \nu_\alpha + d ~\to~ \nu_\alpha+p+n\,,\alpha=e,\mu,\tau\,.
\end{equation}
It was the neutral current measurements from the SNO experiment which confirmed~\cite{SNO:2002tuh} the deficit of $\nu_e$ flux reported by the Homestake experiment and resolved the {\it solar neutrino anomaly}. The deficit can be explained by the phenomenon of neutrino flavor transition or  {\it neutrino oscillation} from $\nu_e$ to $\nu_\mu$ or $\nu_\tau$ while traveling from the Sun to the detector on Earth.  

\underline{\textbf{Borexino}}\\ Borexino~\cite{Borexino:2008gab} is located about 3.8 km underground at the Laboratori Nazionali del Gran Sasso (LNGS) in Italy. The main goal of this experiment was the detection of the monoenergetic (862 keV) $^7$Be solar neutrinos via neutrino-electron elastic scattering. Borexino consists of a 300-ton ultra-pure organic liquid scintillator in a spherical volume, surrounded by a buffer zone to shield it from the external radiation. The liquid scintillator is housed in a nylon vessel, and the detector is submerged in a larger water tank to provide additional shielding against cosmic rays and natural radioactivity. About 2200 photomultiplier tubes (PMTs) are positioned around the scintillator to detect the light flashes produced when neutrinos interact with electrons inside the scintillator. Borexino made precision measurements of neutrinos from various reactions in the Sun,$^7$Be~\cite{Bellini:2011rx}, pp~\cite{BOREXINO:2014pcl, BOREXINO:2018ohr}, pep~\cite{Borexino:2011ufb}, $^8$B~\cite{Borexino:2008fkj}, and CNO cycle~\cite{BOREXINO:2020aww} neutrinos, confirming the solar models and the role of nuclear fusion processes. Borexino also studied neutrino oscillation phenomena, focusing on the flavor transition of electron neutrinos to other neutrino flavors during their journey from the Sun to the Earth. This experiment provided critical evidence supporting the Mikheyev-Smirnov-Wolfenstein (MSW) effect, which explains how neutrino oscillations are influenced by interactions with matter, particularly in the high-density environment of the Sun’s core~\cite{BOREXINO:2018ohr}.

\noindent\textbf{(iii) Reactor antineutrinos:} Electron antineutrinos are artificially produced in Nuclear reactors. The radio isotopes, mainly, $^{235}{\rm U}, ^{238}{\rm U}, ^{239}{\rm Pu}$ and $^{241}{\rm Pu}$ produce lighter unstable fission fragments which are rich in neutrons. The fission fragments undergo radioactive $\beta$-decay producing $\bar\nu_e$ in 1 to 10 MeV energy range.

The reactor antineutrinos are detected via the inverse beta decay process, where the electron antineutrino interacts with the free proton ($p$) in the medium, 
\begin{equation}\label{equ:inverse-beta-decay}
\bar\nu_e+ p ~\to~ e^++n\,.
\end{equation}
the positron quickly annihilates with the electrons inside the detector, producing a prompt signal of two gamma photons, each with 0.511 MeV energy. The neutron is captured by Hydrogen in the case of water detectors, leading to a delayed 2.2 MeV gamma ray signal. The challenges in the case of water detectors involve the long neutron capture time ($\sim200$ $\mu$s) and the low-energy gamma photons, which make these events difficult to distinguish from the background. Gadolinium is doped into water to overcome these difficulties. In this case, the higher gamma energies ($\sim$ 8 MeV) and faster neutron capture time ($\sim30$ $\mu$s) enhance the detection efficiency. 

Various reactor antineutrino experiments like KamLAND~\cite{KamLAND:2002uet}, Daya Bay~\cite{DayaBay:2018yms}, RENO~\cite{RENO:2018dro}, and Double Chooz~\cite{DoubleChooz:2019qbj} have contributed significantly towards measuring the reactor electron antineutrino flux as well as the neutrino oscillation parameters with greater precision. Below, we describe these reactor experiments in more detail.

\underline{\textbf{KamLAND}}\\
KamLAND, the Kamioka Liquid Scintillator Anti-Neutrino Detector, was a scintillator-based reactor experiment in Kamioka, Japan. The detector material is a 1 kt ultra-pure liquid scintillator contained within a nylon balloon suspended in purified mineral oil. It is surrounded by 55 nuclear reactors with an average baseline length of 180 km. KamLAND provided evidence of neutrino oscillation~\cite{KamLAND:2002uet} by confirming the reactor antineutrino disappearance. In combination with the solar neutrino data, it gave the most precise measurements of $\Delta m^2_{21}$ (= $7.59^{+0.21}_{-0.21}\times10^{-5} ~{\rm eV}^2$) and $\tan^2{\theta_{12}}$ (= $0.47^{+0.06}_{-0.05}$)~\cite{KamLAND:2008dgz}.  

\underline{\textbf{Daya Bay, Double Chooz, RENO}} \\
These are short-baseline reactor antineutrino experiments designed to measure $\theta_{13}$ via $\bar\nu_e$ disappearance. Daya Bay is located in China. It consists of eight detector modules, each filled with 20 tons of liquid scintillator that detect electron anineutrinos from six reactor cores situated within a distance of 1.9 km. Double Chooz was built on the site of the earlier Chooz experiment in France. It deployed two identical gadolinium-doped liquid scintillator detectors placed at 400 m and 1050 m from the Chooz nuclear power plant. RENO (Reactor Experiment for
Neutrino Oscillation) also uses two identical gadolinium-doped liquid scintillator detectors of mass 16.5 kt, placed at 294 m and 1383 m from the Hanbit Nuclear power Plant in South Korea.

\noindent The $\bar\nu_e\to\bar\nu_e$ survival probability in vacuum is given by~\cite{Zhan:2009rs}, 
\begin{align}\label{equ:nue_disapp}
P_{\bar\nu_e\to\bar\nu_e}\nonumber= &1- \cos^4\theta_{13}\sin^2{2\theta_{12}}\sin^2\frac{\Delta m^2_{21}L}{4 E}\\\nonumber&
+\cos^2\theta_{12}\sin^2{2\theta_{13}}\sin^2\frac{\Delta m^2_{31}L}{4 E}\\&
+\sin^2\theta_{12}\sin^2{2\theta_{13}}\sin^2\frac{\Delta m^2_{32}L}{4 E}\,.
\end{align}
At short-baseline experiments, where $L\sim1~\rm {km}$, the oscillation is mainly driven by $\Delta m^2_{31}$ only, and the experiments are super-sensitive to the reactor mixing angle, $\theta_{13}$, without any interference from the other mixing angles.
Double Chooz gave the hint of electron antineutrino disappearance and hence non-zero $\theta_{13}$~\cite{DoubleChooz:2011ymz} along with accelerator-based experiments T2K~\cite{T2K:2011ypd} and MINOS~\cite{MINOS:2011amj}. In 2012, Daya Bay~\cite{DayaBay:2012fng} measured the non-zero value $\theta_{13}$ with a significance of $5\sigma$ that was very quickly confirmed by RENO~\cite{RENO:2012mkc}.

The recent reevaluation of reactor antineutrino flux~\cite{Mueller:2011nm, Huber:2011wv} resulted in around a 6\% deficit of the reactor flux observed by Daya Bay, RENO, and Double Chooz, as compared to that theoretically predicted. This discrepancy is known as the {\it Reactor Antineutrino Anomaly (RAA)}~\cite{Mention:2011rk}. This deficit can be explained by the $\bar\nu_e$ disappearance via sterile neutrino with oscillations driven by the mass-squared difference of around 1 eV$^2$. These experiments also observed a distortion in the shape of reactor antineutrino energy spectrum around 5 MeV~\cite{DayaBay:2017jkb, DoubleChooz:2014kuw, RENO:2015ksa}, which further worsened the RAA and is commonly referred to as the {\it Reactor shape anomaly}. However, the latest refinement in the flux calculation models substantially reduced the significance of RAA from that originally reported~\cite{Giunti:2021kab}. 

\underline{\textbf{Upcoming reactor experiment: JUNO}}\\
The Jiangmen Underground Neutrino Observatory (JUNO)~\cite{JUNO:2015zny} is a cutting-edge neutrino experiment under construction in Jiangmen, China, approximately 700 meters underground. It is located at around 53 km from Yangjiang and Taishan nuclear power plants. Yangjiang houses six reactor cores, each with 2.9 Gigawatt (GW) thermal power, and Taishan contains four, each with 4.59 GW thermal power, resulting a total of 35.76 GW. JUNO is designed to be a multipurpose experiment, with its main goal to determine the neutrino mass ordering by studying reactor antineutrino oscillations over a medium baseline of 53 km. From \equ{nue_disapp}, we can see that at medium baselines like JUNO, the oscillation due to the solar mass-squared difference $\Delta m^2_{21}$ also survives along with the atmospheric one, $\Delta m^2_{31}$. These two mass-squared differences interfere differently in the case of normal and inverted neutrino mass orderings. Due to the excellent energy resolution of around $3\%$ at 1 MeV, JUNO will be able to resolve these interference effects, thus resolving the issue of neutrino mass ordering. Additionally, JUNO will significantly enhance the precision of $\theta_{12}$, $\Delta m^2_{21}$, and $\Delta m^2_{31}$. After six years of running, JUNO is expected to measure $\theta_{12}$, $\Delta m^2_{21}$, and $\Delta m^2_{31}$ with relative $1\sigma$ precisions of $0.5\%$, $0.3\%$ and $0.2\%$, respectively~\cite{JUNO:2022mxj}.  

\noindent\textbf{(iv) Geoneutrinos:} Geoneutrinos are naturally produced inside the crust and mantle of Earth via the radioactive decay of Uranium ($^{238}$U), Thorium ($^{232}$Th), and predominantly from Potassium ($^{40}$K). They are electron antineutrinos of MeV scale energy detected via inverse beta decay process (\equ{inverse-beta-decay}). KamLAND was the first to detect the geoneutrinos~\cite{Araki:2005qa}. Borexino also provided complementary information to KamLAND~\cite{Borexino:2010dli}. Due to the extremely low event rates of the geoneutrinos at the present-day experiments, they do not provide a suitable avenue for the neutrino oscillation study.  
%Since geoneutrinos travel small distances before they are detected in the experiments, they do not have significant effect on neutrino oscillation. 

\noindent\textbf{(v) Neutrinos from accelerator:} These are artificially produced neutrinos in high-energy particle accelerators. Protons are accelerated to very high energies before they strike a fixed target made up of carbon, beryllium, or other dense materials. When these high-energy protons hit the target, they produce a shower of secondary particles, including mostly pions ($\pi^+, \pi^-$) and a little amount of kaons ($K^+, K^-$). These secondary particles are then guided inside a decay pipe using magnetic horns to choose the correct electric charge for producing neutrinos or antineutrinos from their decay. The mesons of correct polarity are then allowed to decay in flight inside a long decay pipe to muons and muon neutrinos. The beam dump absorbs the charged muons before they further decay, resulting in a pure muon neutrino beam with a very small contamination of electron neutrinos from kaon or muon decay. The energy of neutrinos from accelerators is in the range of sub-GeV to tens of GeV. 

{\textbf{Short-baseline accelerator-based experiments:}}

\underline{\textbf{LSND}} \\
The Liquid Scintillator Neutrino Detector (LSND)~\cite{LSND:1995lje, Hill:1995gf} was designed to study the $\nu_\mu\to\nu_e$ or $\bar\nu_\mu\to\bar\nu_e$ oscillations using the neutrinos from accelerator situated 30 m away at the Los Alamos Meson decay facility (LAMPF). The $\nu_\mu$ flux was produced from the $\mu^-$ decay in flight (DIF) and $\bar\nu_\mu$ was produced from $\mu^+$ decay at rest (DAR). In 1995, LSND reported candidate events showing the evidence for the $\bar\nu_\mu\to\bar\nu_e$ oscillations~\cite{LSND:1995lje} via the inverse beta decay process, $\bar\nu_e+p\to e^+ + n$ followed by $n+p \to d+\gamma$, that was further strengthened in 1996~\cite{LSND:1996ubh}. It also reported in 1997, the evidence of $\nu_\mu\to\nu_e$ oscillation~\cite{LSND:1997vun} via the charged-current process, $\nu_e +{\rm C} \to e^+ + {\rm X}$. During 1993--1998, LSND reported a total excess of 87.9 $\pm$ 22.4 $\pm$ 6.0 $\bar\nu_e$ events over the expected background~\cite{LSND:2001aii}. This result can not be explained by the standard three-flavor neutrino oscillation. This oscillation anomaly is the so-called {\it LSND anomaly}. It needs active-sterile neutrino oscillation with a mass-squared difference between the active and the sterile state in the range of 0.2--10 eV$^2$.

\underline{\textbf{MiniBooNE}} \\
MiniBooNE~\cite{Bazarko:1999hq, Bazarko:2000id} was a Boster Neutrino Experiment located at Fermilab designed to conclusively test the LSND excess. The neutrino beam was generated from a primary 8 GeV proton beam striking a beryllium target to produce the neutrinos from the successive decay of pions. The neutrinos produced with energy in the range 0.2 GeV to 3 GeV then travel to the detector placed 541 m away. MiniBooNE was running in both neutrino~\cite{MiniBooNE:2012maf} and antineutrino~\cite{MiniBooNE:2010idf, MiniBooNE:2013uba} modes. Magnetic horns are used to separate $\nu_\mu$ and $\bar\nu_\mu$ beams. The $\nu_\mu$ flux peaks around 0.6 GeV and that of $\bar\nu_\mu$ around 0.4 GeV. MiniBooNE reported a combined excess of $460.5\pm99.0$ events at $4.7\sigma$ with $381.2\pm 85.2$ in neutrino mode and $79.3\pm28.6$ in antineutrino mode at low energies in the range 0.2 to 1.25 GeV~\cite{MiniBooNE:2018esg}. This MiniBooNE low energy excess is known as the {\it MiniBooNE anomaly}. It can be explained with active-sterile neutrino oscillations with mass-squared difference in the range of 0.01--1~eV$^2$. MicroBooNE proposed in 2007~\cite{MicroBooNE:2007ivj} is still verifying the possible explanations for the MiniBooNE low energy excess with its total data collected from 2015 to 2020~\cite{MicroBooNE:2021zai}.  

{\textbf{Long-baseline accelerator-based experiments:}}

\underline{\textbf{K2K}} \\
The KEK to Kamioka (K2K)~\cite{K2K:2001nlz, K2K:2002icj} was the first long-baseline experiment with a neutrino travel distance of 250 km. It was designed to confirm the atmospheric neutrino oscillations reported by the Super-Kamiokande~\cite{Super-Kamiokande:1998tou}. The KEK proton synchrotron, operated at 12 GeV, accelerates the protons, which then strike the aluminium target to produce secondary particles, mainly charged pions. The magnetic horns focus the pions. The charged pions then decay to produce neutrinos. The neutrinos thus produced are 98\% pure $\nu_\mu$ with peak energy around 1.3 GeV. K2K collaboration published its final result in 2006~\cite{K2K:2006yov}, analyzing the data from 1999 to 2004. K2K measured the allowed range of $\Delta m^2$ in two-flavor neutrino oscillation framework, to lie between $1.9\times10^{-3}$ eV$^2$ and $3.5\times10^{-3}$ eV$^2$ at 90\% C.L. with the best-fit value $2.8\times 10^{-3}$ eV$^2$ for $\sin^22\theta$ = 1. These results confirmed the $\nu_\mu$ disappearance due to $\nu_\mu\to\nu_\tau$ oscillation, as reported by Super-Kamiokande.   

\underline{\textbf{MINOS/MINOS$+$}} \\
The Main Injector Neutrino Oscillation Search (MINOS)~\cite{MINOS:2014rjg} was one of the first generation long-baseline experiments collecting data from February 2005 till April 2012. During this phase, it used $10.56 \times 10^{20}$ P.O.T. for $\nu_\mu$-dominated beam and $3.36 \times 10^{20}$ P.O.T. for $\bar\nu_\mu$-dominated beam.  This running phase used low energy neutrinos with the peak energy of 3 GeV. Magnetic horns are used to distinguish $\nu_\mu$ from $\bar\nu_\mu$. This experiment used two steel-scintillator detectors -- a  1 kt near detector situated at a distance of 1.04 km from the NuMI beamline at Fermilab and a 5.4 kt far detector positioned 735 km away in Minnesota. After the MINOS phase, it was upgraded to MINOS$+$~\cite{MINOS:2020llm}, which ran from 2013 to June 2016 using $\nu_\mu$-dominated beam. MINOS$+$ used high energy beam of neutrinos, with a peak energy of 7 GeV. The main purpose of this experiment was to probe neutrino oscillation through $\nu_\mu\to\nu_\mu$ and $\bar\nu_\mu\to\bar\nu_\mu$ disappearance channels.

\underline{\textbf{T2K}}\\
Tokai-to-Kamioka (T2K)~\cite{T2K:2011qtm} is a presently running long-baseline experiment that receives neutrino beam from the Japanese Particle Accelerator Center (J-PARC). It uses Super-K as the far detector, which is placed at a distance of 295 km from the source at an off-axis angle of $2.5^\circ$ with respect to the neutrino beam. The off-axis position helps to produce a nearly monochromatic neutrino beam with the flux peaked around 0.6 GeV. The near detectors for this experiment include the on-axis INGRID detector and the off-axis ND280, both placed at around 280 m from J-PARC, which help to reduce the cross section and flux uncertainties at the far detector. 
The main channels that T2K probes are $\nu_\mu\to\nu_e\,,\bar\nu_\mu\to\bar\nu_e\,,\nu_\mu\to\nu_\mu\,,{\rm and}~\bar\nu_\mu\to\bar\nu_\mu$. From 2010 to 2020, T2K has accumulated $1.97\times 10^{21}$ P.O.T. in neutrino mode and $1.63\times 10^{21}$ P.O.T. in antineutrino mode at the far detector~\cite{T2K:2023smv}. The primary goals of T2K include the precision on $\theta_{13}$, the measurement of the CP violation in the leptonic sector, and the determination of the correct octant of $\theta_{23}$ via the $\nu_e~{\rm and}~\bar\nu_e$ appearance study, and the precision measurements of the other oscillation parameters like $\Delta m^2_{31}$ and $\theta_{23}$ via the $\nu_\mu~{\rm and}~\bar\nu_\mu$ disappearance study. The relatively shorter baseline helps to measure the CP phase without much interference from the fake CP asymmetry due to the Earth's matter effect. Recent T2K results~\cite{T2K:2023smv} on oscillation parameter measurements gave the following best-fit values: $\sin^2 \theta_{23} = 0.561^{+0.021}_{-0.032}$\,, $\Delta m^2_{32} =2.494^{+0.041}_{-0.058} \times 10^{-3}~{\rm  eV}^2$, $\delta_{\rm CP} = -1.97^{+0.97}_{-0.70}$\,, and $\sin^2\theta_{13} = 28.0^{+2.8}_
{-6.5}\times 10^{-3}$. 

\underline{\textbf{NO$\nu$A}} \\
The NuMI Off-axis $\nu_e$ Appearance (NO$\nu$A)~\cite{NOvA:2004blv, NOvA:2019cyt} experiment is another currently running long-baseline experiment in the US. It collects neutrinos from the Fermilab's NuMI beam~\cite{Adamson:2015dkw}. It has a $0.8^\circ$ off-axis near detector at 1 km at 100 m underground and a far detector at the same off-axis placed 810 km away in the Minnesota River. The near and far detectors have the fiducial volume of 290 tons and 14 kilotons, respectively. Both detectors are tracking calorimeters where the charged particles are tracked via polyvinyl
chloride (PVC) cells filled with a mineral oil-based liquid scintillator. NO$\nu$A explores $\nu_\mu\to\nu_e\,,\bar\nu_\mu\to\bar\nu_e\,,\nu_\mu\to\nu_\mu\,,{\rm and}~\bar\nu_\mu\to\bar\nu_\mu$ oscillation channels with significant statistics coming from the disappearance channels. The neutrino flux peaks around 2 GeV. In the period, February 2014 to May 2021, NO$\nu$A has accumulated $1.7 
\times 10^{21}$ P.O.T. in neutrino mode and $1.27 \times 10^{21}$ P.O.T. in antineutrino mode~\cite{Shanahan:2021jlp}. The best-fit values of the oscillation parameters obtained using the said exposures assuming normal neutrino mass ordering and $\theta_{23}$ in the upper octant are, $\Delta m^2_{32}=2.41^{+0.07}_{-0.07}\times10^{-3} ~{\rm eV}^2$, $\sin^2{\theta_{23}}=0.57^{+0.03}_{-0.04}$, and $\delta_{\rm CP}=(0.82^{+0.27}_{-0.87})\pi$. The results for inverted neutrino mass ordering are: $\Delta m^2_{31}=-2.45^{+0.07}_{-0.07}\times10^{-3} ~{\rm eV}^2$, $\sin^2{\theta_{23}}=0.56$, and $\delta_{\rm CP}=1.52\pi$~\cite{NOvA:2021nfi, Shanahan:2021jlp}.

\noindent\textbf{(vi) Atmospheric neutrinos:} These are naturally produced neutrinos in Earth's atmosphere. When the primary cosmic rays (mostly protons) interact with the atmospheric nuclei, a shower of secondary particles, pions ($\pi^\pm$) and kaons ($K^\pm$) are produced. The secondary particles further decay to produce neutrinos. Following are the decay chains involved in the process of neutrino and antineutrino generation,

\begin{center}
	\begin{tikzpicture}
	% Positive pion decay
	\node (pi+) at (0,1) {\(\pi^+ \to \mu^+ + \nu_\mu\)};
	\node (decay+) at (2,0) {\(e^+ + \nu_e + \bar{\nu}_\mu\)};
	
	% Negative pion decay
	\node (pi-) at (5,1) {\(\pi^- \to \mu^- + \bar{\nu}_\mu\)};
	\node (decay-) at (7,0) {\(e^- + \bar{\nu}_e + \nu_\mu\)};
	
	% Arrows
	\draw[->] (pi+) |- (decay+); % Right-angle arrow directly from mu+
	\draw[->] (pi-) |- (decay-); % Right-angle arrow directly from mu-
	\end{tikzpicture}
\end{center}

These result in the $\nu_\mu$ to $\nu_e$ flavor ratio at the production to be $2:1$. The atmospheric neutrinos have a wide energy range starting from a few MeV to several TeV. In addition, before their detection, they travel a wide range of path lengths ranging from a few km (downward-going neutrinos) to thousands of km of the order of the diameter of Earth (upward-going neutrinos). Due to the isotropic nature of the cosmic rays, the ratio between the upward and downward-going neutrinos is expected to be unity. Deviation of this ratio from unity may occur at low energies due to the disturbances from the Earth's geomagnetic fields, but at high energies, this ratio should remain unity if there is no oscillation. 
%The flux of neutrinos falls with energy approximately as $E^{-2.7}$ at the high energy region from a few GeV to TeV. 
They are detected using large-volume detectors like Super-Kamiokande, IceCube, and KM3NeT/ORCA that observe Cherenkov light as a result of neutrino interaction in water, ice or other dense material. There is also a proposal of a magnetized Iron Calorimeter detector known as INO-ICAL~\cite{ICAL:2015stm} to detect the atmospheric neutrinos. Below, we briefly discuss the atmospheric neutrino experiments.  

\underline{\textbf{Super-K}} \\
Previously, while discussing solar neutrino experiments, we provided the experimental details of Super-K. One of the significant milestones of this experiment is that in 1998, Super-K reported a deficit of upward-going muon-like events compared to downward-going muon-like events with an up-down ratio of about 0.52 compared to the predicted ratio close to unity~\cite{Super-Kamiokande:1998tou}. This is the famous {\it atmospheric neutrino anomaly}. It could be resolved via $\nu_\mu\to\nu_\tau$ oscillation. In this way, Super-K gave the first-ever evidence of neutrino oscillation which establishes that neutrinos have mass -- a crucial evidence of physics beyond the Standard Model~\cite{Super-Kamiokande:1998tou}. This led to the Nobel prize 2015 to Prof. Takaaki Kajita and Prof. Arthur B. McDonald. The latest measurements of the oscillation parameters by Super-K are~\cite{Super-Kamiokande:2023ahc}: $\sin^2{\theta_{23}}=0.45^{+0.06}_{-0.03}$, $\Delta m^2_{32}=2.40^{+0.07}_{-0.09}\times10^{-3} ~{\rm eV}^2$, and $\delta_{\rm CP}=(-1.89^{+0.87}_{-1.18})\pi$ for normal neutrino mass ordering, and $\sin^2{\theta_{23}}=0.48^{+0.07}_{-0.05}$, $\Delta m^2_{32}=2.40^{+0.05}_{-0.33}\times10^{-3} ~{\rm eV}^2$, and $\delta_{\rm CP}=(-1.89^{+1.32}_{-1.97})\pi$ for inverted neutrino mass ordering. 

\underline{\textbf{IceCube DeepCore}}\\ 
IceCube DeepCore~\cite{IceCube:2011ucd} is a sub-array of the IceCube detector consisting of a more compact array of eight strings placed at the bottom center of the IceCube detector. Its compact configuration reduces the neutrino detection threshold to about 10 GeV, allowing it to be well-suited for atmospheric neutrino oscillation studies. With 9.3 years of data from 2012 to 2021, IceCube DeepCore gives the most precise measurement of the atmospheric oscillation parameters to date, $\Delta m^2_{32} = 2.40^{+0.05}_{-0.04}\times 10^{-3} ~{\rm eV}^2$ and $\sin2 \theta_{23}=0.54^{+0.04}_{-0.03}$ assuming normal neutrino mass ordering~\cite{IceCube:2024xjj}.

\underline{\textbf{KM3NeT/ORCA}}\\ KM3NeT (Cubic Kilometre Neutrino Telescope)~\cite{KM3Net:2016zxf} is a water Cherenkov detector under construction in the deep Mediterranean Sea. KM3NeT consists of three building blocks, each comprising of 115 vertical strings called detection units (DUs). Each DU has 18 spherical Digital Optical Modules (DOMs) arranged in vertical arrays. Each DOM contains 31 photomultiplier tubes (PMTs) to detect Cherenkov light produced by charged particles resulting from neutrino interactions. Multi-PMT DOMs increase the photon collection efficiency and improve energy and directional reconstruction capabilities. KM3NeT/ORCA~\cite{Katz:2013svu}, situated off the coast of Toulon, France, is optimized for the neutrino oscillation study in the energy range of 1 to 100 GeV using atmospheric neutrinos. It contains a single building block of KM3NeT and, thus, is expected to have a total of 115 DUs. The modular design of KM3NeT allows the phase-wise deployment of additional DUs, gradually increasing the detector’s sensitivity and capacity. 
In an intermediate stage comprising of 6 DUs, KM3NeT/ORCA reported the first oscillation measurements; $\sin^2 \theta_{23}=0.50^{+0.10}_
{-0.10}$ and $\Delta m^2_{31}=1.95^{+0.24}_
{-0.22}$ with 355 days of data~\cite{Pestel:2022ewd, Schumann:2023kil}.
 
\underline{\textbf{INO-ICAL}}\\ INO (India-based Neutrino Observatory) is a proposed atmospheric neutrino experiment in the Theni district of Tamil Nadu, India. ICAL~\cite{ICAL:2015stm} is a 50 kt magnetized Iron Calorimeter containing 151 layers of iron plates, with active detector elements, known as resistive plate chambers (RPCs), placed between consecutive layers. The atmospheric neutrinos ($\nu_\mu$, $\bar\nu_\mu$) interact with the nucleus in the iron to produce, through charged-current interactions, muons or hadrons, which are detected by the RPCs. The iron plates are magnetized to generate a uniform magnetic field of about 1.5 Tesla, which allows the identification of $\mu^+$ from $\mu^-$, separately. This capability enables the experiment to distinguish between neutrinos and antineutrinos. These RPCs provide excellent timing resolution and position accuracy, allowing precise reconstruction of the energy, direction, and charge of the particles. ICAL will help to improve the precision measurements of the atmospheric oscillation parameters, $\Delta m^2_{32}$ and $\theta_{23}$. Its ability to distinguish neutrinos from antineutrinos makes it a powerful tool for resolving the neutrino mass ordering issue by observing the Earth's matter effect in neutrinos and antineutrinos separately.

\noindent\textbf{(v) Astrophysical neutrinos:} These are high-energy neutrinos produced by some of the most energetic processes in the universe. These neutrinos are created in distant astrophysical sources such as supernovae, gamma-ray bursts (GRBs), active galactic nuclei (AGN), and blazar jets. Inside these extragalactic sources, the charged particles, like protons, are accelerated to ultra-relativistic speeds and interact with matter or radiation, leading to the production of very high energy pions, which subsequently decay into neutrinos. These neutrinos provide a unique window into our universe at high-energies universe and are crucial for multi-messenger astronomy. High-energy astrophysical neutrinos typically have energies in the range of TeV to PeV and even beyond. These astrophysical neutrinos travel vast distances without being deflected by the galactic medium. Hence, detection of these neutrinos can shed light on the high-energy events occurring in astrophysical sources. Below are some of the experiments designed for detecting these astrophysical neutrinos. 

\underline{\textbf{IceCube}} \\
IceCube~\cite{IceCube:2016zyt} is a cubic-kilometer ice detector present in Antarctica, South Pole. It is designed to study the neutrinos of ultra-high energies from astrophysical sources. It comprises of 5160 digital optical modules (DOMs) distributed along 86 vertical strings, each containing 60 DOMs. The horizontal spacing between a pair of DOMs is 125 meters, while the vertical separation between them is 17 meters. Out of these 86 strings, towards the bottom central part of the detector, eight compact strings are placed with a tighter vertical spacing of 7 meters and a horizontal spacing of 40 to 70 meters. This compact nature helps to reduce the detection threshold to about 10 GeV, making it suitable for studying neutrino oscillation with atmospheric neutrinos. This part of the detector is called the IceCube DeepCore~\cite{IceCube:2011ucd, IceCube:2016zyt}. When the high energy neutrinos enter into the detector, they interact with the nucleons/partons present inside the ice, predominantly via deep-inelastic scattering, producing secondary charged particles that give rise to Cherenkov radiation. This Cherenkov light is then captured by the photomultiplier tubes (PMTs) present inside the DOM. The light signal inside the DOMs are then converted to electric signals and sent to the processors present at the detector surface. There, it is further analyzed to extract the information on the energy, direction, and flavor of the neutrinos. The events recorded in IceCube are of three major types: track, cascade, and double cascade. The muons from $\nu_\mu$-CC interactions give elongated track-like events. A subdominant contribution to the track-like events also comes from the $\nu_\tau$-CC interactions as the final state tau leptons can decay to muons. The cascade-like events are from the $\nu_e,\nu_\tau$-CC or NC events from all flavors. The $\nu_\tau$-CC interactions can give double cascade events, one from the shower of hadrons produced at the interaction vertex and the other from the decay of the final state tau leptons to hadrons. The double cascade events are unique signature of the $\nu_\tau$ events. IceCube has so far detected 164 high-energy starting events (HESE) with 7.5 years of data~\cite{IceCube:2023sov}.

\underline{\textbf{IceCube-Gen2}}\\
IceCube-Gen2~\cite{IceCube-Gen2:2020qha, IC-gen2} is the proposed next-generation upgrade to the IceCube detector. It will have almost eight times larger volume than IceCube. This huge volume with advanced optical modules and better angular resolution will enable IceCube-Gen2 to have better sensitivity to neutrino interaction and better reconstruction of the neutrino energy, direction, and flavor beyond the TeV-PeV range. 

\newpage
\underline{\textbf{KM3NeT/ARCA}}\\
KM3NeT/ARCA~\cite{KM3Net:2016zxf} located near Sicily, Italy, is optimized for detecting high-energy astrophysical neutrinos from the sources such as active galactic nuclei (AGN), gamma-ray bursts, and supernovae. It would contain two building blocks of KM3NeT. The large detector volume and sparse distribution of optical modules make ARCA sensitive to neutrinos in the TeV–PeV energy range and beyond, enabling the identification of astrophysical neutrino sources with improved angular resolution. Recently, KM3NeT/ARCA has detected an ultra-high energy cosmic neutrino of energy around 220 PeV~\cite{KM3NeT:2025npi}. 

\section{Present status of the oscillation parameters in three-neutrino framework}
The phenomenon of mass-induced neutrino oscillation in three-flavor framework involves six fundamental parameters, three mixing angles, $\theta_{12}\,,\theta_{13}\,,{\rm and}\,\,\theta_{23}$, one  CP-violating Dirac phase, $\delta_{\rm CP}$, and two independent mass-squared differences, $\Delta m^2_{31}$ and $\Delta m^2_{21}$. The KamLAND~\cite{KamLAND:2013rgu} reactor antineutrino experiment along with the solar experiments like Homestake~\cite{Cleveland:1998nv}, Gallex~\cite{Kaether:2010ag}, SAGE~\cite{SAGE:2009eeu}, Super-K~\cite{Super-Kamiokande:2005wtt, Super-Kamiokande:2008ecj, Super-Kamiokande:2010tar, Super-Kamiokande:2023jbt}, SNO~\cite{SNO:2011hxd, sno_solar_neutrino2024}, and Borexino~\cite{Borexino:2008fkj, Bellini:2011rx} have immensely contributed towards the precision measurement of the solar oscillation parameters, $\Delta m^2_{21}$ and $\theta_{12}$~\cite{Esteban:2024eli}. The long-baseline experiments, K2K~\cite{K2K:2001nlz, K2K:2002icj}, MINOS~\cite{MINOS:2013utc, MINOS:2013xrl}, T2K~\cite{T2K:2023mcm, t2k_neutrino2024}, NO$\nu$A~\cite{Shanahan:2021jlp} along with the atmospheric neutrino experiments, IceCube DeepCore~\cite{IceCube:2019dyb, IceCube:2019dqi, IceCube:2024xjj} and Super-K~\cite{Super-Kamiokande:2023ahc}, KM3NeT/ORCA~\cite{Pestel:2022ewd, Schumann:2023kil} and reactor experiments, Daya-Bay, RENO and Double-Chooz have measured the atmospheric mass-squared difference, $\Delta m^2_{31}$. The combined efforts from long-baseline and atmospheric experiments have measured the atmospheric mixing angle $\theta_{23}$ with a relative $1\sigma$ precision of about $5\%$~\cite{Esteban:2024eli}. The reactor data from Double-Chooz~\cite{DoubleChooz:2019qbj, Soldin:2024fgt, Thiago_neutrino2020}, Daya Bay~\cite{DayaBay:2022orm}, and RENO~\cite{RENO:2018dro, Yoo_neutrino2020} have measured the reactor mixing angle $\theta_{13}$ with utmost precision. 
The preliminary information on $\delta_{\rm CP}$ is mainly coming from the long-baseline sector by studying the $\nu_\mu\to\nu_e$ and $\bar\nu_\mu\to\bar\nu_e$ appearance data and by studying the sub-GeV neutrinos of atmospheric sample of Super-K~\cite{Super-Kamiokande:2023ahc}. 

Various global fit groups~\cite{NuFIT, deSalas:2020pgw, Capozzi:2021fjo} are working on providing precision measurements of various oscillation parameters by combining the data from neutrino oscillation experiments all over the world. Here, we provide the current status of the oscillation parameters based on the global fit to the world oscillation data, as given in NuFit 6.0~\cite{Esteban:2024eli}. Table~\ref{tab:nufit-6} gives the current best-fit values of the oscillation parameters along with the uncertainties involved in their measurements. The relative $1\sigma$ precision{\footnote{The relative $1\sigma$ precision is given by the formula,
		\begin{equation}
		1\sigma~{\rm precision} = \frac{3\sigma~{\rm upper~ limit}-3\sigma~{\rm lower~limit}}{6\times{\rm best{\tiny-}fit~value}}\times 100\%
		\end{equation}}} 
on the measurements of the oscillation parameters are given in the last column of table~\ref{tab:nufit-6}. As can be seen from the table, most of the oscillation parameters have been measured quite precisely with the a relative $1\sigma$ precision below $\sim5\%$. However, at present, the Dirac CP phase, $\delta_{\rm CP}$, is poorly constrained as compared to all other oscillation parameters. The NuFIT 6.0 analysis depicts that even if the individual dataset from long-baseline experiments, T2K and NOvA, show a preference for normal neutrino mass ordering, the global fit does not strongly prefer anyone from normal and inverted neutrino mass orderings.
	
\begingroup
\renewcommand{\arraystretch}{1.5} % Adjust row height	
	\begin{table}[t!]
		\centering
		\begin{center} 
			\begin{tabular}{|c|c|c|c|}
				\hline
				Parameter &  Best-fit value~$\pm~1\sigma$ & 3$\sigma$ range &  $1\sigma$ precision (\%) \\
				\hline
				${\sin^2\theta_{12}}$ & $0.308^{+0.012}_{-0.011}$ & 0.275--0.345  & 3.79 \\
				\hline
				\mr{2}{*}{${\sin^2\theta_{13}}$} & $0.02215^{+0.00056}_{-0.00058}$ & 0.02030--0.02388 & {2.69} \\ 
				& ($0.02231^{+0.00056}_{-0.00056}$) & (0.02060--0.02409) & (2.49) \\ 
				\hline
				\mr{2}{*}{${\sin^2\theta_{23}}$} & $0.470^{+0.017}_{-0.013}$ & 0.435--0.585 & 5.32  \\
				& ($0.550^{+0.012}_{-0.015}$) & (0.440--0.584) & (4.36) \\
				\hline
				\mr{2}{*}{$\delta_{\rm CP}~[^{\circ}]$} & $212^{+26}_{-41}$ & 124--364 & 18.87 \\
				& ($274^{+22}_{-25}$) & (201--335) & (8.15)\\
				\hline
				\mr{2}{*}{$\frac{\Delta{m^2_{21}}}{10^{-5} \, \rm{eV}^2}$} & \mr{2}{*}{$7.49^{+0.19}_{-0.19}$} & \mr{2}{*}{6.92--8.05}  & \mr{2}{*}{2.51} \\
				& & & \\
				\hline
				\mr{2}{*}{$\frac{\Delta{m^2_{31}}}{10^{-3} \, \rm{eV}^2}$} & $2.513^{+0.021}_{-0.019}$  & 2.451--2.578 & 0.84 \\
				& (-$2.484^{+0.020}_{-0.020}$)  & (-2.547--(-2.421)) & (0.84)\\
				\hline
			\end{tabular}
			\caption{The best-fit values and the associated uncertainties of the three neutrino oscillation parameters according to the global fit to oscillation data, NuFit 6.0~\cite{Esteban:2024eli}. The values outside parentheses are for normal neutrino mass ordering (NMO); values inside, for inverted mass ordering (IMO).}
			\label{tab:nufit-6}
		\end{center}
	\end{table} 
\endgroup
%%%%%%%%%%%%%%%%%%%%%%%%%%%%%%%%%%%%%%%%%%%%%%%%%%%%%%%%%%%%%%
\section{Unsolved issues in $3\nu$ oscillation framework and future roadmap for their resolution}
%%%%%%%%%%%%%%%%%%%%%%%%%%%%%%%%%%%%%%%%%%%%%%%%%%%%%%%%%%%%%%

\noindent{\textbf {Octant of $\theta_{23}$:}} Though the value of the atmospheric mixing angle $\theta_{23}$ has been measured quite precisely, still the $3\sigma$ uncertainties in this parameter allows both the higher ($\theta_{23}>45^\circ$) and lower octant ($\theta_{23}<45^\circ$) values along with the maximal mixing ($\theta_{23}=45^\circ$) of $\theta_{23}$ (see table~\ref{tab:nufit-6}). We still do not know if this angle is maximal or non-maximal. If non-maximal, to what octant does it belong? However, there are hints towards non-maximal $\theta_{23}$ from the global fit studies~\cite{NuFIT, deSalas:2020pgw, Capozzi:2021fjo}. Determining the octant of $\theta_{23}$ will help verify certain mass models~\cite{Mohapatra:2006gs, Albright:2006cw, Albright:2010ap, King:2013eh}. The experiments probing $\theta_{23}$ with $\nu_\mu\to\nu_\mu$ disappearance data can only measure $\theta_{23}$. However, they can not resolve its octant as the $\nu_\mu\to\nu_\mu$ disappearance probability is proportional to $\sin^22\theta_{23}$ at the leading order (see eqs.~\ref{equ:disapp_vac}, \ref{equ:disapp_mat}). Since the $\nu_\mu\to\nu_e$ appearance probability is proportional to $\sin^2\theta_{23}$ at the leading order (see eqs.~\ref{equ:app_vac}, \ref{equ:app_mat}), this channel is crucial in fixing the correct octant of $\theta_{23}$ if it turns out to be non-maximal in Nature. Long-baseline (LBL) experiments are crucial in resolving the octant degeneracy, since $\nu_\mu\to\nu_e$ is one of the main oscillation channels they probe~\cite{Antusch:2004yx, Hagiwara:2006nn, Chatterjee:2013qus, Agarwalla:2013hma}. Recent studies reveal that the next-generation long-baseline experiments, DUNE and T2HK, hold strong potential towards excluding the maximal $\theta_{23}$~\cite{Agarwalla:2021bzs, Agarwalla:2024kti}.

\noindent{\textbf {Value of the leptonic CP phase, $\delta_{\rm CP}$:}}
The leptonic Dirac CP phase $\delta_{\rm CP}$ is a cyclic variable and can, in principle, take any value in the range [$0^\circ:360^\circ$]. The values $0^\circ$ and $180^\circ$ imply CP conservation; values other than these would imply CP violation in the lepton sector. The values, $90^\circ$ and $270^\circ$ lead to maximal CP-violation. Establishing CP violation in the lepton sector may shed light in explaining the observed matter-antimatter asymmetry of the Universe. As we can see from table~\ref{tab:nufit-6}, $\delta_{\rm CP}$ is the poorest measured parameter with the present global neutrino oscillation data. Though it excludes one of the CP-conserving values of $\delta_{\rm CP}=0^\circ$ at $3\sigma$ C.L., the other CP-conserving value $\delta_{\rm CP}=180^\circ$ is still allowed for NMO. Since $\delta_{\rm CP}$ changes sign while going from neutrino to antineutrino, the LBL experiments can probe it using the combined data from $\nu_\mu\to\nu_e$ and $\bar\nu_\mu\to\bar\nu_e$ oscillation channels. The presently running LBL experiments T2K and NO$\nu$A have predominantly contributed to the measurement of $\delta_{\rm CP}$. However, there exists tension in their dataset. The preferred values of $\delta_{\rm CP}$ by NO$\nu$A and T2K in case of NMO do not agree with each other~\cite{NOvA:2021nfi, T2K:2021xwb, alex_neutrino2020, Patrick_neutrino2020, t2k_neutrino2024, nova_neutrino2024}. This tension is expected to be resolved with the accumulation of more data from these experiments. If the tension persists, it can hint towards the presence of some new physics. For an instance, in \Refes~\cite{Chatterjee:2020kkm, Chatterjee:2020yak, Chatterjee:2024kbn}, authors have explored the possibility of new physics being a solution to the disagreement between T2K and NO$\nu$A. Next-generation LBL experiments will significantly contribute towards establishing the leptonic CP violation~\cite{Abe:2015zbg, Nath:2015kjg, Machado:2015vwa, DUNE:2020jqi, Agarwalla:2022xdo}.

\noindent{\textbf {Determination of the neutrino mass ordering:}}
The solar neutrino experiments have measured both the magnitude and sign of the solar mass-squared difference, $\Delta m^2_{21}$ (positive, $m_2>m_1$). The magnitude of $\Delta m^2_{31}$ is measured quite precisely with the data from reactor, atmospheric, and LBL experiments. However, we still do not know the sign of $\Delta m^2_{31}$. This leaves us with the two possible neutrino mass ordering, normal mass ordering (NMO, $\Delta m^2_{31}>0$) and inverted mass ordering (IMO, $\Delta m^2_{31}<0$). While neutrino oscillation in vacuum is blind to the neutrino mass ordering, the matter effects can distinguish between the two mass orderings. The presently running experiments, Super-K~\cite{Super-Kamiokande:2023ahc}, IceCube DeepCore~\cite{PradoRodriguez:2023sor} and KM3NeT/ORCA~\cite{KM3NeT:2021ozk} with significantly larger matter effects are playing pivotal role in determining the neutrino mass ordering. The NuFIT 6.0 analysis indicates that while individual datasets from the LBL experiments, T2K and NOvA, show a preference for NMO, the global fit does not strongly favor one ordering over the other. Specifically, the combined analysis using the long-baseline, reactor, and IceCube DeepCore atmospheric data results in an almost equally good fit for both the orderings. Only when the atmospheric dataset from Super-K and IceCube DeepCore are included, the global fit shows a preference for NMO, with a $\Delta \chi^2$ of 6.1. The upcoming medium-baseline reactor experiment JUNO with six years of data is expected to determine the neutrino mass ordering with $3\sigma$ C.L. by exploiting the interference effect of the solar and atmospheric mass-square differences~\cite{Heinz:2024dyt}. The mass ordering determination is one of the major goals for the next-generation long-baseline experiments like DUNE and T2HK.

%==========================================================================
\section{Next-generation long-baseline neutrino oscillation experiments}
%==========================================================================

\textbf{(i) DUNE}
\vspace{0.1in}

DUNE (Deep Underground Neutrino Experiment)~\cite{DUNE:2020lwj, DUNE:2020jqi, DUNE:2021cuw, DUNE:2021mtg} is a next-generation LBL neutrino oscillation experiment expected to start taking data around 2030. It will use a liquid-argon time-projection chamber (LArTPC) detector with a net volume of 40~kton. Its neutrino beam will travel a distance of 1285~km from Fermilab to the Homestake Mine. Neutrinos will be produced by a 1.2 MW beam of 120 GeV protons delivering $1.1 \cdot 10^{21}$ protons-on-target (P.O.T.) per year.  It will produce a wide-band, on-axis beam of neutrinos with energies of 0.1--20~GeV and peaking around 2.5~GeV.  While simulating neutrino events in DUNE, we follow the latest configuration details from \Refe~\cite{DUNE:2021cuw}. DUNE will run in both neutrino and antineutrino modes, 5 years in each mode, for a total run-time of 10 years. We use this run time for our simulation work in chapters~\ref{C3} and~\ref{C4} of this thesis. However, in chapter~\ref{C5}, we consider a run time of 7 years, divided equally into 3.5 years each in neutrino and antineutrino modes. We bin events between $E_{\rm rec} = 0.5$ and 8 GeV with a uniform bin width of 0.125~GeV; between 8 and 10~GeV, with a width of 1~GeV; and between 10 and 18~GeV, with a width of 2~GeV. For all values of the CP- violating phase $\delta_{\rm CP}$, DUNE is expected to resolve the neutrino mass ordering at a significance of $5\sigma$  with only two years of its run leveraging the large matter effect that the neutrinos of high-energy and longer travel-distance will face~\cite{DUNE:2020jqi}. It will also measure CP violation at $5\sigma$ with 7 years of data if $\delta_{\rm CP}$ has the maximal value of $-\pi/2$~\cite{DUNE:2020jqi}.\\

\noindent\textbf{(ii) T2HK/T2HKK}
\vspace{0.1in}

T2HK (Tokai-to-Hyper-Kamiokande)~\cite{Abe:2015zbg, Hyper-Kamiokande:2016srs, Hyper-Kamiokande:2018ofw} is another next-generation long-baseline oscillation experiment in Japan. It will use a water Cherenkov detector with a fiducial volume of 187~kt. Neutrinos will be produced at the J-PARC facility~\cite{McDonald:2001mc} by a 1.3-MW beam of 80 GeV protons delivering $2.7 \times 10^{22}$~P.O.T.~per year.  The ensuing neutrino beam will be narrow-band, and will peak around 0.6~GeV, will travel 295~km to the detector at the Tochibara Mines in Japan, placed $2.5^\circ$ off-axis.  The off-axis position helps tune the beam such that the flux peaks at the first oscillation maximum. For our analysis, we follow the configuration details from \Refe~\cite{Hyper-Kamiokande:2016srs}.  T2HK will run 2.5 years in neutrino mode and 7.5 years in antineutrino mode, adhering to the proposed 1:3 ratio between the modes. We bin events uniformly between $E_{\rm rec} = 0.1$ and 3~GeV with a bin width of 0.1~GeV.

T2HKK will have two far detectors --- one is the Japanese detector (JD), same as the above discussed T2HK, and the other is a proposed detector in Korea~\cite{Seo:2019dpr} (Korean detector, KD). We call this combination JD+KD. KD is assumed to be placed 1100 km away from J-PARC at the same off-axis angle as JD, which is $2.5^\circ$ from the beamline. KD will perform at the second oscillation maximum ($\sim 0.7$ GeV) for 1100 km baseline. This setup offers enhanced sensitivity to mass ordering and leptonic CP violation. Working at the second oscillation maxima would be helpful in probing $\delta_{\rm CP}$, and the longer baseline of 1100 km provides a larger matter effect, contributing to mass ordering determination. With both JD and KD detectors covering different oscillation maxima, T2HKK aims to provide a more robust measurements of CP violation and other parameters, thus advancing the precision of three-flavor neutrino oscillation parameters.

Besides the far detectors discussed above, two near detectors have also been proposed. The first one is ND280~\cite{T2K:2019bbb} which is an off-axis detector located 280 m from the neutrino source and $2.5^\circ$ away from the beamline. ND280, situated close to the source, will help measure the unoscillated neutrino flux and the neutrino cross sections. The second near detector is the Intermediate Water Cherenkov Detector (IWCD)~\cite{Drakopoulou:2017qdu,Wilson:2020trq, Andreopoulos:2016rqc} which has a mass of 1 kt and is assumed to be located at a distance of 1 km from J-PARC. IWCD can be moved vertically to have any off-axis angle while collecting the data. However, in the present work, we consider IWCD at the same off-axis angle as JD and KD. Unlike the other near detector, which helps to reduce the flux and crosssection uncertainties at the far detectors, IWCD has its independent physics program such as the sterile neutrino search, either standalone or in combination with ND280~\cite{Hyper-Kamiokande:2018ofw}.

%%%%%%%%%%%%%%%%%%%%%%%%%%%%%%%%%%%%%%%%%%%%%%%%%%%%%%%%%
\section{Summary}
%%%%%%%%%%%%%%%%%%%%%%%%%%%%%%%%%%%%%%%%%%%%%%%%%%%%%%%%%
This chapter discusses in detail the theoretical formalism of neutrino oscillation. We discuss the neutrino oscillation both in two-flavor and three-flavor pictures and derive oscillation probability expressions for the $\nu_\mu\to\nu_e$ appearance and $\nu_\mu\to\nu_\mu$ disappearance channels. Then, we briefly discuss the sources of neutrinos and the completed, ongoing, or planned experiments to detect neutrinos from various sources and their contributions to neutrino oscillation studies. We also discuss various short-baseline (SBL) anomalies like LSND, MiniBooNE, Gallium, and Reactor antineutrino anomalies coming from the accelerator, Gallium, and reactor neutrino experiments, respectively. These anomalous data cannot be accommodated in the standard three-flavor oscillation framework and point towards the existence of a light sterile neutrino where the active-sterile neutrino oscillation with a mass-squared difference of $\sim$ 1~eV$^2$ may resolve these anomalies. In light of these experimental  anomalies, we study, as a part of this thesis work, the active-sterile neutrino oscillation in the context of next-generation long-baseline experiments (see chapter~\ref{C5}). We provide the current status of the oscillation parameters based on the latest global fit analysis, NuFIT 6.0. We then discuss the issues yet to be solved in the three-flavor scenario. 

The precise determination of the oscillation parameters by several world-class experiments makes the neutrino oscillation experiments sensitive to any subleading new physics effects. In the subsequent chapters, we explore three new physics scenarios: long-range neutrino interaction, Lorentz invariance violation, and active-sterile neutrino oscillation in the context of next-generation long-baseline neutrino oscillation experiments.
%%%%%%%
%%%%%%%
%%%%%%%
%%%%%%%
\blankpage
%%%%%%%%%%%%%%%%%%%%%%%%%%%% Chapter-3 %%%%%%%%%%%%%%%%%%%%%%%%%%%%%%
%%%%%%%%%%%%%%%%%%%%%%%%%%%%%%% CHAPTER - 3 %%%%%%%%%%%%%%%%%%%%%%%%%%%%%%%%%%%%
\chapter{A plethora of long-range neutrino interactions probed by DUNE and T2HK}
\label{C3} 
%%%%%%%%%%%%%%%%%%%%%%%%%%%%%%%%%%%%%%%%%%%%%%%%%%%%%%%%%%%%%%%%%%%%%%%%%%%%%%%%
%=================================================
\section{Introduction}
\label{sec:intro}
%=================================================

In the search for physics beyond the Standard Model, neutrino flavor transitions provide versatile and exacting probes.  One large class of proposed models of new neutrino physics posits the existence of new flavor-dependent neutrino interactions.  These are interactions beyond the standard weak ones that, by affecting $\nu_e$, $\nu_\mu$, and $\nu_\tau$ differently, could modify the transitions between them relative to the standard expectation.  Because the new interaction is likely feeble, the modifications are likely difficult to spot.  Yet, if the range of the new interaction is long~\cite{He:1990pn,Foot:1990uf, Foot:1990mn, He:1991qd, Foot:1994vd}, then vast repositories of matter located far from the neutrinos may source a large matter potential that could affect flavor transitions appreciably, even if the new interaction is significantly more feeble than weak interactions.

%=================================================
\begin{figure}
	\centering
	\includegraphics[width=0.85\textwidth]{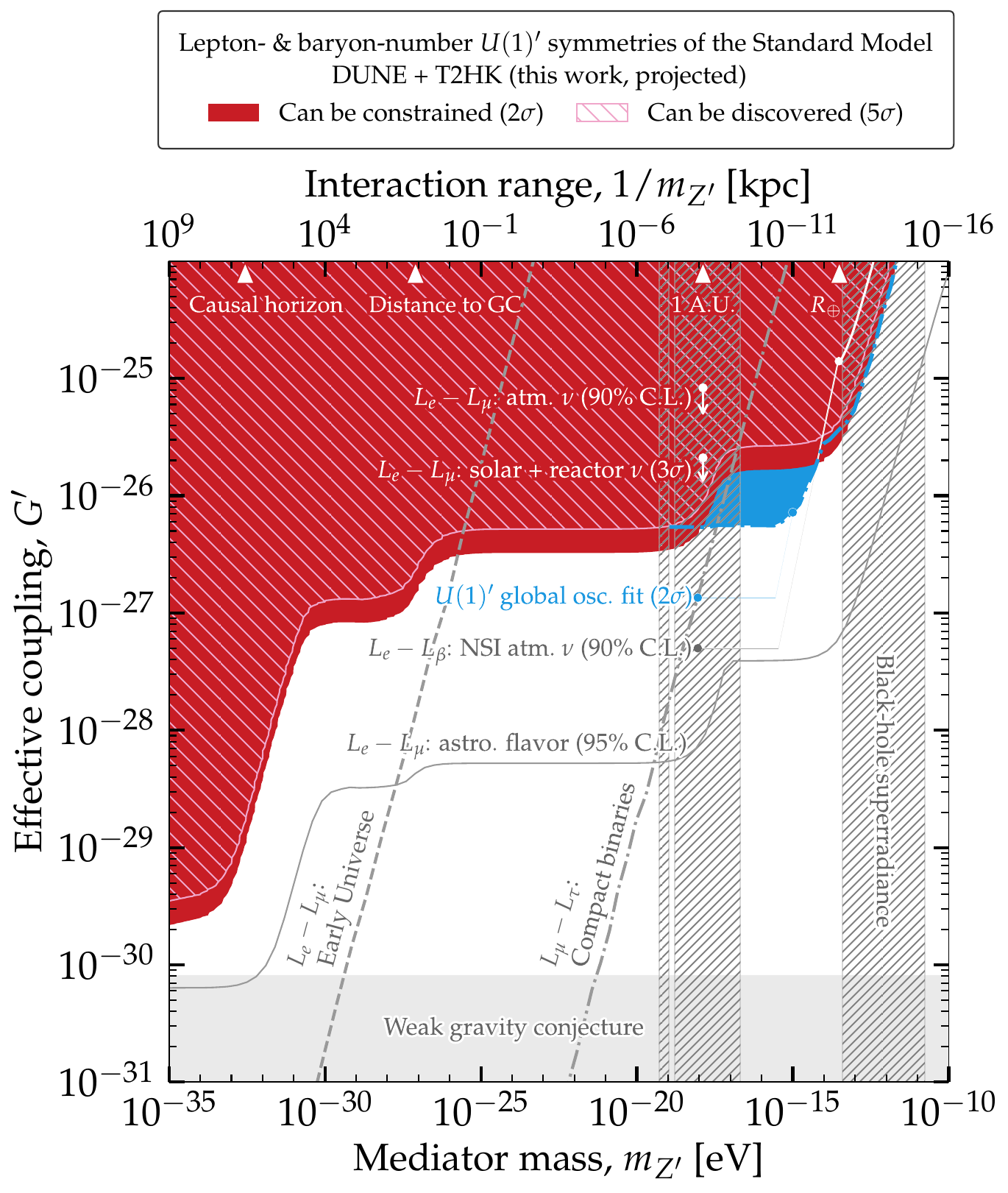}
	\caption{\textit{Overview of projected constraints and discovery prospects of long-range neutrino interactions achieved by combining DUNE and T2HK.}  Results are on the effective coupling of the new gauge boson, $Z^\prime$, that mediates the interaction, across all the candidate $U(1)^\prime$ symmetries that we consider could induce long-range interactions (table~\ref{tab:charges}), and for 10 years of operation of each experiment.  For this figure, we assume that the true neutrino mass ordering is normal. Existing limits are from a recent global oscillation fit~\cite{Coloma:2020gfv}, shown also across all symmetries, and, for specific symmetries, from atmospheric neutrinos~\cite{Joshipura:2003jh}, solar and reactor neutrinos~\cite{Bandyopadhyay:2006uh}, and non-standard interactions~\cite{Super-Kamiokande:2011dam, Ohlsson:2012kf, Gonzalez-Garcia:2013usa}.  The estimated sensitivity from present flavor-composition measurements of high-energy astrophysical neutrinos in IceCube is from~\Refe~\cite{Agarwalla:2023sng}.  Indirect limits~\cite{Wise:2018rnb} are from black-hole superradiance~\cite{Baryakhtar:2017ngi}, the early Universe~\cite{Dror:2020fbh}, compact binaries~\cite{KumarPoddar:2019ceq}, and the weak gravity conjecture~\cite{Arkani-Hamed:2006emk}, assuming a lightest neutrino mass of $0.01$~eV. See sections~\ref{sec:intro}, \ref{sec:constraints_pot}, and  \ref{sec:discovery} for details.  \textit{DUNE and T2HK may constrain long-range interactions more strongly than ever before, or discover them, regardless of which $U(1)^\prime$ symmetry is responsible for inducing them.}}    
	\label{fig:constraints_discovery_dune_t2hk-NMO}
\end{figure}
%=================================================

So far, there is no evidence for such long-range neutrino interactions, but there are stringent constraints on them inferred from observations of atmospheric~\cite{Joshipura:2003jh}, solar~\cite{Grifols:2003gy, Bandyopadhyay:2006uh,Gonzalez-Garcia:2006vic}, accelerator~\cite{Heeck:2010pg, Heeck:2018nzc}, and high-energy astrophysical~\cite{Bustamante:2018mzu, Agarwalla:2023sng} neutrinos, and from their combination~\cite{Davoudiasl:2011sz, Farzan:2016wym, Wise:2018rnb, Dror:2020fbh, Coloma:2020gfv, Alonso-Alvarez:2023tii}. Other constraints do not involve neutrinos, \eg, gravitational fifth-force searches~\cite{Adelberger:2009zz, Salumbides:2013dua}, tests of the equivalence principle~\cite{Schlamminger:2007ht}, black-hole superradiance~\cite{Baryakhtar:2017ngi}, the orbital period of compact binary systems~\cite{KumarPoddar:2019ceq}, and the perihelion precession of planets~\cite{KumarPoddar:2020kdz}; we show some of them in \figu{constraints_discovery_dune_t2hk-NMO}.  Reference~\cite{Singh:2023nek} (also, \Refe~\cite{Wise:2018rnb}) contains a brief review of existing limits, including some shown here in figs.~\ref{fig:constraints_discovery_dune_t2hk-NMO}, \ref{fig:all_symmetry}, and \ref{fig:discovery_all_DUNE+T2HK}. For completeness, in appendix~\ref{sec:other-constraints}, we briefly discuss the other studies constraining the long-range interaction.

Long-baseline neutrino oscillation experiments, where the distance between the source and detector is of hundreds of kilometers or more, are particularly well-suited for searching for new neutrino interactions that may affect flavor transitions.  The high precision of their detectors and their well-characterized neutrino beams facilitate identifying subtle deviations from standard expectations~\cite{DUNE:2020fgq,Arguelles:2022tki,NOvA:2024lti}.  In combination with other experiments, they have placed stringent limits on long-range neutrino interactions~\cite{Honda:2007wv,Chatterjee:2015gta,Khatun:2018lzs, Coloma:2020gfv,Mishra:2024riq}.  

In the coming 10--20 years, new long-baseline experiments, larger, using more advanced detection and reconstruction techniques, and more intense neutrino beams, hold an opportunity for important progress.  We prepare to seize it by forecasting the capability of two of the leading long-range neutrino oscillation experiments, the Deep Underground Neutrino Experiment (DUNE)~\cite{DUNE:2021cuw} and Tokai-to-Hyper-Kamiokande (T2HK)~\cite{Hyper-Kamiokande:2016srs, Hyper-Kamiokande:2018ofw}, presently under construction, to constrain, discover, and characterize new flavor-dependent neutrino interactions, and we interpret it in the context of long-range neutrino interactions.

We introduce new neutrino interactions by gauging the accidental global $U(1)^\prime$ symmetries of the Standard Model that involve combinations of lepton numbers, $L_e$, $L_\mu$, and $L_\tau$, and baryon number, $B$; see \Refes~\cite{He:1990pn, Foot:1990uf, Foot:1990mn, He:1991qd, Foot:1994vd} for early works and \Refe~\cite{Langacker:2008yv} for a review.  Gauging one of the several candidate symmetries (more on this later) introduces a new neutral vector gauge boson, $Z^\prime$, whose mass and coupling strength are a priori unknown, and which induces a new Yukawa potential sourced by electrons, neutrons, or protons, depending on the symmetry.  Also depending on the symmetry, the new interaction affects only $\nu_e$, $\nu_\mu$, or $\nu_\tau$, or a combination of them, and modifies flavor transitions differently.  The lighter the mediator, the longer the range of the interaction.  We focus on masses between $10^{-10}$~eV and $10^{-35}$~eV, corresponding to an interaction range between meters and Gpc.  (The complementary case for heavy mediators, studied in the context of contact neutrino interactions, was first studied in \Refes~\cite{Wolfenstein:1977ue,Valle:1987gv,Guzzo:1991hi}; see also \Refe~\cite{Coloma:2020gfv}.)

There are three core ingredients to our analysis; we sketch them below and expand on them later.  First, as in \Refes~\cite{Bustamante:2018mzu, Singh:2023nek, Agarwalla:2023sng}, we use the long-range matter potential sourced by vast repositories of matter in the local and distant Universe: the Earth, Moon, Sun, Milky Way, and the cosmological matter distribution.  Previous calculations of this potential~\cite{Bustamante:2018mzu, Agarwalla:2023sng}, limited to lepton-number symmetries (more on this momentarily) used only the distributions of electrons and neutrons; our new analysis extends that to include also protons.  Second, as in \Refe~\cite{Singh:2023nek}, we base our analysis on detailed simulations of DUNE and T2HK, which grounds our results in realistic detection capabilities.  Third, motivated by \Refe~\cite{Coloma:2020gfv} --- which, unlike us, used present-day oscillation data --- we explore a plethora of candidate $U(1)^\prime$ symmetries that could introduce long-range neutrino interactions, each affecting neutrino oscillations differently (see table~\ref{tab:charges}).  Doing this extends the first forecasts for DUNE and T2HK reported in \Refe~\cite{Singh:2023nek}, which were limited to three candidate symmetries, $L_e-L_\mu$, $L_e-L_\tau$, and $L_\mu-L_\tau$, and allows us to establish whether the sensitivity claimed therein was limited to those three cases, or applies broadly to other symmetries.  {\textit{While the above ingredients have been accounted for before separately, or in other contexts, the novelty and strength of our analysis lies in combining them.}

Figure~\ref{fig:constraints_discovery_dune_t2hk-NMO} conveys the essence of our findings; we elaborate on them later.  It summarizes our forecasts on constraining and discovering long-range interactions across the fourteen candidate symmetries that we consider (table~\ref{tab:charges}); the results for individual symmetries are comparable among them and we show them later (figs.~\ref{fig:all_symmetry} and \ref{fig:discovery_all_DUNE+T2HK}).  Figure~\ref{fig:constraints_discovery_dune_t2hk-NMO} shows that DUNE and T2HK may place the strongest constraints on long-range interactions, especially for mediators lighter than $10^{-18}$~eV, and discover them, even if they are subdominant.  This reaffirms the outlook first reported in \Refe~\cite{Singh:2023nek}.  (The sensitivity from flavor measurements of high-energy astrophysical neutrinos~\cite{Agarwalla:2023sng} (see also \Refes~\cite{Bustamante:2018mzu, Ackermann:2022rqc}) could be comparable but, for now, it is subject to large astrophysical uncertainties not captured in \figu{constraints_discovery_dune_t2hk-NMO}.)  

The novel perspective revealed by our results is that {\textit{DUNE and T2HK may constrain or discover new neutrino interactions with matter --- including long-range ones --- regardless of which symmetry, out of the candidates we consider, induces them.}  When searching for new interactions, the sensitivity of DUNE and T2HK is not limited to spotting a handful of specific modifications to the flavor transitions, but extends to a broad range of them.  Pivoting on this, we show later that, in some cases, \textit{DUNE and T2HK may identify or narrow down which candidate symmetry is responsible for inducing the new interaction; see \figu{confusion-matrix}.} 

This chapter is organized as follows.  Section~\ref{sec:formalism} introduces new neutrino-matter interactions due to various $U(1)^\prime$ symmetries and the long-range matter potential they induce.  Section~\ref{sec:probability_event_rates} illustrates their effect on the neutrino oscillation probability and the event spectra in DUNE and T2HK.  Section~\ref{sec:results} contains our main results: the constraints on the new matter potential, its discovery prospects, their interpretation as being due to long-range interactions, and the separation between different candidate symmetries.  Section~\ref{sec:conclusion} summarizes and concludes.  Appendices \ref{app:U1-charges}--\ref{app:discovery} contain additional details and results.

%=================================================
\section{Long-range neutrino interactions}
\label{sec:formalism}
%=================================================

%=================================================
\subsection{New neutrino-matter interactions from $U(1)^\prime$ symmetries}
\label{sec:formalism_lagrangians}
%=================================================

%=================================================
\begin{figure}
	\centering
	\includegraphics[width=\textwidth]{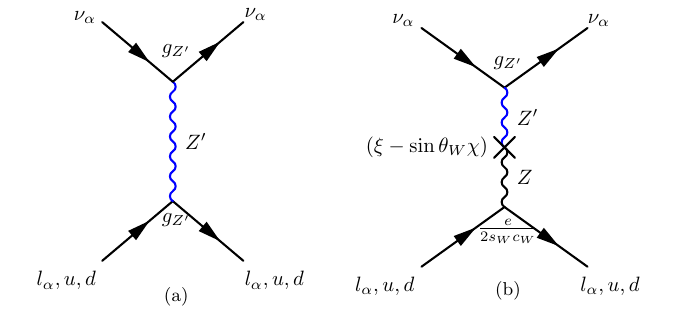}
	\caption{\textit{Feynman diagrams of the new neutrino-matter interactions that we consider.}  The interaction Lagrangian is \equ{full_lagrangian}. Diagram (a) represents the new interaction mediated by a new $Z^\prime$ neutral vector boson, with coupling constant $g_{Z^\prime}$.   Diagram (b) represents the mixing between $Z$ and $Z^\prime$.  In our analysis, we account for the contribution of diagram (a) for all $U(1)^\prime$ symmetries except for $L_\mu-L_\tau$, for which diagram (a) is replaced by diagram (b).  See section~\ref{sec:formalism_lagrangians} for details.}
	\label{fig:feynman_diagram}
\end{figure}
%=================================================

The Standard Model (SM) contains accidental global $U(1)$ symmetries that involve $L_e$, $L_\mu$, $L_\tau$, and $B$. Gauging them individually introduces anomalies. However, certain combinations of them can be gauged anomaly-free, either within the SM particle content or by adding right-handed neutrinos~\cite{Araki:2012ip, Allanach:2018vjg}. Since neutrino oscillations are not affected by flavor-universal gauge symmetries --- say, $B-L$ --- we focus on $U(1)^\prime$ symmetries that are flavor-dependent, namely (table~\ref{tab:charges}), $B-3L_e$, $B-3L_\mu$, $B-3L_\tau$, $B-L_e-2L_\tau$, $B_y+L_\mu+L_\tau$, $B-\frac{3}{2}(L_\mu+L_\tau)$, $L-3L_e$, $L-3L_\mu$, $L-3L_\tau$, $L_e-\frac{1}{2}(L_\mu+L_\tau)$, $L_e+2L_\mu+2L_\tau$, $L_e-L_\mu$, $L_e-L_\tau$, and $L_\mu-L_\tau$.  This is the same list of fourteen candidate symmetries explored in \Refes~\cite{Davoudiasl:2011sz, Araki:2012ip, Coloma:2020gfv, delaVega:2021wpx}.  (Here, $B_y \equiv B_1-yB_2-(3-y)B_3$~\cite{Farzan:2015doa, Coloma:2020gfv}, where $B_1$, $B_2$, and $B_3$ are the baryon numbers of quarks of the first, second, and third generation, respectively, and $y$ is an arbitrary constant that we set to $y=0$ because we consider neutrino interactions with first-generation quarks only.) Their rich phenomenology has been discussed in, \eg,  \Refes~\cite{Ma:1997nq,Lee:2010hf,Davoudiasl:2011sz,Lee:2011uh,Araki:2012ip,Farzan:2015doa,Kownacki:2016pmx,Heeck:2018nzc,Coloma:2020gfv,delaVega:2021wpx,Barman:2021yaz,AtzoriCorona:2022moj,DeRomeri:2024dbv}. We expand on them later. Each one, after being promoted to a local gauge symmetry, generates new flavor-dependent neutrino-matter interactions mediated by a new vector boson, $Z^\prime$.  (In principle, the stringent constraints on new interactions of charged leptons could render the possibility of new neutrino interactions unfeasible, but this limitation can be circumvented by suitable model building; see, \eg, \Refes~\cite{Farzan:2016wym, Joshipura:2019qxz, Almumin:2022rml}.)

Figure~\ref{fig:feynman_diagram} shows the neutrino-matter interactions that we consider between the three active neutrinos, $\nu_e$, $\nu_\mu$, and $\nu_\tau$, and electrons ($e$), and up ($u$) and down ($d$) quarks inside protons and neutrons. Apart from the standard $W^{\pm}$- and $Z$-boson mediated interactions, which we do not show explicitly, for a given $U(1)^\prime$ symmetry, the effective interaction Lagrangian is
\begin{equation}
	\label{equ:full_lagrangian}
	\mathcal{L}
	= \mathcal{L}_{Z^\prime}
	+
	\mathcal{L}_{\rm mix} \;.
\end{equation}

The first term on the right-hand side of \equ{full_lagrangian} describes the new neutral-current flavor-dependent neutrino-matter interactions~\cite{He:1990pn, He:1991qd, Heeck:2010pg,Coloma:2022dng}, mediated by $Z^\prime$, whose mass, $m_{Z^\prime}$, and coupling strength, $g_{Z^\prime}$, are a priori unknown, \ie,
\begin{multline}
	\label{equ:Lagrangian}
	\mathcal{L}_{Z'}
	= -g_{Z^\prime}\big( a_u\, \bar{u}\gamma^\alpha u
	+ a_d\, \bar{d}\gamma^\alpha d
	+ a_e\, \bar{e}\gamma^\alpha e
	\\
	+ b_e \,\bar{\nu}_e\gamma^\alpha P_L \nu_e
	+ b_\mu\,\bar{\nu}_\mu\gamma^\alpha P_L \nu_\mu
	+ b_\tau\, \bar{\nu}_\tau \gamma^\alpha P_L \nu_\tau \big)  Z'_\alpha\,,
\end{multline}
where $a_e$, $a_u$, and $a_d$ are the $U(1)^\prime$ charges of the electron, up quark, and down quark, and $b_e$, $b_\mu$, and $b_\tau$ are the charges of $\nu_e$, $\nu_\mu$, and $\nu_\tau$. Appendix~\ref{app:U1-charges} contains the values of the $U(1)^\prime$ charges of the symmetries that we consider.

The above neutrino-matter interactions can be generated upon extending the SM gauge group $SU(3)_{C}\times SU(2)_{L}\times U(1)_{Y}$ by the maximal abelian gauge group $U(1)^\prime=U(1)_{B-L}\times U(1)_{L_{\mu}-L_{\tau}}\times U(1)_{L_\mu-L_e}$ by adding three right-handed neutrinos to the SM particle content and imposing the condition that all new couplings are vector-like and the $U(1)^\prime$ charges of the quarks are flavor-universal. Then, any subset of the $U(1)^\prime$ hypercharge $c_\textsc{bl} (B-L) + c_{\mu\tau} (L_\mu-L_\tau)+ c_{\mu e} (L_\mu-L_e)$ can be gauged in an anomaly-free way~\cite{Araki:2012ip, Heeck:2018nzc, Allanach:2018vjg, Coloma:2020gfv}, with $a_u = a_d = c_\textsc{bl} / 3$, $a_e = b_e = -(c_\textsc{bl} +
c_{\mu e})$, $b_\mu = -c_\textsc{bl} + c_{\mu e} + c_{\mu\tau}$, and
$b_\tau = -(c_\textsc{bl} + c_{\mu\tau})$.

Table~\ref{tab:charges} lists all the $U(1)^\prime$ symmetries that we consider.  We group them according to the texture of the new matter potential that they introduce, $\mathbf{V}_{\rm LRI}$ (section~\ref{sec:yukawa_interaction}); later, we interpret this potential as being due to long-range interactions (LRI).  Different textures affect neutrino oscillations differently; we elaborate on this in section~\ref{sec:prob_var_pot}.  Reference~\cite{Singh:2023nek} explored long-range interactions due to $L_e-L_\mu$, $L_e-L_\tau$, and $L_\mu-L_\tau$ under a similar analysis that we perform here, but when computing their effect on neutrino oscillations (section~\ref{sec:prob_var_pot}) used values of the neutrino mixing parameters from \Refe~\cite{Capozzi:2021fjo}.  We revisit them here using mixing parameters from the NuFIT~5.1 global fit to oscillation data~\cite{Esteban:2020cvm, NuFIT} instead.

%=================================================
\begin{table}[t!]
	\centering
	\footnotesize
	\begin{adjustbox}{width=\linewidth}
		\renewcommand{\arraystretch}{1.4}
		\begin{tabular}{|c|c|c|c|c|}
			\hline
			\mr{3}{*}{\shortstack{Texture 
					\\[0.4em] of $\mathbf{V}_{\rm LRI}$}} &
			\mr{3}{*}{$U(1)^\prime$ symmetry} &  
			\mc{3}{c|}{New matter potential, $\mathbf{V}_{\rm LRI} = \textrm{diag}(V_{{\rm LRI},e}, V_{\rm LRI,\mu}, V_{\rm LRI,\tau})$}
			\\ 
			&
			& 
			\mr{2}{*}{\shortstack{Texture to place limits,
					\\[0.4em] $\mathbf{V}_{\rm LRI} = V_{\rm LRI} \cdot \textrm{diag}(\ldots)$}}
			&
			\mr{2}{*}{\shortstack{General form
					\\[0.4em] of $V_{\rm LRI}$, \equ{pot_total_general}}} 
			&
			\mr{2}{*}{\shortstack{Form of $V_{\rm LRI}$ to convert limits on it 
					\\[0.4em] into limits on $G^\prime$ {\it vs.}~$m_{Z^{\prime}}$, \equ{pot_total_simplified}}}
			\\
			& 
			& 
			&
			&
			\\
			\hline
			%%%%%%
			\mr{6}{*}{$\left(\begin{array}{ccc} \bullet &  &  \\  & 0 & \\ &  & 0 \end{array}\right)$} &
			%%%%%%
			$B-3L_{e}$ &
			$\textrm{diag}(1, 0, 0)$ &
			$9V_{e}-3(V_{p}+V_{n})$ &
			$3(V^{\oplus}_{e}+V^{\leftmoon}_{e}+V^{\rm MW}_{e})+\frac{21}{4}V^{\astrosun}_{e}+\frac{39}{7}V^{\rm cos}_{e}$ 
			\\
			\cline{2-5} 
			%%%%%%
			&
			$L-3L_e$ &
			$\textrm{diag}(1, 0, 0)$ &
			$6 V_e$ &
			$6(V^{\oplus}_{e}+V^{\leftmoon}_{e}+V^{\rm MW}_{e}+V^{\astrosun}_{e}+V^{\rm cos}_{e})$ 
			\\
			\cline{2-5}
			&
			$B-\frac{3}{2}(L_\mu+L_\tau)$ &
			$\textrm{diag}(1, 0, 0)$ &
			$\frac{3}{2}(V_p+V_n)$ &
			$3(V^{\oplus}_{e}+V^{\leftmoon}_{e}+V^{\rm MW}_{e})+\frac{15}{8}V^{\astrosun}_{e}+\frac{12}{7}V^{\rm cos}_{e}$
			\\
			\cline{2-5}
			&
			$L_e-\frac{1}{2}(L_\mu+L_\tau)$ &
			$\textrm{diag}(1, 0, 0)$ &
			$\frac{3}{2}V_e$ &
			$\frac{3}{2}(V^{\oplus}_{e}+V^{\leftmoon}_{e}+V^{\rm MW}_{e}+V^{\astrosun}_{e}+V^{\rm cos}_{e})$ 
			\\
			\cdashline{2-5}
			&
			$L_e+2L_\mu+2L_\tau$ &
			$\textrm{diag}(-1, 0, 0)$ &
			$V_e$ &
			$V^{\oplus}_{e}+V^{\leftmoon}_{e}+V^{\rm MW}_{e}+V^{\astrosun}_{e}+V^{\rm cos}_{e}$ 
			\\
			\cline{2-5}
			&
			$B_{y}+L_\mu+L_\tau$ &
			$\textrm{diag}(-1, 0, 0)$ &
			$V_p+V_n$ &
			$2(V^{\oplus}_{e}+V^{\leftmoon}_{e}+V^{\rm MW}_{e})+\frac{5}{4}V^{\astrosun}_{e}+\frac{8}{7}V^{\rm cos}_{e}$
			\\
			\hline
			%%%%%%
			\mr{4}{*}{$\left(\begin{array}{ccc} 0 &  &  \\  & \bullet & \\ &  & 0 \end{array}\right)$} &
			%%%%%%
			\mr{2}{*}{$B-3L_\mu$} &
			\mr{2}{*}{$\textrm{diag}(0, -1, 0)$} &
			\mr{2}{*}{$3(V_p+V_n)$} &
			\mr{2}{*}{$6(V^{\oplus}_{e}+V^{\leftmoon}_{e}+V^{\rm MW}_{e})+\frac{15}{4}V^{\astrosun}_{e}+\frac{24}{7}V^{\rm cos}_{e}$} 
			\\
			& & & &
			\\
			\cline{2-5}
			%%%%%%
			&
			\mr{2}{*}{$L-3L_\mu$} &
			\mr{2}{*}{$\textrm{diag}(0, -1, 0)$} &
			\mr{2}{*}{$3V_e$} &
			\mr{2}{*}{$3 (V^{\oplus}_{e}+V^{\leftmoon}_{e}+V^{\rm MW}_{e}+V^{\astrosun}_{e}+V^{\rm cos}_{e})$}
			\\ 
			& & & &
			\\
			%%%%%%
			\hline
			\mr{4}{*}{$\left(\begin{array}{ccc} 0 &  &  \\  & 0 & \\ &  & \bullet  \end{array}\right)$} &
			%%%%%%
			\mr{2}{*}{$B-3L_\tau$} &
			\mr{2}{*}{$\textrm{diag}(0, 0, -1)$} &
			\mr{2}{*}{$3 (V_p + V_n)$} &
			\mr{2}{*}{$6(V^{\oplus}_{e}+V^{\leftmoon}_{e}+V^{\rm MW}_{e})+\frac{15}{4}V^{\astrosun}_{e}+\frac{24}{7}V^{\rm cos}_{e}$}
			\\ 
			& & & &
			\\
			\cline{2-5}
			&
			\mr{2}{*}{$L-3L_\tau$} &
			\mr{2}{*}{$\textrm{diag}(0, 0, -1)$} &
			\mr{2}{*}{$3 V_e $} &
			\mr{2}{*}{$3 (V^{\oplus}_{e}+V^{\leftmoon}_{e}+V^{\rm MW}_{e}+V^{\astrosun}_{e}+V^{\rm cos}_{e})$}
			\\ 
			& & & &
			\\
			%%%%%%
			\hline
			\mr{3}{*}{$\left(\begin{array}{ccc} \bullet &  &  \\  & \bullet & \\ &  & 0 \end{array}\right)$} &
			%%%%%%
			\mr{3}{*}{$L_e-L_\mu$} &
			\mr{3}{*}{$\textrm{diag}(1, -1, 0)$} &
			\mr{3}{*}{$V_e$} &
			\mr{3}{*}{$V^{\oplus}_{e}+V^{\leftmoon}_{e}+V^{\rm MW}_{e}+V^{\astrosun}_{e}+V^{\rm cos}_{e}$}
			\\
			& & & &
			\\
			& & & &
			\\
			%%%%%%
			\hline
			\mr{3}{*}{$\left(\begin{array}{ccc} \bullet &  &  \\  & 0 & \\ &  & \bullet \end{array}\right)$} &
			%%%%%%
			\mr{3}{*}{$L_e-L_\tau$} &
			\mr{3}{*}{$\textrm{diag}(1, 0, -1)$} &
			\mr{3}{*}{$V_e$} &
			\mr{3}{*}{$V^{\oplus}_{e}+V^{\leftmoon}_{e}+V^{\rm MW}_{e}+V^{\astrosun}_{e}+V^{\rm cos}_{e}$}
			\\
			& & & &
			\\
			& & & &
			\\
			%%%%%%
			\hline
			\mr{4}{*}{$\left(\begin{array}{ccc} 0 & &  \\  & \bullet & \\ &  & \bullet \end{array}\right)$} &
			%%%%%%
			\mr{2}{*}{$L_\mu-L_\tau$} &
			\mr{2}{*}{$\textrm{diag}(0, 1, -1)$} &
			\mr{2}{*}{$-V_e+V_p+V_n$} &
			\mr{2}{*}{$V^{\oplus}_{e}+V^{\leftmoon}_{e}+V^{\rm MW}_{e}+\frac{1}{4}V^{\astrosun}_{e}+\frac{1}{7}V^{\rm cos}_{e}$}
			\\
			& & & &
			\\
			\cline{2-5}
			&
			%%%%%%
			\mr{2}{*}{$B-L_e-2L_\tau$} &
			\mr{2}{*}{$\textrm{diag}(0, 1, -1)$} &
			\mr{2}{*}{$-V_e+V_p+V_n$} &
			\mr{2}{*}{$V^{\oplus}_{e}+V^{\leftmoon}_{e}+V^{\rm MW}_{e}+\frac{1}{4}V^{\astrosun}_{e}+\frac{1}{7}V^{\rm cos}_{e}$}
			\\
			& & & &
			\\
			\hline
		\end{tabular}
	\end{adjustbox}
	\caption{
		\textit{$U(1)^\prime$ gauge symmetries considered in our analysis and the new matter potential they induce.}  We group symmetries according to the texture of the new matter potential, $\mathbf{V}_{\rm LRI}$, that they induce; equal or similar textures lead to equal or similar sensitivity (sections~\ref{sec:constraints_pot}--\ref{sec:confusion-theory}).  Elements of  $\mathbf{V}_{\rm LRI}$ marked with $\bullet$ represent nonzero entries.  When computing constraints and discovery prospects on the new matter potential, and the distinguishability between competing candidate symmetries, we use for it the form $\mathbf{V}_{\rm LRI} = V_{\rm LRI} \cdot \textrm{diag}(\dots)$, where the texture of the diagonal matrix is indicated in the table for each symmetry, after subtracting terms proportional to the identity.  The general expression for $V_{\rm LRI}$, sourced by electrons, protons, and neutrons, regardless of their source, is from \equ{pot_total_general}.  We use knowledge of the relative abundance of electrons, protons, and neutrons in the Earth ($\oplus$), Moon ($\leftmoon$), Sun ($\odot$), Milky Way (MW), and in the cosmological matter distribution (cos) to convert limits obtained on $V_{\rm LRI}$ into limits on the mass and coupling strength of the $Z^\prime$ mediator, $m_{Z^\prime}$ and $G^\prime$ (sections~\ref{sec:constraints_pot} and \ref{sec:discovery}).  See sections~\ref{sec:intro} and \ref{sec:hamiltonians} for details.
		\label{tab:charges}
	}
\end{table}
%=================================================

The second term on the right-hand side of \equ{full_lagrangian} describes the mixing between the neutral gauge bosons, $Z$ and $Z^\prime$~\cite{Babu:1997st,Heeck:2010pg,Joshipura:2019qxz}, \ie, $\mathcal{L}_{ZZ^\prime} \supset(\xi-\sin\theta_W\chi) Z'_{\mu}Z^{\mu}$, where $\chi$ is the kinetic mixing angle between the two gauge bosons and $\xi$ is the rotation angle between physical states and gauge eigenstates. This induces a four-fermion interaction of neutrinos with matter given by
\begin{equation}
	\label{equ:Vmutau}
	\mathcal{L}_{\rm mix}
	=
	-g_{Z^\prime}\frac{e}{\sin\theta_W \cos\theta_W}(\xi-\sin\theta_W\chi)
	J'_\sigma J_3^\sigma \;,
\end{equation}
where $J^\prime_\sigma = \bar{\nu}_\mu \gamma_\sigma P_L\nu_\mu-\bar{\nu}_\tau \gamma_\rho P_L\nu_\tau$ and $J_3^\rho = -\frac{1}{2}\bar{e}\gamma^\sigma P_L e+\frac{1}{2}\bar{u}\gamma^\rho P_L u-\frac{1}{2}\bar{d}\gamma^\rho P_L d$, $e$ is the unit electric charge, $\theta_W$ is the Weinberg angle, and $P_L$ is the left-handed projection operator. In this case, the contributions from electrons and protons cancel each other out, leaving only neutrons to source the new matter potential.
Because the value of the mixing factor $(\xi-\sin\theta_W\chi)$ is not known (though there are upper limits on it~\cite{Schlamminger:2007ht, Adelberger:2009zz, Heeck:2010pg}), we place bounds instead on the effective coupling $g_{Z^\prime} (\xi-\sin\theta_W\chi)$. In our analysis, we consider the contribution of the mixing potential only for the symmetry $L_\mu-L_\tau$, since in this case there are no muons and taus to source the matter potential via $\mathcal{L}_{Z^\prime}$ that would otherwise be dominant due to the abundance of baryons and electrons. 

%=================================================
\subsection{Long-range interaction potential}
\label{sec:yukawa_interaction}
%=================================================

When the new neutrino interactions stem from purely leptonic symmetries, the matter potential is sourced only by electrons, given the dearth of naturally occurring muons and taus. (In the case of $L_\mu-L_\tau$, neutrons contribute through $Z$--$Z^\prime$ mixing; see above.) When they stem from symmetries that blend baryon and lepton numbers, the potential is sourced by electrons, neutrons, and protons, depending on the specific symmetry.

For a given $U(1)^\prime$ symmetry out of our candidates (table~\ref{tab:charges}), the Yukawa potential, mediated by $Z^\prime$, that is experienced by a neutrino situated a distance $d$ away from an electron ($f = e$), a proton ($f = p$), or a neutron ($f = n$) is
\begin{equation}
	V_{Z^\prime, f}
	=
	G^{\prime 2}
	\frac{1}{4\pi d}
	e^{-m_{Z^\prime}\,d} \;,
	\label{equ:potential_zprime}
\end{equation}
where the interaction range is $1/m_{Z^\prime}$; beyond this distance, the potential is suppressed.  Under the $L_\mu-L_\tau$ symmetry, we ignore the contribution of \equ{potential_zprime}; instead, we consider that a neutrino experiences only a potential due to the mixing between $Z$ and $Z^\prime$ (section~\ref{sec:formalism_lagrangians}), sourced by a neutron, \ie,
\begin{equation}
	V_{ZZ^\prime, n}
	=
	G^{\prime 2}
	\frac{e}{\sin\theta_W\cos\theta_W}
	\frac{1}{4 \pi d}
	e^{-m_{Z^\prime}d} \;.
	\label{equ:potential_zzprime}
\end{equation}
In eqs.~(\ref{equ:potential_zprime}) and (\ref{equ:potential_zzprime}), the effective coupling strength is
\begin{equation}
	G^\prime
	=
	\left\{
	\begin{array}{lll}
		g_{Z^\prime} & , & ~{\rm for}~\nu~{\rm interaction~via}~ Z^\prime \\
		\sqrt{g_{Z^\prime} (\xi-\sin \theta_W \chi)} & , & ~{\rm for}~\nu~{\rm interaction~via}~ Z-Z^\prime ~{\rm mixing}\\ 
	\end{array}
	\right.\;.
	\label{equ:G_prime}
\end{equation}
Combining eqs.~(\ref{equ:potential_zprime})--(\ref{equ:G_prime}) yields the potential
\begin{equation}
	V_f
	=
	\left\{
	\begin{array}{ll}
		V_{Z^\prime, f} & ,~
		\textrm{for~all~symmetries~but~}L_\mu-L_\tau
		\\
		V_{ZZ^\prime, n} & ,~
		\textrm{for~}L_\mu-L_\tau~\textrm{~and~}f = n \\
		0 & ,~
		\textrm{otherwise}  
	\end{array}
	\right. \;.
\end{equation}

Following \Refe~\cite{Bustamante:2018mzu}, we focus on long-range interactions, with ultra-light mediators with masses $m_{Z^\prime} = 10^{-35}$--$10^{-10}$~eV that result in interaction ranges from a few hundred meters to Gpc; see also \Refe~\cite{Wise:2018rnb}. Based on the methods introduced in \Refe~\cite{Bustamante:2018mzu} (see also \Refe~\cite{Wise:2018rnb}) and developed in \Refes~\cite{Agarwalla:2023sng, Singh:2023nek}, we estimate the total potential sourced by the electrons, protons, and neutrons in nearby and distant celestial objects --- the Earth ($\oplus$), Moon ($\leftmoon$), Sun ($\astrosun$), and the Milky Way (MW) --- and by the cosmological distribution of matter (cos) in the local Universe., \ie,
\begin{equation}
	\label{equ:pot_total}
	V_f
	(m_{Z^\prime}, G^\prime)
	=
	\left.
	\left(
	V_f^\oplus
	+
	V_f^{\leftmoon}
	+
	V_f^{\astrosun}
	+
	V_f^{\rm MW}
	+ 
	V_f^{\rm cos}
	\right)
	\right\rvert_{m_{Z^\prime}, G^\prime}
	\;.
\end{equation}
Hence, the potential experienced by $\nu_\alpha$ ($\alpha = e, \mu, \tau$) is
\begin{equation}
	\label{equ:pot_total_general}
	V_{{\rm LRI}, \alpha}
	(m_{Z^\prime}, G^\prime)
	=
	b_\alpha
	\sum_{f = e, p, n}
	\kappa_f
	V_f
	(m_{Z^\prime}, G^\prime)
	\;,
\end{equation}
where $b_\alpha$ is the $U(1)^\prime$ charge of the neutrino (table~\ref{tab:charges2}).  For all symmetries but $L_\mu - L_\tau$, $\kappa_f \equiv a_f$ is the $U(1)^\prime$ charge of an electron, $a_e$, a proton, $a_p = 2a_u + a_d$, or a neutron, $a_n = 2a_d + a_u$ (table~\ref{tab:charges2}).  For $L_\mu - L_\tau$, $\kappa_f = y_f$ is instead their weak hypercharge, and only neutrons contribute, with $y_n = 2y_d + y_u$.
The value of $m_{Z^\prime}$ determines the relative sizes of the contributions of the different celestial objects to the total potential.  We defer to \Refes~\cite{Bustamante:2018mzu, Singh:2023nek} for details on the calculation of these  contributions; below, we sketch it.

We make the assumption that the matter responsible for generating this potential is electrically neutral, \ie, it contains equal abundance of electrons and protons ($N_e = N_p$), and isoscalar, \ie, it contains equal abundance of protons and neutrons ($N_p = N_n$), except for the Sun~\cite{Heeck:2010pg} and the cosmological matter distribution~\cite{Hogg:1999ad, Steigman:2007xt, Planck:2015fie}, as follows.  We treat the Sun ($N_{e,\astrosun} = N_{p,\astrosun} \sim 10^{57}$, $N_{n,\astrosun} = N_{e,\astrosun}/4$) and the Moon ($N_{e,\leftmoon} = N_{p,\leftmoon} = N_{n,\leftmoon} \sim 5 \cdot 10^{49}$) as point sources of electrons, protons, and neutrons, and the Earth ($N_{e, \oplus} = N_{p, \oplus} = N_{n, \oplus} \sim 4 \cdot 10^{51}$), the Milky Way ($N_{e, {\rm MW}} = N_{p, {\rm MW}} \approx N_{n, {\rm MW}} \sim 10^{67}$), and the cosmological matter ($N_{e,\mathrm{cos}} = N_{p,\mathrm{cos}} \sim 10^{79}$, $N_{n,\mathrm{cos}} \sim 10^{78}$) as continuous distributions. 

For the contribution of matter inside the Earth, we adopt the approximation of computing the average potential that acts on the neutrinos at their point of detection, as in \Refes~\cite{Bustamante:2018mzu, Agarwalla:2023sng, Singh:2023nek}.  We do not calculate the changing potential as the neutrinos traverse inside the Earth; see \Refes~\cite{ Smirnov:2019cae, Coloma:2020gfv} for such detailed treatment. Our approximation holds well for mediator mass below $10^{-14}$~eV, for which the interaction range is longer than the radius of the Earth (\figu{constraints_discovery_dune_t2hk-NMO}), so that all of the electrons, protons, and neutrons inside it contribute to the potential regardless of their position relative to the neutrino trajectory.

The above approximations allow us to simplify the calculation of the total potential, \equ{pot_total_general}.  First, for each celestial body, we compute the potential sourced by the electrons in it.  Then, for a choice of  symmetry, we compute the potential sourced by protons and neutrons by rescaling the electron potential by their abundance relative to electrons, \ie,
\begin{eqnarray}
	\label{equ:pot_total_simplified}
	V_{{\rm LRI}, \alpha}
	&=&
	b_\alpha
	\left[
	\left(
	\kappa_e
	+
	\kappa_p
	\frac{N_{p, \oplus}}{N_{e, \oplus}}
	+
	\kappa_n
	\frac{N_{n, \oplus}}{N_{e, \oplus}}
	\right)
	V_e^\oplus
	+
	(\oplus \to \leftmoon)
	+
	(\oplus \to \astrosun)
	+
	(\oplus \to {\rm MW})
	\right. 
	\nonumber
	\\
	&&
	\left.
	\qquad
	+~
	(\oplus \to {\rm cos})
	\right] \;.
\end{eqnarray}
By following this procedure, we need only compute explicitly the potential due to electrons --- which may be computationally taxing~\cite{Bustamante:2018mzu, Agarwalla:2023sng, Singh:2023nek} --- rather than the potential due to electrons, protons, and neutrons separately.  Table~\ref{tab:charges} shows \equ{pot_total_simplified} evaluated for each of our candidate symmetries.  Later, in section~\ref{sec:constraints_pot}, we use these expressions to convert the limits we place on the new matter potential into limits on $G^\prime$ as a function of $m_{Z^\prime}$.

%=================================================
\section{Neutrino oscillation probabilities and event rates}
\label{sec:probability_event_rates}
%=================================================

%=================================================
\subsection{Neutrino interaction Hamiltonian}
\label{sec:hamiltonians}
%=================================================

The Hamiltonian that describes neutrinos traveling through matter is, in the flavor basis,
\begin{equation}
	\label{equ:hamiltonian_tot}
	\mathbf{H}
	=
	\mathbf{H}_{\rm vac}
	+
	\mathbf{V}_{\rm mat}
	+
	\mathbf{V}_{\rm LRI} \;.
\end{equation}
The first term on the right-hand side is responsible for the oscillation of neutrinos in vacuum.  For neutrinos with energy $E$, it is
\begin{equation}
	\label{equ:hamiltonian_vac}
	\mathbf{H}_{\rm vac}
	=
	\frac{1}{2 E}
	\mathbf{U}~
	{\rm diag}(0, \Delta m^2_{21}, \Delta m^2_{31})
	~\mathbf{U}^{\dagger} \;,
\end{equation}
where $\mathbf{U}$ is the Pontecorvo-Maki-Nakagawa-Sakata (PMNS) matrix, parameterized in terms of three mixing angles, $\theta_{23}$, $\theta_{13}$, and $\theta_{12}$, and one CP-violation phase, $\delta_{\rm CP}$, $\Delta m^2_{31} \equiv m^2_{3}-m^2_{1}$, and $\Delta m^2_{21} \equiv m^2_{2}-m^2_{1}$, with $m_i$ ($i = 1, 2, 3$) the mass of the neutrino mass eigenstate $\nu_i$.  

Table~\ref{tab:params_value1} shows the values of the oscillation parameters that we use in our analysis.  Later (section~\ref{sec:expt-details}), when producing mock event samples for DUNE and T2HK, we adopt as true values of the oscillation parameters their best-fit values of the recent NuFIT~5.1~\cite{Esteban:2020cvm, NuFIT} global fit to oscillation data.  When forecasting limits or discovery potential of long-range interactions, we allow their values to float as part of our statistical methods (section~\ref{sec:stat_methods}).  

%=================================================
\begin{table}[t!]
	\centering
	\begin{center} 
		\begin{tabular}{|c|c|c|c|}
			\hline
			Parameter &  Best-fit value & 3$\sigma$ range &  Statistical treatment \\
			\hline
			${\theta_{12}}~[^{\circ}]$ & 33.45 & 31.27--35.87  & Fixed to best fit \\
			\hline
			\mr{2}{*}{${\theta_{13}}~[^{\circ}]$} & 8.62 & 8.25--8.98 & \mr{2}{*}{Fixed to best fit} \\ 
			& (8.61) & (8.24--9.02) & \\ 
			\hline
			\mr{2}{*}{${\theta_{23}}~[^{\circ}]$} & 42.1 & 39.7--50.9 & \mr{2}{*}{Minimized over $3\sigma$ range} \\
			& (49.0) & (39.8--51.6) & \\
			\hline
			\mr{2}{*}{$\delta_{\rm CP}~[^{\circ}]$} & 230 & 144--350 & \mr{2}{*}{Minimized over $3\sigma$ range} \\
			& (278) & (194--345) & \\
			\hline
			\mr{2}{*}{$\frac{\Delta{m^2_{21}}}{10^{-5} \, \rm{eV}^2}$} & \mr{2}{*}{7.42} & \mr{2}{*}{6.82--8.04}  & \mr{2}{*}{Fixed to best fit} \\
			& & & \\
			\hline
			\mr{2}{*}{$\frac{\Delta{m^2_{31}}}{10^{-3} \, \rm{eV}^2}$} & 2.51  & 2.430--2.593 & \mr{2}{*}{Minimized over $3\sigma$ range} \\
			& (-2.41)  & (-2.506--(-2.329)) & \\
			\hline
		\end{tabular}
		\caption{\textit{Best-fit values and allowed ranges of the oscillation parameters used in our analysis.}  The values are from the NuFIT~5.1 global fit to oscillation data~\cite{Esteban:2020cvm, NuFIT}.  Values outside parentheses are for normal neutrino mass ordering; values inside, for inverted mass ordering.  To produce the illustrative figs.~\ref{fig:dune_prob_events} and \ref{fig:t2hk_prob_events}, we fix all parameters to their best-fit values.}
		\label{tab:params_value1}
	\end{center}
\end{table}
%=================================================

The second term on the right-hand side of \equ{hamiltonian_tot} is the potential from the standard CC coherent forward $\nu_e$-$e$ scattering, \ie,
\begin{equation}
	\label{equ:v_mat}
	\mathbf{V}_{\rm mat}
	=
	{\rm diag}(V_{\rm CC}, 0, 0) \;,
\end{equation}
where $V_{\rm CC} = \sqrt{2} G_F n_e$, $G_F$ is the Fermi constant, and $n_e$ is the number density of electrons along the trajectory of the neutrinos.  This term contributes only during neutrino propagation inside Earth, where electron densities are high.  Since we do not compute the changing potential as the neutrino propagates (section~\ref{sec:yukawa_interaction}), and since the neutrino beams in DUNE and T2HK travel exclusively inside the crust of the Earth, where the matter density is fairly uniform, we use the average matter density along the neutrino trajectory from production to detection, $\rho_\textrm{avg}$, to approximate the potential, \ie,  $V_{\rm CC} \approx 7.6 \cdot Y_{e} \cdot 10^{-14} \left(\frac{\rho_{\text{avg}}}{\mathrm{g}~\mathrm{cm}^{-3}}\right) \; \mathrm{eV}$,where $Y_e \equiv n_e / (n_p + n_n)$ is the density of electrons relative to that of protons, $n_p$, and neutrons,  $n_n$.  We estimate $\rho_\textrm{avg}$ using the Preliminary Reference Earth Model~\cite{Dziewonski:1981xy}, which yields $2.848$~g~cm$^{-3}$ and 2.8~g~cm$^{-3}$ for DUNE and T2HK, respectively.  The potential above is for neutrinos; for antineutrinos, it flips sign, \ie, $\mathbf{V}_{\rm mat} \to - \mathbf{V}_{\rm mat}$.

The third term on the right-hand side of \equ{hamiltonian_tot} is the contribution from the new neutrino-matter interactions, \ie,
\begin{equation}
	\label{equ:lri_pot}
	\mathbf{V}_{\rm LRI}
	=
	{\rm diag}(V_{{\rm LRI}, e}, V_{{\rm LRI},\mu}, V_{{\rm LRI},\tau}) \;,
\end{equation}
where, for a specific choice of $U(1)^\prime$ symmetry, and for given values of $G^\prime$ and $m_{Z^\prime}$, $V_{{\rm LRI}, \alpha}$ is computed using \equ{pot_total_simplified}.  When computing limits and discovery prospects of the new matter potential, we use for $\mathbf{V}_{\rm LRI}$ instead the textures in table~\ref{tab:charges}; see sections~\ref{sec:constraints_pot} and \ref{sec:discovery}.  The potential above is for neutrinos; for antineutrinos, it flips sign, \ie, $\mathbf{V}_{\rm LRI} \to - \mathbf{V}_{\rm LRI}$.

The relative sizes of the standard and new contributions to the total Hamiltonian, \equ{hamiltonian_tot}, determine the range of values of the new matter potential to which DUNE and T2HK are sensitive.  On the one hand, in the absence of new interactions, \ie, when $\mathbf{V}_{\rm LRI} = 0$, oscillations are driven by standard vacuum and matter effects.  Only the coherent forward scattering of $\nu_e$ on electrons inside the Earth modifies the oscillation parameters.  On the other hand, if the new matter potential is the dominant contribution, \ie, when $\mathbf{V}_{\rm LRI} \gg \mathbf{H}_{\rm vac} + \mathbf{V}_{\rm mat}$, oscillations are suppressed because $\mathbf{V}_{\rm LRI}$ is diagonal.  

In-between, when the new interactions contribute comparably to the standard contributions, \ie, when $\mathbf{V}_{\rm LRI} \approx \mathbf{H}_{\rm vac} + \mathbf{V}_{\rm mat}$, the new matter potential introduces a resonance that enhances the values of the oscillation parameters and affects the oscillation probabilities significantly.  For DUNE, the standard contribution to the Hamiltonian is roughly $10^{-13}$--$10^{-12}$~eV, depending on the specific neutrino energy; for T2HK, which has lower energies, it is slightly higher, roughly $10^{-12}$--$10^{-11}$~eV.  For the new matter potential to induce resonant flavor conversions --- and thus to boost the detectability of the new interactions --- it must be within this range.  Below, we show that this is indeed the case, by computing oscillation probabilities including the new interactions.

%=================================================
\begin{figure}
	\centering
	\includegraphics[width=\textwidth]{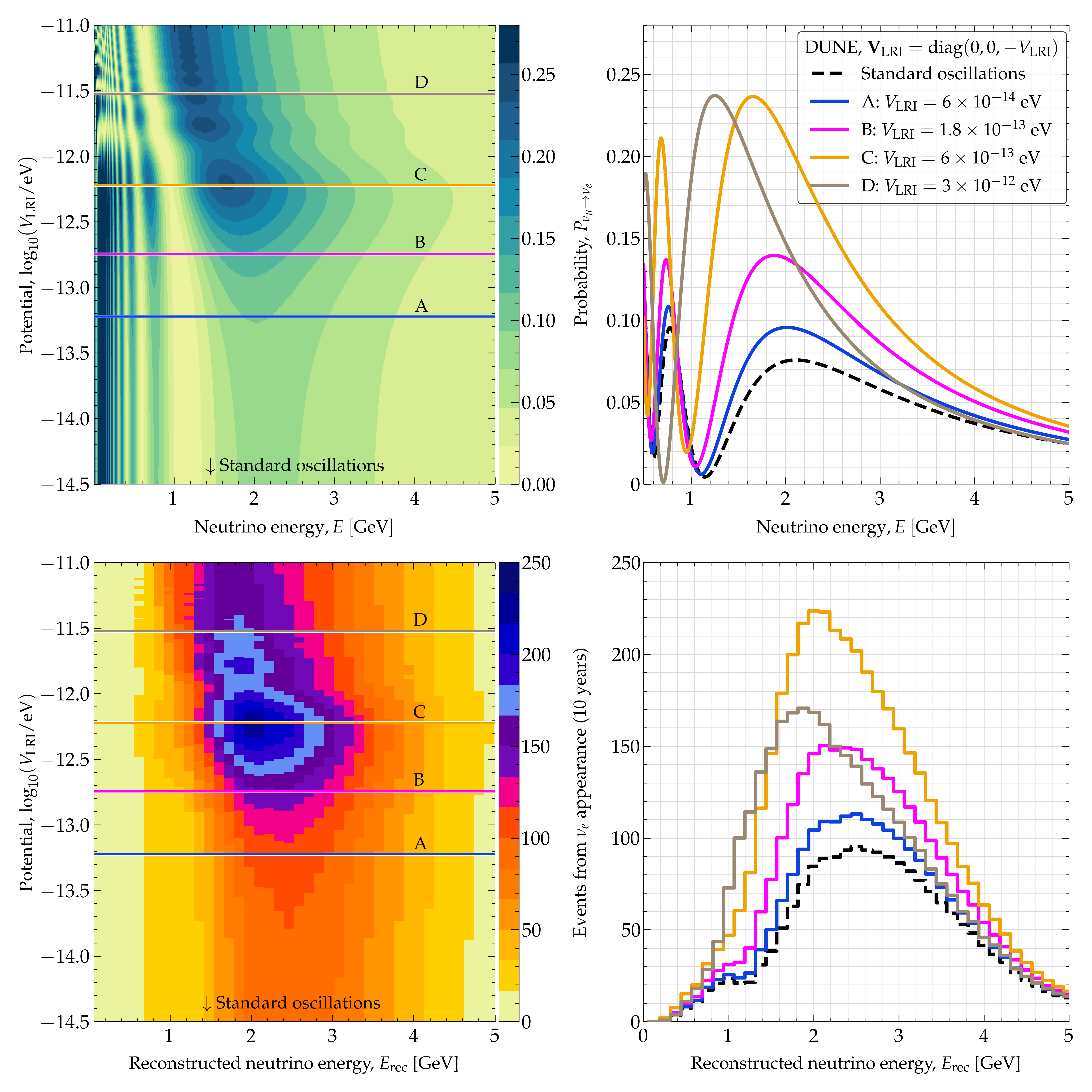}
	\caption{\textit{Oscillation probabilities (top) and event distributions (bottom) in the presence of a new matter potential in DUNE.}  In this figure, we show examples computed assuming a matter potential matrix of the form $\mathbf{V}_{\rm LRI} = \textrm{diag}(0,0,-V_{\rm LRI})$, with varying value of $V_{\rm LRI}$, as would be induced by the $B - 3L_\tau$ and $L - 3L_\tau$ symmetries (table~\ref{tab:charges}). \textit{Top left:} $\numu \to \nue$ probability as a function of the neutrino energy, $E$, and new matter potential, $V_{\rm LRI}$. \textit{Top right:} Oscillation probability computed for choices A--D of the potential, showing the change in amplitude and phase compared to standard oscillations.  \textit{Bottom left:}  Total number, \ie, signal plus background, of $\numu\to\nue$ appearance events after 10 years of run-time (5~yr in $\nu$ and $\bar{\nu}$ modes each), as a function of reconstructed neutrino energy, $E_{\rm rec}$, and $V_{\rm LRI}$.  \textit{Bottom right:}  Event spectra computed for choices A--D of the potential.  See section~\ref{sec:formalism} for details and \figu{t2hk_prob_events} for analogous results for T2HK.  \textit{In DUNE, resonant effects may appear if the new matter potential $V_{\rm LRI} \approx 10^{-13}$--$10^{-12}$~eV.}
		\label{fig:dune_prob_events}} 
\end{figure}
%=================================================

%=================================================
\begin{figure}
	\centering
	\includegraphics[width=\textwidth]{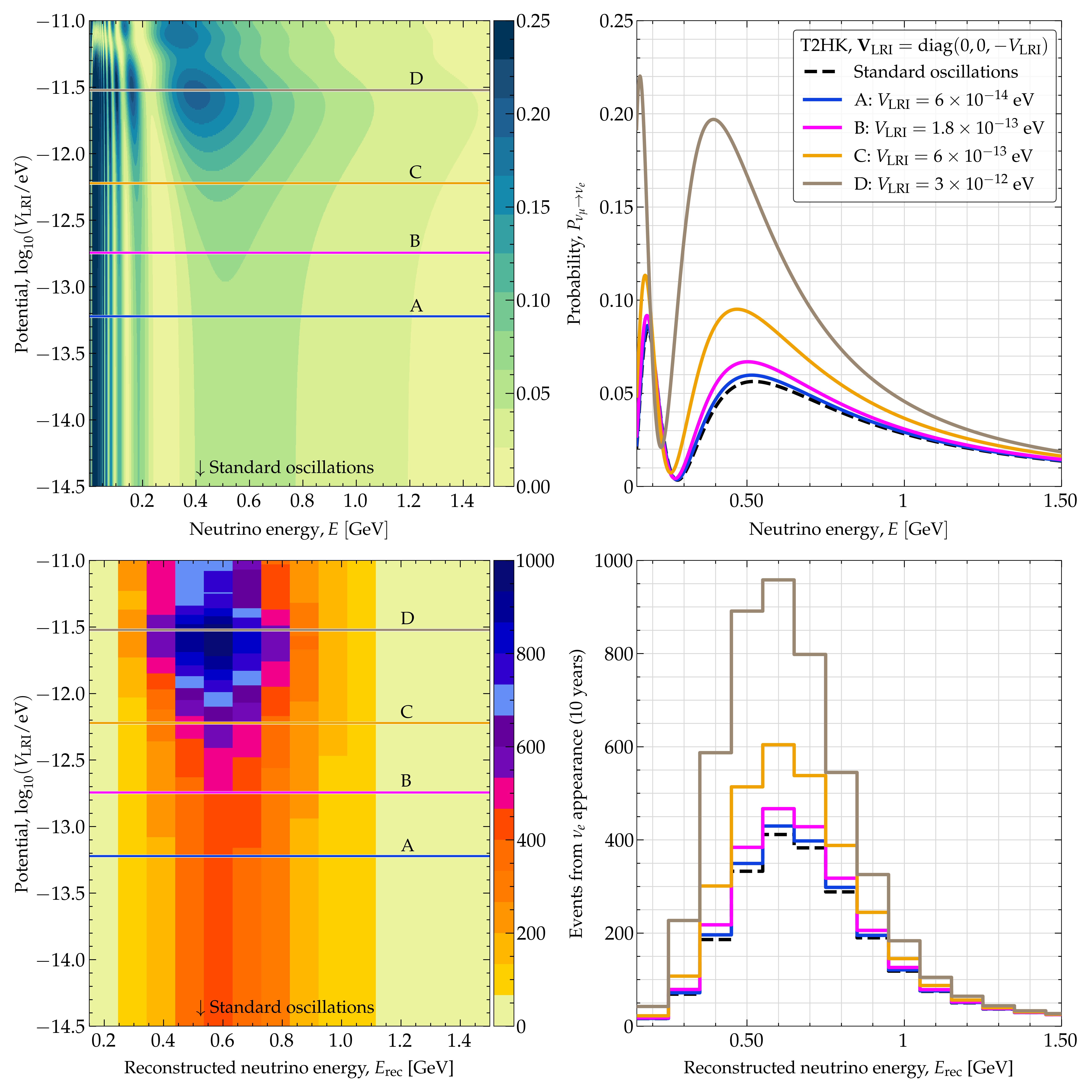}
	\caption{\textit{Oscillation probabilities (top) and event distributions (bottom) in the presence of a new matter potential in T2HK.}  Same as \figu{dune_prob_events}, but for T2HK.  The four illustrative choices of $V_{\rm LRI}$, A--D, are the same as in \figu{dune_prob_events}.  For the event rates, we use 10~years of run-time (2.5~yr in $\nu$ mode and 7.5~yr in $\bar{\nu}$ mode).  See section~\ref{sec:formalism} for details.  \textit{In T2HK, resonant effects may appear if the new matter potential $V_{\rm LRI} \approx 10^{-12}$--$10^{-11}$~eV.}
		\label{fig:t2hk_prob_events}} 
\end{figure}
%=================================================

%===============================================
\subsection{Neutrino oscillation probabilities}
\label{sec:prob_var_pot}
%===============================================

The $\nu_\alpha \to \nu_\beta$ transition probability under neutrino-matter interactions, including standard and new contributions, and governed by the Hamiltonian in \equ{hamiltonian_tot} is~\cite{Kimura:2002hb,Akhmedov:2004ny}
\begin{equation}
	\label{equ:osc_prob_gen}
	P_{\nu_{\alpha}\to \nu_{\beta}}
	=
	\left\vert
	\sum^3_{i=1}
	\tilde{U}_{\alpha i}
	\exp
	\left[
	-\frac{\Delta \tilde{m}^2_{i1} L}{2E}
	\right]
	\tilde{U}^{^{\ast}}_{\beta i}
	\right\vert^2 \;,
\end{equation}
where $L$ is the distance that the neutrino traverses from production to detection, $\tilde{m}^2_{i}/2E$ are the eigenvalues of the Hamiltonian, and $\Delta \tilde{m}^2_{ij} \equiv \tilde{m}_{i}^2 - \tilde{m}_{j}^2$. The matrix $\tilde{\mathbf{U}}$ diagonalizes the Hamiltonian; it is parameterized like the PMNS matrix but depends on the mixing parameters modified by matter effects, $\tilde{\theta}_{23}$, $\tilde{\theta}_{13}$, $\tilde{\theta}_{12}$, and $\tilde{\delta}_{\rm CP}$.  The values of the modified oscillation parameters deviate from their values in vacuum increasingly with rising neutrino energy; the magnitude of the deviation and its dependence with energy are different for the different symmetries.  Appendix~\ref{app:param_run} shows this explicitly.  We compute the oscillation parameters numerically (see below); \Refes~\cite{Barger:1980tf, Zaglauer:1988gz, Ohlsson:1999um, Akhmedov:2004ny, Agarwalla:2013tza, Chatterjee:2015gta, Agarwalla:2015cta, Khatun:2018lzs, Agarwalla:2021zfr} provide approximate analytical expressions for them, some of which we use in our discussion below.

To generate our results, including in figs.~\ref{fig:dune_prob_events} and \ref{fig:t2hk_prob_events}, we compute  the $\nu_\mu \to \nu_e$ and $\bar{\nu}_\mu \to \bar{\nu}_e$ \textit{appearance} probabilities and the $\nu_\mu \to \nu_\mu$ and $\bar{\nu}_\mu \to \bar{\nu}_\mu$ \textit{disappearance} probabilities to which DUNE and T2HK are sensitive.  We do this numerically using {\sc GLoBES}~\cite{Huber:2004ka, Huber:2007ji}, with a version of the {\sc snu} matrix-diagonalization library~\cite{Kopp:2006wp, Kopp:2007ne} modified by us to include the new matter potential.  Later (section~\ref{sec:expt-details}), we also use these tools to compute event rates. 

Figures~\ref{fig:dune_prob_events} and \ref{fig:t2hk_prob_events} show the $\nu_\mu \to \nu_e$ probability computed in the presence of a new matter potential, evaluated, respectively, for the baselines and energy ranges of DUNE and T2HK.  For the new matter potential matrix, we adopt the illustrative choice of $\mathbf{V}_{\rm LRI} = \mathrm{diag}(0, 0, -V_{\rm LRI})$, which has the texture of the potential induced by the $B - 3L_\tau$ and $L - 3L_\tau$ symmetries (table~\ref{tab:charges}), and vary the value of $V_{\rm LRI}$.  However, our  observations below hold also for other choices of the texture of the potential.  As anticipated above, the effects of the new interaction on the oscillation probabilities (and the event distributions) are significant when the new matter potential is comparable to the standard term in the Hamiltonian.  Broadly stated, the closer in size the new and standard contributions are, the closer to resonant are the effects induced by the new matter potential.  Appendix~\ref{app:prob_plots} shows oscillation probabilities for our other candidate symmetries. 

Exclusively for the purpose of understanding the effect of the new interactions on the probabilities, we use approximate analytical expressions for the $\nu_\mu \to \nu_e$ and $\nu_\mu \to \nu_\mu$ probabilities.  For the $\nu_\mu \to \nu_e$ probability, we expand  \equ{osc_prob_gen} under the approximation that $\tilde{\theta}_{12}$ saturates to $90^\circ$~\cite{Chatterjee:2015gta, Agarwalla:2021zfr}, which occurs early with rising energy in the presence of standard and new matter effects (\figu{theta_run}).  This yields~\cite{Agarwalla:2021zfr}
\begin{equation}
	\label{equ:app_mat}
	P_{\nu_{\mu} \to \nu_{e}}
	\approx
	\sin^{2}\tilde{\theta}_{23} ~ 
	\sin^{2}(2\tilde{\theta}_{13}) ~
	\sin^{2}
	\bigg[
	1.27~
	\frac{ (\Delta \tilde{m}^{2}_{32} / {\rm eV^2}) (L / {\rm km}) }
	{E / {\rm GeV}}
	\bigg] \,.
\end{equation}
For the $\nu_\mu \to \nu_\mu$ probability, we expand \equ{osc_prob_gen}, assuming $\tilde{\theta}_{12} = 90^{\circ}$, and keep the first two terms from eq.~(3.27) of~\cite{Khatun:2018lzs} (see also \Refe~\cite{Agarwalla:2021zfr}).  This yields
\begin{equation}
	P_{\nu_{\mu}\to\nu_{\mu}}
	\approx
	1
	-
	\sin^2 (2\tilde{\theta}_{23})~
	\cos^2 \tilde\theta_{13}~
	\sin^2
	\bigg[
	1.27~
	\frac{(\Delta \tilde{m}^{2}_{31} / {\rm eV^2}) (L/{\rm km} )}
	{(E / {\rm GeV})}
	\bigg] \;.
	\label{equ:surv_mat}
\end{equation}

In DUNE (\figu{dune_prob_events}), on account of its baseline, the leading vacuum contribution over most of the relevant energy range is $\propto \Delta m_{31}^2 / (2 E)$; see \equ{hamiltonian_vac}.  In the $\mu$-$\tau$ sector, which determines the modification of the $\tilde{\theta}_{23}$ and $\tilde{\theta}_{13}$  angles that drive the $\nu_\mu \to \nu_e$ probability, the vacuum contribution dominates over the matter potential at energies roughly below 6~GeV.  

From roughly 1~GeV to 2~GeV, resonant features induced by the new interaction on the probability are possible when the new matter potential is roughly of the same size as the standard contributions (see above).  The exact relation between these quantities, which we do not show explicitly but compute numerically and implicitly, stems from the conditions imposed on them in order to achieve the resonant enhancement of the probability.  In addition to increasing the probability amplitude, the new interaction shifts the position of the first oscillation maximum to slightly lower energies, due to the growth of $\Delta \tilde{m}^{2}_{31}$ and $\Delta \tilde{m}^{2}_{32}$ with energy (\figu{delm_run}).  Appendix~\ref{app:param_run} expands on this.  (Below about 0.5~GeV, the effects of the new interaction on the probability are driven instead by $\tilde{\theta}_{12}$, on account of its rapid growth with rising energy (\figu{theta_run}).  However, because of the paucity of the DUNE neutrino beam at these energies (section~\ref{sec:expt-details}), these effects contribute little to our analysis.)

Figure~\ref{fig:dune_prob_events} shows that, between 1~GeV and 2~GeV, for a given value of $V_{\rm LRI}$, the probability is enhanced at values of the energy for which the resonance condition is satisfied.  Higher values of $V_{\rm LRI}$ require larger matching energies to trigger the resonance, on account of the $\propto 1/E$ dependence of the vacuum term.  For a fixed value of $V_{\rm LRI}$, the modification with energy of the $\nu_\mu \to \nu_e$ probability is driven by the growth with energy of $\tilde{\theta}_{23}$ and $\tilde{\theta}_{13}$ (\figu{theta_run}).  Overall, for the illustrative symmetry in \figu{dune_prob_events}, this makes DUNE sensitive to $V_{\rm LRI} \approx (\mathbf{H}_{\rm vac})_{\tau\tau} \in [3.8 \cdot 10^{-14}, 1.4 
\cdot10^{-12}]$~eV (see bottom panel of \figu{lrf_bounds_NH_selective}), given the reconstructed neutrino energy range of 0.5--18 GeV in DUNE (section~\ref{sec:expt-details}). Values of $V_{\rm LRI}$ significantly smaller than that are unable to match the standard contribution and, therefore, trigger no resonance. 

The above behavior is not limited to the illustrative symmetry in \figu{dune_prob_events}, but applies to all of the symmetries that we consider.  Indeed, although the specific elements of the standard contribution that the long-range potential must match are different for different symmetries, later we find that our upper limits on $V_{\rm LRI}$ are contained within the above range; most are within $10^{-14}$--$10^{-13}$~eV; see \figu{constraints_on_pot_dune_t2hk-NMO}.  

In T2HK (\figu{t2hk_prob_events}), the results are similar as in DUNE.  However, because the T2HK neutrino beam has lower energies, the vacuum contribution is larger than in DUNE and, therefore, the values of the new matter potential to which T2HK is sensitive are higher, \ie, $V_{\rm LRI} \in [2.3 \cdot 10^{-13}, 6.8 \cdot 10^{-12}]$~eV, corresponding to the reconstructed neutrino energy range of 0.1--3 GeV; see \figu{lrf_bounds_NH_selective}.

%=================================================
\subsection{Event rates in DUNE and T2HK}
\label{sec:expt-details}
%=================================================

Long-baseline neutrino experiments, with precisely characterized neutrino beams, are excellent platforms to perform precision tests of the standard oscillation paradigm and to search for physics beyond it.  Like \Refe~\cite{Singh:2023nek}, we gear our forecasts of the sensitivity to new neutrino-matter interactions to two of the leading long-baseline experiments under construction, DUNE and T2HK.  Both experiments plan to have near and far detectors; in our analysis, we focus exclusively on the latter, where the effects of oscillations are more apparent; however, the near detectors also have interesting probing capabilities~\cite{Melas:2023olz}.

The neutrino beams are produced as mainly $\nu_\mu$ or $\bar{\nu}_\mu$, with a small contamination of $\nu_e$ and $\bar{\nu}_e$.  The experiments will look for the appearance of $\nu_e$ and $\bar{\nu}_e$ and the disappearance of $\nu_\mu$ and $\bar{\nu}_\mu$. Hence, there are four oscillation channels that we use in our analysis:  $\nu_{\mu}\to\nu_e$, $\bar\nu_\mu\to \bar\nu_e$, $\nu_{\mu}\to\nu_\mu$, and $\bar\nu_\mu\to \bar\nu_\mu$.  Detection of a sought signal is primarily via CC neutrino interactions of $\nu_e$, $\bar{\nu}_e$, $\nu_\mu$, and $\bar{\nu}_\mu$.  While the majority of the background is contributed by the NC events triggered by neutrinos of all flavors, there is also a small contribution from CC events triggered by $\nu_{\tau}$ and $\bar{\nu}_{\tau}$ born from oscillations.  Following \Refes~\cite{Hyper-Kamiokande:2016srs, DUNE:2021cuw}, we assume 2\% and 5\% appearance and 5\% and 3.5\% disappearance systematic uncertainties when computing the signal event rates in DUNE and T2HK, respectively. For the background contribution, for T2HK we assume 10\% systematic uncertainties for all kinds of background events and, for DUNE, 5--20\% depending on the background channel.  For details, see table 7 of \Refe~\cite{Singh:2023nek}.  Since the far detectors cannot distinguish neutrinos from antineutrinos, we add the ``wrong-sign'' contamination events as part of the signal.

Just as for the oscillation probabilities, we compute the expected rate of neutrino-induced events detected by DUNE and T2HK using {\sc GLoBES}~\cite{Huber:2004ka, Huber:2007ji}, with the new neutrino-matter interactions included by modifying the {\sc snu} library~\cite{Kopp:2006wp, Kopp:2007ne}.  We compute events binned in reconstructed neutrino energy, $E_{\rm rec}$, \ie, the energy inferred by analyzing the properties of the particles created in the neutrino interaction.  

DUNE~\cite{DUNE:2020lwj, DUNE:2020jqi, DUNE:2021cuw, DUNE:2021mtg} will use a liquid-argon time-projection chamber detector with a net volume of 40~kton. Its neutrino beam will travel 1285~km from Fermilab to the Homestake Mine. Neutrinos will be produced by a 1.2-MW beam of 120-GeV protons delivering $1.1 \cdot 10^{21}$ protons-on-target (P.O.T.) per year.  It will produce a wide-band, on-axis beam of neutrinos with energies of 0.5--110~GeV and peaking around 2.5~GeV.  When simulating neutrino detection in DUNE, we follow the configuration details from \Refe~\cite{DUNE:2021cuw}.  DUNE will run in neutrino and antineutrino modes, 5 years in each, for a total run-time of 10 years.  To make our forecasts conservative, we use only the fiducial volume and beam power of the completed form of DUNE and ignore the smaller contribution from runs during its construction~\cite{DUNE:2021tad}.   We bin events between $E_{\rm rec} = 0.5$ and 8 GeV with a uniform bin width of 0.125~GeV; between 8 and 10~GeV, with a width of 1~GeV; and between 10 and 18~GeV, with a width of 2~GeV.

T2HK~\cite{Abe:2015zbg, Hyper-Kamiokande:2016srs, Hyper-Kamiokande:2018ofw} will use a water Cherenkov detector with a fiducial volume of 187~kton. Neutrinos will be produced at the J-PARC facility~\cite{McDonald:2001mc} by a 1.3-MW beam of 80-GeV protons delivering $2.7 \cdot 10^{22}$~P.O.T.~per year.  The ensuing neutrino beam will be narrow-band, will peak around 0.6~GeV, will travel 295~km to the detector at the Tochibara Mines in Japan, and arrive $2.5^\circ$ off-axis.  For our analysis, we follow the configuration details from \Refe~\cite{Hyper-Kamiokande:2016srs}.  T2HK will run 2.5 years in neutrino mode and 7.5 years in antineutrino mode, adhering to the proposed 1:3 ratio between the modes.  We bin events uniformly between $E_{\rm rec} = 0.1$ and 3~GeV with a bin width of 0.1~GeV.

Figures~\ref{fig:dune_prob_events} and \ref{fig:t2hk_prob_events} illustrate the event spectra from $\nu_e$ appearance expected in DUNE and T2HK, computed using the same illustrative potential of $\mathbf{V}_{\rm LRI} = \textrm{diag}(0, 0, -V_{\rm LRI})$ used for the probabilities in these figures.  The features in the event spectra reflect the features in the oscillation probabilities (section~\ref{sec:prob_var_pot}). The event distribution starts deviating from the standard-oscillation expectation at $V_{\rm LRI} \gtrsim 10^{-14}~{\rm eV}$, and the deviation grows with $V_{\rm LRI}$.  Because T2HK has a larger detector than DUNE, its event rate is 3--4 times higher.  Yet, because DUNE reaches higher energies than T2HK,  it is sensitive to smaller values of $V_{\rm LRI}$, since the sensitivity to $V_{\rm LRI}$ is $\propto 1/E$; see section~\ref{sec:prob_var_pot}.  Appendix~\ref{app:prob_plots} shows event spectra for our other candidate symmetries. 

The above interplay between the experiments has important consequences for their sensitivity to new neutrino interactions; they become apparent later in our analysis, \eg, in section~\ref{sec:constraints_pot}.  The constraints on and discovery potential of the new interactions are driven by DUNE, on account of it being sensitive to the smallest values of $V_{\rm LRI}$.  However, in order to reach high statistical significance in our claims --- \eg, for discovery or distinguishing between competing candidate symmetries --- the contribution of T2HK is key, since it can reach the larger values of $V_{\rm LRI}$ that are needed to make those claims.

%=================================================
\section{Limits and discovery potential}
\label{sec:results}
%=================================================

Based on the above calculation of oscillation probabilities and event rates, we forecast the sensitivity of T2HK and DUNE to the new neutrino-matter interactions induced by our candidate symmetries.  First, we forecast constraints on the new matter potential and then convert them into constraints on the mass and coupling strength of $Z^\prime$ (sections~\ref{sec:stat_methods} and \ref{sec:constraints_pot}).  Second, we forecast discovery prospects (sections~\ref{sec:stat_methods} and \ref{sec:discovery}).  Third, we forecast prospects of distinguishing between different symmetries (sections~\ref{sec:stat_methods} and \ref{sec:confusion-theory}).

In the main text, we show results as figures (figs.~\ref{fig:lrf_bounds_NH_selective}--\ref{fig:confusion-matrix}).  In the appendices, we show some of them in tables (tables~\ref{tab:upper_limits_potential_NMO_organized} and \ref{tab:discovery_strength}).   In \Refe~\cite{GitHub_lrf-repo}, we provide them as digitized files.

%=================================================
\subsection{Statistical methods}
\label{sec:stat_methods}
%=================================================

When computing the constraints and discovery prospects, we treat each symmetry individually.  We follow the same statistical methods as in \Refe~\cite{Singh:2023nek}; below, we sketch it, and defer to \Refe~\cite{Singh:2023nek} for details.  When computing prospects for distinguishing between symmetries, we extend these methods to make pairwise comparisons between symmetries.  Throughout, we fix $\theta_{13}$ and $\theta_{12}$ to their present-day best-fit values (table~\ref{tab:params_value1}).  For $\theta_{13}$, this is because the current precision on its value is already small, of 2.8\%~\cite{DayaBay:2022orm}.  For $\theta_{12}$, it is because it has only a small impact on our results (eqs.~(\ref{equ:app_mat}) and  (\ref{equ:surv_mat})).

The experiments are blind to the origin of the new matter potential.  They are only sensitive to how and by how much it influences neutrino oscillations, \ie, to the texture of the matter potential $\mathbf{V}_{\rm LRI}$, as shown in table~\ref{tab:charges}, and to the size of the parameter $V_{\rm LRI}$ on which it depends.  In our statistical analysis, for each choice of symmetry, we adopt its corresponding $\mathbf{V}_{\rm LRI}$ potential texture.  As a consequence, symmetries that have equal or similar  texture yield equal or similar sensitivity to $V_{\rm LRI}$; see, \eg, \figu{constraints_on_pot_dune_t2hk-NMO}.  Only afterwards, when we convert the resulting sensitivity to $V_{\rm LRI}$ into sensitivity on $G^\prime$ and $m_{Z^\prime}$, do the particular $U(1)^\prime$ charges of each symmetry and the knowledge of the matter content in celestial bodies play a role in yielding different sensitivity between equally or similarly textured symmetries; see, \eg, \figu{all_symmetry}.  We show this explicitly below.

\smallskip

\textbf{\textit{Constraints on new neutrino interactions.---}}  When forecasting constraints, we generate the \textit{true} event spectrum under standard oscillations by fixing $V_{\rm LRI}^{\rm true} = 0$.  We compare it to \textit{test} event spectra computed using non-zero values of $V_{\rm LRI}$.  In experiment $e = \{{\rm T2HK},~{\rm DUNE}\}$, we bin the spectra in $N_e$ bins of $E_{\rm rec}$; see section~\ref{sec:expt-details} for a description of the binning.  In the $i$-th bin, we compare the true {\it vs.}~test numbers of events from each detection channel $c = \{{\rm app}~\nu,~{\rm app}~\bar{\nu},~{\rm disapp}~\nu,~{\rm disapp}~\bar{\nu}\}$, \ie,  $N_{e,c,i}^\textrm{true}$ {\it vs.}~$N_{e,c,i}^\textrm{test}$, via the Poisson $\chi^2$ function~\cite{Baker:1983tu, Cowan:2010js, Blennow:2013oma, Singh:2023nek}
\begin{eqnarray}
	\chi_{e,c}^{2}
	(V_{\rm LRI}, \boldsymbol{\theta}, o)
	=
	&&
	\underset{\left\{\xi_{s}, \{\xi_{b, c, k}\}\right\}}{\mathrm{min}} 
	\left\{
	2\sum^{N_e}_{i=1}
	\left[
	N_{e, c, i}^{{\rm test}}
	(V_{\rm LRI}, \boldsymbol{\theta}, o, \xi_s, \{\xi_{b,c,k}\})
	\right. \right.
	\nonumber \\
	&&
	\left. \left.
	-
	N_{e, c, i}^{{\rm true}}
	\left( 
	1
	+
	\ln
	\frac{N_{e, c, i}^{{\rm test}}
		(V_{\rm LRI}, \boldsymbol{\theta}, o, \xi_s, 
		\{\xi_{b, c, k}\})}
	{N_{e, c, i}^{{\rm true}}}
	\right)
	\right]
	+
	\xi^{2}_{s}
	+
	\sum_k \xi^{2}_{b, c, k} 
	\right\} \;, \;\,\,\,\,\,\,\,\,\,\,
	\label{equ:chi2_per_channel}
\end{eqnarray}
where $\boldsymbol{\theta} \equiv \{ \sin^2\theta_{23}, \delta_{\rm CP}, \vert \Delta m_{31}^2\vert \}$ are the oscillation parameters that we vary (the other parameters are fixed, see above) and $o = \{ \text{NMO, IMO} \}$ is the choice of neutrino mass ordering, which can be normal (NMO) or inverted (IMO).  On the right-hand side of \equ{chi2_per_channel}, $\xi_{s}$ and $\xi_{b, c, k}$ represent, respectively, systematic uncertainties on the signal rate and the $k$-th background contribution to the detection channel $c$; these uncertainties are identical for neutrinos and antineutrinos.  We treat them as in \Refe~\cite{Singh:2023nek}.  The right-hand side of \equ{chi2_per_channel} is profiled over the systematic uncertainties; the last two terms are pull terms that keep the values of the systematic uncertainties under control when minimizing over them.  

In \equ{chi2_per_channel}, the event spectra contain both signal and background contributions.  The true number of events is computed using $V_{\rm LRI}^{\rm true} = 0$, and for choices of $\boldsymbol{\theta}^{\rm true}$ and $o^{\rm true}$, \ie,
\begin{equation}
	N_{e, c, i}^{{\rm true}}
	= 
	N_{e, c, i}^{s, {\rm true}}
	+ 
	N_{e, c, i}^{b, {\rm true}} \;,
\end{equation}
where $N_{e, c, i}^{s, {\rm true}}$ and $N_{e, c, i}^{b, {\rm true}}$ are, respectively, the number of signal ($s$) and background ($b$) events; the latter is summed over all sources of background that affect this detection channel.  Similarly, the test number of events is
\begin{equation}
	\label{equ:num_test}
	N_{e, c, i}^{{\rm test}}
	(V_{\rm LRI}, \boldsymbol{\theta}, o, \xi_s, \left\{\xi_{b,c,k}\right\})
	=
	N^s_{e,c,i}(V_{\rm LRI}, \boldsymbol{\theta}, o)
	(1+\pi_{e,c}^s\xi_s)
	+
	\sum_k
	N^{b}_{e,c,k,i}(\boldsymbol{\theta}, o)
	\left(
	1+\pi_{e,c,k}^b\xi_{b,c,k}
	\right) \;,
\end{equation}
where $\pi_{e,c}^s$ and $\pi_{e,c,k}^b$ are normalization errors on the signal and background rates (refer to Table~\ref{tab:normalization_err}). See \Refe~\cite{Singh:2023nek} for more details.

%%%%%%%%%%%%%%%%%%%%%%%%%%%%%%%%%%%%%%%%%%
\begin{table}[t!]
	\resizebox{\columnwidth}{!}{%
		\centering
		\begin{tabular}{|c|c|c|c|c|c|c|c|c|}
			\hline
			\multirow{3}{*}{Experiment}  & \multicolumn{8}{c|}{Normalization errors~[\%]}  \\
			& \multicolumn{4}{c|}{Signal, $\pi_{e,c}^s$} & \multicolumn{4}{c|}{Background, $\pi_{e,c,k}^b$}\\
			&  App.~$\nu$ & App.~$\bar{\nu}$ & Disapp.~$\nu$ & Disapp.~$\bar{\nu}$ & $\nu_{e}$, $\bar{\nu}_{e}$ CC & $\nu_{\mu}$, $\bar{\nu}_{\mu}$ CC & $\nu_{\tau}$, $\bar{\nu}_{\tau}$ CC & NC \\ 
			\hline
			DUNE & 2 & 2 & 5 & 5 & 5 & 5 & 20 & 10\\
			T2HK & 5 & 5 & 3.5 & 3.5 & 10 & 10 & -- & 10\\
			\hline
	\end{tabular}}
	\caption{Normalization errors used in the calculation of event rates in DUNE and T2HK, including signal and background detection channels. We show them separately for the neutrino ($\nu$) and antineutrino ($\bar{\nu}$) modes, and for appearance (``App.'') and disappearance (``Disapp.'') channels. The errors, sourced from \cite{Hyper-Kamiokande:2016srs, DUNE:2021cuw}, are used in \equ{num_test}.}
	\label{tab:normalization_err}
\end{table}
%%%%%%%%%%%%%%%%%%%%%%%%%%%%%%%%%%%%%%%%%%

For T2HK or DUNE, separately and together, we compute the total $\chi^2$ by adding the contributions of all the channels $c$, \ie,
\begin{eqnarray}
	\label{equ:chi2_dune}
	\chi_{\rm DUNE}^{2}
	(V_{\rm{LRI}}, \boldsymbol{\theta}, o)
	&=&
	\sum_c \chi_{{\rm DUNE}, c}^{2}
	(V_{\rm{LRI}}, \boldsymbol{\theta}, o)
	\;, \\
	\label{equ:chi2_t2hk}
	\chi_{\rm T2HK}^{2}
	(V_{\rm{LRI}}, \boldsymbol{\theta}, o)
	&=&
	\sum_c \chi_{{\rm T2HK}, c}^{2}
	(V_{\rm{LRI}}, \boldsymbol{\theta}, o) 
	\;, \\
	\label{equ:chi2_dune_t2hk}
	\chi^2_{{\rm DUNE}+{\rm T2HK}}
	(V_{\rm{LRI}}, \boldsymbol{\theta}, o)
	&=&
	\chi^2_{\rm DUNE}
	(V_{\rm{LRI}}, \boldsymbol{\theta}, o)
	+
	\chi^2_{\rm T2HK}
	(V_{\rm{LRI}}, \boldsymbol{\theta}, o)
	\;.
\end{eqnarray}
We treat the contributions of different channels as uncorrelated.

We compute the sensitivity to $V_{\rm LRI}$ by comparing the minimum value of the above functions, $\chi^2_{e, {\rm min}}$, which is obtained when evaluating them at $V_{\rm LRI} = V_{\rm LRI}^{\rm true} = 0$, $\boldsymbol{\theta} = \boldsymbol{\theta}^{\rm true}$, and $o = o^{\rm true}$, against test values of these parameters.  In the main text, we fix $\boldsymbol{\theta}^{\rm true}$ to its best-fit value under normal ordering (table~\ref{tab:charges}) and $o^{\rm true}$ to NMO.  In appendix~\ref{app:other}, we fix them to inverted ordering instead; our conclusions do not change.  Since we are interested in obtaining limits only on $V_{\rm LRI}$, we profile over $\boldsymbol{\theta}$ and $o$.  This yields the test statistic that we use to place constraints on $V_{\rm LRI}$; \eg, for DUNE, it is
\begin{equation}
	\label{equ:delta_chi2_dune}
	\Delta\chi^2_{ {\rm DUNE}, {\rm con}}(V_{\rm LRI})
	=
	\underset
	{\{ \boldsymbol{\theta}, o\}}{\mathrm{min}} 
	\left[
	\chi_{\rm DUNE}^{2}
	(V_{\rm LRI}, \boldsymbol{\theta}, o) 
	-
	\chi_{{\rm DUNE}, {\rm min}}^{2}
	\right]
	\;,
\end{equation}
and similarly for T2HK and DUNE + T2HK.  When profiling, we follow the same procedure as in \Refe~\cite{Singh:2023nek}.  When profiling over $\sin^2 \theta_{23}$, $\delta_{\rm CP}$, and $\vert \Delta m_{31}^2 \vert$, we vary each of them within their present-day $3\sigma$ allowed ranges~\cite{Esteban:2020cvm}.  We assume no correlations between them, since these are expected to disappear in the near future; see, \eg, \Refe~\cite{Song:2020nfh}. 
In principle, varying the values of the oscillation parameters over ranges different than the ones we have used could change our results. However, we have found that the test statistic aligns closely with the chosen true values, so we do not expect significant changes were we to use wider ranges for the oscillation parameters. Using \equ{delta_chi2_dune}, we report upper limits on $V_{\rm LRI}$ with $2\sigma$ and $3\sigma$ significance, for 1 degree of freedom (d.o.f.).

\smallskip

\textbf{\textit{Discovery of new neutrino interactions.---}}  When forecasting discovery prospects, we follow a similar procedure as when forecasting constraints, with some changes.  In $\chi^2_{e,c}(V_{\rm LRI}, \boldsymbol{\theta}, o)$ in \equ{chi2_per_channel}, the true event spectrum is instead computed using the nonzero value $V_{\rm LRI}^{\rm true} = V_{\rm LRI}$ from the left-hand side and, like before, for choices of $\boldsymbol{\theta}^{\rm true}$ and $o^{\rm true}$ (NMO in the main text and IMO in appendix~\ref{app:other}), \ie, $N_{e,c,i}^{\rm true} \to N_{e,c,i}^{\rm true}(V_{\rm LRI}^{\rm true} = V_{\rm LRI})$ in \equ{chi2_per_channel}.  The test spectrum is instead computed under standard oscillations, \ie, $N_{e, c, i}^{{\rm test}} (V_{\rm LRI}, \boldsymbol{\theta}, o, \xi_s, \{\xi_{b,c,k}\}) \to N_{e, c, i}^{{\rm test}} (V_{\rm LRI} = 0, \boldsymbol{\theta}, o, \xi_s, \{\xi_{b,c,k}\})$ in \equ{chi2_per_channel}.  Like before we profile over $\boldsymbol{\theta}$ and $o$ to build the test statistic that we use to compute the significance with which oscillations with a new matter potential $V_{\rm LRI}$ would be discovered; \eg, for DUNE,
\begin{equation}
	\label{equ:delta_chi2_disc}
	\Delta\chi^2_{ {\rm DUNE}, \rm {disc}}(V_{\rm LRI})
	=
	\underset
	{\{ \boldsymbol{\theta}, o\}}{\mathrm{min}} 
	\left[\chi_{ {\rm DUNE}, {\rm min}}^{2}
	-
	\chi_{\rm DUNE}^{2}
	(V_{\rm LRI}, \boldsymbol{\theta}, o) 
	\right]
	\;,
\end{equation}
and similarly for T2HK and DUNE + T2HK.  This test statistic measures the separation between the observed event distribution, which includes the new matter potential, and standard oscillations.  Using \equ{delta_chi2_disc}, we report the values of $V_{\rm LRI}$ for which LRI would be discovered at $3\sigma$ and $5\sigma$, for 1 d.o.f.  Reference~\cite{Singh:2023nek} shows complementary results on jointly measuring the values of $V_{\rm LRI}$ and of the mixing parameters $\theta_{23}$ and $\delta_{\rm CP}$.

\smallskip

\textbf{\textit{Distinguishing between symmetries.---}}Symmetries that introduce new  matter potentials with different textures have qualitatively different effects on the oscillation probabilities (\figu{probability}).  Hence, we explore whether, in the event of discovery of evidence of a new neutrino interaction, we may identify which symmetry is responsible for it, or narrow down the possibilities to a subset of candidate symmetries.

We proceed similarly as before.  Out of the set of candidate symmetries  (table~\ref{tab:charges}), the true symmetry responsible for the new potential observed is SA, and SB is an alternative one.  We modify \equ{chi2_per_channel} by changing $N_{e,c,i}^{\rm true} \to  N_{e,c,i}^{\rm true}(V_{\rm LRI}) \vert_{\rm SA}$ and $N_{e, c, i}^{{\rm test}}
(V_{\rm LRI}, \boldsymbol{\theta}, o, \xi_s, \allowbreak \{\xi_{b,c,k}\}) \to N_{e, c, i}^{{\rm test}} (V_{\rm LRI}, \boldsymbol{\theta}, o, \xi_s, \{\xi_{b,c,k}\}) \vert_{\rm SB}$.  We show only results assuming NMO for $\boldsymbol{\theta}^{\rm true}$ and $o^{\rm true}$.  The test statistic that we use to distinguish SA from SB is
\begin{equation}
	\label{equ:delta_chi2_conf}
	\Delta\chi^2_{ {\rm DUNE}, {\rm dist}}(V_{\rm LRI}) \vert_{{\rm SA}, {\rm SB}}
	=
	\underset
	{\{ \boldsymbol{\theta}, o\}}{\rm{min}} 
	\left[\chi^{2}_{{\rm DUNE}, {\rm min}}(V_{\rm LRI}) \vert_{\rm SA}
	-
	\chi_{\rm DUNE}^{2}
	(V_{\rm LRI}, \boldsymbol{\theta}, o) \vert_{\rm SB}
	\right]
	\;,
\end{equation}
and similarly for T2HK and DUNE + T2HK.  Using \equ{delta_chi2_disc}, we report the significance, for 1~d.o.f., with which all pairs of SA and SB can be distinguished, via confusion matrices produced for illustrative values of $V_{\rm LRI}$.  

%=================================================
\begin{figure}
	\centering
	\includegraphics[width=0.75\textwidth]{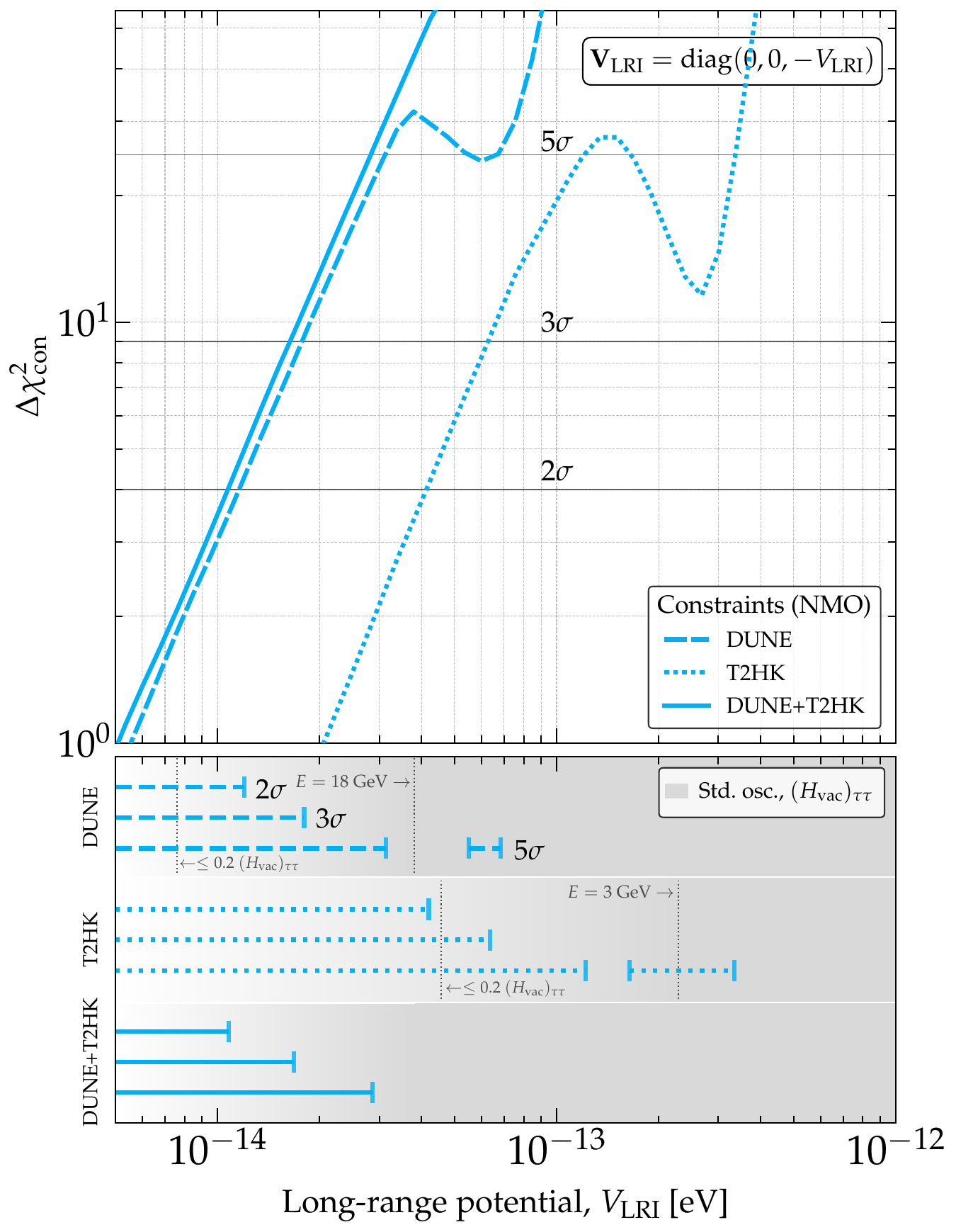}
	\caption{\textit{Projected test statistic used to constrain the new matter potential induced by a $U(1)^\prime$ symmetry.} For this plot, as illustration, we show limits on a potential of the form $\mathbf{V}_{\rm LRI} = \textrm{diag}(0, 0, -V_{\rm LRI})$ for neutrinos and $-\mathbf{V}_{\rm LRI}$ for antineutrinos, like in figs.~\ref{fig:dune_prob_events} and \ref{fig:t2hk_prob_events}, as would be introduced by symmetries $L - 3L_\tau$ or $B - 3L_\tau$ (table~\ref{tab:charges}); fig.~\ref{fig:lrf_bounds_NH} shows results for all the symmetries.  The test statistic is \equ{delta_chi2_dune}.  Results are for DUNE and T2HK separately and combined.  The true neutrino mass ordering is assumed to be normal; fig.~\ref{fig:lrf_bounds_IH} shows that results under inverted mass ordering are similar.  See sections~\ref{sec:stat_methods} and \ref{sec:constraints_pot} for details.  The experiments are sensitive to values of $V_{\rm LRI}$ that are comparable to the standard-oscillation terms in the Hamiltonian; for the choice of $\mathbf{V}_{\rm LRI}$ texture in this figure, this is $(\mathbf{H}_{\rm vac})_{\tau\tau}$, which is $\propto 1/E$. Constraints on $V_{\rm LRI}$ lie around 20\% of the value of $(\mathbf{H}_{\rm vac})_{\tau\tau}$ evaluated at the highest energy in each experiment.  \textit{Combining DUNE and T2HK strengthens the constraints, especially at high values of $V_{\rm LRI}$, by removing the degeneracies between $V_{\rm LRI}$ and $\theta_{23}$ and $\delta_{\rm CP}$ that plague each experiment individually (see also \Refe~\cite{Singh:2023nek}).}
		\label{fig:lrf_bounds_NH_selective}} 
\end{figure}
%=================================================

%=================================================
\subsection{Constraints on new neutrino interactions}
\label{sec:constraints_pot}
%=================================================

%=================================================
\begin{figure}
	\centering
	\includegraphics[width=\textwidth]{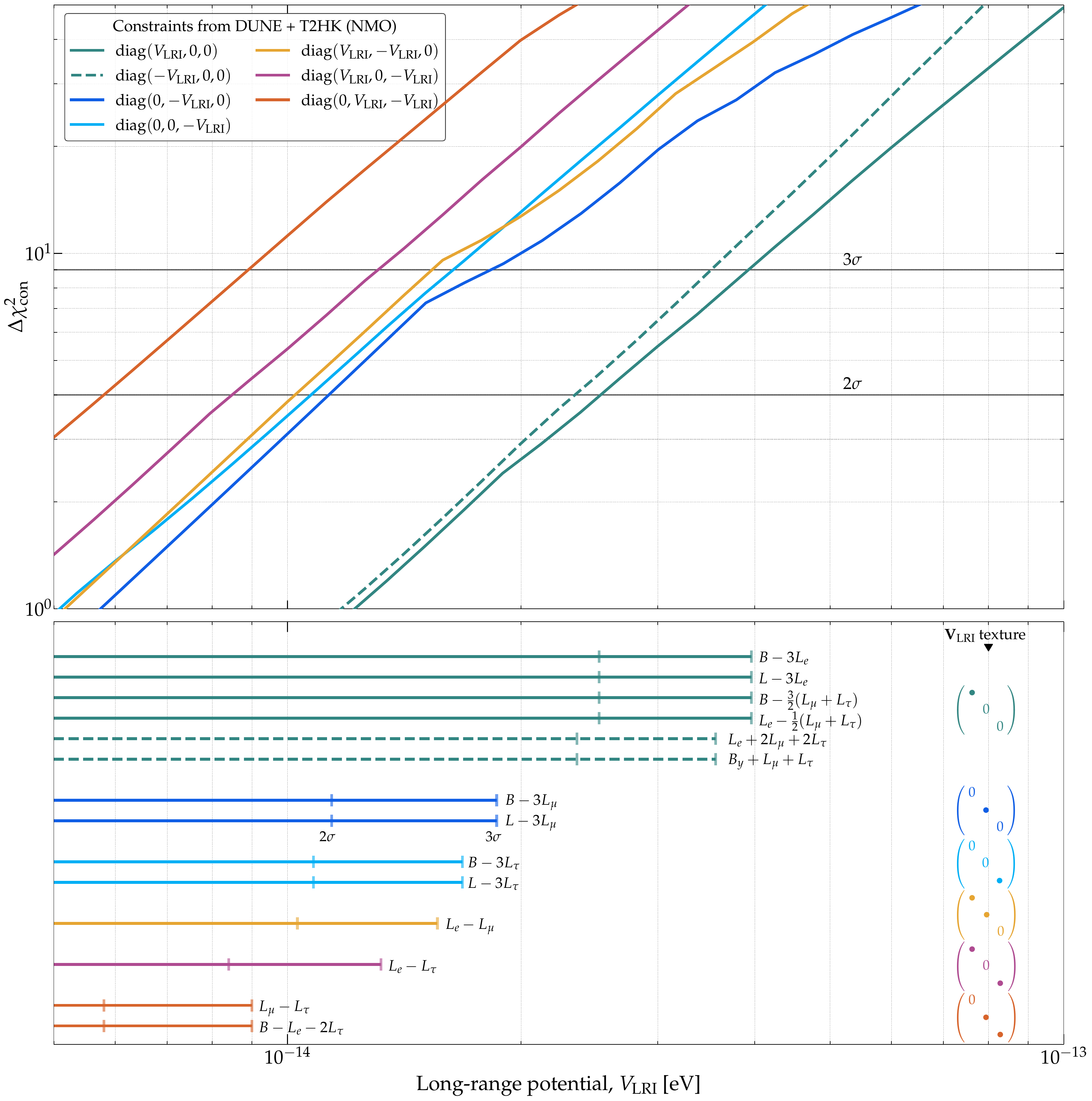}
	\caption{\textit{Projected test statistic (top) used to place upper limits (bottom) on the new matter potential induced by our candidate $U(1)^\prime$ symmetries.}  The true neutrino mass ordering is assumed to be normal; results under inverted ordering are similar (\figu{constraints_on_pot_dune_t2hk-IMO}).  Results are for DUNE and T2HK combined; the test statistic is \equ{delta_chi2_dune} computed for their combination.  See sections~\ref{sec:stat_methods} and \ref{sec:constraints_pot} for details.  The symmetries are grouped according to the texture of the new matter potential that they introduce, $\mathbf{V}_{\rm LRI}$ in table~\ref{tab:charges}.  The numerical values of the limits are in table~\ref{tab:upper_limits_potential_NMO_organized}.  \textit{Symmetries that induce equal or similar potential texture yield equal or similar upper limits on $V_{\rm LRI}$.  All limits lie near the value of the standard-oscillation terms in the Hamiltonian (see \figu{lrf_bounds_NH_selective}), since this triggers resonances in the oscillation probabilities (section~\ref{sec:hamiltonians}).}
		\label{fig:constraints_on_pot_dune_t2hk-NMO}} 
\end{figure}
%=================================================

Figure~\ref{fig:lrf_bounds_NH_selective} shows our resulting projected constraints on $V_{\rm LRI}$ assuming, for illustration, a matter potential with the texture $\mathbf{V}_{\rm LRI} = \textrm{diag}(0,0,-V_{\rm LRI})$, as introduced by the symmetries $L-3L_\tau$ and $B-3L_\tau$, just as in figs.~\ref{fig:dune_prob_events} and \ref{fig:t2hk_prob_events}.  (We show results for other symmetries later.)  Our findings reiterate one of the key results first reported by \Refe~\cite{Singh:2023nek}: DUNE and T2HK can each separately place upper limits on $V_{\rm LRI}$ --- with DUNE placing stronger constraints due to its higher energies ({\it cf.}~figs.~\ref{fig:dune_prob_events}~{\it vs.}~\ref{fig:t2hk_prob_events}), as anticipated in section~\ref{sec:expt-details}.   However, the limits that they can place individually weaken at high values of $V_{\rm LRI}$, due to degeneracies between $V_{\rm LRI}$, $\theta_{23}$, and $\delta_{\rm CP}$; in \figu{lrf_bounds_NH_selective}, this shows up as dips in the test statistic; see \Refe~\cite{Singh:2023nek} for details.  Combining DUNE and T2HK lifts these degeneracies: T2HK lifts the degeneracies due to $\theta_{23}$ and $\delta_{\rm CP}$, while DUNE fixes the neutrino mass ordering, \ie, the sign of $\Delta m_{31}^2$.  As anticipated (section~\ref{sec:prob_var_pot}), the limits on $V_{\rm LRI}$ are comparable to the size of standard-oscillation terms in the Hamiltonian.

Our results extend those of \Refe~\cite{Singh:2023nek}, which had shown the above interplay between DUNE and T2HK only for the symmetries $L_e-L_\mu$, $L_e-L_\tau$, and $L_\mu-L_\tau$.  We find that the same complementarity is present for all the other candidate symmetries that we consider (\figu{lrf_bounds_NH}), with some variation depending on whether the mass ordering is normal or inverted ({\it cf.}~figs.~\ref{fig:lrf_bounds_NH} {\it vs.}~\ref{fig:lrf_bounds_IH}), stemming from differences in the signs of the standard and new matter potentials for neutrinos and antineutrinos (section~\ref{sec:hamiltonians}), and in the run times for each in T2HK (section~\ref{sec:expt-details}).

Figure~\ref{fig:constraints_on_pot_dune_t2hk-NMO} (also, table~\ref{tab:upper_limits_potential_NMO_organized}) shows the upper limits on $V_{\rm LRI}$ for all the symmetries that we consider.  Like in table~\ref{tab:charges}, we group symmetries according to the texture of the matter potential, $\mathbf{V}_{\rm LRI}$, that they induce, since the effects of new interactions on the neutrino oscillation probabilities depend on the texture of the potential matrix $\boldsymbol{V}_{\rm LRI}$, and on the size of its elements, regardless of the source of the potential.  Because of this, the limits on $V_{\rm LRI}$ in \figu{constraints_on_pot_dune_t2hk-NMO} are equal or similar for symmetries that have equal or similar textures for $\mathbf{V}_{\rm LRI}$; {\it cf.}~table~\ref{tab:charges} and \figu{constraints_on_pot_dune_t2hk-NMO}.  Thus, our results broaden the perspectives first put forward by \Refe~\cite{Singh:2023nek}: \textit{DUNE and T2HK may constrain the new matter potential to a level comparable to the standard-oscillation terms, roughly $10^{-14}$--$10^{-13}$~eV, regardless of what is the $U(1)^\prime$ symmetry responsible for inducing the new interaction.}

%=================================================
\begin{figure}
	\centering
	\includegraphics[width=0.85\textwidth]{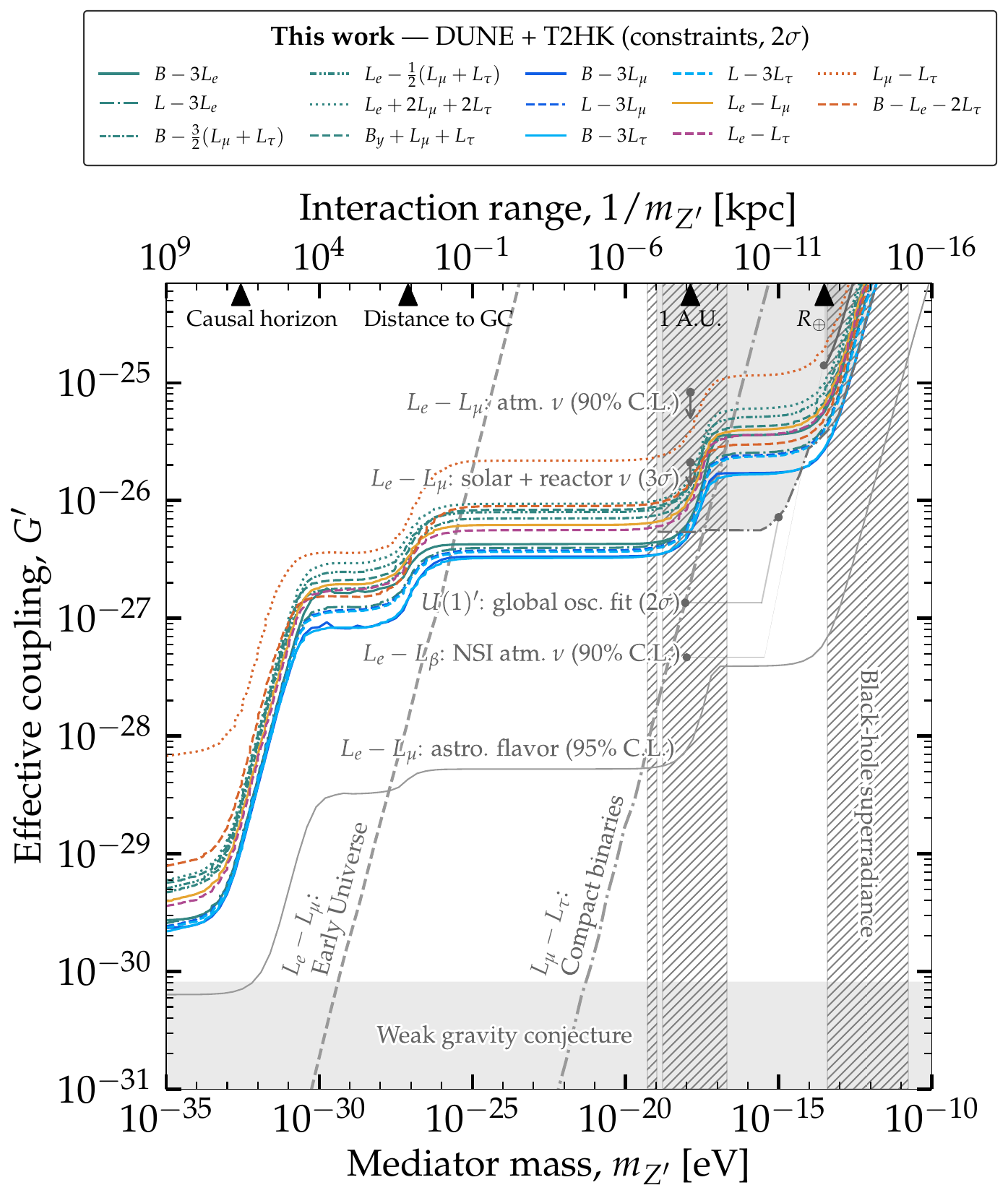}
	\caption{\textit{Projected upper limits on the effective coupling of the new gauge boson, $Z^\prime$, that mediates flavor-dependent long-range neutrino interactions.}  Results are for DUNE and T2HK, combined, after 10 years of operation, and for each of our candidate $U(1)^\prime$ symmetries (table~\ref{tab:charges}).  For this figure, we assume that the true neutrino mass ordering is normal. For each symmetry, the limits on the coupling, $G^\prime$, as a function of the mediator mass, $m_{Z^\prime}$, are converted from the limits on $V_{\rm LRI}$ in \figu{constraints_on_pot_dune_t2hk-NMO} (also in table~\ref{tab:upper_limits_potential_NMO_organized}) using the expressions for $V_{\rm LRI}$ in table~\ref{tab:charges}.  The existing limits are the same as in \figu{constraints_discovery_dune_t2hk-NMO}.  See sections~\ref{sec:stat_methods} and \ref{sec:constraints_pot} for details.  \textit{DUNE and T2HK may constrain long-range interactions more strongly than ever before, regardless of which $U(1)^\prime$ symmetry is responsible for inducing them, especially for mediators lighter than $10^{-18}$~ eV}.}
	\label{fig:all_symmetry}
\end{figure}
%=================================================

The strongest limits can be placed when the new matter potential affects primarily the $\mu$-$\tau$ sector, \ie, when $\mathbf{V}_{\rm LRI} = \textrm{diag}(0, V_{\rm LRI}, -V_{\rm LRI})$, as would be induced by symmetries $B-L_{e}-2L_{\tau}$ and $L_\mu - L_\tau$; see table~\ref{tab:charges}.  A potential of this form affects primarily the $\nu_\mu \to \nu_\mu$ and $\bar{\nu}_\mu \to \bar{\nu}_\mu$ disappearance probabilities. Because in DUNE and T2HK the disappearance channels have the highest event rates (\figu{event_spectra}), the effects of the new matter potential in this case can be detected more easily, leading to stronger limits on $V_{\rm LRI}$.  In contrast, the weakest limits can be placed when the new matter potential affects primarily the electron sector, \ie, when the only non-zero entry is $(\mathbf{V}_{\rm LRI})_{ee}$, as would be induced by symmetries $B - 3L_e$, $L - 3L_e$, $B-\frac{3}{2} (L_\mu + L_\tau)$, $B_y + L_\mu + L_\tau$, $L_{e} + 2L_\mu + 2L_\tau$, and $L_e - \frac{1}{2} (L_\mu + L_\tau)$.  A potential of this form affects primarily the $\nu_\mu \to \nu_e$ and $\bar{\nu}_\mu \to \bar{\nu}_e$ appearance probabilities.  Because in DUNE and T2HK the appearance channels have lower event rates, the limits on $V_{\rm LRI}$ in this case are weaker.

Figure~\ref{fig:all_symmetry} shows the limits on $V_{\rm LRI}$ converted into limits on $G^\prime$ as a function of $m_{Z^\prime}$.  To convert them, we use knowledge of the distribution of electrons, protons, and neutrons in the Earth, Moon, Sun, the Milky Way, and the cosmological matter distribution, and the long-range potential that they source, as introduced in section~\ref{sec:yukawa_interaction}.  In practice, for each symmetry, we take the limit on $V_{\rm LRI}$ from \figu{constraints_on_pot_dune_t2hk-NMO} and equate it to the expression for the simplified potential in table~\ref{tab:charges}, which depends on $m_{Z^\prime}$ and $G^\prime$, and which contains the contribution of the celestial bodies weighed by their  abundance of electrons, protons, and neutrons.  Then, for each value of $m_{Z^\prime}$, we find the upper limit on $G^\prime$ that we report in \figu{all_symmetry}.  In \figu{constraints_discovery_dune_t2hk-NMO}, the region constrained is the envelope of all the individual curves in \figu{all_symmetry}.

%=================================================
\begin{figure}
	\centering
	\includegraphics[width=1\textwidth]{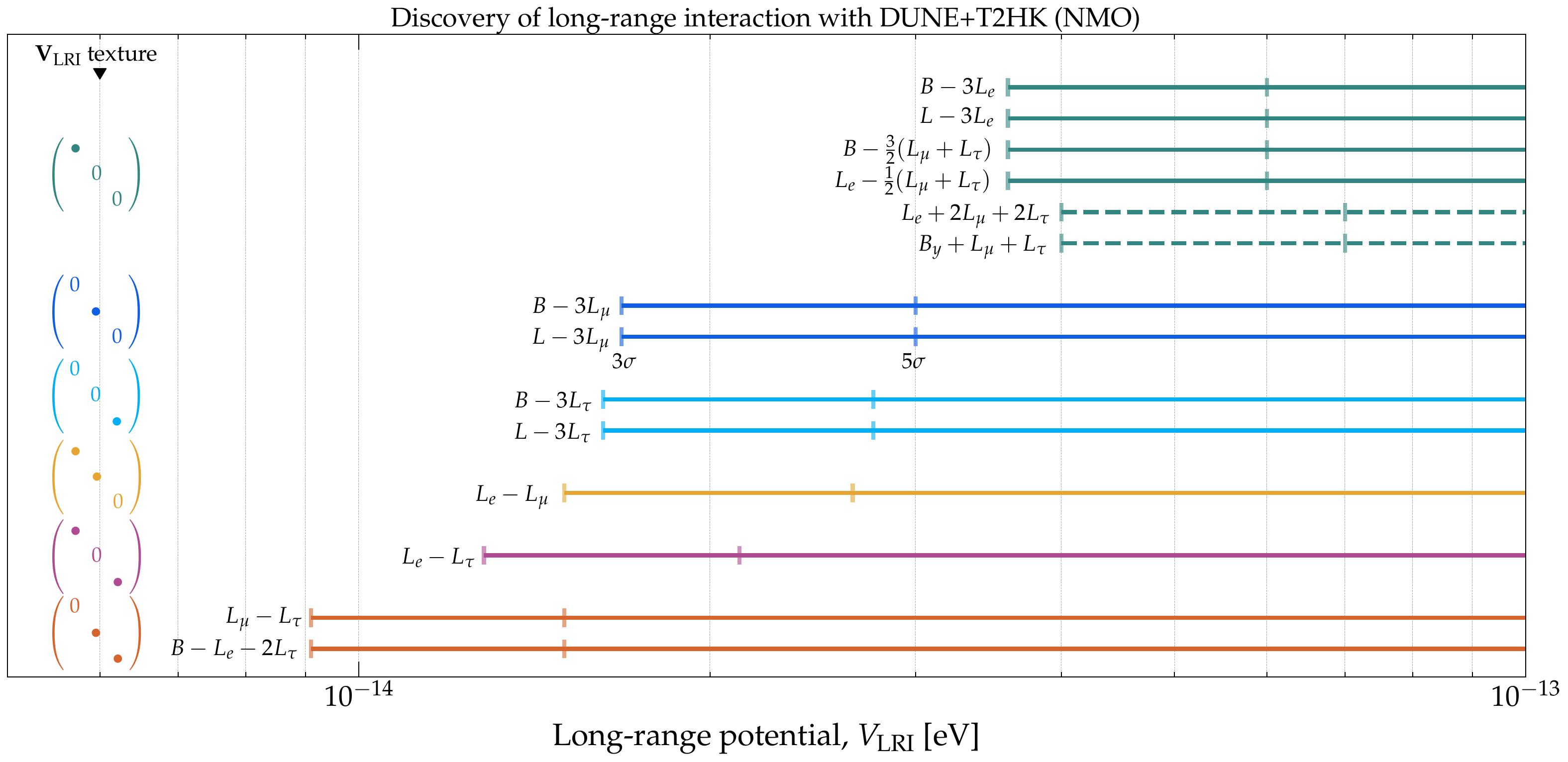}
	\caption{\textit{Projected discovery prospects of the new matter potential induced by our candidate $U(1)^\prime$ symmetries.}  The true neutrino mass ordering is assumed to be normal.  Results are for DUNE and T2HK combined; the test statistic is \equ{delta_chi2_disc} computed for their combination.   See sections~\ref{sec:stat_methods} and \ref{sec:discovery} for details.  The symmetries are grouped according to the texture of the new matter potential that they introduce, $\mathbf{V}_{\rm LRI}$ in table~\ref{tab:charges}.  The numerical values of the results are in table~\ref{tab:discovery_strength}.  \textit{Symmetries that induce equal or similar potential texture have equal or similar discovery prospects.  Discoverable ranges of $V_{\rm LRI}$ lie near the value of the standard-oscillation terms in the Hamiltonian, since this triggers resonances in the oscillation probabilities (section~\ref{sec:hamiltonians}).}}
	\label{fig:discovery_3s_5s_dune_t2hk}
\end{figure}
%=================================================

The limits in \figu{all_symmetry} exhibit step-like transitions occurring at different values of $m_{Z^\prime}$.  As explained in \Refe~\cite{Bustamante:2018mzu} (see also \Refes~\cite{Agarwalla:2023sng, Singh:2023nek} and section~\ref{sec:yukawa_interaction}), the transitions occur when the interaction range, $1/m_{Z^\prime}$, reaches the distance to a new celestial body.  Because different bodies have different abundances of electrons, protons, and neutrons, the tightest limits on $V_{\rm LRI}$ from \figu{constraints_on_pot_dune_t2hk-NMO} do not necessarily translate into the tightest limits on $G^\prime$ in \figu{all_symmetry}.  Once again, our results broaden the perspectives first put forward by \Refe~\cite{Singh:2023nek}: \textit{regardless of which of our candidate $U(1)^\prime$ symmetries is responsible for inducing long-range neutrino interactions, DUNE and T2HK may outperform existing limits on the coupling strength of the new $Z^\prime$ mediator.}  (See section~\ref{sec:intro} for an explanation of why the limits coming from flavor measurements of high-energy astrophysical neutrinos in \figu{all_symmetry} are in reality no match, for now, for DUNE and T2HK.)

References~\cite{Dolgov:1995hc, Blinnikov:1995kp, Joshipura:2003jh, Grifols:2003gy, Bustamante:2018mzu} indicated that if the relic neutrino background consists of equal numbers of $\nu_e$ and $\bar{\nu}_e$ it may partially screen out the long-range matter potential sourced by distant electrons by inducing corrections to the mass of the $Z^\prime$.  We have not considered this effect in our analysis, but, like \Refe~\cite{Bustamante:2018mzu}, we point out that it would affect the sensitivity to coupling strengths $G^\prime \lesssim 10^{-29}$, for which the Debye length of this effect, \ie, the distance at which it becomes appreciable, is about a factor-of-10 smaller~\cite{Joshipura:2003jh} than the interaction length to which we are sensitive in \figu{all_symmetry}.  A recent recalculation of the magnitude of the screening in \Refe~\cite{Chauhan:2024qew} suggests that it might have a stronger impact on the constraints on long-range interactions; however, a detailed assessment of this possibility within our analysis lies beyond the scope of the present work.

%=================================================
\begin{figure}
	\centering
	\includegraphics[width=0.85\textwidth]{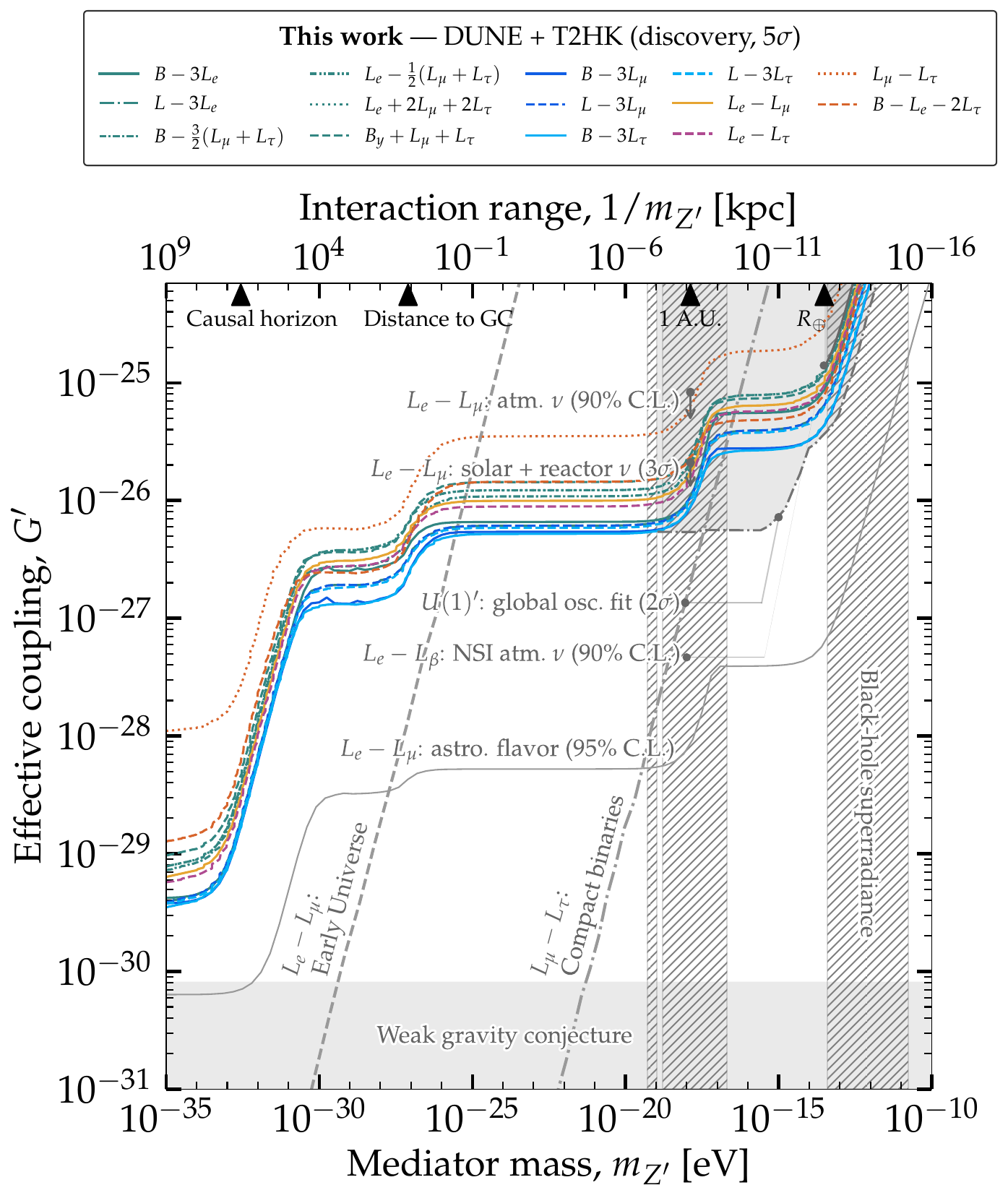}
	\caption{\textit{Projected discovery prospects of the effective coupling of the new gauge boson, $Z^\prime$, that mediates flavor-dependent long-range neutrino interactions.}  Results are for DUNE and T2HK, combined, after 10 years of operation, and for each of our candidate $U(1)^\prime$ symmetries (table~\ref{tab:charges}).  For this figure, we assume that the true neutrino mass ordering is normal. For each symmetry, the discovery prospects of the coupling, $G^\prime$, as a function of the mediator mass, $m_{Z^\prime}$, are converted from the discovery prospects on $V_{\rm LRI}$ in \figu{discovery_3s_5s_dune_t2hk} (also in table~\ref{tab:discovery_strength}) using the expressions for $V_{\rm LRI}$ in table~\ref{tab:charges}.  The existing limits are the same as in \figu{constraints_discovery_dune_t2hk-NMO}.  See sections~\ref{sec:stat_methods} and \ref{sec:discovery} for details.  \textit{DUNE and T2HK may discover long-range interactions, if they induce a matter potential comparable to the standard-oscillation terms of the Hamiltonian, regardless of which $U(1)^\prime$ symmetry is responsible for inducing them.}}
	\label{fig:discovery_all_DUNE+T2HK}
\end{figure}
%=================================================

%=================================================
\subsection{Discovery of new neutrino interactions}
\label{sec:discovery}
%=================================================

Figure~\ref{fig:discovery_3s_5s_dune_t2hk} (also, table~\ref{tab:discovery_strength}) shows, for each symmetry, the projected range of values of $V_{\rm LRI}$ that would result in the discovery of the presence of a new matter potential with a statistical significance of $3\sigma$ or $5\sigma$.  The ranges of values that can be discovered are similar to the ranges of values that can be constrained (\figu{constraints_on_pot_dune_t2hk-NMO}), since in both cases it is the size of the standard-oscillation term in the Hamiltonian that determines the sensitivity.  Like when placing constraints, symmetries whose matter potentials have equal or similar texture yield equal or similar discovery prospects.  Symmetries that affect primarily the $\nu_\mu \to \nu_\mu$ and $\bar{\nu}_\mu \to \bar{\nu}_\mu$ disappearance probabilities require smaller values of $V_{\rm LRI}$ to be discovered, due to the event rates being largest in the disappearance channels (\figu{event_spectra}).

Figure~\ref{fig:discovery_all_DUNE+T2HK} shows the associated discoverable regions of $G^\prime$ as a function of $m_{Z^\prime}$, converted from the discoverable intervals of $V_{\rm LRI}$ via the expressions for the potential sourced by celestial bodies in table~\ref{tab:charges}, just like we did for the constraints.  The results for discovery exhibit the same step-like transitions as for constraints, and the hierarchy of discoverability of the symmetries in \figu{discovery_all_DUNE+T2HK} is, as expected, the same as that of the constraints in \figu{all_symmetry}.  In \figu{constraints_discovery_dune_t2hk-NMO}, the discoverable region is the envelope of all the individual curves  in \figu{discovery_all_DUNE+T2HK}.

The results in figs.~\ref{fig:discovery_3s_5s_dune_t2hk} and \ref{fig:discovery_all_DUNE+T2HK} are the first reported discovery prospects in DUNE and T2HK of the full list of candidate $U(1)^\prime$ gauge symmetries in table~\ref{tab:charges}.  \textit{DUNE and T2HK may discover long-range interactions, regardless of what is the $U(1)^\prime$ symmetry responsible for inducing them}, provided the new matter potential is roughly within $10^{-14}$--$10^{-13}$~eV, \ie, comparable to the standard-oscillation terms.

%=================================================
\subsection{Distinguishing between symmetries}
\label{sec:confusion-theory}
%=================================================

Finally, we forecast how well, in the event of discovery of a new neutrino interaction, DUNE and T2HK could identify which of our candidate $U(1)^\prime$ symmetries is responsible for it.  

Figure~\ref{fig:confusion-matrix} shows this via confusion matrices.  They depict the statistical separation between pairs of symmetries, one true and one test, computed using the test statistic in \equ{delta_chi2_conf} for the combination of DUNE and T2HK.  We show results assuming two illustrative values of the new matter potential, $V_{\rm LRI} = 10^{-14}$~eV and $6 \cdot 10^{-14}$~eV.  The higher the potential is, the more prominent the effects of the new interaction are, and the easier it becomes to contrast event distributions due to competing symmetries.  The confusion matrices are nearly, but not fully, symmetric, since the true and test event spectra are treated differently (section~\ref{sec:stat_methods}).

The separation is clearer between symmetries whose matter potential matrices, $\mathbf{V}_{\rm LRI}$, have different textures; see table~\ref{tab:charges}.  Conversely, the separation is blurred between symmetries whose matter potentials have similar texture, \eg, between $B - 3 L_e$ and $L_e + 2 L_\mu + 2 L_\tau$, and it is null between symmetries whose matter potentials have equal texture, \eg, between $B - 3 L_e$ and $L - 3 L_e$.  This is a fundamental limitation; it persists regardless of the value of $V_{\rm LRI}$.  As when constraining (section~\ref{sec:constraints_pot}) and discovering (section~\ref{sec:discovery}) a new interaction, symmetries that affect primarily the disappearance probabilities, \ie, $L_\mu-L_\tau$ and $B-L_e-2L_\tau$, introduce features into the event rate that may be more easily spotted due to higher event rates, and are thus more easily distinguished from alternative symmetries.  Conversely, symmetries that affect primarily the appearance probabilities are less easily distinguished from alternative symmetries.

%=================================================
\begin{figure}
	\centering
	\includegraphics[width=1\textwidth]{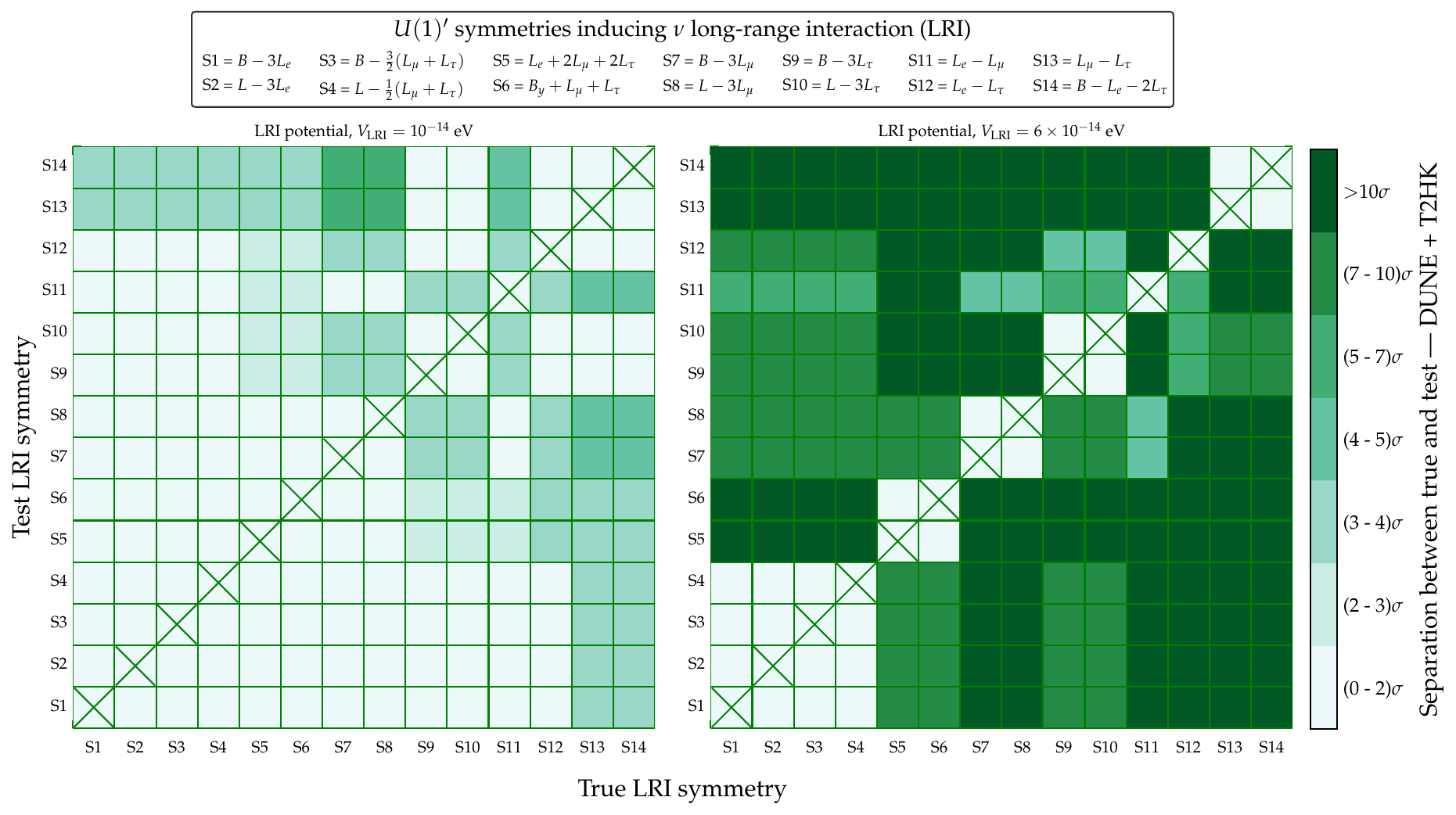}
	\caption{\textit{Confusion matrices showing the degree of separation between true and test $U(1)^\prime$ symmetries.}  The separation is evaluated using the test statistic in \equ{delta_chi2_conf}, and expressed in terms of number of standard deviations, $\sigma$, between the symmetries.  Results are for DUNE and T2HK combined, after 10 years of operation, assuming normal neutrino mass ordering.  We show results for two illustrative values of the new matter potential, $V_{\rm LRI} =  10^{-14}$~eV (\textit{left}) and $6 \cdot 10^{-14}$~eV (\textit{right}).  See sections~\ref{sec:stat_methods} and \ref{sec:confusion-theory} for details.  \textit{Distinguishing between competing symmetries may be feasible, especially for higher values of $V_{\rm LRI}$ and when the texture of the long-range matter interaction potential of each symmetry (table~\ref{tab:charges}) is different.}}
	\label{fig:confusion-matrix}
\end{figure}
%=================================================

%=================================================
\section{Summary and conclusions}
\label{sec:conclusion}
%=================================================

The growing precision achieved by neutrino oscillation experiments endows them with the capability to look for new neutrino interactions with matter that could modify the transitions between $\nu_e$, $\nu_\mu$, and $\nu_\tau$, revealing long-sought physics beyond the Standard Model.  The possibility that the interaction has a long range --- which we focus on --- is particularly compelling.  In this case, neutrinos on Earth could experience a large potential sourced by vast repositories of matter in the local and distant Universe --- the Earth, Moon, Sun, Milky Way, and the cosmological distribution of matter --- thereby enhancing our chances of probing the new interaction.

We have constructed the new interaction by gauging accidental global, anomaly-free $U(1)$ symmetries of the Standard Model that involve combinations of lepton and baryon numbers.  Doing this introduces a new neutral gauge boson, $Z^\prime$, that acts as mediator and induces a matter potential sourced by electrons, neutrons, or protons, depending on the specific symmetry considered.  The interaction range is inversely proportional to the mediator mass, which, along with its coupling strength, is a priori unknown; they are to be determined experimentally.   We have explored masses between $10^{-35}$~eV and $10^{-10}$~eV, corresponding to an interaction range between Gpc and hundreds of meters.

We have studied the prospects of constraining, discovering, and identifying the symmetry responsible for new neutrino interactions in two leading next-generation long-baseline oscillation experiments, DUNE and T2HK, expected to start operations within the next decade.  An initial study~\cite{Singh:2023nek} showed that they could outperform the present-day sensitivity to long-range interactions because of their large sizes, advanced detectors, and intense neutrino beams.  However, that study explored only three out of the many possible candidate symmetries that could induce new interactions, and ones involving exclusively lepton numbers.  Since different symmetries affect flavor transitions differently, it remained to be determined whether the sensitivity claimed in \Refe~\cite{Singh:2023nek} applies broadly.

We have addressed this, motivated by present-day comprehensive searches for long-range interactions~\cite{Coloma:2020gfv}, by exploring a plethora of possible candidate symmetries (table~\ref{tab:charges}) which induce a new matter potential that affects only $\nu_e$, $\nu_\mu$, or $\nu_\tau$, or combinations of them.  To make our forecasts realistic, we base them on detailed simulations of DUNE and T2HK, including accounting for multiple detection channels, energy resolution, backgrounds, and planned operation times of their neutrino beams in $\nu$ and $\bar{\nu}$ modes.

Our conclusions cement and broaden earlier promising perspectives. 
Although the different symmetries have diverse effects on oscillations, we find that \textit{regardless of which symmetry is responsible for inducing new neutrino-matter interactions, including long-range ones, DUNE and T2HK may constrain them more strongly than ever before (figs.~\ref{fig:constraints_discovery_dune_t2hk-NMO}, \ref{fig:constraints_on_pot_dune_t2hk-NMO},  and \ref{fig:all_symmetry})  or may discover them (figs.~\ref{fig:constraints_discovery_dune_t2hk-NMO}, \ref{fig:discovery_3s_5s_dune_t2hk}, and \ref{fig:discovery_all_DUNE+T2HK}).}  The experiments are predominantly sensitive to a new matter potential whose size is comparable to the standard-oscillation potential (\figu{lrf_bounds_NH_selective}), since this induces resonant effects on the oscillation probabilities.  Also, they are predominantly sensitive to new interactions that affect the disappearance channels, $\nu_\mu \to \nu_\mu$ and $\bar{\nu}_\mu \to \bar{\nu}_\mu$, since they have higher event rates, making it easier to spot subtle effects.  

In addition, for the first time, we report that \textit{it may be possible to identify the symmetry responsible for the new interaction, or to narrow down the possibilities (\figu{confusion-matrix})}, especially if the new matter potential is relatively large.  There is, however, an unavoidable limitation to disentangling the effects of two competing symmetries whose effects on the flavor transitions are equal or similar. Nevertheless, in all cases, combining events detected by DUNE and T2HK is key to lifting degeneracies between standard and new oscillation parameters that limit the sensitivity of each experiment individually.

Overall, our results demonstrate that the reach of DUNE and T2HK to probe new neutrino interactions is not only deep, but also broad in its scope.
\blankpage 
%%%%%%%%%%%%%%%%%%%%%%%%%%%% Chapter-4 %%%%%%%%%%%%%%%%%%%%%%%%%%%%%%
%%%%%%%%%%%%%%%%%%%%%%%%%%%%%%% CHAPTER - 4 %%%%%%%%%%%%%%%%%%%%%%%%%%%%%%%%%%%%
\chapter{Constraining Lorentz invariance violation with next-generation long-baseline experiments}
\label{C4}
%=============================%
\section{Introduction and motivation}
\label{sec:introduction}
%=============================% 
It is a well-established fact that the Lorentz symmetry is an exact symmetry of Nature. As a consequence, the Standard Model of particle physics conserves Lorentz symmetry. However, there exist several models, unifying SM and general relativity, that violate the Lorentz and CPT symmetry at the Planck scale ($\sim10^{19}$ GeV). In this chapter, we focus on Lorentz Invariance violation (LIV). This can be realized at a low energy scale accessible to the current experiments under Standard Model Extension (SME) framework. Various neutrino experiments are at the forefront to test the Lorentz Invariance Violation at a low energy scale. For example, in an attempt to understand the excess of $\nu_e$ signal events in the $\nu_{\mu}$ beam, LSND collaboration~\cite{LSND:2005oop}, searched for the possible signature of LIV in the context of neutrino oscillation. They did not find any signature of LIV and put strong constraints on relevant LIV parameters.

Several other experiments have made efforts to search for LIV, which include MINOS~\cite{MINOS:2008fnv, MINOS:2010kat, MINOS:2012ozn}, MiniBooNE~\cite{MiniBooNE:2011pix}, Double Chooz~\cite{DoubleChooz:2012eiq}, Super-K~\cite{Super-Kamiokande:2014exs}, IceCube~\cite{IceCube:2010fyu} and T2K~\cite{Abe:2017eot}. None of these experiments found any positive signal of LIV and set competitive bounds on various LIV parameters. In addition to the aforementioned studies by the experimental collaborations, there are various independent works towards the exploration of LIV with accelerator neutrinos in long-baseline experiments~\cite{Dighe:2008bu, Barenboim:2009ts, Rebel:2013vc, Diaz:2015dxa, deGouvea:2017yvn, Barenboim:2017ewj, Barenboim:2018lpo, Barenboim:2018ctx, Majhi:2019tfi, Fiza:2022xfw, Majhi:2022fed}, reactor antineutrinos in short-baseline experiments~\cite{Giunti:2010zs}, atmospheric neutrinos~\cite{Datta:2003dg, Chatterjee:2014oda, SinghKoranga:2014mxh, Sahoo:2021dit, Sahoo:2022rns}, solar neutrinos~\cite{Diaz:2016fqd}, and high-energy neutrinos from astrophysical sources~\cite{Hooper:2005jp, Tomar:2015fha, Liao:2017yuy}. Recently, the KATRIN experiment, using the data from the first scientific run, placed limits on some of the oscillation-free LIV parameters that can't be probed by the time-of-flight or neutrino oscillation experiments~\cite{KATRIN:2022qou}. Note that due to $SU(2)_{L}$ gauge invariance in the electroweak sector of the SME Lagrangian, the bounds on the LIV parameters obtained from the neutrino experiments and the charged-lepton sector may be related to each other~\cite{Crivellin:2020oov}. An exhaustive list of the constraints on all the CPT-violating and CPT-conserving LIV parameters can be found in Ref.~\cite{Kostelecky:2008ts}.

In the present work, we mainly focus on the capability of the upcoming long-baseline experiments, DUNE and T2HK in isolation and combination, to constrain the LIV parameters. We derive the sensitivities of these experiments to place competitive limits on the off-diagonal CPT-violating LIV parameters ($a_{\alpha\beta}$ where $\alpha, \beta = e, \mu, \tau$ and $\alpha\neq\beta$) and for the first time, the off-diagonal CPT-conserving LIV parameters ($c_{\alpha\beta}$ where $\alpha, \beta = e, \mu, \tau$ and $\alpha\neq\beta$). We study the impact of these LIV parameters and their associated phases at the probability level. Then we shift our attention to explore the possible degeneracies between the standard oscillation parameters ($\theta_{23}$ and $\delta_{\rm CP}$) and the above-mentioned LIV parameters. Finally, we derive the expected constraints on these LIV parameters using the upcoming long-baseline experiments, DUNE and T2HK in standalone mode and also in combination, considering their state-of-the-art simulation details. To understand various interesting features of our numerical results, we derive simple and compact analytical expressions of the oscillation probabilities for both appearance and disappearance channels.

This chapter is organized as follows. In section~\ref{sec:LIV}, we discuss the formalism of neutrino oscillation in the presence of Lorentz Invariance Violation and the effects of various LIV parameters on the appearance and disappearance probabilities. In section~\ref{sec:LBL}, we give the details of the long-baseline experimental setups considered for our work and discuss the expected synergies between DUNE and T2HK in various aspects. Also, this section discusses the effect of LIV parameters at the event level. We dedicate section~\ref{sec:RnA} to describe the numerical technique used for our analyses. We present our results in section~\ref{sec:results}, where we show the correlations among different LIV parameters, the atmospheric mixing angle ($\theta_{23}$), and the Dirac CP phase ($\delta_{\rm CP}$) and finally the expected bounds on the CPT-conserving and CPT-violating LIV parameters. We summarize our results and give our concluding remarks in section~\ref{sec:SnC}. In appendix~\ref{appndx}, we compare the numerical (exact) and analytical (approximate) probabilities. In appendix~\ref{appndx-B}, we discuss the neutrino appearance and disappearance event spectra for DUNE and T2HK in standard interaction case and with new physics.
%=============================%
\section{Neutrino oscillation in presence of Lorentz invariance violation}
\label{sec:LIV}
%============================%
\subsection{Theoretical formalism of LIV}
The Lorentz invariance has been considered to be an inalienable part in both the Standard Model of particle physics and General Relativity for the global as well as the local variables. However, a few proposed models in String Theory~\cite{Polyakov:1987ez, Kostelecky:1988zi, Kostelecky:1989jp, Kostelecky:1990pe, Kostelecky:1991ak, Kostelecky:1995qk, Kostelecky:1999mu, Kostelecky:2000hz} and Loop Quantum Gravity~\cite{Gambini:1998it, Alfaro:2002xz, Sudarsky:2002ue, Amelino-Camelia:2002aqz, Ng:2003jk} allow the Lorentz invariance violation while attempting a unification of gravity with the SM gauge fields at the Planck scale ($M_P \sim 10^{19}$ GeV). Here, we consider the mechanism proposed in the string theory, which spontaneously breaks the CPT and Lorentz symmetries at a higher dimension $(>4)$ of space-time. One can realize such a violation of Lorentz and CPT symmetries in the realistic four-dimensional space-time by introducing new interactions in the minimal SME framework in the form of tiny perturbations. In this minimal SME framework, the strength of such interactions are expected to be suppressed by  $\sim1/M_P$~\cite{Colladay:1998fq, Kostelecky:2003fs, Colladay:1996iz, Kostelecky:2000mm, Kostelecky:2003cr, Bluhm:2005uj}, manifesting the effect of Planck scale physics at low energy. The impacts of such LIV interactions can be experienced by the fundamental particles in a broad category of experiments via coherent, interference, or extreme effects.

By virtue of mass-induced neutrino flavor oscillations, the neutrinos are sensitive to the LIV effects while propagating through space-time. Under the minimal SM extension (SME), the Lagrangian density of the induced renormalizable and gauge-invariant LIV interaction terms for the left-handed neutrinos can be expressed as~\cite{Kostelecky:2011gq, KumarAgarwalla:2019gdj, Antonelli:2020nhn, Sahoo:2021dit, Sahoo:2022rns}:
\begin{align}
	\mathcal{L}_{\rm LIV} & = -\frac{1}{2}\left[a^{\mu}_{\alpha\beta}\,\overline{\psi}_\alpha\,\gamma_{\mu}\,P_L\,\psi_\beta - i c^{\mu\nu}_{\alpha\beta}\,\overline{\psi}_\alpha\,\gamma_{\mu}\,\partial_\nu P_L\,\psi_\beta \right] + h.c.\,, 
	\label{equ:LIV-1}
\end{align}
where $P_L$ is the projection operator, $a^{\mu}$ and $c^{\mu\nu}$ are the CPT-violating and CPT-conserving LIV parameters, respectively. Here, $(\mu,\,\nu)$ are space-time indices, and $\alpha,\,\beta$ are the neutrino-flavor indices. The CPT-violating LIV parameter changes its sign under CPT transformation while the CPT-conserving one does not (see Refs.~\cite{Sahoo:2021dit, Kostelecky:2003cr}). Now, considering a realistic scenario where the neutrinos can propagate through the Earth matter, the effective Hamiltonian of an ultra-relativistic left-handed neutrino, in a three neutrino mixing scenario, can be expressed in the flavor basis as ~\cite{Kostelecky:2011gq, Sahoo:2021dit, Sahoo:2022rns}:
\begin{align}
	\mathcal{H}_{\rm eff} & = \frac{1}{2E}\,U\,\Delta m^2\,U^\dagger
	+ \frac{1}{E}\big(a^\mu_{L} p_\mu
	- c^{\mu\nu}_{L} p_\mu p_\nu \big) + \sqrt{2}G_FN_e\tilde{I}.
	\label{equ:LIV-2}
\end{align} 
The first term in the above equation, $U$ represents the three-neutrino unitary mixing matrix, also known as PMNS matrix, and the $\Delta m^2$ part contains two independent mass-squared splittings in the form of a diagonal-matrix: ${\rm diag}(0,\,\Delta{m}^2_{21},\,\Delta{m}^2_{31})$. The second term shows the strength of induced potential with left-handed neutrino due to LIV, and $p$ is the four-momenta. The third term defines the effective matter-potential induced due to the elastic charged-current scattering between $\nu_e$ and electron. Here, $G_F$ is an electroweak coupling constant, also known as the Fermi constant, $N_e$ is the number density of the ambient electrons present in matter, and $\tilde{I}$ is a diagonal matrix of the form diag$(1,\,0,\,0)$. The scalar part of the last term can be parameterized with matter density as $\sqrt{2}G_FN_e$ $\approx$ $7.6\,\times 10^{-23}\cdot Y_e \cdot \rho\left(\rm g/cm^3\right)$ GeV, where $Y_e$ is the relative electron number density in the ambient matter having an average Earth-matter density $\rho$.

In this study, we only focus on the time-like component ($\mu,\,\nu\,=\,0$) of LIV parameters in an isotropic space-time coordinate. From here onwards, we will consider $(a^0_L)_{\alpha\beta} \equiv a_{\alpha\beta}$ and $(c^{00}_L)_{\alpha\beta} \equiv c_{\alpha\beta}$.  Using the Sun-centered celestial-equatorial coordinate (see Ref.~\cite{Kostelecky:2008ts}) as an approximated inertial frame of reference, the \equ{LIV-2} can be written in the following fashion:

\begin{align}
	\mathcal{H}_{\rm eff} = \; \frac{1}{2E} 
	U\left(\begin{array}{ccc}
		0 & 0 & 0                 \\
		0 & \Delta m^{2}_{21} & 0 \\
		0 & 0 & \Delta m^{2}_{31} \\
	\end{array}\right)U^{\dagger}
	+&\left(\begin{array}{ccc}
		a_{ee} & a_{e\mu} & a_{e\tau} \\
		a^*_{e\mu} & a_{\mu\mu} & a_{\mu\tau} \\
		a^*_{e\tau} & a^*_{\mu\tau} & a_{\tau\tau}
	\end{array} \right) \nonumber \\
	-&\frac{4}{3} E
	\left(
	\begin{array}{ccc}
		c_{ee} & c_{e\mu} & c_{e\tau} \\
		c^*_{e\mu} & c_{\mu\mu} & c_{\mu\tau} \\
		c^*_{e\tau} & c^*_{\mu\tau} & c_{\tau\tau}
	\end{array}
	\right) 
	+ \sqrt{2}G_{F}N_{e}
	\left(\begin{array}{ccc}
		1 & 0 & 0 \\ 
		0 & 0 & 0 \\ 
		0 & 0 & 0
	\end{array}\right).
	\label{equ:LIV-3}
\end{align}
For the case of right-handed antineutrino $U \to U^*$, $a_{\alpha\beta} \to -a_{\alpha\beta}^*$, $c_{\alpha\beta} \to c_{\alpha\beta}^*$ and $\sqrt{2}G_FN_e \to -\sqrt{2}G_FN_e$. Note that an extra fraction $4/3$ appears due to the choice of isotropic coordinates. It is essential to note that due to the origin of the LIV parameters from the spontaneous symmetry breaking and the hermiticity of the underlying theory, $a_{\alpha\beta}$ and $c_{\alpha\beta}$ must be real valued quantities in space-time. However, while implementing them in an effective Hamiltonian in flavor space, the off-diagonal components of these parameters can be treated as complex, due to hermiticity of the effective Hamiltonian (\equ{LIV-3}).

To have an estimate of the strength of LIV parameters which may affect the outcome of the long-baseline experiments under consideration, we compare the first three terms in the neutrino propagation Hamiltonian as shown in \equ{LIV-3}. The first term governs the neutrino oscillation in vacuum, whereas the second and third terms are the contribution from CPT-violating and CPT-conserving LIV, respectively. For typical long-baseline experiments with neutrino energy in the GeV range, the relevant scale for standard atmospheric neutrino oscillation is $\Delta m^2_{31}/{2E}\sim 10^{-22}$ GeV. So, in order to have noticeable effects from LIV, the strength of the parameters in the second and third terms should be around the same order as the standard neutrino oscillation scale. So, we observe that both CPT-violating ($a_{\alpha\beta}$) and CPT-conserving ($E\times c_{\alpha\beta}$) LIV parameters should have the strength of the order $10^{-22}$ GeV to have visible effects over standard neutrino oscillation in the matter.

\begin{table}[h!]
	\centering
	\begin{center}
		\begin{adjustbox}{width=1\textwidth}
			\begin{tabular}{|c| c| c| c|}
				\hline \hline
				\multicolumn{4}{|c|}{Existing constraints on CPT-violating LIV parameters} \\ \hline
				Experiments & $a_{e\mu} ~[ \rm GeV ]$ &  $a_{e\tau} ~[ \rm GeV ]$ & $a_{\mu\tau} ~[ \rm GeV ]$ \\ \hline
				\multirow{2}{*}{Super-K (95\% C.L.)} & $\mathrm{Re}(a_{e\mu}) < 1.8\times10^{-23}$  & $\mathrm{Re}(a_{e\tau}) < 4.1\times10^{-23}$ & $\mathrm{Re}(a_{\mu\tau}) < 0.65\times10^{-23}$ \\
				
				& $\mathrm{Im}(a_{e\mu}) < 1.8\times10^{-23}$  & $\mathrm{Im}(a_{e\tau}) < 2.8\times10^{-23}$ & $\mathrm{Im}(a_{\mu\tau}) < 0.51\times10^{-23}$ \\
				\hline
				\multirow{2}{*}{IceCube (99\% C.L.)} &\multirow{2}{*}{--} & \multirow{2}{*}{--} & $|\mathrm{Re}(a_{\mu\tau})| < 0.29\times10^{-23}$ \\ 
				&\multirow{2}{*}{--} & & $|\mathrm{Im}(a_{\mu\tau})| < 0.29\times10^{-23}$ \\ 
				\hline 
				\hline
				\multicolumn{4}{|c|}{Existing constraints on CPT-conserving LIV parameters} \\ \hline 
				
				Experiments & $c_{e\mu}$ &  $c_{e\tau}$ & $c_{\mu\tau}$ \\ \hline
				
				\multirow{2}{*}{Super-K (95\% C.L.)} & $\mathrm{Re}(c_{e\mu}) < 8.0\times10^{-27}$  & $\mathrm{Re}(c_{e\tau}) < 9.3\times10^{-25}$ & $\mathrm{Re}(c_{\mu\tau}) < 4.4\times10^{-27}$ \\
				
				& $\mathrm{Im}(c_{e\mu}) < 8.0\times10^{-27}$  & $\mathrm{Im}(c_{e\tau}) < 1.0\times10^{-24}$ & $\mathrm{Im}(c_{\mu\tau}) < 4.2\times10^{-27}$ \\
				
				\hline
				
				\multirow{2}{*}{IceCube (99\% C.L.)} &\multirow{2}{*}{--} & \multirow{2}{*}{--} & $|\mathrm{Re}(c_{\mu\tau})| < 0.39\times10^{-27}$ \\ 
				&\multirow{2}{*}{--} & & $|\mathrm{Im}(c_{\mu\tau})| < 0.39\times10^{-27}$
				\\\hline\hline
				
			\end{tabular}
		\end{adjustbox}
	\end{center}
	\mycaption{Existing constraints on the off-diagonal CPT-violating and CPT-conserving LIV parameters from Super-K~\cite{Super-Kamiokande:2014exs} and IceCube~\cite{IceCube:2017qyp}.}
	\label{tab:existing_bounds}
\end{table}

Following the above-discussed formalism, various neutrino experiments have given bounds on the CPT-violating and CPT-conserving LIV parameters. In particular, a recent publication by the IceCube collaboration~\cite{IceCube:2017qyp}, where the analysis has been performed in an effective two-flavor oscillation scenario, presented the most stringent bounds on the CPT-violating and CPT-conserving LIV parameters in $\mu-\tau$ sector. Apart from this, there are also limits on both CPT-violating and CPT-conserving LIV parameters using the atmospheric neutrino data\footnote{Though the main focus of this paper is to explore the potential of DUNE and T2HK with their long-baseline setups, it is important to mention that these experiments will also collect huge atmospheric data which can be used to probe the LIV parameters under consideration in this paper. For instance, in Ref.~\cite{Hyper-Kamiokande:2018ofw}, the T2HK collaboration has already estimated the bounds on both CPT-violating and CPT-conserving LIV parameters which are 3 to 4 times more stringent than Super-K. This improvement is attributed to the larger fiducial volume of T2HK than Super-K. The DUNE collaboration has derived future limits on the LIV parameters associated with the Lorentz-violating operators of higher dimension (mass dimension $>$ 4) in Ref.~\cite{DUNE:2020ypp} with its atmospheric sample.} from Super-K~\cite{Super-Kamiokande:2014exs}. In table~\ref{tab:existing_bounds}, we tabulate the existing limits on various off-diagonal LIV parameters from these two experiments. Note that in our work, we represent the off-diagonal LIV parameters as $|R|\cdot e^{i\phi}$, where $|R|$ denotes the magnitude of the LIV parameter and $\phi$ is the complex phase associated with it.

Several existing studies, for example, T2K near detector~\cite{Abe:2017eot}, MiniBooNE~\cite{MiniBooNE:2011pix}, MINOS near detector~\cite{MINOS:2008fnv}, MINOS far detector~\cite{MINOS:2010kat}, explore the possibility of sidereal time variation\footnote{Sidereal time is measured with respect to fixed distant stars rather than the Sun. The sidereal day is the time taken by Earth for one complete rotation ($360^\circ$), measured with respect to a fixed star, which is around 23 hours 56 minutes. However, the solar day is 24 hours long. The longer solar day is due to the simultaneous orbiting of Earth around Sun and the rotation around itself.}} of neutrino flavor transition, considering the rotation of Earth in the LIV background. But these studies did not get any signature of the sidereal time dependence of neutrino oscillation probability and put limits on various combinations of LIV parameters. The sidereal dependence of the transition probability requires the presence of the spatial anisotropic LIV parameters~\cite{Diaz:2009qk}. However, since in the present work, we consider only the time-like components of the LIV parameters (one-at-a-time) in an isotropic coordinate system, the limits obtained from our work will not be directly applicable to the parameters that are probed in the above-mentioned studies.

%=============================%
\subsection{Analytical expressions of the oscillation probabilities with LIV}
%=============================%

The presence of LIV terms in the neutrino Hamiltonian would affect the neutrino propagation through a medium, consequently modifying the neutrino flavor transition probability. So, it is possible to probe LIV in various neutrino oscillation experiments. In order to have an analytical understanding of the neutrino flavor transition probabilities in the presence of LIV, we follow the approach given in Refs.~\cite{Kikuchi:2008vq, Agarwalla:2016fkh, KumarAgarwalla:2019gdj}, where authors use perturbation theory to calculate the neutrino evolution matrix in various BSM scenarios like the presence of neutral current NSI and CPT-violating LIV parameters. We use $\alpha~(\equiv \Delta m^2_{21}/\Delta m^2_{31})$, $\sin^2\theta_{13}$, and LIV parameters $a'_{\alpha\beta}~(\equiv a_{\alpha\beta}/\sqrt{2}G_F N_e)$ and $c'_{\alpha\beta}~(\equiv c_{\alpha\beta}E/\sqrt{2}G_F N_e)$ ($\alpha,\beta = e,\mu,\tau$) as the expansion parameters. In this work, we mainly probe the off-diagonal CPT-violating and CPT-conserving LIV parameters $a_{\alpha\beta}$ and $c_{\alpha\beta}$ ($\alpha,\beta=e,\mu,\tau;\,\alpha\neq\beta$), respectively.

\vspace{0.3cm}
$\bullet$  \textbf{$\nu_\mu\to\nu_e$ appearance channel:}
\vspace{0.3cm}

The expression for the $\nu_\mu\to\nu_e$ transition probability, considering terms up to first-order in the above mentioned expansion parameters, can be written as~\cite{KumarAgarwalla:2019gdj},
\begin{align}\label{equ:pme_liv}
	\centering
	P_{\mu e} \simeq P_{\mu e}(\text{SI}) + P_{\mu e}(a_{e\beta}/c_{e\beta}) + \mathcal{O}(\alpha^2, \alpha \sin^2\theta_{13},a'^2_{e\beta},c'^2_{e\beta},a'^2_{\mu\tau},c'^2_{\mu\tau}),\,\,\,\,\,\,\,\beta=\mu,\tau.
\end{align}
The first term in right-hand-side (RHS) is the standard $\nu_\mu\to\nu_e$ appearance probability in the absence of any new physics parameters,
\begin{align}\label{equ:p_si}
	P_{\mu e}(\text{SI}) 
	&\simeq \mathbb{X} + \mathbb{Y}\cos(\delta_{\text{CP}} + \Delta),
\end{align}
where,
\begin{align}\label{equ:si_coeff}
	&\mathbb{X} = \sin^22\theta_{13}\sin^2\theta_{23} \frac{\sin^{2}\big[(1-\hat{A})\Delta \big]}{(1-\hat{A})^{2}};\nonumber \\
	&\mathbb{Y} = \alpha \sin2\theta_{12} \sin2\theta_{13} \sin2\theta_{23} \frac{\sin \hat{A}\Delta}{\hat{A}}  \frac{\sin \big[(1-\hat{A})\Delta\big]}{1-\hat{A}}; \nonumber \\
	&\hat{A} = \frac{2\sqrt{2}G_{F}N_{e}E}{\ldm}, \qquad  \Delta = \frac{\ldm L}{4E}.
\end{align}
The second term in RHS of \equ{pme_liv} are the contributions from the LIV parameters $a_{e\beta}/c_{e\beta}$ ($\beta = \mu,\tau$). For CPT-violating LIV cases, the expression of this term is;
\begin{align}\label{equ:p_cptv_liv}
	&P_{\mu e}(a_{e\beta}) 
	\simeq 
		2|a_{e\beta}|L\sin\tym\sin2\tzm \sin\Delta\big[\mathbb{Z}_{e\beta}\sin(\delta_{\rm CP} + \varphi_{e\beta}) +
		\mathbb{W}_{e\beta}\cos(\delta_{\rm CP} + \varphi_{e\beta})\big],\nonumber\\
\end{align}
and for CPT-conserving case,
\begin{align}\label{equ:p_cptc_liv}
	&P_{\mu e}(c_{e\beta}) 
	\simeq\frac{-8}{3}|c_{e\beta}|EL\sin\tym\sin2\theta_{23}\sin\Delta \big[\mathbb{Z}_{e\beta}\sin(\delta_{\rm CP} + \varphi_{e\beta}) +
		\mathbb{W}_{e\beta}\cos(\delta_{\rm CP} + \varphi_{e\beta})\big],\nonumber\\
\end{align}
where,
\begin{align}\label{equ:liv_coeff}
	&\mathbb{Z}_{e\beta} = 
	\begin{cases}
		- c_{23}  \sin \Delta, & \text{if}\ \beta=\mu. \\
		s_{23}  \sin \Delta, & \text{if}\ \beta=\tau.
	\end{cases} \nonumber \\
	&\mathbb{W}_{e\beta} = 
	\begin{cases}
		c_{23}  \big(\frac{s_{23}^{2}\sin \Delta}{c_{23}^{2}\Delta} + \cos \Delta \big), & \text{if}\ \beta=\mu. \\
		s_{23}  \big(\frac{\sin \Delta}{\Delta} - \cos \Delta \big), & \text{if}\ \beta=\tau.
	\end{cases} 
\end{align}
Note that the other off-diagonal LIV parameter $\amt/\cmt$ does not appear in the first-order terms. However, it may be present in higher-order terms and has a relatively smaller impact on the appearance probabilities. Note that for the appearance probability in the antineutrino case, one needs to apply $a_{\alpha\beta}\rightarrow-a^{*}_{\alpha\beta}$, $c_{\alpha\beta} \to c_{\alpha\beta}^*$, and $\hat{A}\to -\hat{A}$ in eqs.~(\ref{equ:p_si})--(\ref{equ:liv_coeff}). In appendix~\ref{appndx}, we show the validity of the approximate analytical expression of the appearance probability derived in this section by comparing it with the exact probability calculated numerically.

\begin{figure}[h!]
	\centering
	\includegraphics[width=\textwidth]{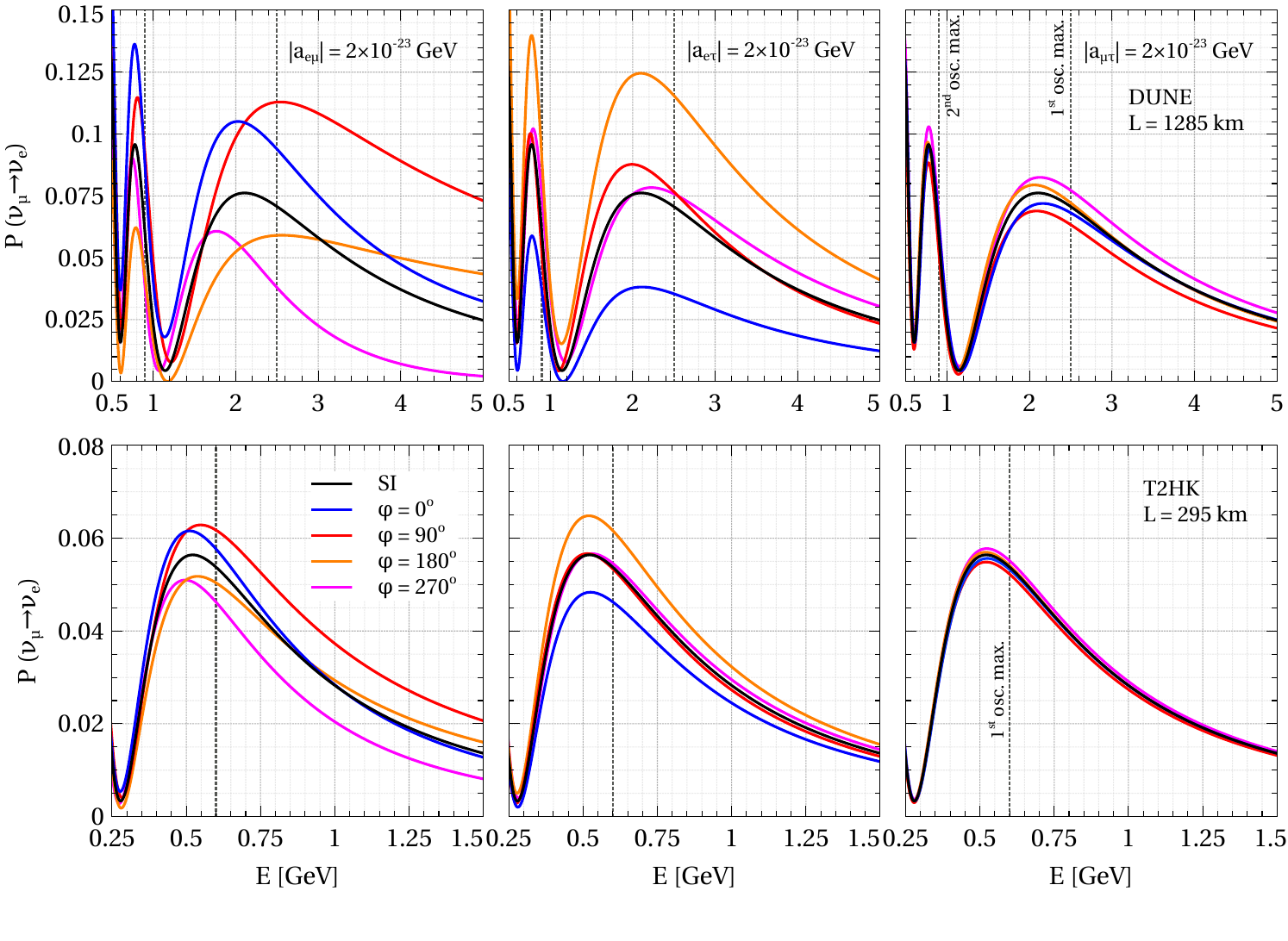}
	\vspace*{-10mm}
	\mycaption{$\nu_\mu\rightarrow\nu_e$ appearance probability as a function of energy in the presence of off-diagonal CPT-violating LIV parameters $a_{e\mu}$ (left column), $a_{e\tau}$ (middle column), and $a_{\mu\tau}$ (right column). The top and bottom rows correspond to the baselines of DUNE ($L=1285$ km) and T2HK ($L=295$ km), respectively. The black line in each panel shows the SI case (no LIV) and four colored lines correspond to four benchmark values of the phases associated with the LIV parameters: $0^\circ$, $90^\circ$, $180^\circ$, and $270^\circ$ with LIV strength $|a_{\alpha\beta}|=2.0\times10^{-23}$ GeV ($\alpha,\beta=e,\mu,\tau;\alpha\neq\beta$). The vertical grey-dashed lines in each panel show the energies at the first and second oscillation maxima. The values of the standard oscillation parameters used in this plot are mentioned in table~\ref{tab:params_value}.}
	\label{fig:app_prob}
\end{figure}

\begin{table}[htb!]
	\centering
	 \begin{adjustbox}{width=0.7\textwidth}
	\begin{tabular}{|c|c|c|c|c|c|c|}
		\hline
		\hline
		$\theta_{12}$ & $\theta_{13}$ & $\theta_{23}$ & $\delta_{\text{CP}}$ & $\Delta{m^2_{21}}~[\rm{eV}^2]$ & $\Delta{m^2_{31}}~[\rm{eV}^2]$ & Mass Ordering\\
		\hline
		$33.45^\circ$ & $8.62^\circ$ & $42.1^\circ$ & $230^\circ$ & $7.42\times10^{-5}$ & $2.51\times10^{-3}$ & NMO\\
		\hline
		\hline
	\end{tabular}
   \end{adjustbox}
	\caption{Benchmark values of the oscillation parameters used in our analysis~\cite{Esteban:2020cvm}. We consider normal mass ordering (NMO) throughout this work which corresponds to $m_1<m_2<m_3$.}
	\label{tab:params_value}
\end{table}

\begin{figure}[h!]
	\centering
	\includegraphics[width=\textwidth]{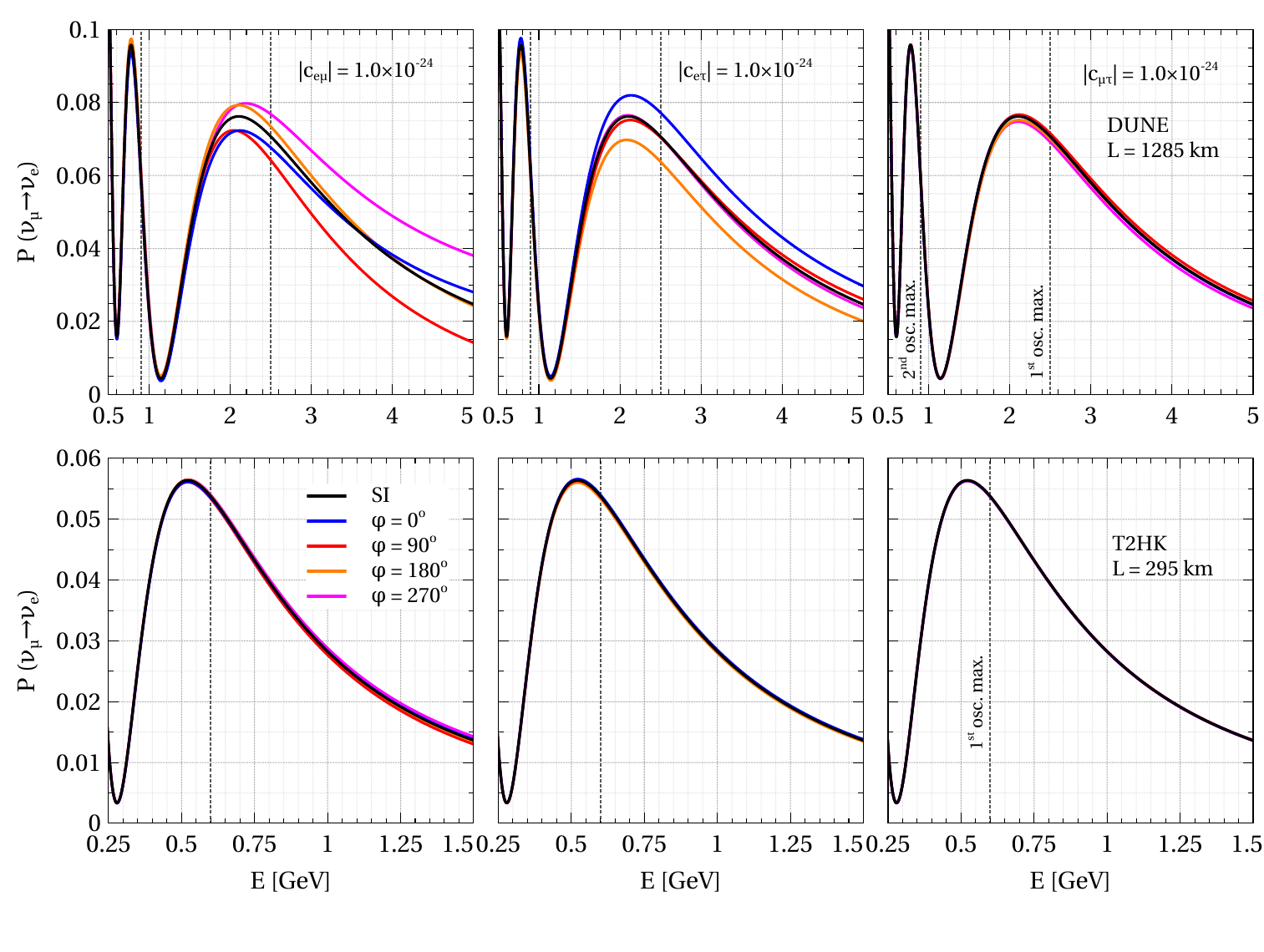}
	\vspace*{-10mm}
	\mycaption{$\nu_\mu\rightarrow\nu_e$ appearance probability as a function of energy in the presence of off-diagonal CPT-conserving LIV parameters $c_{e\mu}$ (left column), $c_{e\tau}$ (middle column), and $c_{\mu\tau}$ (right column). The top and bottom rows correspond to the baselines of DUNE ($L=1285$ km) and T2HK ($L=295$ km), respectively. The black line in each panel shows the SI case (no LIV) and four colored lines correspond to four benchmark values of the phases associated with the LIV parameters: $0^\circ$, $90^\circ$, $180^\circ$, and $270^\circ$ with LIV strength $|c_{\alpha\beta}|=1.0\times10^{-24}$ ($\alpha,\beta=e,\mu,\tau;\alpha\neq\beta$). The vertical grey-dashed lines in each panel show the energies at first and second oscillation maxima. The values of the standard oscillation parameters used in this plot are mentioned in table~\ref{tab:params_value}.}
	\label{fig:cptc_app_prob}
\end{figure}

In \figu{app_prob}, we plot $\nu_\mu\rightarrow\nu_e$ oscillation probability as a function of neutrino energy for the baseline $L=1285$ km (top row) and $L=295$ km (bottom row) in the standard interaction (SI) case where there is no LIV and in the presence of the CPT-violating LIV parameters. To plot exact oscillation probability, we use GLoBES software~\cite{Huber:2004ka, Huber:2007ji} and modify the probability calculator accordingly to introduce LIV. The values of the standard oscillation parameters used to calculate the probability are given in table~\ref{tab:params_value}. The left, middle, and right columns correspond to the appearance probability in the presence of  $a_{e\mu}$, $a_{e\tau}$, and $a_{\mu\tau}$ one-at-a-time, respectively with strength $|a_{\alpha\beta}|=2\times10^{-23}$ GeV. In each panel, the solid black curve shows the SI case. Four colored curves correspond to the probabilities in the presence of CPT-violating LIV parameter for the four chosen values of the associated phase, namely, 0, $90^\circ$, $180^\circ$, and $270^\circ$. It is clear from the figure that the impact of $\amt$ is marginal compared to $\aem$ and $\aet$. This behavior is obvious from our analytical expression in \equ{pme_liv} where $\amt$ does not appear at the leading order, contrary to the other two off-diagonal CPT-violating LIV parameters. In the presence of $\aem$, the appearance probability shows a significant deviation from the SI case depending on the value of the associated phase. Near first oscillation maxima, probability is maximum when $\phi_{e\mu}=90^\circ$ and minimum when $\phi_{e\mu}=270^\circ$. It can be explained using \equ{p_cptv_liv}, which shows the contribution from the CPT-violating LIV parameters to the oscillation probability. In case of $\aem$, the sign of $\mathbb{Z}$ ($\mathbb{W}$) in \equ{p_cptv_liv} is negative (positive) near the first oscillation maxima. Therefore, around the first oscillation maxima, the appearance probability is maximum (minimum) when the multiplicative factors associated with $\mathbb{Z}$ and $\mathbb{W}$ are negative (positive) and positive (negative), respectively. Since $\delta_{\rm CP}=230^\circ$, this happens at $\phi_{e\mu} = 90^\circ$ ($270^\circ$). However, in the middle panels, we observe that in the presence of $\aet$, the appearance probability is maximum (minimum) at $\phi_{e\tau}=180^\circ$ ($0^\circ$). This happens because now both $\mathbb{Z}$ and $\mathbb{W}$ are positive. So, the maximum (minimum) probability corresponds to the value of $\phi_{e\tau}$ for which both $\sin(\delta_{\rm CP} + \varphi_{e\beta})$ and $\cos(\delta_{\rm CP} + \varphi_{e\beta})$ in \equ{p_cptv_liv} are positive (negative). This occurs at $\phi_{e\tau} = 180^\circ$ ($0^\circ$) considering our benchmark value of $\delta_{\text{CP}}$.
Even though all these features can be observed both in the top and the bottom rows, we notice that the spread of the oscillation probability due to phase is significantly lesser in the bottom row, where we consider a comparatively smaller baseline ($L=295$ km). It is because of $L$ dependency in \equ{p_cptv_liv}, which adds the contribution from LIV parameters.

In \figu{cptc_app_prob}, we plot the appearance probability in the presence of three off-diagonal CPT-conserving LIV parameters, $\cem$ (left column), $\cet$ (middle column), and $\cmt$ (right column) with strength of $1.0\times10^{-24}$.
Here also, we use $L=1285$ km (top row) and $L=295$ km (bottom row). For DUNE ($L=1285$ km), in the presence of $c_{e\beta}$ ($\beta=\mu,\tau$), we observe that the impact of the associated phases is in the opposite order compared to the corresponding CPT-violating LIV parameters in \figu{app_prob}. It is because of the opposite sign of the LIV contributing term in the case of CPT-conserving LIV, as shown in \equ{p_cptc_liv}. As expected, $c_{\mu\tau}$ has almost a negligible effect on the probability in the case of DUNE. However, for T2HK ($L=295$ km), all three off-diagonal CPT-conserving LIV parameters have almost no impact on the oscillation probabilities because of the smaller baseline and comparatively lower neutrino energy.

\begin{figure}[h!]
	\centering	
	\includegraphics[width=\textwidth]{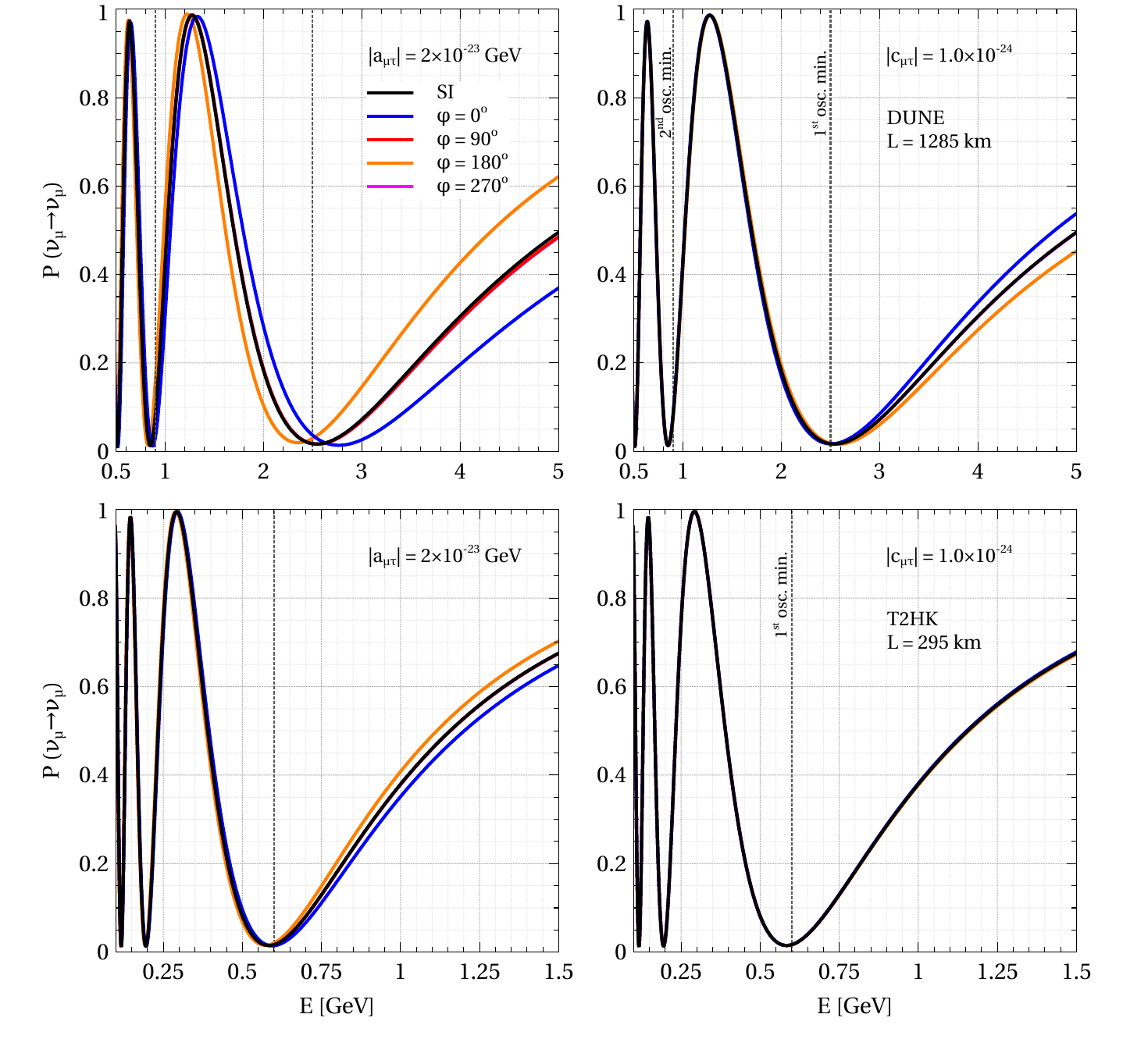}
	\vspace*{-10mm}
	\mycaption{$\nu_\mu\rightarrow\nu_\mu$ disappearance probability as a function of energy in the presence of off-diagonal LIV parameters $a_{\mu\tau}$ (left column), $c_{\mu\tau}$ (right column). The top and bottom rows correspond to the baselines of DUNE ($L=1285$ km) and T2HK ($L=295$ km), respectively. The black line in each panel shows the SI case (no LIV) and four colored lines correspond to four benchmark values of the phases associated with the LIV parameters: $0^\circ$, $90^\circ$, $180^\circ$, and $270^\circ$ with LIV strength $|a_{\alpha\beta}|=2.0\times10^{-23}$ GeV, $|c_{\alpha\beta}|=1.0\times10^{-24}$ ($\alpha,\beta=e,\mu,\tau;\alpha\neq\beta$). The vertical grey-dashed lines in each panel show the energies at first and second oscillation maxima. The values of the standard oscillation parameters used in this plot are mentioned in table~\ref{tab:params_value}.}
	\label{fig:disapp_prob}
\end{figure} 

\vspace{0.25cm}
$\bullet$  \textbf{$\nu_\mu\to\nu_\mu$ disappearance channel:}
\vspace{0.25cm}

Now we discuss the impact of the CPT-violating and the CPT-conserving LIV parameters on the $\nu_\mu\to\nu_\mu$ disappearance probability.
Following the same strategy as the appearance channel, we derive the compact analytical expression for $\nu_\mu\to\nu_\mu$ disappearance probability,
\begin{align}\label{equ:pmm}
	P_{\mu\mu} \simeq P_{\mu\mu}(\text{SI}) + P_{\mu\mu}(\amt/\cmt) +\mathcal{O}(\alpha^2, \alpha \sin^2\theta_{13},a'^2_{e\beta},c'^2_{e\beta},a'^2_{\mu\tau},c'^2_{\mu\tau}),\,\,\,\,\,\,\,\beta=\mu,\tau.
\end{align}
The first term in the RHS is the standard disappearance probability without any new physics contribution,
\begin{align}
	P_{\mu\mu}(\text{SI}) = P_{\mu\mu}(\text{vacuum},\text{two flavor})+\alpha\mathbb{P}+\sin^2\theta_{13}\mathbb{Q}+\alpha\sin\theta_{13}\mathbb{R},
\end{align}
where,
\begin{align}
	P_{\mu\mu}(\text{vacuum},\text{two flavor}) = 1-\sin^22\theta_{23}\sin^2\Delta.
\end{align}
$\mathbb{P}$, $\mathbb{Q}$, and $\mathbb{R}$ are defined as follows,
\begin{align}
	&\mathbb{P} = \cos^2\theta_{12}\sin^22\theta_{23}\sin2\Delta,\\
	&\mathbb{Q} = -4\sin^2\theta_{23}\frac{\sin^2(\hat{A}-1)\Delta}{(\hat{A}-1)^2}-\frac{2}{\hat{A}-1}\sin^22\tzm\left(\sin\Delta\cos \hat{A}\Delta\frac{\sin(\hat{A}-1)\Delta}{\hat{A}-1}-\frac{\hat{A}}{2}\Delta\sin2\Delta\right),\\
	&\mathbb{R}= 2\sin2\theta_{12}\sin2\theta_{23}\cos\delta_{\rm CP}\cos\Delta\frac{\sin \hat{A}\Delta}{\hat{A}}\frac{\sin(\hat{A}-1)\Delta}{\hat{A}-1}.
\end{align}
Note that in the above expressions, we have also neglected the term with $\alpha\sin\tym\cos2\tzm$, since it is of the same order as $\alpha \sin^2\theta_{13}$. The contribution from the LIV parameters\footnote{Only off-diagonal LIV parameter that appears at the first order in the disappearance probability is $\amt/\cmt$. This has already been discussed in Refs.~\cite{Kopp:2007ne,Kikuchi:2008vq} in case of NSI.} is given by,
\begin{align}\label{equ:Pmm_liv}
	P_{\mu\mu}(\amt/\cmt) = \frac{\sin^22\tzm}{2}\left[2\sin^2\tym\Delta-\mathbb{S}\right] \sin2\Delta,
\end{align}
where,
\begin{align}\label{equ:S}
	\mathbb{S} = \begin{cases}
		2L\sin2\tzm|a_{\mu\tau}|\cos\phi_{\mu\tau}, & \text{CPT-violating LIV}. \\
		-\frac{8}{3}E L \sin2\tzm|c_{\mu\tau}|\cos\phi_{\mu\tau}, & \text{CPT-conserving LIV}.
	\end{cases} 
\end{align}
In appendix~\ref{appndx}, we compare the disappearance probability calculated using the analytical expressions derived above with the same calculated numerically.

In \figu{disapp_prob}, we show the $\nu_\mu\to\nu_\mu$ disappearance probability as a function of energy for the baseline $L=1285$ km (top row) and $L=295$ km (bottom row) in SI case as well as in the presence of LIV parameters. In the left column, we show the impact of CPT-violating LIV parameter $\amt$, whereas the right column corresponds to the CPT-conserving LIV parameter $c_{\mu\tau}$. As discussed earlier, the other two off-diagonal parameters do not appear at the leading order and are expected to have a negligible impact on disappearance probability. From the left panels, we observe that the phase associated with $\amt$, shows a significant impact with positive (negative) deviation for $\phi_{\mu\tau}=180^\circ$ ($0^\circ$) from the SI case. We can explain this feature using the LIV contributing terms in the analytical expression (refer to \equ{pmm}). It is clear that when $\mathbb{S}$ (shown in \equ{S}) is negative (positive), the disappearance probability is larger (smaller) than its corresponding SI value. Similar to the appearance probability, here also, we observe the impact of the phases become smaller for a smaller baseline, as shown in lower panels, because the LIV contributing term is proportional to the baseline length $L$. In the right column, we show the impact of the CPT-conserving LIV parameters. Here, we observe that for $L=1285$ km, disappearance probability is maximum (minimum) for $\phi_{\mu\tau}=0^\circ$ ($180^\circ$), which shows an opposite behavior compared to the CPT-violating LIV parameters. This attributes to the opposite sign of $\mathbb{S}$ in \equ{S}, in the case of the CPT-conserving LIV parameters. 

%=============================%
\section{Long-baseline experiments: DUNE and T2HK}
\label{sec:LBL}
\subsection{Essential features of the experimental setups}
%=============================%
Accelerator-based neutrino oscillation experiments are playing a very important role in resolving issues in the standard $3\nu$ paradigm and exploring various BSM physics in the neutrino sector. Precise information about the neutrino flux, cross section, and baseline make these experiments unique. In this work, we probe LIV in the context of next-generation long-baseline experiments DUNE and T2HK. DUNE is a future long-baseline experiment with an on-axis, high-intensity wide-band neutrino beam produced at Fermilab~\cite{DUNE:2020lwj, DUNE:2020jqi, DUNE:2021cuw, DUNE:2021mtg}. The detector would be a 40 kt Liquid argon time projection chamber (LArTPC) placed underground at Homestake mine, 1285 km from the source. On the other hand, T2HK~\cite{Abe:2015zbg, Hyper-Kamiokande:2018ofw} is another next-generation long-baseline experiment with off-axis, narrow band beam produced at the J-PARC proton synchrotron facility. The beam would be detected at Hyper-kamiokande, a 187 kt water Cherenkov detector placed at a distance of 295 km from the source with an off-axis angle $2.5^\circ$. In table~\ref{tab:exp_details}, we give the other relevant information on these two experiments.

\begin{table}
	\centering
 \begin{adjustbox}{width=0.8\textwidth}
	\begin{tabular}{|c|c|c|}
		\hline \hline
		& DUNE & T2HK\\ 
		\hline
		Detector Mass& 40 kt LArTPC & 187 kt WC   \\
		\hline
		Baseline & 1285 km & 295 km \\
		\hline
		Proton Energy & 120 GeV & 80 GeV \\
		\hline
		Beam type & Wide-band, on-axis & Narrow-band, off-axis ($2.5^{\circ}$)\\
		\hline
		Beam power & 1.2 MW & 1.3 MW \\
		\hline
		P.O.T./year& $1.1\times10^{21}$ & $2.7\times10^{21}$\\ 
		\hline
		Run time ($\nu+\bar{\nu}$) & 5 yrs + 5 yrs & 2.5 yrs + 7.5 yrs  \\
		\hline
		Normalization error & 2\% (app.) 5\% (disapp.) & 5\% (app.), 3.5\% (disapp.)\\
		\hline\hline
	\end{tabular}
 \end{adjustbox}
	\mycaption{Major features of LBL experiments, DUNE~\cite{DUNE:2020lwj} and T2HK~\cite{Abe:2015zbg} used in our simulation.}
	\label{tab:exp_details}
\end{table}

As mentioned earlier, the DUNE setup uses an on-axis, wide-band neutrino beam. This allows DUNE to explore both the first and second oscillation maxima for the baseline 1285 km, which are around 2.5 GeV and 0.9 GeV, respectively. On the other hand, T2HK has an off-axis narrow band beam with an energy peak around 0.6 GeV, which is the first oscillation maxima for T2HK. The monochromatic beam will give the advantage of high statistics at the first oscillation maxima, where the impact of various physics can be significant. 
DUNE will have equal runtime for neutrino and antineutrino mode, which will give a larger number of expected neutrino events than the antineutrino, since the neutrino has almost three times the cross section of the antineutrino. For T2HK, antineutrino runtime is three times the neutrino runtime in order to compensate for suppression in the cross section. So, depending on the physics under probe, different ratios of the neutrino and antineutrino events can be useful. The longer baseline of DUNE allows it to have larger matter effect compared to T2HK. Apart from the complementarity at the neutrino flux, baseline, and runtime, the detector properties of the two setups are different.
As proposed by the collaborations, DUNE is going to have total systematic uncertainties of 2.5\% in the appearance channel and 5\% in the disappearance channel~\cite{DUNE:2020ypp, DUNE:2020jqi, DUNE:2021cuw} and for T2HK, it is 5\% and 3.5\%~\cite{Hyper-KamiokandeWorkingGroup:2014czz, Abe:2015zbg}, respectively. Lower systematics in the appearance channel in DUNE can allow it to have comparatively larger sensitivity to some physics that have a larger impact on the appearance channel. A similar argument can be given for T2HK in the disappearance channel. Note that the mentioned values of the systematic uncertainties are estimated values. In future, these values may change, which would affect the results presented in the later sections.

Each oscillation channel in both the experiments have backgrounds. For DUNE, the appearance channel has background from intrinsic $\nu_e$ beam contamination and misidentified $\nu_\mu$, $\nu_\tau$, and neutral current (NC) events. For the disappearance channel, backgrounds are misidentified $\nu_\tau$ and NC events. Similarly, for T2HK, backgrounds in the appearance channel come from the $\nu_e$ beam contamination and misidentified $\nu_\mu$ and NC events. Backgrounds for the disappearance channel come from the misidentified $\nu_e$ and NC events.
%=============================%
\subsection{Expected event rates in the presence of LIV}
\label{subsec:LBL-Evt-Sim}
%=============================%

As mentioned earlier, in this work, we consider two experimental configurations, DUNE and T2HK, for our analysis. We calculate the expected event rates of these two configurations using the GLoBES software~\cite{Huber:2004ka, Huber:2007ji}. For the oscillation analysis with LIV parameters, we use GLoBES-extension $snu.c$~\cite{Kopp:2007ne}.

In table~\ref{tab:total_events}, we give the expected $\nu_e/\bar{\nu}_e$ and $\nu_\mu/\bar{\nu}_\mu$ event rates from DUNE and T2HK in SI case and in the presence of various LIV parameters. Assumed configurations of the experiments are discussed in detail in section~\ref{sec:LBL} (see table~\ref{tab:exp_details}). While generating the events, the strength of the CPT-violating (CPT-conserving) LIV parameters is considered to be $2.0\times10^{-23}$ GeV ($1.0\times10^{-24}$), one at-a-time. Here, we consider the phase associated with the off-diagonal LIV parameters to be zero ($\phi_{\alpha\beta}=0^\circ$; $\alpha, \beta = e, \mu, \tau$ and $\alpha\neq\beta$) for a demonstration purpose.

	\begin{table}[h!]		
		\centering
		\begin{adjustbox}{width=1\textwidth}
			\begin{tabular}{|ccc|*{8}{c|}}		
				\hline\hline
				
				\multicolumn{3}{|c|}{\multirow{2}{*}{}} & \multicolumn{2}{|c}{$\nu_e$ appearance} & \multicolumn{2}{|c|}{$\bar\nu_e$ appearance}& \multicolumn{2}{|c}{$\nu_\mu$ disappearance} & \multicolumn{2}{|c|}{$\bar\nu_\mu$ disappearance}\\ 
				
				\cline{4-11}
				
				\multicolumn{3}{|c|}{} & DUNE & T2HK & DUNE & T2HK & DUNE & T2HK & DUNE & T2HK \\
				
				\hline
				
				\multicolumn{3}{|c|}{SI} & 1614 & 1613  & 292 & 727 & 15624 & 9577  & 9052 & 9074 \\
				
				\cline{1-11}
				
				\hline
				
				\multicolumn{3}{|c|}{$|a_{e\mu}|$ (= $2\times10^{-23}$ GeV)} & 2276 & 1731 & 666 & 901  & 15446 & 9581  & 8891 & 9042\\
				\hline
				\multicolumn{3}{|c|}{$|a_{e\tau}|$ (= $2\times10^{-23}$ GeV)} & 817 & 1387  & 226 & 697 & 15613 & 9565   & 9063 & 9084\\
				\hline
				\multicolumn{3}{|c|}{$|a_{\mu\tau}|$ (= $2\times10^{-23}$ GeV)} & 1567 & 1598  & 303 & 735 & 14404 & 9366  & 9237 & 9373\\
				\hline
				\multicolumn{3}{|c|}{$|c_{e\mu}|$ (= $1.0\times10^{-24}$)} & 1757 & 1610 & 480 & 735  & 15315 & 9575  & 8800 & 9071\\
				\hline
				\multicolumn{3}{|c|}{$|c_{e\tau}|$ (= $1.0\times10^{-24}$)} & 1792 & 1623  & 296 & 724 & 15634 & 9578   & 9056 & 9074\\
				\hline
				\multicolumn{3}{|c|}{$|c_{\mu\tau}|$ (= $1.0\times10^{-24}$)} & 1620 & 1614  & 295 & 727 & 16263 & 9600  & 9411 & 9099\\
				
				\hline
				\hline
			\end{tabular}
		\end{adjustbox}
		\mycaption{Total signal rate for the $\nu_{e}$ appearance channel and  $\nu_{\mu}$ disappearance channel both in neutrino and antineutrino mode for DUNE and T2HK setups in SI case as well as in the presence off-diagonal LIV parameters. The relevant features of these facilities are given in table~\ref{tab:exp_details}. The strength of the CPT-violating (CPT-conserving) LIV parameters is taken to be $2\times10^{-23}$ GeV ($1.0\times10^{-24}$). The phases associated with the off-diagonal LIV parameters are considered to be zero. The values of the standard oscillation parameters used to calculate event rate are quoted in table~\ref{tab:params_value}.}
	
	\label{tab:total_events}
\end{table}

We make following observations from table~\ref{tab:total_events}:
\begin{itemize}

\item In the presence of $a_{e\mu}$ ($a_{e\tau}$), $\nu_e$ event deviates from the SI case by 41\% (49\%) for DUNE and by 7.3\% (14\%) for T2HK. 

\item Similarly, we observe that the presence of LIV parameter $a_{\mu\tau}$ changes the expected $\nu_{\mu}$ disappearance event rates by 7.8\% for DUNE and 2.2\% for T2HK from the SI case. In the presence of other LIV parameters, changes in the event rates are very small ($\leq 1\%$).

\item In presence of $c_{e\mu}$ ($c_{e\tau}$), the $\nu_{e}$ appearance event rates changes by $\sim$9\% (11\%) for DUNE, but for T2HK, the changes are very minute ($<$ 1\%). However $c_{\mu\tau}$ changes the $\nu_{\mu}$ event rates by 4\% for DUNE and 0.2\% for T2HK.

\end{itemize}

All these observations are consistent with the results seen at the probability level in the previous section. Though, in this section, we give only the total signal event rates for DUNE and T2HK, however, we perform a binned study while presenting the sensitivity results  in section~\ref{sec:results}. For demonstration purpose, we show in appendix~\ref{appndx-B}, the $\nu_e$ appearance and $\nu_\mu$ disappearance signal event spectra expected to be seen by DUNE and T2HK in SI case as well as in presence of LIV (SI+LIV). Similar plots can be made for antineutrinos. Note that while calculating the final sensitivity results, we take bin contributions from appearance and disappearance channels in both neutrino and antineutrino modes.

\begin{table}[h!]
	\begin{center}
		\begin{tabular}{|c|c|c|c|}
			\hline\hline
			Experiments & $E_{\rm rec}$ (GeV)& Bin width (GeV) & Total bins \\
			\hline\hline
			DUNE & \makecell[c]{$0.5\,-\, 8$ \\ $8\,-\,10$ \\ $10\,-\,18$} & 
			\makecell[c]{$0.125$ \\ $1.0$ \\ $2.0$} & 
			$\left.\begin{tabular}{l}
			60 \\ 2 \\ 4
			\end{tabular}\right\}$  $66$ \\
			\hline
			T2HK & $0.1\,-\,3.0$ & 0.1 & 29\\
			\hline\hline
		\end{tabular}
		\mycaption{The binning scheme used in our analysis for both DUNE and T2HK. The binning is performed over the reconstructed energy ($E_{\rm rec}$).\label{tab:binning}}
	\end{center}
\end{table}

%=============================%
\section{Numerical analysis}
\label{sec:RnA}
%=============================%
One of the major goals of this work is to study the ability of DUNE and T2HK to constrain the off-diagonal CPT-violating and CPT-conserving LIV parameters. To estimate the sensitivity of a given experiment towards the LIV parameters, we use the following form of the Poissonian $\chi^2$:
\begin{equation}
\chi^2 (\vec{\lambda}, \xi_{s}, \xi_{b}) = \min_{\vec\rho, \xi_{s}, \xi_{b}}~\Big[{{2\sum_{i=1}^{n}}\big(y_{i}-x_{i}-x_{i} \text{ln}\frac{y_{i}}{x_{i}}\big) + \xi_{s}^2 + \xi_{b}^2}~\Big],
\end{equation} 
which gives the median sensitivity of the experiment where $n$ is the total number of reconstructed energy bins. The binning scheme adopted in this work for the simulation of DUNE and T2HK are given in table~\ref{tab:binning}.
\begin{equation}
y_{i} = N^{th}_{i}
(\vec\rho)~[1 + \pi^{s}\xi_{s}] + N^{b}_{i}(\vec\rho)~[1+\pi^b\xi_{b}],
\end{equation}
\noindent
where $N^{th}_{i}$ is the expected number of signal events in the $i$-th bin with the set of oscillation parameters $\vec\lambda$ = \{$\theta_{12}, \theta_{13}, \theta_{23}, \Delta{m}^2_{21}, \Delta{m}^2_{31}, \delta_{\rm CP}, a_{\alpha\beta}, \phi^{a}_{\alpha\beta}, c_{\alpha\beta}, \phi^{c}_{\alpha\beta}$\}. $\phi^{a}_{\alpha\beta}$ and $\phi^{c}_{\alpha\beta}$ are the phases associated with the off-diagonal CPT-violating and CPT-conserving LIV parameters, respectively. $N^{b}_{i}$ is the number of background events in the $i$-th energy bin. The systematic pulls on the signal and background are denoted by the variables $\xi_{s}$ and $\xi_{b}$, respectively. We marginalize the $\chi^2$ over the set of parameters $\vec\rho$ and also over the systematic pulls ($\xi_{s}$ and $\xi_{b}$) in the fit. The variables $\pi^s$ and $\pi^b$ stand for the normalization error on the signal and background. The values of the normalization errors for DUNE and T2HK are listed in table~\ref{tab:norm_err}. $x_{i} = N^{obs}_{i} + N^{b}_{i}$ embodies the prospective data from the experiment, where $N^{obs}_{i}$ is the number of charged current (CC) signal events and $N^b_{i}$, as mentioned before, is the number of background events. 

We quantify our results in terms of the statistical significance given by Poissonian $\Delta{\chi}^2$ defined as,
\begin{equation}
	\Delta{\chi}^2 = \min_{\vec\rho, \xi_{s}, \xi_{b}}\Big[\chi^2 (a_{\alpha\beta}/c_{\alpha\beta} \neq 0) - \chi^2 (a_{\alpha\beta}/c_{\alpha\beta} = 0)\Big],\,\,\,\,\,\alpha,\beta=e,\mu,\tau;\alpha\neq\beta.
\end{equation}
The first term in the right-hand side of the above equation is obtained when we fit the prospective data from the experiment with the theory in the presence of LIV, and the second term is calculated by fitting with the standard case with no LIV present in theory. Due to suppression in the statistical fluctuations, we can take $\chi^2 (a_{\alpha\beta}/c_{\alpha\beta} = 0)$ $\sim$ 0 while obtaining the median sensitivity of the experiment using frequentist approach~\cite{Blennow:2013oma}. Here $\vec\rho$ is the set of parameters over which the $\chi^2$ is marginalized, and $\xi_{s}$ and $\xi_{b}$ are the systematic pulls on the signal and background, respectively.
For our analysis, we use the true values of the standard oscillation parameters as given in table~\ref{tab:params_value}. In theory, we keep the two mixing angles $\theta_{12}$, $\theta_{13}$ and two mass-splittings $\Delta m^2_{21}$, $\Delta m^2_{31}$ fixed at the same values. We do not marginalize over the present uncertainty in the magnitude of $\Delta m^2_{31}$, because the global oscillation data attain a relative 1$\sigma$ precision of 1.1\% in the measurement of $\Delta m^2_{31}$. JUNO, with six years of data, may further improve the measurement of this parameter to 0.2\%~\cite{JUNO:2022mxj}. Also, we do not marginalize over the neutrino mass ordering since there is a hint towards the normal mass ordering from the global oscillation data~\cite{deSalas:2020pgw, Esteban:2020cvm, Capozzi:2021fjo}. Moreover, T2K~\cite{T2K:2023smv}, NO$\nu$A~\cite{Carceller:2023kdz}, Super-K~\cite{Posiadala-Zezula:2022vzn}, IceCube-DeepCore~\cite{Mead:2023spo}, and ORCA~\cite{KM3NeT:2023ncz} are going to collect more data in the coming days which will further strengthen the estimation of mass ordering. Also, it is expected that with preliminary data coming from $\sim$ 3 years run of DUNE would be able to settle the issue of neutrino mass ordering at very high confidence level~\cite{DUNE:2020ypp}. For the correlation between LIV parameters and $\delta_{\mathrm{CP}}$, we marginalize over the $\theta_{23}$ in its allowed $3\sigma$ range~\cite{Esteban:2020cvm} and the phase $\phi$ associated with the off-diagonal LIV parameters in the range [0, $360^\circ$]. Similarly, for the LIV parameter and $\theta_{23}$ correlation analysis, we marginalize over $\delta_{\mathrm{CP}}$ in its allowed $3\sigma$ range and $\phi$ in its entire permitted range.
While deriving the limits on the off-diagonal LIV parameters, we marginalize over $\theta_{23}\in[39.7^\circ:50.9^\circ]$, $\delta_{\rm CP}\in[144^\circ:350^\circ]$, and $\phi\in[0^\circ:360^\circ]$.

\begin{table}[t!]
	\resizebox{\columnwidth}{!}{%
		\centering
		\begin{tabular}{|c|c|c|c|c|c|c|c|c|}
			\hline\hline
			\multirow{3}{*}{Expts.}  & \multicolumn{8}{c|}{Normalization errors~[\%]}  \\
			\cline{2-9}
			& \multicolumn{4}{c|}{Signal ($\pi^s$)} & \multicolumn{4}{c|}{Background ($\pi^b$)}\\ \cline{2-9}
			&  $\nu$ App. & $\bar{\nu}$ App. & $\nu$ Disapp.& $\bar{\nu}$ Disapp. & $\nu_{e}$, $\bar{\nu}_{e}$ CC & $\nu_{\mu}$, $\bar{\nu}_{\mu}$ CC & $\nu_{\tau}$, $\bar{\nu}_{\tau}$ CC & NC \\ 
			\hline
			DUNE & 2 & 2 & 5 & 5 & 5 & 5 & 20 & 10\\
%			&&&&&&&&\\
			T2HK & 5 & 5 & 3.5 & 3.5 & 10 & 10 & -- & 10\\
			\hline\hline
	\end{tabular}}
	\caption{The values of the normalization errors associated to the signal and background event rates in DUNE and T2HK for neutrino and antineutrino in both appearance and disappearance channels. These values are taken from Ref.~\cite{DUNE:2021cuw} for DUNE and Ref.~\cite{Hyper-Kamiokande:2018ofw} for T2HK.}
	\label{tab:norm_err}
\end{table}

%============================%
\section{Our results}
\label{sec:results}
%============================%
In this section, we present our results in two parts. First, we discuss the correlations between the LIV parameters and the most unsettled standard oscillation parameters, $\delta_{\mathrm{CP}}$ and $\theta_{23}$. This helps us to find if there is any degeneracy between the LIV parameters and the standard oscillation parameters. In the second part, we present the expected constraints on the LIV parameters from DUNE, T2HK individually, and their combination DUNE+T2HK.

%=============================%
\subsection{Correlations in test ($\delta_{\rm CP} - |a_{\alpha\beta}|/|c_{\alpha\beta}|$) and test ($\theta_{23} - |a_{\alpha\beta}|/|c_{\alpha\beta}|$) planes}
\label{subsec:correlation}
%=============================%

\begin{figure}[h!]
	\centering
	\includegraphics[width=\textwidth]{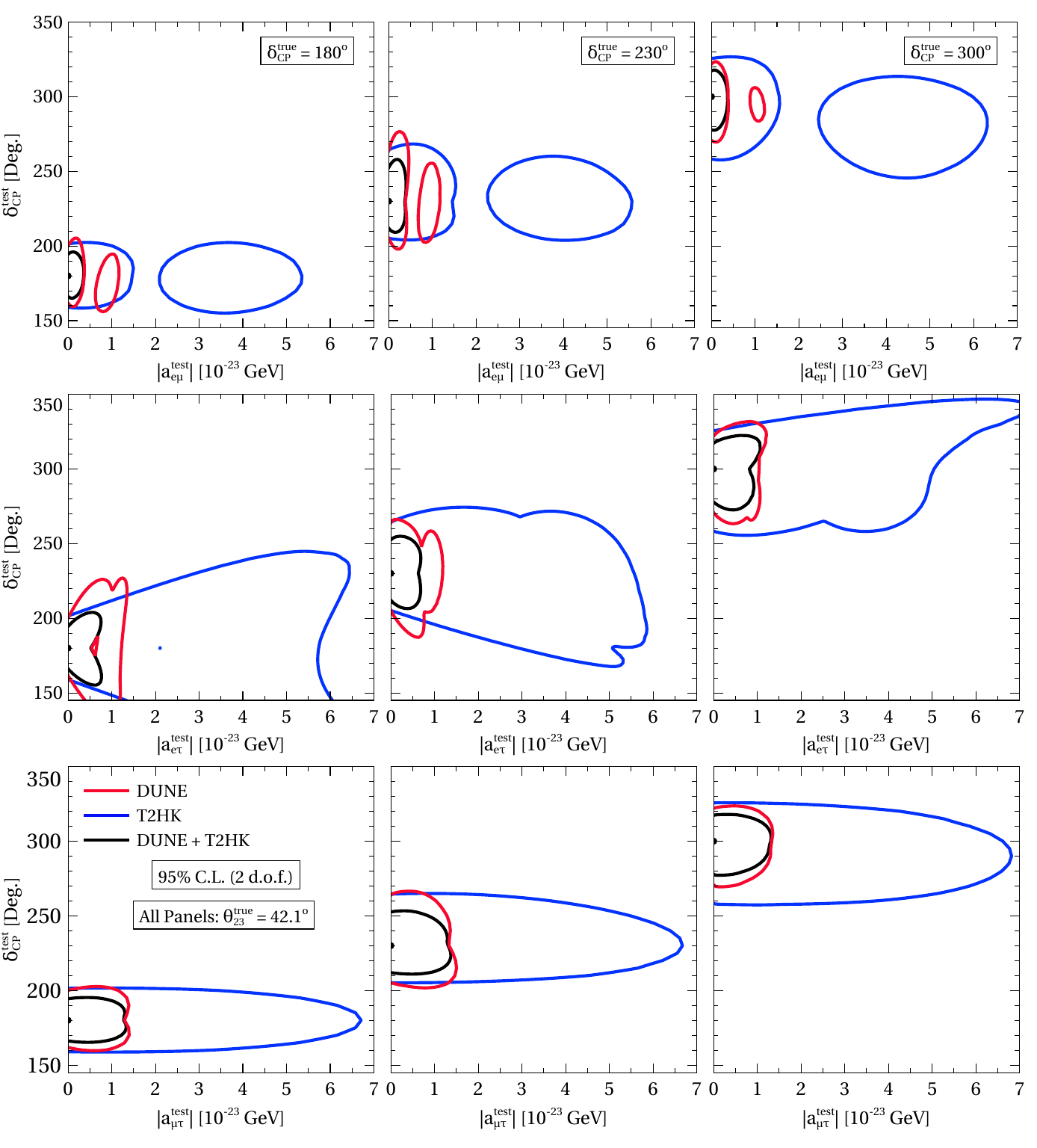}
	\vspace*{-10mm}
	\mycaption{95\% C.L. (2 d.o.f.) contours in the $\delta_{\rm CP}-|a_{e\mu}|$ (top row), $\delta_{\rm CP}-|a_{e\tau}|$ (middle row), and $\delta_{\rm CP}-|a_{\mu\tau}|$ (bottom row) planes for DUNE, T2HK, and DUNE+T2HK. Three benchmark values of $\delta_{\mathrm{CP}}$ considered in the data are $180^\circ$ (left column), $230^\circ$ (middle column), and $300^\circ$ (right column), as shown by the black dots in each panel. True values of other oscillation parameters are given in table~\ref{tab:params_value}. In the fit, we marginalize over $\theta_{23}$ in its allowed $3\sigma$ range and new phase $\phi$ in its entire allowed range.}
	\label{fig:correlation_dcp_cptv}
\end{figure}

\begin{figure}[h!]
	\centering
	\includegraphics[width=\textwidth]{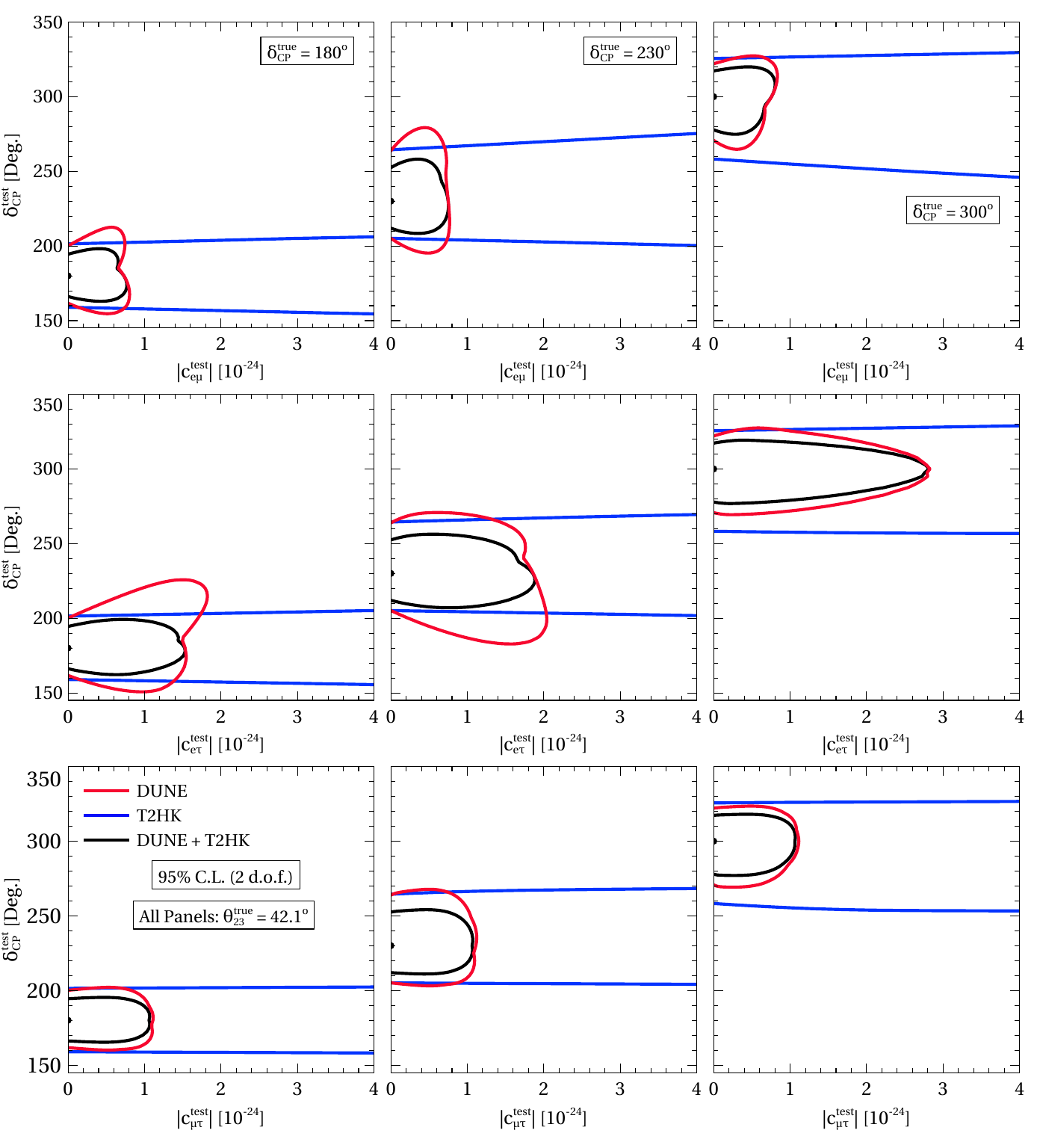}
	\vspace*{-10mm}
	\mycaption{95\% C.L. (2 d.o.f.) contours in the $\delta_{\rm CP}-|c_{e\mu}|$ (top row), $\delta_{\rm CP}-|c_{e\tau}|$ (middle row), and $\delta_{\rm CP}-|c_{\mu\tau}|$ (bottom row) planes for DUNE, T2HK, and DUNE+T2HK. Three benchmark values of $\delta_{\mathrm{CP}}$ considered in the data are $180^\circ$ (left column), $230^\circ$ (middle column), and $300^\circ$ (right column), as shown by black dots in each panel. True values of other oscillation parameters are given in table~\ref{tab:params_value}. In the fit, we marginalize over $\theta_{23}$ in its allowed $3\sigma$ range and new phase $\phi$ in its entire allowed range.}
	\label{fig:correlation_dcp_cptc}
\end{figure}

\begin{figure}[h!]
	\centering
	\includegraphics[width=\textwidth]{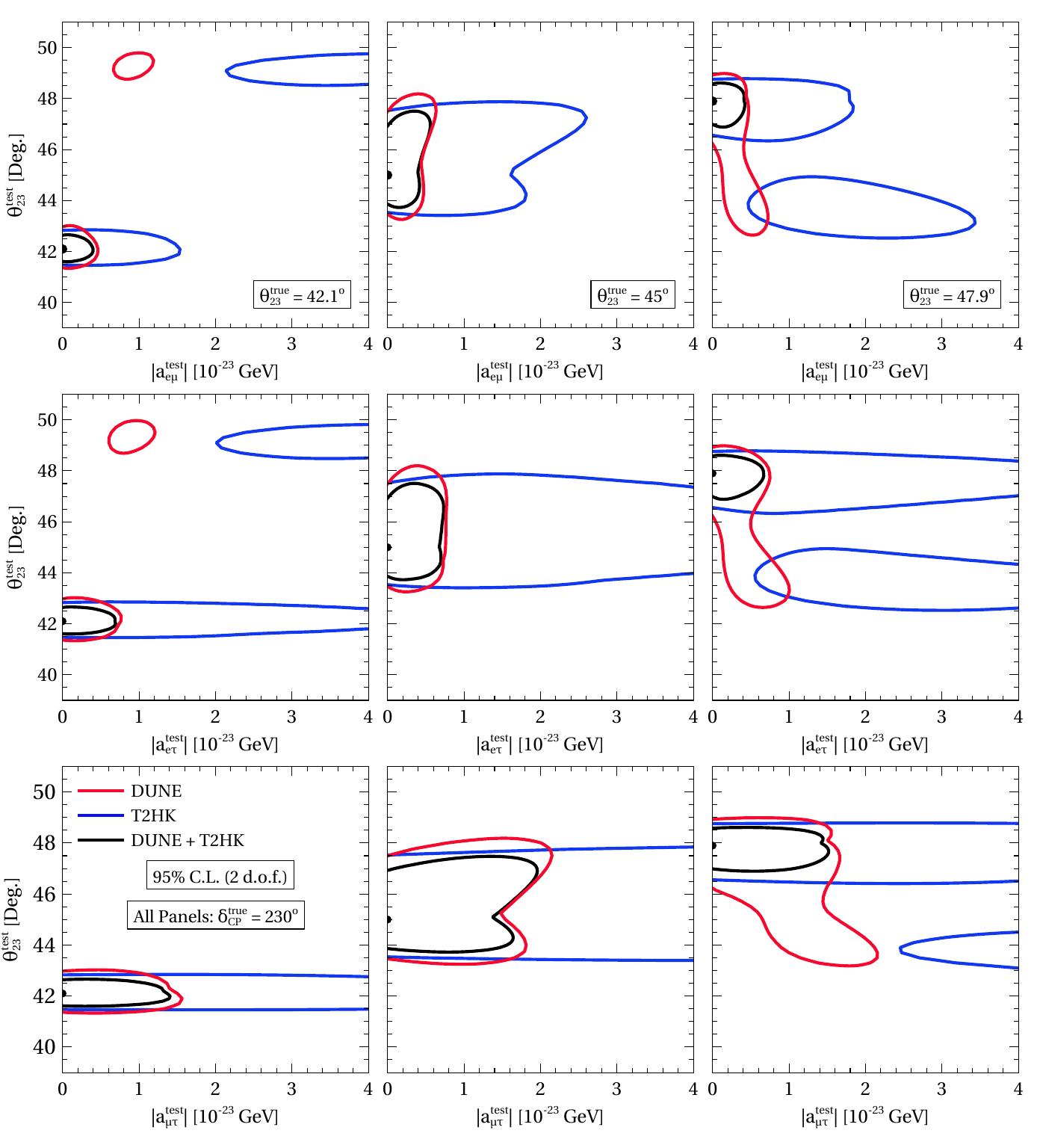}
	\vspace*{-10mm}
	\mycaption{ 95\% C.L. (2 d.o.f.) contours in the $\tzm-|a_{e\mu}|$ (top row), $\tzm-|a_{e\tau}|$ (middle row), and $\tzm-|a_{\mu\tau}|$ (bottom row) planes for DUNE, T2HK, and DUNE+T2HK. Three benchmark values of $\tzm$ considered in the data are $42.1^\circ$ (left column), $45^\circ$ (middle column), and $47.9^\circ$ (right column), as shown by black dots in each panel. True values of other oscillation parameters are given in table~\ref{tab:params_value}. In the fit, we marginalize over $\delta_{\mathrm{CP}}$ in its allowed $3\sigma$ range and new phase $\phi$ in its entire allowed range.}
	\label{fig:correlation_th23_cptv}
\end{figure}

\begin{figure}[h!]
	\centering
	\includegraphics[width=\textwidth]{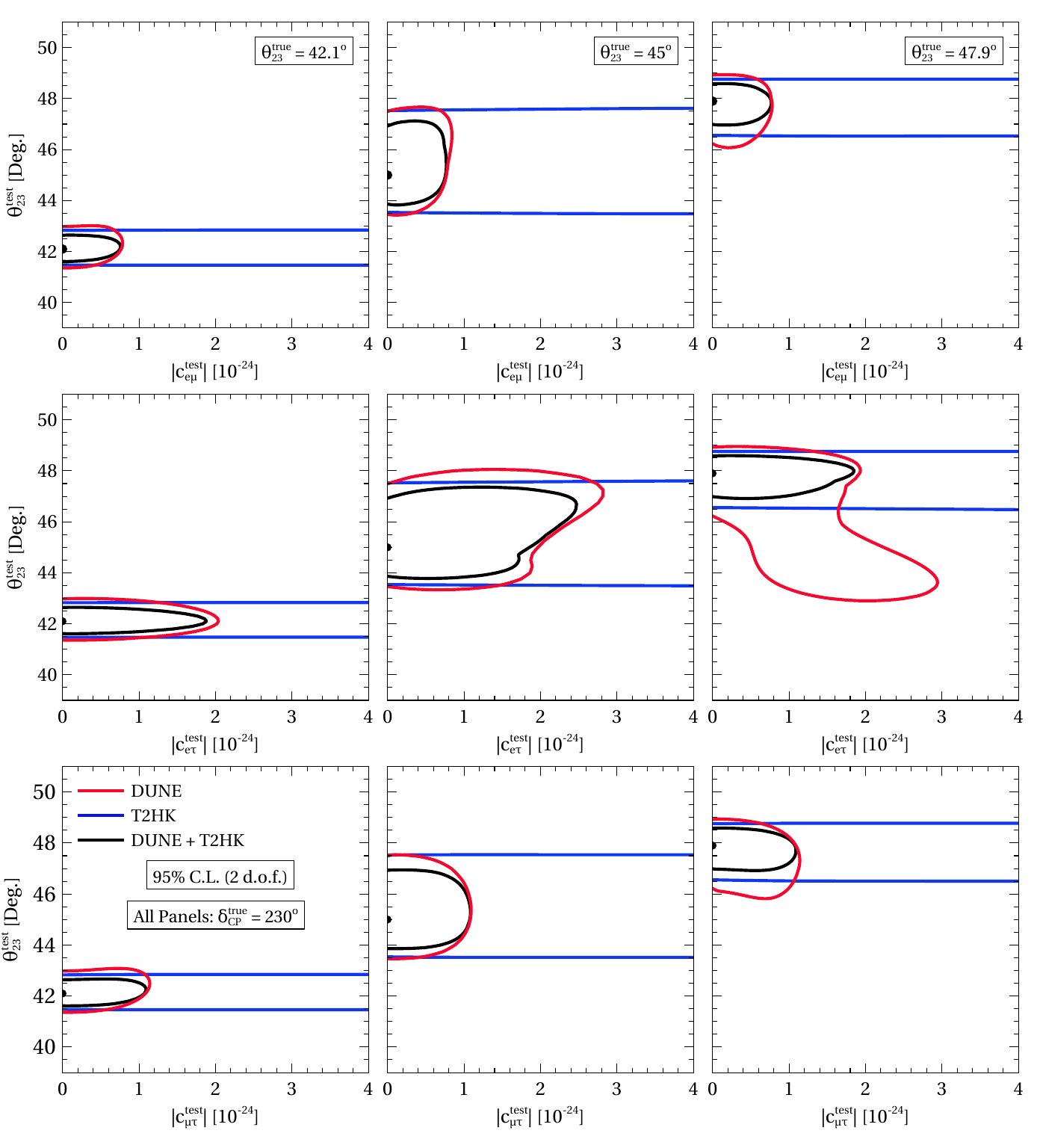}
	\vspace*{-10mm}
	\mycaption{95\% C.L. (2 d.o.f.) contours in the $\tzm-|c_{e\mu}|$ (top row), $\tzm-|c_{e\tau}|$ (middle row), and $\tzm-|c_{\mu\tau}|$ (bottom row) planes for DUNE, T2HK, and DUNE+T2HK. Three benchmark values of $\tzm$ considered in the data are $42.1^\circ$ (left column), $45^\circ$ (middle column), and $47.9^\circ$ (right column), as shown by black dots in each panel. True values of other oscillation parameters are given in table~\ref{tab:params_value}. In the fit, we marginalize over $\delta_{\mathrm{CP}}$ in its allowed $3\sigma$ range and new phase $\phi$ in its entire allowed range.}
	\label{fig:correlation_th23_cptc}
\end{figure}

In \figu{correlation_dcp_cptv}, we show the correlations between the CPT-violating LIV parameters and the standard CP phase $\delta_{\rm CP}$. In the fit, we marginalize over $\theta_{23}$ in its allowed $3\sigma$ range~\cite{Esteban:2020cvm} and the phase $\phi$ associated with the off-diagonal LIV parameters in the range [0, $360^\circ$]. As mentioned earlier, all the other standard oscillation parameters are fixed at their best-fit values given in table~\ref{tab:params_value}, both in data and theory. 
Top, middle, and bottom rows correspond to non-zero $a_{e\mu}$, $a_{e\tau}$, and $a_{\mu\tau}$, respectively, where we consider these LIV parameters 
one at-a-time in the fit. We take three different choices for the value of $\delta_{\rm CP}$ in the data that are allowed in the current $3\sigma$ limits, namely, $180^\circ$ (left panels), $270^\circ$ (middle panels), $300^\circ$ (right panels) as shown by the black dot in each panel. The red, blue, and black curves in each plot correspond to DUNE, T2HK, and the combination DUNE+T2HK, respectively. Each contour represents the allowed regions at 95\% C.L. (2 d.o.f.). We observe from the figure that for all the LIV parameters and all choices of the $\delta_{\mathrm{CP}}$ in the data, the allowed regions in $\delta_{\rm CP}-|a_{\alpha\beta}|$ planes are significantly small for DUNE compared to T2HK. One can understand it from the analytical expression of the oscillation probabilities discussed in section~\ref{sec:LIV}. From \equ{p_cptv_liv} and \equ{Pmm_liv}, we see that contribution from the CPT-violating LIV parameters is directly proportional to $L$. So, DUNE being an experiment with a comparatively longer baseline, shows better sensitivity to the CPT-violating LIV parameters. Hence, it has smaller allowed regions in  $\delta_{\rm CP}-|a_{\alpha\beta}|$ plane as compared to T2HK.

In case of $a_{e\mu}$, we notice a non-trivial degenerate solution in $\delta_{\rm CP}-|a_{e\mu}|$ plane which is centered around a non-zero value of  $|a_{e\mu}|$. This happens for both DUNE and T2HK at $|a_{e\mu}|\approx1\times 10^{-23}$ GeV and  $|a_{e\mu}|\approx4\times 10^{-23}$ GeV, respectively. This mainly occurs due to the degeneracy between $\theta_{23}$ and the complex phases ($\delta_{\rm CP}~\rm and~ \phi$) in the LIV contributing term in the appearance channel (see \equ{p_cptv_liv}), which plays a major role in constraining this parameter. For some combination of $\theta_{23}$ and $\delta_{\mathrm{CP}}+\phi$, this term minimizes at some non-zero value of $|a_{e\mu}|$ resulting in degeneracy with the standard oscillation case. As a result, we observe an allowed region at around that value of $|\aem|$. However, when we take the combined setup DUNE+T2HK, this degeneracy disappears. This happens because the values of $L$ and $E$ in the LIV contributing terms are now different for these two experiments, which helps in lifting the degeneracy. We also observe that upon combining these two setups, the allowed regions shrink further. In the case of $|\aet|$ (middle row) and $|\amt|$ (bottom row), we do not observe such degenerate solutions at 95\% C.L. (2 d.o.f.). In both cases, DUNE+T2HK shows a small improvement in the sensitivity as compared to DUNE. 

In \figu{correlation_dcp_cptc}, we repeat the above analysis for three off-diagonal CPT-conserving LIV parameters, $c_{e\mu}$ (top row), $c_{e\tau}$ (middle row), and $c_{\mu\tau}$ (bottom row). It is clear from the figure that DUNE shows a noticeable correlation between $c_{\alpha\beta}$ ($\alpha,\beta=e,\mu,\tau;\alpha\neq\beta$) and $\delta_{\mathrm{CP}}$. However, for T2HK, there is almost no correlation between those two parameters in all three cases. One can explain it from the fact that CPT-conserving parameters have negligible impact on both appearance and disappearance probabilities for T2HK as shown in the bottom row of \figu{cptc_app_prob} and bottom right panel of \figu{disapp_prob}. As discussed earlier,
this happens because of $L\times E$ dependencies in LIV contributing terms in the CPT-conserving case (see \equ{p_cptc_liv} and \equ{Pmm_liv}). Since T2HK has a shorter baseline and access to low energy neutrino beam compared to DUNE, it shows almost no sensitivity to CPT-conserving LIV parameters. 

In \figu{correlation_th23_cptv}, we show the correlation between the CPT-violating LIV parameters $a_{\alpha\beta}$ ($\alpha,\beta=e,\mu,\tau$; $\alpha\neq\beta$) and $\theta_{23}$. We consider three values of $\theta_{23}$ in data, namely, $42.1^\circ$ (left column) in the lower octant, 45$^\circ$ (middle column) maximal mixing case, and $47.9^\circ$ (right column) in the upper octant\footnote{We choose $\theta_{23}=42.1^\circ$ in lower octant as it is the current best-fit value from the global-fit of the oscillation parameters~\cite{Esteban:2020cvm}. For simulation, we marginalize over $\delta_{\mathrm{CP}}$ in its allowed $3\sigma$ range and the corresponding LIV phases in their entire allowed range. We consider the corresponding value ($47.9^\circ$) in the upper octant.}. We observe that for both DUNE and T2HK, the best result is obtained when the true value of $\theta_{23}$ is in the lower octant, where the allowed region is relatively small compared to the other two cases. 
Similar to $\delta_{\mathrm{CP}}-a_{\alpha\beta}$ correlation, DUNE performs significantly better compared to T2HK for all three choices of true $\theta_{23}$.
Here also, we observe degenerate allowed regions at non-zero values of $a_{e\mu}$ and $a_{e\tau}$ that appear at the opposite octant of $\theta_{23}$ for both DUNE and T2HK. This happens because of the degeneracy between $\theta_{23}$, $\delta_{\mathrm{CP}}$, and $\phi$ in the LIV contributing terms in the oscillation probabilities. On adding the data from DUNE and T2HK, the allowed regions become smaller, and interestingly, the degenerate regions appearing for the individual setups vanish, as shown by the black contour in each panel. In \figu{correlation_th23_cptc}, we show the same for CPT-conserving LIV parameters. Here also, the best results are obtained when $\theta_{23}$ is in lower octant for the individual setups. We observe that DUNE shows noticeable correlations in $\theta_{23}-|c_{\alpha\beta}|$ planes, but T2HK shows almost no correlation with the CPT-conserving LIV parameters. As mentioned before, T2HK has almost no sensitivity to the CPT-conserving LIV parameters because of its smaller baseline and lower neutrino energy.
However, we observe a slight improvement in the allowed regions when we combine the data from DUNE and T2HK.

%=============================%
\subsection{Constraints on CPT-violating and CPT-conserving LIV parameters}
\label{subsec:constraints}
%=============================%

In the previous section, we have discussed the correlations of the LIV parameters with the most uncertain standard oscillation parameters, $\delta_{\mathrm{CP}}$ and $\theta_{23}$ in the context of DUNE, T2HK, and their combination. In this section, we present the limits on the off-diagonal LIV parameters that would be obtained by these three setups. As discussed earlier, in our simulation, we marginalized over $\theta_{23}$, $\delta_{\text{CP}}$, and the phase associated with the off-diagonal LIV parameters in the fit (see section~\ref{sec:RnA} for details). 

\begin{figure}[h!]
	\centering
	\includegraphics[width=\textwidth]{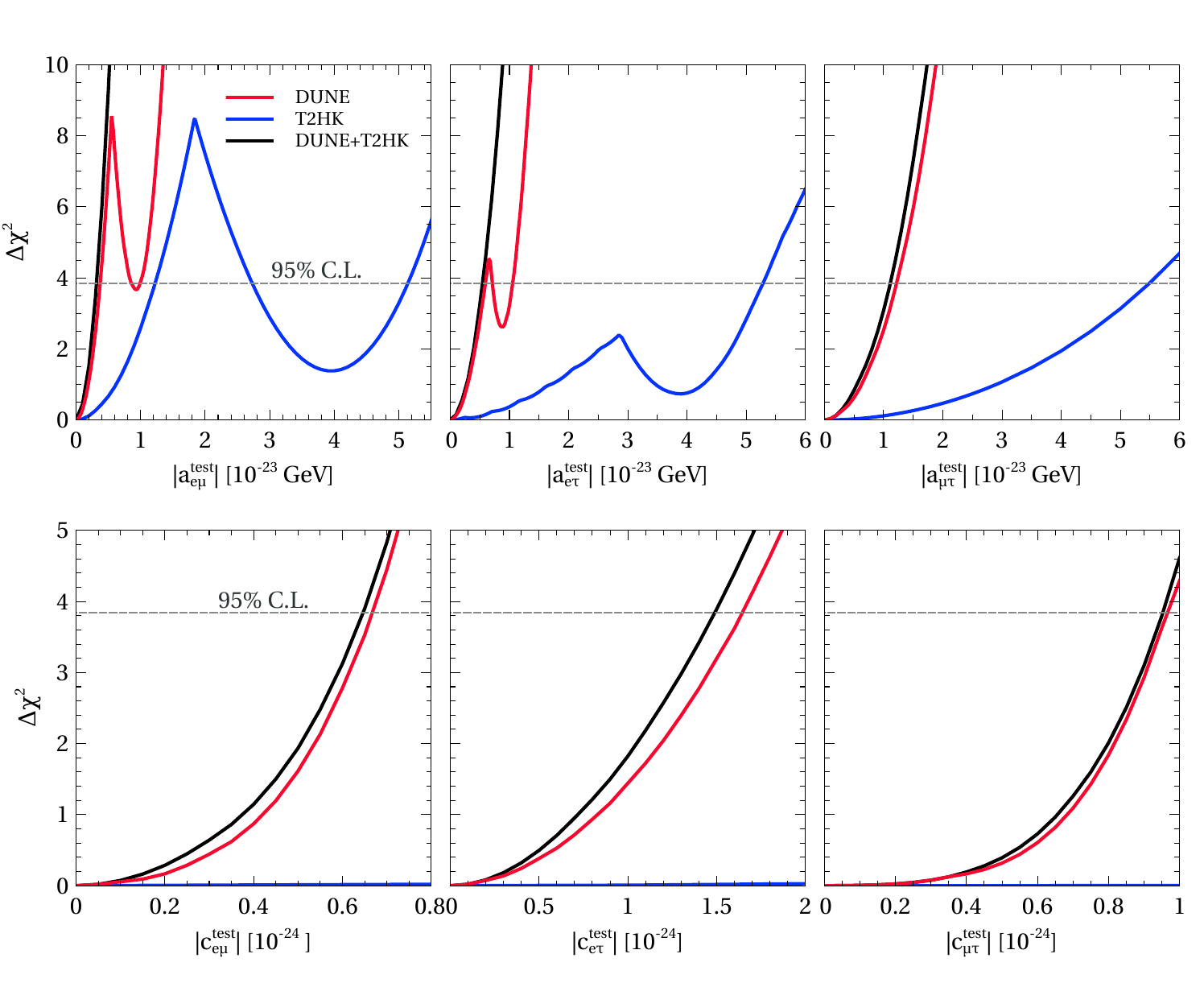}
	\vspace*{-10mm}
	\mycaption{ The expected limits on the CPT-violating (top row) and CPT-conserving (bottom row) LIV parameters
		from DUNE (red curves), T2HK (blue curves), and DUNE+T2HK (black curves).
		The upper (lower) panels show the $\Delta\chi^2$ for the off-diagonal CPT-violating (CPT-conserving) LIV parameters considering one-at-a-time. The true values of $\theta_{23}$ and $\delta_{\mathrm{CP}}$ are kept at their best fit values given in table~\ref{tab:params_value}. We marginalize over $\theta_{23}$ and $\delta_{\mathrm{CP}}$ in their 3$\sigma$ allowed range in the fit. Apart from $\theta_{23}$ and $\delta_{\mathrm{CP}}$, we also marginalize over the associated LIV phases in their total allowed range [$0^{\circ}$, $360^{\circ}$].}
	\label{fig:cptv_liv_bounds}
\end{figure}

In \figu{cptv_liv_bounds}, we show $\Delta \chi^2$ as a function of the off-diagonal CPT-violating (top row) and CPT-conserving (bottom row) LIV parameters. The red, blue, and black lines in each panel correspond to the sensitivity of DUNE, T2HK, and DUNE+T2HK setups, respectively. The top left panel correspond to $|a_{e\mu}|$, where we see that DUNE shows better sensitivity compared to T2HK at 95\% C.L.. Here, we find that for both DUNE and T2HK, there are local minima of $\Delta\chi^2$ around $1\times10^{-23}$ GeV and $4\times10^{-23}$ GeV, respectively. This feature can be explained using the correlations of the LIV parameters with standard oscillation parameters $\theta_{23}$ and $\delta_{\mathrm{CP}}$ discussed in section~\ref{subsec:correlation}. We observe that there are degenerate allowed regions in $\delta_{\mathrm{CP}}-|a_{e\mu}|$ and $\theta_{23}-|a_{e\mu}|$ planes (see figs.~\ref{fig:correlation_dcp_cptv} and \ref{fig:correlation_th23_cptv}) around the same values of $|\aem|$ ($\approx1\times10^{-23}$ GeV for DUNE and $\approx 4\times10^{-23}$ GeV for T2HK), where the local minima occur. Since this parameter is mainly constrained by the appearance channel, it hints towards a degeneracy between the appearance probability in absence of any new physics and the same in presence of the LIV for some combination of $\theta_{23}, \delta_{\mathrm{CP}}$ and new phase $\phi$. However, this degeneracy vanishes as we combine the data from DUNE and T2HK, giving a more stringent limit on $|\aem|$. In the top middle panel, we show the sensitivity for $|\aet|$. Here also, the local minima of $\Delta\chi^2$ are observed for the individual setup DUNE and T2HK, which again occur due to degeneracies between $\theta_{23}$, $\delta_{\mathrm{CP}}$, and $\phi$ (see top middle panel of \figu{correlation_dcp_cptv} and \ref{fig:correlation_th23_cptv}). Adding the data from the two experiments solves the issue of local minima.
Top right panel shows the constraints on $|\amt|$ for the three setups. We observe that DUNE gives significantly better limits for $|\amt|$ as compared to T2HK. Unlike $|\aem|$ and $|\aet|$, we do not observe any local minima of $\Delta\chi^2$ for $|\amt|$. It happens because $|\amt|$ is mainly constrained by the disappearance channel, where such degeneracy among $\theta_{23}$ and the CP phases ($\delta_{\rm CP}$ and $\phi$) does not occur.
In the lower panels of \figu{cptv_liv_bounds}, we show the constraints on CPT-conserving LIV parameters. As it is clear from the oscillation probability plots in \figu{cptc_app_prob} (see bottom row) and \figu{disapp_prob} (see bottom right panel), T2HK has almost no sensitivity on the CPT-conserving LIV parameters. However, when the data from T2HK and DUNE are added, sensitivities are slightly improved for all the three off-diagonal parameters.

In table~\ref{tab:constraints_a}, we list the expected constraints on the off-diagonal CPT-violating and CPT-conserving LIV parameters at 95\% C.L. The second and third columns show the limits from DUNE and T2HK, respectively. The fourth column is the ultimate limit on LIV parameters from the combination of DUNE and T2HK.
Note that for the bounds on $|\aem|$ and $|\aet|$, we consider the most conservative scenarios, $\ie$ the largest value of $|a_{e\beta}|$ ($\beta = \mu,~\tau$), which reaches 95\% C.L. value. For $|a_{e\beta}|$, the obtained constraints from DUNE are almost five times better than that of T2HK. Also, combining the data from DUNE and T2HK, the limits improved further by a factor of $\approx 3$ for $|\aem|$  and $\approx 2$ for $|\aet|$ compared to DUNE. For $|\amt|$ also, constraints from DUNE outperform T2HK approximately by a factor of four. However, DUNE+T2HK setup improves the bounds on $|\amt|$ by only $\approx12\%$ compared to the standalone DUNE. In the case of CPT-conserving LIV parameters, the constraints from DUNE are incomparable to that of T2HK, as the latter shows almost no sensitivity to CPT-conserving LIV parameters. However, the combined setup DUNE+T2HK shows a marginal improvement in the limits compared to DUNE only.

\begin{table}[h!]
	\begin{center}
		\begin{adjustbox}{width=0.8\textwidth}
			\begin{tabular}{|c|c|c|c|c|}
				\hline\hline
				&DUNE& T2HK & DUNE+T2HK & T2K+NO$\nu$A\\ 
				\hline
				$|a_{e\mu}|~[10^{-23}~\rm {GeV}]$ &  $<$ 1.0 & $<$ 5.15 & $<$ 0.32 & $<$ 6.1\\ 
				\hline
				$|a_{e\tau}|~[10^{-23}~\rm {GeV}]$& $<$ 1.05 & $<$ 5.3 & $<$ 0.55& $<$ 7.0\\
				\hline
				$|a_{\mu\tau}|~[10^{-23}~\rm {GeV}]$& $<$ 1.26 & $<$ 5.5 & $<$ 1.1& $<$ 8.3\\
				\hline
				$|c_{e\mu}|~[10^{-24}]$ &  $<$ 0.66 &  $<$ 17.1  &   $<$ 0.64 & $<$ 11.0 \\ 
				\hline
				$|c_{e\tau}|~[10^{-24}]$& $<$ 1.65 &  $<$ 71.1 &   $<$ 1.49& $<$ 37.5\\
				\hline
				$|c_{\mu\tau}|~[10^{-24}]$& $<$ 0.97 &  $<$ 42.4 &   $<$ 0.95 & $<$ 29.0\\
				\hline\hline
			\end{tabular}
		\end{adjustbox}
		\mycaption{Expected bounds on the off-diagonal CPT-violating and CPT-conserving LIV parameters at 95\% C.L. (1 d.o.f.) using DUNE, T2HK, and the combination of DUNE and T2HK. Last column shows the results using the combination of T2K and NO$\nu$A with their full exposures.  }
		\label{tab:constraints_a}
	\end{center}
\end{table}

For a comparison with currently running long-baseline experiments T2K and NO$\nu$A, in the last column, we provide expected bounds from the combination of T2K and NO$\nu$A considering their full exposure. We assume a total exposure of 84.4 kt$\cdot$MW$\cdot$yrs for T2K~\cite{T2K:2001wmr, T2K:2011qtm, T2K:2014xyt} with five years of total runtime divided equally in neutrino and antineutrino modes. For NO$\nu$A~\cite{Ayres:2002ws, NOvA:2004blv, NOvA:2007rmc, Patterson:2012zs}, we consider total exposure of 58.8 kt$\cdot$MW$\cdot$yrs with six years of runtime with three years each in neutrino and antineutrino mode. We observe that DUNE puts significantly better constraints on both the CPT-violating and CPT-conserving LIV parameters than T2K+NO$\nu$A setup because the former has a larger baseline and better systematic uncertainties. However, the constraints on CPT-violating LIV parameters from T2K+NO$\nu$A setup are close to that of T2HK. Since NO$\nu$A has a comparatively larger baseline ($L=810$ km), it has the upper hand in putting stringent bounds on the CPT-violating parameters. However, T2HK has less systematic uncertainties to compensate for its small baseline. Similarly, for the CPT-conserving LIV parameters, the limits from T2K+NO$\nu$A setup are of the same order as T2HK, where the former gives slightly better constraints. It is because, apart from the larger baseline, the energy of the neutrino beam is also prominent for NO$\nu$A, which plays a vital role in constraining CPT-conserving LIV parameters.

The fourth column of table~\ref{tab:constraints_a} shows the ultimate constraints on the off-diagonal CPT-violating and CPT-conserving LIV parameters that would be set from the combination of two next-generation long-baseline experiments DUNE and T2HK at 95\% C.L. In table~\ref{tab:existing_bounds}, we show the existing bounds on some CPT-violating and CPT-conserving LIV parameters from the atmospheric neutrino  experiments, Super-K and IceCube. Comparing these with our results in table~\ref{tab:constraints_a}, we observe that DUNE alone would be able to give better constraints of $\aem$ and $\aet$ as compared to the existing bounds from Super-K. Combining DUNE and T2HK can improve the constraints for $a_{e\beta}$ ($\beta=\mu,~\tau$) by almost one order of magnitude. One reason for this is the fact that $a_{e\beta}$ ($\beta=\mu,~\tau$) are mainly constrained by the appearance channel, which is the most important oscillation channel for a next-generation long-baseline experiment like DUNE that has a considerably larger baseline. For the atmospheric neutrino experiment like Super-K, the major channel is $\nu_\mu\to\nu_\mu$ disappearance channel, in which these two LIV parameters do not appear in the leading order. For $\amt$, projected results from the DUNE+T2HK setup are of the same order as Super-K, with the later having a slightly better limit. In the case of CPT-conserving LIV parameters, we observe that the existing limits from Super-K are at least one order better for $\cem$ and $\cmt$ compared to our results for DUNE+T2HK. However, for $\cet$, the expected limits from DUNE+T2HK are comparable with Super-K. This mainly happens because the contributions from CPT-conserving LIV parameters to oscillation probabilities in both the oscillation channels are proportional to both neutrino energy and the baseline. Atmospheric neutrino experiment like Super-K probes a significantly larger range of neutrino energies and baselines as compared to the LBL experiment like DUNE. So we expect Super-K to have better limits on the CPT-conserving LIV parameters.

%=============================%
\section{Summary and conclusions}
\label{sec:SnC}
%=============================%

In the past few decades, data from outstanding neutrino oscillation experiments that are either completed or currently operational has almost settled the issue of measuring standard three-flavor neutrino oscillation parameters with excellent precision. Apart from resolving a few remaining issues in the three-neutrino paradigm, another major goal of the next-generation neutrino oscillation experiments will be to search for various physics beyond the standard model, which will open up a new era in particle physics. 
With that motivation, in this work, we probe the Lorentz invariance violation and its impact on neutrino flavor transition in the context of the two most anticipated upcoming long-baseline experiments, DUNE and T2HK. The Lorentz invariance violation can be realized in low energy effective field theories where the LIV interaction terms in the lagrangian come as multiplication of Lorentz violating coefficients and Lorentz violating operators of arbitrary mass dimensions. The coefficients of the dimension-three and dimension-four operators are, respectively, CPT-violating and CPT-conserving. In this work, for the first time, we have explored the CPT-conserving LIV parameters in the context of long-baseline experiments. Here, we focus on the isotropic components of the off-diagonal CPT-violating and CPT-conserving LIV parameters. The presence of non-zero CPT-violating and CPT-conserving parameters modify the neutrino propagation Hamiltonian and hence the oscillation probabilities, making them worth studying in neutrino oscillation experiments.

To have an analytical understanding about the impact of various LIV parameters on the neutrino oscillation probability, we use the perturbative approach to derive simple approximate analytical expressions for the $\nu_\mu\to\nu_e$ appearance and $\nu_\mu\to\nu_\mu$ disappearance probabilities. Here, we keep the terms up to first order in $\alpha$, $\sin^2\theta_{13}$, and LIV parameters $a_{\alpha\beta}/c_{\alpha\beta}$ ($\alpha,\beta = e,\mu,\tau;\alpha\neq\beta$). We find that for the appearance channel, $a_{e\mu}/c_{e\mu}$ and $a_{e\tau}/c_{e\tau}$ appear at the leading order, whereas for the disappearance channel, only $a_{\mu\tau}/c_{\mu\tau}$ presents. Our analytical expressions explain various features of oscillation probabilities shown in figs.~\ref{fig:app_prob}--\ref{fig:disapp_prob}, where we plot the exact oscillation probabilities numerically. We explain how the impact of LIV on oscillation probabilities depends on the values of phases associated with the off-diagonal LIV parameters using the LIV contributing terms in the oscillation probabilities given in eqs.~(\ref{equ:p_cptv_liv}),~(\ref{equ:p_cptc_liv}), and (\ref{equ:Pmm_liv}). Also, we find that the LIV contributing terms in our analytical expressions for CPT-violating and CPT-conserving LIV parameters are proportional to $L$ and $L\times E$, respectively, both in the appearance and disappearance channels. As a result, DUNE, with a larger baseline and higher energy of the neutrino beam, shows significantly larger sensitivity to the LIV parameters than T2HK. As shown in lower panels of \figu{cptc_app_prob}, T2HK shows negligible sensitivity to the CPT-conserving LIV parameters.

Using the configuration of the DUNE and T2HK as tabulated in table~\ref{tab:exp_details}, we calculate the expected total event rates from the two setups in the standard case and in the presence of off-diagonal CPT-violating and CPT-conserving LIV parameters considered one-at-a-time (see table~\ref{tab:total_events}). As expected from the probability analysis, we observe a major change in the event rate from the SI case in the presence of $a_{e\mu}$ ($a_{e\tau}$), which shows 41\% (49\%) deviation in the event rates for DUNE and 7.3\%(14\%) for T2HK when we consider $|a_{e\beta}|=2\times10^{-23}$ GeV ($\beta=\mu,\tau$). In the disappearance channel, the presence of $a_{\mu\tau}$ leads to modifications in the event rates by 7.8\% for DUNE and 2.2\% for T2HK. The CPT-conserving LIV parameters for which we consider the strength $|c_{\alpha\beta}|=1\times 10^{-24}$ show comparatively small changes in the event rates with a maximum 11\% changes for DUNE and $<1\%$ for T2HK.

We discuss the correlation between various LIV parameters and most uncertain oscillation parameters $\theta_{23}$ and $\delta_{\mathrm{CP}}$ (see section~\ref{subsec:correlation}). To demonstrate this, we show allowed regions at 95\% C.L. (2 d.o.f.) in  $\delta_{\mathrm{CP}}-|a_{\alpha\beta}|/|c_{\alpha\beta}|$ (see \figu{correlation_dcp_cptv} and \figu{correlation_dcp_cptc})  and $\theta_{23}-|a_{\alpha\beta}|/|c_{\alpha\beta}|$ (see \figu{correlation_th23_cptv} and \figu{correlation_th23_cptc}) planes. In all cases, DUNE shows a more constrained allowed region in the above-mentioned planes, as expected from the probability plots. We notice almost no correlation between the CPT-conserving LIV parameters and standard oscillation parameters ($\theta_{23}$ and $\delta_{\rm CP}$) for T2HK. For $\delta_{\mathrm{CP}}-|a_{e\mu}|$ case, we observe some allowed regions centered around a non-zero value of $|a_{e\mu}|$, both for DUNE and T2HK. It happens due to degeneracy between the $\theta_{23}$ and the phases $\delta_{\mathrm{CP}}$ and $\phi$, which minimizes the contribution from LIV at some non-zero value of the $|a_{e\mu}|$. Also, in the case of $\theta_{23}-|a_{e\mu}|$ correlation, we observe such degenerate regions in opposite octant to the value of $\theta_{23}$ considered in the data. This also happens due to degeneracy between these three parameters ($\theta_{23}$, $\delta_{\mathrm{CP}}$ and $\phi$). In both the cases, the degenerate allowed regions vanish upon combining the data from DUNE and T2HK. Also, there is improvement in allowed regions in both $\delta_{\mathrm{CP}}-|a_{\alpha\beta}|$ and $\theta_{23}-|a_{\alpha\beta}|$ planes for the DUNE+T2HK setup. For the CPT-conserving case, this improvement is marginal. 
 
From the discussion on the correlation between the LIV parameters and standard oscillation parameters $\delta_{\mathrm{CP}}$ and $\theta_{23}$ in DUNE and T2HK, we get an idea about the sensitivity of these two setups to the off-diagonal CPT-violating and CPT-conserving LIV parameters. Here, we present the limits (see \figu{cptv_liv_bounds} and table~\ref{tab:constraints_a}) on the LIV parameters that DUNE, T2HK, and their combination are expected to place with their full exposures. For CPT-violating LIV parameters, DUNE shows almost five times better constraints than T2HK.
Also, for $|a_{e\beta}|$ ($\beta = \mu,\tau$), we observe a local minimum in $\Delta\chi^2$ which results in deterioration in the constraints from DUNE and T2HK. It happens due to the degeneracy between $\theta_{23}$ and the phases $\delta_{\mathrm{CP}}$ and $\phi$ which is clear from the correlation plots. 
For the combination DUNE+T2HK, limits improve significantly for $|a_{e\beta}|$ ($\beta = \mu, \tau$) as the above discussed degeneracies vanish. For $a_{\mu\tau}$, DUNE outperforms T2HK, and their combination results in a small improvement in the limits. T2HK shows almost no sensitivity for the CPT-conserving LIV parameters. As a result, T2HK produces incomparably worse constraints relative to DUNE. To compare the limits from the future LBL experiments with the currently operating ones, we show the expected constraints by combining T2K and NO$\nu$A setups with their full exposures (see the last column of table~\ref{tab:constraints_a}). We observe for CPT-violating LIV parameters, though the bounds from T2K+NO$\nu$A are worse than DUNE, it is comparable to that of T2HK. It is mainly because of the larger baseline of NO$\nu$A than T2HK that has better systematic uncertainties than the former. For the CPT-conserving parameters, T2K+NO$\nu$A gives slightly better constraints than T2HK, again because the former has a longer baseline and higher energy of the neutrino beam. We also compare our results with the existing limits on CPT-violating and CPT-conserving parameters listed in table~\ref{tab:existing_bounds} from Super-K. We find that for CPT-violating parameters, especially $a_{e\beta}$ ($\beta=\mu,\tau$), projected limits at 95\% C.L. for the DUNE and T2HK combination is better compared to Super-K. It happens because the limits on $a_{e\beta}$ are mainly driven by the appearance channel, which is a major oscillation channel for LBL experiments, whereas the atmospheric neutrino experiment like Super-K mainly probe disappearance channel. For CPT-conserving parameters, Super-K shows almost two-order better constraints compared to the DUNE+T2HK setup. It is because of the large energy of the neutrinos in atmospheric neutrino experiments.

In this work, we discuss the impact of LIV on neutrino flavor transition probabilities in the context of next-generation long-baseline experiments DUNE and T2HK and explore the ability of these two setups to constrain the CPT-violating and CPT-conserving LIV parameters. We conclude that although these two setups have good complementarity in their configurations, a smaller baseline and lower energy of the neutrino beam for T2HK make it less favorable to probe LIV as compared to DUNE. However, combining DUNE and T2HK can improve the limits on LIV parameters up to a certain extent. We hope that our present study can be an important addition to the several interesting beyond the Standard Model scenarios which can be probed in the next-generation long-baseline neutrino oscillation experiments.

%%%%%%%%%%%%%%%%%%%%%%%%%%%%%%%%%%%
\blankpage 
%%%%%%%%%%%%%%%%%%%%%%%%%%% Chapter-5 %%%%%%%%%%%%%%%%%%%%%%%%%%%%%%%
%%%%%%%%%%%%%%%%%%%%%%% CHAPTER - 5 %%%%%%%%%%%%%%%%%%%%
\chapter{Active-sterile neutrino oscillations in long-baseline experiments over a wide mass-squared range}
\label{C5} 
%%%%%%%%%%%%%%%%%%%%%%%%%%%%%%%%%%%%%%%%%%%%%%%%%%%%%%%%%%
%======================================
\section{Introduction and motivation}
%======================================

The three flavor oscillation framework involves three mixing angles ($\theta_{12}, \theta_{13}, \theta_{23}$), two distinct mass-squared splittings ($ \Delta m^{2}_{21} =m^{2}_{2}-m^{2}_{1}\,,  ~\Delta m^{2}_{31} = m^{2}_{3}-m^{2}_{1}$), and one CP phase $\delta_{13}$ (equivalent to $\delta_{\rm CP}$). Marvelous data from various world-class experiments such as KamLAND~\cite{KamLAND:2013rgu}, Daya Bay~\cite{DayaBay:2018yms}, RENO~\cite{RENO:2018dro}, Homestake~\cite{Cleveland:1998nv}, Gallex~\cite{Kaether:2010ag}, SAGE~\cite{SAGE:2009eeu}, SNO~\cite{SNO:2011hxd, sno_solar_neutrino2024}, Borexino~\cite{Borexino:2008fkj, Bellini:2011rx}, Super-K~\cite{Super-Kamiokande:2010tar, Super-Kamiokande:2017yvm, Super-Kamiokande:2019gzr, Super-Kamiokande:2004orf}, IceCube DeepCore~\cite{IceCube:2017lak}, ANTARES~\cite{ANTARES:2018rtf}, K2K~\cite{K2K:2001nlz, K2K:2002icj}, MINOS~\cite{MINOS:2013utc}, T2K~\cite{T2K:2019bcf}, and NO$\nu$A~\cite{NOvA:2021nfi} have been conclusive in establishing the three neutrino oscillation framework with the measurement of the oscillation parameters with satisfactory precision. 

Despite being very successful in explaining most of the neutrino oscillation data from various experiments, the standard three flavor oscillation framework fails to accommodate some of the anomalous results that come from various short-baseline (SBL) neutrino experiments such as LSND~\cite{LSND:1995lje, LSND:2001aii} that observed an excess of $\bar{\nu}_{e}$-like events in the $\bar{\nu}_{\mu}$ beam, the Gallium radioactive source experiments GALLEX~\cite{Kaether:2010ag} and SAGE~\cite{SAGE:2009eeu} which indicated the deficit in $\nu_e$ events, and the SBL reactor antineutrino experiments~\cite{Mention:2011rk} which reported a deficit in the observed $\bar{\nu}_{e}$ events compared to the event rates expected from the reactor antineutrino fluxes estimated by Huber~\cite{Huber:2011wv} and the Mueller group~\cite{Mueller:2011nm}. Moreover, MiniBooNE experiment, which was designed to verify the LSND anomaly, has reported a 4.8$\sigma$ excess of ~$\nu_e/\bar\nu_{e}$ charged-current quasi-elastic (CCQE) events in ~$\nu_\mu/\bar\nu_{\mu}$ beam at low energies~\cite{MiniBooNE:2008yuf, MiniBooNE:2018esg, Arguelles:2019xgp}. All these anomalous results can be explained via active-sterile neutrino oscillations driven by a new and comparatively large mass-squared splitting ($\Delta m^{2} \sim$1 eV$^2$) with respect to the standard atmospheric and solar mass-squared splittings. Such a large mass-squared splitting can be incorporated into the oscillation scenario by extending the standard 3$\nu$ oscillation to the so-called 3+1 scheme with one sterile neutrino. The later involves one new mass-squared splitting $\Delta m^2_{41}\,(\equiv m^2_4-m^2_1)$, three new mixing angles ($\theta_{14}, \theta_{24}$, and $\theta_{34}$), and two new CP phases ($\delta_{14}$ and $\delta_{34}$). The combined analysis of NEOS and DANSS~\cite{Gariazzo:2018mwd} spectral ratios in the 3+1 framework favors the SBL $\bar\nu_e$ oscillations with a statistical significance of 3.7$\sigma$ with $\Delta m^2_{41}\approx 1.3~\text{eV}^2$. The recent reexamination of the data from the ILL~\cite{Cogswell:2018auu} experiment claims a $3\sigma$ evidence of active-sterile oscillations with $\Delta m^2_{41}\approx 1~\text{eV}^2$. Very recently, the BEST~\cite{Barinov:2022wfh}  experiment, which probes the Gallium anomaly, has reported the $\nu_e\rightarrow\nu_s$ oscillations at 4$\sigma$ C.L. with $\Delta m^2_{41}\approx 3.3~\text{eV}^2$. A $3.5\sigma$ indication of active-sterile oscillations has been reported by Neutrino-4~\cite{NEUTRINO-4:2018huq} experiment with $\Delta m^2_{41}\approx 7.26~\text{eV}^2$ and $\sin^{2}2\theta_{14}\approx 0.38$. Even if there are positive indications for the short-baseline $\barparen{e}\rightarrow\barparen{e}$ disappearance and $\barparen\mu\rightarrow\barparen{e}$ appearance, the non-observation of $\barparen\mu\rightarrow\barparen\mu$ disappearance events~\cite{MINOS:2017cae, IceCube:2016rnb} leads to a strong appearance-disappearance tension~\cite{Giunti:2013aea, Kopp:2013vaa, Conrad:2012qt, Gariazzo:2015rra, Giunti:2019aiy, Boser:2019rta} in the 3+1 framework. Despite having a similar experimental setup as LSND, the KARMEN experiment found no evidence of electron-like excess~\cite{Eitel:2002jr}. The STEREO experiment, which was looking for the electron antineutrino survival events, reported null results for the sterile oscillation hypothesis and rejected the best-fit of the reactor antineutrino anomaly at 97.5\% C.L.~\cite{STEREO:2018rfh}. The measurement of the energy spectrum at PROSPECT experiment rejects the best-fit point of the reactor antineutrino anomaly at 2.5$\sigma$ C.L.~\cite{PROSPECT:2020sxr}. Moreover, the reactor antineutrino anomaly is almost resolved with the updated and refined calculations of the reactor electron antineutrino fluxes~\cite{Estienne:2019ujo, Kopeikin:2021ugh, Giunti:2021kab}. The recent analysis of the global $\nu_e$ and $\bar\nu_e$ disappearance data leaves Gallium Anomaly in a visible tension with the reactor antineutrino anomaly, which seems to be getting resolved in light of recent flux model refinements~\cite{Giunti:2022btk}. MicroBooNE experiment, designed to probe the MiniBooNE anomaly, disfavors the $\nu_e$ interpretation of the MiniBooNE low-energy excess (MiniBooNE LEE)~\cite{MicroBooNE:2021ktl}. However, the MicroBooNE experiment does not probe the entire sterile neutrino parameter space allowed by the MiniBooNE experiment~\cite{Arguelles:2021meu}. Furthermore, the $\nu_e$ interpretation of the MiniBooNE LEE is not studied in a model-independent way~\cite{Arguelles:2021meu}. The authors of \Refe~\cite{MiniBooNE:2022emn} claim that even though the 3+1 active-sterile oscillation model is not the quintessential scheme to explain the MiniBooNE LEE, it gives a better fit to MiniBooNE data compared to the no-oscillation scenario. In reference~\cite{Denton:2021czb}, the search for $\nu_e$ survival events using the MicroBooNE data gives a hint for the active-sterile oscillations at 2.2$\sigma$. However, very recently, the MicroBooNE collaboration published their result~\cite{MicroBooNE:2022sdp}, considering both the appearance and disappearance events. They do not find any signal of active-sterile neutrino oscillation and hence, provide constraints on the sterile neutrino parameter space. All these results suggest that the active-sterile oscillation hypothesis is under tension within the global picture of neutrino oscillation data and needs more investigations to be verified conclusively. The upcoming Fermilab Short-Baseline Neutrino (SBN) Program~\cite{MicroBooNE:2015bmn} and JSNS$^2$ experiment at the J-PARC facility~\cite{JSNS2:2013jdh, Ajimura:2020qni} are expected to provide definitive tests for the MiniBooNE and LSND anomalies, respectively. Future experiments: JUNO-TAO~\cite{JUNO:2020ijm}, PROSPECT-II~\cite{PROSPECT:2021jey}, and IsoDAR~\cite{Alonso:2022mup} will be able to probe most of the allowed regions by the reactor antineutrino and the gallium anomalies. Recent reviews on the search for active-sterile neutrino oscillations can be found in the Refs.~\cite{Diaz:2019fwt, Boser:2019rta, Palazzo:2020tye, Dasgupta:2021ies, Acero:2022wqg}.

All the SBL anomalies discussed so far can be explained by the active-sterile oscillations with a mass-squared difference of around 1 $\text{eV}^2$. Additionally, there exist a handful of motivations to look for the existence of sterile neutrinos over a wide range of mass-squared differences. For example, the absence of the expected upturn\footnote{{The upturn is a prediction of the large mixing angle LMA solution~\cite{Smirnov:2003da, deHolanda:2002dko} to the solar neutrino problem. The $\nu_e$ disappearance probability is driven by matter effect in relatively higher energy, $\gtrsim$ 5 MeV, where, $P_{\nu_{e} \to \nu_{e}} \approx \sin^2\theta_{12}\sim0.3$, however at lower energy, $\lesssim$ 2 MeV, vacuum oscillation takes over ($P_{\nu_{e} \to \nu_{e}} \approx 1-\frac{1}{2}\sin^22\theta_{12}\sim0.58$). In both the cases, the probability becomes energy independent, and thus an upturn in the survival probability happens in the intermediate energies.}} in the solar neutrino spectrum at low energy below 8 MeV can be explained with the help of a super-light sterile neutrino with a value of $\Delta m^2_{41}\sim(0.7-2)\times10^{-5}~\text{eV}^2$~\cite{deHolanda:2003tx, deHolanda:2010am, BhupalDev:2012jvh, Liao:2014ola, Divari:2016jos}. The data from $\theta_{13}$ sensitive reactor antineutrino experiments can be used to place constraints on very light neutrinos having their masses below 0.1 eV~\cite{Palazzo:2013bsa, Esmaili:2013yea, Girardi:2014wea, DayaBay:2014fct, DayaBay:2016qvc}. The possible existence of a light sterile neutrino with a mass of a few eV can help to understand the neutron deficit during the nucleosynthesis of heavy elements inside supernova~\cite{Caldwell:1999zk, Tamborra:2011is, Wu:2013gxa}. A keV scale sterile neutrino can simultaneously explain the dark matter and the baryon asymmetry of the Universe in the Neutrino Minimal Standard Model ($\nu$MSM)\footnote{For a detailed review on the keV scale sterile neutrino dark matter, see \Refe~\cite{Drewes:2016upu}.}~\cite{Asaka:2005an, Asaka:2005pn, Boyarsky:2009ix, Drewes:2016upu}. So, it is logical to probe the active-sterile oscillations over a wide range of the sterile neutrino mass. There is a plethora of studies in this direction~\cite{Orchanian:2011qq, Berryman:2015nua, IceCube:2016rnb,  MINOS:2016viw,  DayaBay:2016lkk, DayaBay:2016qvc, Choubey:2016fpi, Coelho:2017cwp, Choubey:2017ppj, Terliuk:2017jyw, Leitner:2017jco, Thakore:2018lgn, Todd:2018bzx, ANTARES:2018rtf, T2K:2019efw, Jones:2019nix, KATRIN:2020dpx, MINOS:2020iqj,  Hu:2020uvx, IceCube:2020tka, Licciardi:2021hyi, Arguelles:2021gwv, Coloma:2021uhq, NOvA:2021smv, KM3NeT:2021uez, Andriamirado:2021qjc, Chattopadhyay:2022hkw}.

In this work, we explore the potential of DUNE, T2HK (the Japanese Detector, JD), and T2HKK (the combination of Japanese and the Korean Detectors, JD+KD), in probing the sterile neutrino parameter space for a wide range of the sterile mass-squared difference ($\Delta m^2_{41}$) starting from $10^{-5}$ eV$^2$ to 100 eV$^2$. We discuss the CP violation discovery potential and the CP reconstruction capabilities of these experiments in the 3+1 scheme. The near detectors of these future long-baseline experimental facilities will help to probe the allowed regions by the SBL anomalies. We consider the Intermediate Water Cherenkov Detector (IWCD)~\cite{Drakopoulou:2017qdu} for the JD/KD setup and both the near and far detectors of DUNE, in deriving the exclusion contours on the sterile neutrino parameter space in the 3+1 active-sterile neutrino oscillation scenario. We find that these setups will probe most of the SBL-allowed parameter regions. The CP-violating phases can be observed only through the interference between two frequencies. In the SBL setups, only one frequency ($\Delta m^2_{41}$) is relevant. So, due to the absence of any interference effect, the SBL setups are completely blind to the presence CP phases. However, in the long-baseline (LBL) setups, as first pointed out in~\cite{Klop:2014ima}, the interference between any two distinct frequencies (out of the three, i.e., the solar ($\Delta m^2_{21}$), atmospheric ($\Delta m^2_{31}$), and the sterile ($\Delta m^2_{41}$) frequencies) may show up. Hence, the LBL experiments can probe both the standard and the new sterile CP phases, providing a complementarity between the LBL and SBL experiments in the context of sterile neutrino search. Previously, the sensitivity to the CP phases in the 3+1 scheme has been studied in the references~\cite{Agarwalla:2016mrc, Agarwalla:2018nlx, Agarwalla:2016xxa, KumarAgarwalla:2019blx}. Recently, similar studies have been carried out in \Refe~\cite{Singha:2022btw}, using the possible LBL experiment options at KM3NeT and in \Refe~\cite{Sharma:2023jzg} in the context of the proposed MOMENT experiment that uses a muon beam for neutrino production instead of the conventional pion beams. Unlike the previous studies that only explore the CP sensitivity for a eV-scale sterile neutrino, the present work focuses on the CP sensitivities at three different scales; when the sterile frequency ($\Delta m^2_{41}$) is equal to (i) 1 eV$^2$, (ii) $\Delta m^2_{21}$, and (iii) $\Delta m^2_{31}$. We explore the CP violation discovery potential and reconstruction capabilities of CP phases of DUNE, JD, and JD+KD setups at these three different scales of $\Delta{m}^2_{41}$.

This chapter is organized in the following fashion. In section~\ref{sec:basics}, we discuss the parameterization of the mixing matrix in the 3+1 scheme. In the same section, we also provide the approximate $\nu_{\mu}\rightarrow\nu_e$ transition probability expressions to extract the information on CP violation sensitivity of the experiments from the probability plots.  Section~\ref{sec:experimental details} gives the details regarding the experimental setups and simulation method used for our analysis. Section~\ref{sec:event level discussion} deals with the bi-event plots and associated discussion. In section~\ref{sec: CPV discovery}, we present our results on the CP violation discovery potential of JD, JD+KD, and DUNE. In section~\ref{sec: CP reconstruction}, we discuss the CP phase reconstruction capabilities of these experiments. Section~\ref{sec: exclusion contours} is devoted in determining the parameter space excluded by the long-baseline experiments under consideration if there is no sterile neutrino in Nature. We summarize the results and draw our conclusions in section~\ref{sec: conclusion}.
%\newpage
%========================Parameter Table====================================

\begin{table}[h!]
	\begin{center}
		{
			\begin{tabular}{|c|c|c|}
				\hline\hline
				%	\mr{2}{*}{\bf Parameter} & \mr{2}{*}{\bf True Value} & \mr{2}{*}{\bf Marginalization Range} \\
				{\bf Oscillation parameters} & {\bf Benchmark values} & {\bf Marginalization ranges} \\
				%& &  \\
				\hline\hline
				{${\theta_{12}/^{\circ}}$} & {33.46} & {Not marginalized} \\
				%& &  \\
				\hline
				{${\theta_{13}/^{\circ}}$} & {8.47} & {Not marginalized} \\ 
				%& &  \\
				\hline
				{${\theta_{23}/^{\circ}}$} & {45} & {[39.23, 50.77]} \\
				%\mr{2}{*}{[0.4, 0.6]} \\
				%& &  \\
				\hline
				%	\mr{2}{*}{${\theta_{14}}/^{\circ}$} & \mr{2}{*}{0} & \mr{2}{*}{[0, 10]} \\
				{${\theta_{14}/^{\circ}}$} & {= ${\theta_{13}}$} & {[0, 10]} \\
				%& &  \\  
				\hline
				%	\mr{2}{*}{${\theta_{24}}/^{\circ}$} & \mr{2}{*}{0} & \mr{2}{*}{[0, 10]} \\
				{${\theta_{24}/^{\circ}}$} & {= ${\theta_{13}}$} & {[0, 10]} \\
				%& &  \\ 
				\hline
				{${\theta_{34}}/^{\circ}$} & {0} & {Not marginalized}  \\
				%& &  \\  
				\hline
				%	\mr{2}{*}{$\delta_{13}/^{\circ}$} & \mr{2}{*}{- 90} & \mr{2}{*}{[- 180, 180]} \\
				{$\delta_{13}/^{\circ}$} & {[- 180, 180]} & {[- 180, 180]} \\
				%& &  \\
				\hline
				%	\mr{2}{*}{$\delta_{14}/^{\circ}$} & \mr{2}{*}{0} & \mr{2}{*}{[- 180, 180]} \\
				{$\delta_{14}/^{\circ}$} & {[- 180, 180]} & {[- 180, 180]} \\
				%& &  \\
				\hline
				{$\delta_{34}/^{\circ}$} & {0} & {Not marginalized} \\
				%& &  \\  
				\hline
				{${\Delta{m^2_{21}}}/{10^{-5} \, \rm{eV}^2}$} & {7.50} & {Not marginalized} \\
				%& &  \\
				\hline
				{${\Delta{m^2_{31}}}/{10^{-3} \, \rm{eV}^2}$ (NMO)} & {2.475}  & {Not marginalized} \\
				%& & \\
				\hline
				%	\mr{2}{*}{$\frac{\Delta{m^2_{41}}}{\rm{eV}^2}$} & \mr{2}{*}{0} & \mr{2}{*}{Not marginalized} \\
				{${\Delta{m^2_{41}}}/{\rm{eV}^2}$} & {$10^{-5}-10^{2}$} & {Not marginalized} \\
				%& &  \\
				\hline\hline
			\end{tabular}
		}
		\mycaption{Benchmark values of the oscillation parameters and their marginalization ranges used in our analysis in the 3+1 scheme. These values are in agreement with the present best-fit values obtained by various global fit studies~\cite{Esteban:2020cvm, deSalas:2020pgw, Capozzi:2021fjo, Gariazzo:2017fdh,  Dentler:2018sju}.  Normal mass ordering (NMO) is considered throughout this work. We do not marginalize over the neutrino mass ordering.}
		\label{tab:params_value_str}
	\end{center}
\end{table}

%=====================================================
\section{Oscillation probabilities in the 3+1 scheme}
\label{sec:basics}
%=====================================================

In the presence of a light sterile neutrino, the mass and the flavor bases are connected via a 4$\times$4 unitary matrix, which is a product of six rotational matrices~\cite{Agarwalla:2016xxa, Klop:2014ima},
\begin{equation}
\label{equ:U}
U =   \tilde R_{34}R_{24}\tilde R_{14}R_{23}\tilde R_{13}R_{12}\,,
\end{equation} 
%------------------------------------------------------------------------------------
where, $R_{ij}$ ($\tilde R_{ij}$) is a real (complex) $4\times4$ rotation matrix in the ($i,j$) plane containing the $2\times2$ matrix,
%------------------------------------------------------------------------------------
\begin{eqnarray}
\label{equ:R_ij_2dim}
R^{2\times2}_{ij} =
\begin{pmatrix}
c_{ij} &  s_{ij}  \\
- s_{ij}  &  c_{ij}
\end{pmatrix}
\,\,\,{\rm and}\,\,\,   
\tilde R^{2\times2}_{ij} =
\begin{pmatrix}
c_{ij} &  \tilde s_{ij}  \\
- \tilde s_{ij}^*  &  c_{ij}
\end{pmatrix}
\,,   
\end{eqnarray}
%------------------------------------------------------------------------------------
in the  $(i,j)$ sub-block, with the following definitions,
%------------------------------------------------------------------------------------
\begin{eqnarray}
c_{ij} \equiv \cos \theta_{ij} \,\,, s_{ij} \equiv \sin \theta_{ij}\,\,, {\rm and}\,\,  \tilde s_{ij} \equiv s_{ij} e^{-i\delta_{ij}}.
\end{eqnarray}
%------------------------------------------------------------------------------------
The  parameterization in \equ{U} has the following interesting properties~\cite{Klop:2014ima}:
\begin{itemize}
	\item When there is no mixing with the fourth state (i.e. $\theta_{14} = \theta_{24} = \theta_{34} =0$), 
	one can get back to the standard $3\nu$ matrix in its usual parameterization.
	\item For small values of $\theta_{13}$ and the mixing angles 
	involving $\nu_4$, one has $|U_{e3}|^2 \simeq s^2_{13}$, $|U_{e4}|^2 = s^2_{14}$, 
	$|U_{\mu4}|^2  \simeq s^2_{24}$ and $|U_{\tau4}|^2 \simeq s^2_{34}$, 
	with a clear physical interpretation of the new mixing angles. 
	\item The leftmost positioning of the matrix $\tilde R_{34}$ guarantees that 
	the vacuum $\nu_{\mu} \to \nu_{e}$ transition probability is independent of $\theta_{34}$ and of the associated CP phase $\delta_{34}$.\\
\end{itemize}
\vspace{-0.5cm}
The lepton mixing-matrix elements in the 3+1 scheme are given by,
\begin{align}
&U_{e1}=c_{14} c_{13} c_{12}\nonumber\,,\\
&
U_{\mu1}= (-s_{24} s_{14} c_{13}  e^{i\delta_{14}}-c_{24}s_{23}s_{13} e^{i\delta_{13}})c_{12}
-c_{24}c_{23}s_{12}\nonumber\,,\\
&U_{\tau 1}=c_{12}\left[ -s_{34} c_{24} s_{14}c_{13} e^{-i(\delta_{34}-\delta_{14})}
-s_{13} e^{i\delta_{13}}(c_{34}c_{23}-s_{34}s_{24}s_{23} e^{-i\delta_{34}})\right]\nonumber\,\\
&\hspace{1cm}
+s_{12}(s_{34}s_{24}c_{23} e^{-i\delta_{34}}+c_{34} s_{23})\nonumber\,,\\	
&U_{s1}=c_{12} \left[-c_{34}c_{24}s_{14} c_{13} e^{i\delta_{14}}
+s_{13} e^{i\delta_{13}}(s_{34}c_{23} e^{i\delta_{34}}+c_{34}s_{24}s_{23})\right]\nonumber\,\\
&\hspace{1cm}
-s_{12}(-c_{34}s_{24}c_{23}+s_{34}s_{23} e^{i\delta_{34}})\nonumber\,,\\
&U_{e2}=c_{14}c_{13}s_{12}\nonumber\,,\\
&U_{\mu 2}=c_{24}c_{23}c_{12}-s_{12}(s_{24}s_{14}c_{13}e^{i\delta_{14}}
+c_{24}s_{23}s_{13}e^{i\delta_{13}})\nonumber\,,\\
&U_{\tau 2}=-c_{12}(s_{34} s_{24} c_{23} e^{-i\delta_{34}}+c_{34}s_{23})\nonumber\\
&\hspace{1cm}
-s_{12}\left[s_{34}c_{24}s_{14}s_{13} e^{-i(\delta_{34}-\delta_{14})}
+s_{13} e^{i\delta_{13}} (c_{34} c_{23}-s_{34}s_{24}s_{23} e^{-i\delta_{34}})\right]\nonumber\,,\\
& U_{s2}=c_{12}\left(-c_{34}s_{24}c_{23}+s_{34}s_{23} e^{i\delta_{34}}\right)\nonumber\\
&\hspace{1cm}
-s_{12}\left[ c_{34}c_{24}s_{14}c_{13}e^{i\delta_{14}}-s_{13}e^{i\delta_{13}}\left(
s_{34}c_{23}e^{i\delta_{34}}+c_{34}s_{24}s_{23}\right)\right]\nonumber\,,\\
&U_{e3}=c_{14}s_{13} e^{-i\delta_{13}}\nonumber\,,\\
&U_{\mu 3}=c_{24}s_{23}c_{13}-s_{24}s_{14}s_{13}e^{i(\delta_{14}-\delta_{13})}\nonumber\,,\\
&U_{\tau 3}=c_{13}\left(c_{34}c_{23}-s_{34}s_{24}s_{23}e^{-i\delta_{34}}\right)-s_{34}c_{24}s_{14}s_{13}e^{-i(\delta_{34}
	-\delta_{14}+\delta_{13})}\nonumber\,,\\
&U_{s3}=-c_{13}\left(s_{34}c_{23}e^{i\delta_{34}}
+c_{34}s_{24}s_{23}\right)
-c_{34}c_{24}s_{14}s_{13}e^{i(\delta_{14}-\delta_{13})}\nonumber\,,\\
&U_{e4}=s_{14} e^{-i\delta_{14}}\nonumber\,,\\
&U_{\mu 4}=s_{24}c_{14}\nonumber\,,\\
&U_{\tau 4}=s_{34}c_{24}c_{14}e^{-i\delta_{34}}\nonumber\,,\\
&U_{s4}=c_{34}c_{24}c_{14}\,.
\end{align}
\noindent
The general expression for neutrino oscillation probability from a flavor $\nu_\alpha$ to $\nu_\beta$ in vacuum, is given by (refer to \equ{general-prob}),
\begin{eqnarray} \label{equ:prob_gen}
\nonumber \pab = \delta_{\alpha\beta}-4\sum_{k>j}Re(U^{*}_{\alpha k}U_{\beta k}
U_{\alpha j}U^{*}_{\beta j})\sin^2\Delta_{kj}
+ 2\sum_{k>j}Im(U^{*}_{\alpha k}U_{\beta k}U_{\alpha j}U^{*}_{\beta j})\sin2\Delta_{kj}\,,
\end{eqnarray}
where,
\begin{equation}
\Delta_{kj} = 1.27 \times \Delta m^{2}_{kj}~[\textrm{eV}^{2}] \times L~[\textrm{km}]/E~[\textrm{GeV}]\,.
\end{equation}  
Here, $L$ and $E$ correspond to the baseline
length and the neutrino energy, respectively. Using the elements of mixing matrix $U$, we obtain the most general expression for the transition probability in vacuum~\cite{Gandhi:2015xza}, 
\begin{align} \label{equ:pme_general}
\nonumber P_{\mu e}^{4\nu} &= \sin^22\tmef\big[ \cos^2\theta_{13}\cos^2\theta_{12}\sin^2\Delta_{41}+\cos^2\theta_{13}\sin^2\theta_{12}\sin^2\Delta_{42}+\sin^2\theta_{13}\sin^2\Delta_{43}\big]\\
\nonumber &+  (a^2\sin^22\tmet - \frac{1}{4}\sin^22\ty\sin^22\tmef)
\big[\cos^2\tx\sin^2\Delta_{31}+\sin^2\tx\sin^2\Delta_{32} \big]\\
\nonumber &+ a^2b \cos\da \sin2\tmet 
\big[\cos2\tx\sin^2\Delta_{21}+\sin^2\Delta_{31}-\sin^2\Delta_{32} \big]\\
\nonumber &+ ab \cos\db \sin2\tmef 
\big[\cos2\tx\cos^2\ty\sin^2\Delta_{21}-\sin^2\ty(\sin^2\Delta_{31}-\sin^2\Delta_{32}) \\
\nonumber&+ \sin^2 \Delta_{41} - \sin^2 \Delta_{42} \big]\\
\nonumber &+ a \cos(\da - \db) \sin2\tmet \sin2\tmef
\Big[-\frac{1}{2}\sin^22\tx\cos^2\ty\sin^2\Delta_{21} \\ 
\nonumber &+ \cos2\ty(\cos^2\tx\sin^2\Delta_{31}+\sin^2\tx\sin^2\Delta_{32})
+\cos^2\theta_{12}\sin^2\Delta_{41}+\sin^2\theta_{12}\sin^2\Delta_{42}-\sin^2\Delta_{43}\Big]\\
\nonumber &+
(a^2b^2-\frac{1}{4}a^2\sin^22\tx\sin^22\tmet-\frac{1}{4}
\cos^4\ty\sin^22\tx\sin^22\tmef) \sin^2\Delta_{21}\\
\nonumber &- \frac{1}{2} a^2 b \sin\da \sin2\tmet 
\big[\sin2\Delta_{21}-\sin2\Delta_{31}+\sin2\Delta_{32} \big]\\
\nonumber &- \frac{1}{2} ab \sin\db \sin2\tmef 
\big[\cos^2\ty\sin2\Delta_{21}+\sin^2\ty(\sin2\Delta_{31}-\sin2\Delta_{32})
-\sin2\Delta_{41}+\sin2\Delta_{42}\big]\\
\nonumber&+ \frac{1}{2} a \sin(\da - \db) \sin2\tmet \sin2\tmef 
\big[\cos^2\tx\sin2\Delta_{31}+\sin^2\tx\sin2\Delta_{32}\\
&-\cos^2\theta_{12}\sin2\Delta_{41} - \sin^2\theta_{12}\sin2\Delta_{42}+\sin2\Delta_{43} \big]\,,
\end{align}
where, we have used the following shorthand notations,
\begin{eqnarray} 
\sin2\theta_{\mu e}^{3\nu}=\sin2\ty\sin\tz\,,\\
b=\cos\theta_{13}\cos\theta_{23}\sin2\tx\,,\\
\sin2\theta_{\mu e}^{4\nu}=\sin2\theta_{14}\sin\theta_{24}\,,\\
a=\cos\theta_{14}\cos\theta_{24}\,.
\end{eqnarray}
From the general expression of the oscillation probability, we further derive the approximate analytical expressions for probabilities in the 3$\nu$-scenario and in the 3+1 scheme for three different choices of the sterile frequency $\Delta{m}^2_{41}$. One of these choices corresponds to a very high value of $\Delta m^2_{41}$, which induces rapid oscillations that will be averaged out, and the other two choices represent the two degenerate cases where the sterile frequency becomes equal to the atmospheric frequency (\ie~$\Delta m^2_{41}$ $\sim$ $\Delta m^2_{31}$) and to the solar frequency (\ie~$\Delta m^2_{41}$ $\sim$ $\Delta m^2_{21}$), respectively. The respective probabilities for the above-mentioned scenarios are given in eqs.~(\ref{equ:palazzo_3n})--(\ref{equ:palazzo_sol}), up to fourth order in small parameter $\epsilon$, where, $s_{13}, s_{14}, s_{24}\simeq 0.15\sim\epsilon$ and $\alpha= \Delta m^2_{21}/\Delta m^2_{31}\simeq \pm\,0.03\sim\epsilon^2$. The corresponding full probabilities to all orders in $\epsilon$ are given in appendix~\ref{appendix:full_prob}. We make the further replacements $\Delta_{31}\rightarrow\Delta, \Delta_{21}\rightarrow\alpha\Delta, \Delta_{32}\rightarrow(1-\alpha)\Delta$.
\begin{eqnarray} \label{equ:palazzo_3n}
\nonumber P_{\mu e}^{3\nu}  & \simeq&  4 s_{23}^2 s^2_{13}  \sin^2{\Delta}\\
\nonumber
&+& 4 c_{12}^2 c_{23}^2 s_{12}^2 (\alpha \Delta)^2\\
\nonumber
&+& 8 s_{13} s_{12} c_{12} s_{23} c_{23} (\alpha \Delta)\sin \Delta \cos({\Delta + \delta_{13}})\\
&-& 4s_{13}^2 s_{23}^2 s_{12}^2 (\alpha\Delta) \sin(2\Delta)\,,
\end{eqnarray} 
\begin{eqnarray}  \label{equ:palazzo_avg}
\nonumber P_{\mu e}^{4\nu}\big|_{\text{high}~\Delta m^2_{41}}   & \simeq& (1 - s^2_{14} - s^2_{24}) P_{\mu e}^{3\nu} \\
\nonumber
&+& 4 s_{14} s_{24} s_{13} s_{23} \sin\Delta \sin (\Delta + \delta_{13} - \delta_{14}) \\
\nonumber
&-& 4 s_{14} s_{24} c_{23} s_{12} c_{12} (\alpha \Delta) \sin \delta_{14}\\
&+& 2 s_{14}^2 s^2_{24} \,,
\end{eqnarray} 
\begin{eqnarray}  \label{equ:palazzo_atm}
\nonumber P_{\mu e}^{4\nu}\big|_{\Delta m^2_{41}\,\simeq\,\Delta m^2_{31}}   & \simeq& (1 - s^2_{14} - s^2_{24}) P_{\mu e}^{3\nu} \\
\nonumber
&+& 8 s_{14} s_{24} s_{13} s_{23} \cos(\delta_{13}-\delta_{14})\sin^{2}{\Delta} \\
\nonumber
&-& 8 s_{14} s_{24} c_{23} s_{12} c_{12} (\alpha\Delta)\sin\delta_{14}\sin^{2}{\Delta}\\
&+& 4 s^2_{14} s^2_{24} \sin^2\Delta\,,
\end{eqnarray} 
\begin{eqnarray}  \label{equ:palazzo_sol}
\hspace{-3cm}P_{\mu e}^{4\nu}\big|_{\Delta m^2_{41}\,\simeq\,\Delta m^2_{31}}   & \simeq& (1 - s^2_{14} - s^2_{24}) P_{\mu e}^{3\nu} \,.
\end{eqnarray} 

Note that for compactness, we denote, $\cos\theta_{ij}\equiv{c_{ij}}$ and $\sin\theta_{ij}\equiv{s_{ij}}$, where $i,j=\{1, 2, 3, 4\}$. Here, we provide the probability expressions in vacuum to understand the behavior of our numerical results qualitatively\footnote{The analytical expressions for the oscillation probabilities in matter in the 3+1 scheme over a wide mass-squared range can be found in Ref.~\cite{Chattopadhyay:2022hkw}, where the sterile mass-ordering sensitivity of DUNE has been studied in detail.}. The accuracy of these expressions is shown in the appendix~\ref{validation_prob}, where the oscillation probabilities obtained using this set of equations are compared with those obtained numerically using the General Long-baseline Experiment Simulator (GLoBES)~\cite{Huber:2004ka,Huber:2007ji}. As discussed in previous literature~\cite{Klop:2014ima, Gandhi:2015xza}, the CP phase dependency of the transition probability appears in the interference terms between any two of the three distinct frequencies, namely, $\Delta m^2_{31}$, $\Delta m^2_{21}$, and $\Delta m^2_{41}$. The third term in \equ{palazzo_3n} is due to the interference between the solar and atmospheric frequencies, which brings the standard CP phase ($\delta_{13}$) into the picture. Similarly, the second and third terms in eqs.~(\ref{equ:palazzo_avg}) and (\ref{equ:palazzo_atm}) appear as a result of the interference between the atmospheric and sterile frequencies and the solar and sterile frequencies, respectively. The second term in eqs.~(\ref{equ:palazzo_avg}) and (\ref{equ:palazzo_atm})  contains both $\delta_{13}$ as well as the sterile CP phase $\delta_{14}$, whereas the third term depends on $\delta_{14}$ only. However, \equ{palazzo_sol} does not contain any interference terms between the standard and the sterile frequencies up to the fourth order in $\epsilon$.

\begin{figure}[h!]
	\centering
	\includegraphics[width=\textwidth]{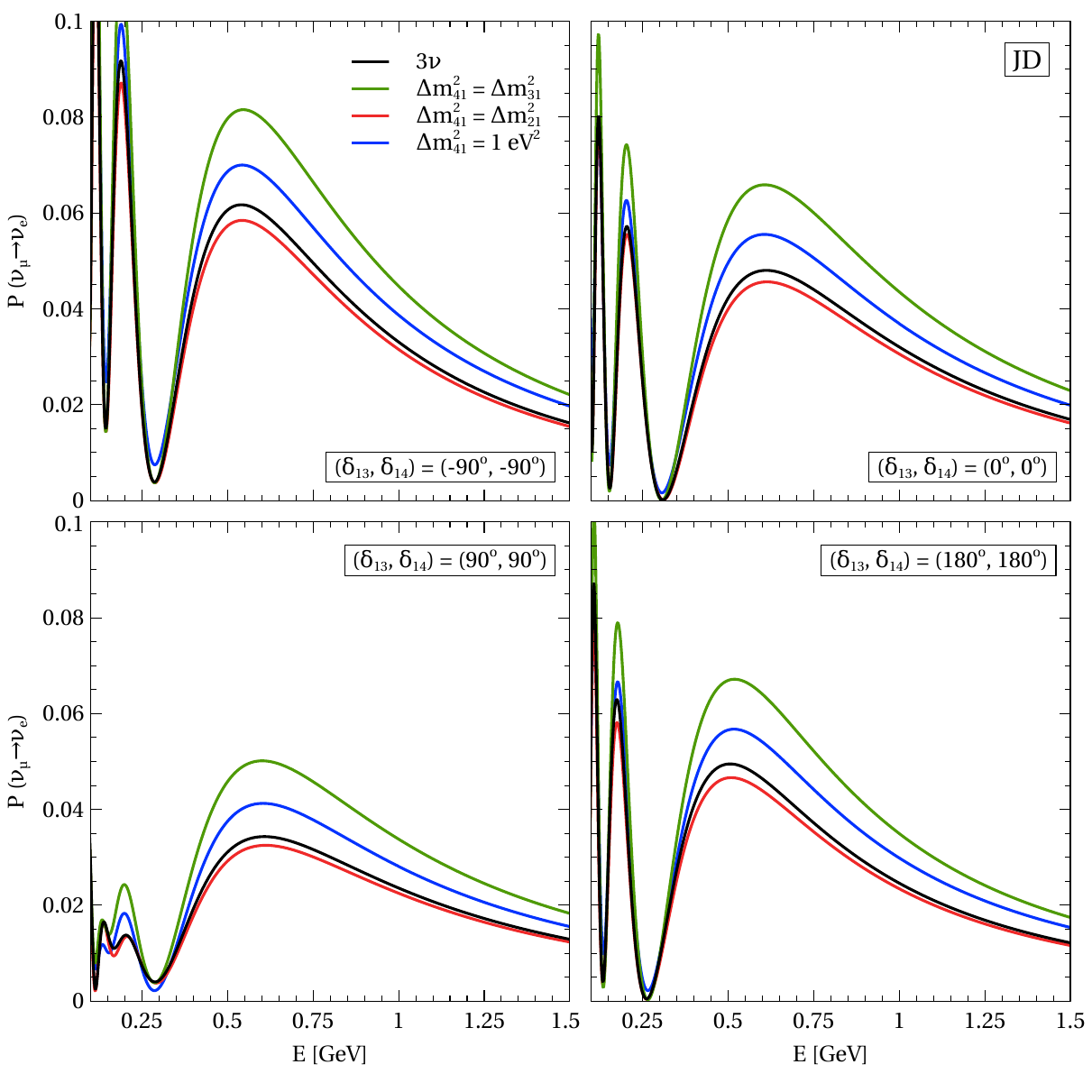}
	\vspace*{-10mm}
	\mycaption {$\nu_{\mu}\rightarrow\nu_{e}$ transition probability as a function of neutrino energy for JD with a baseline of 295 km. The four panels correspond to four different combinations of CP phases ($\delta_{13}$, $\delta_{14}$) and in each panel, we show the probabilities for the standard 3$\nu$-case and 3+1 scheme with three benchmark choices of $\Delta{m}^2_{41}$ as labeled in the figure. The values of the other oscillation parameters are fixed at their benchmark values as given in table~\ref{tab:params_value_str}. We assume normal mass ordering.\label{fig: JD_prob}}
\end{figure}

\begin{figure}[h!]
	\centering
	\includegraphics[width=\textwidth]{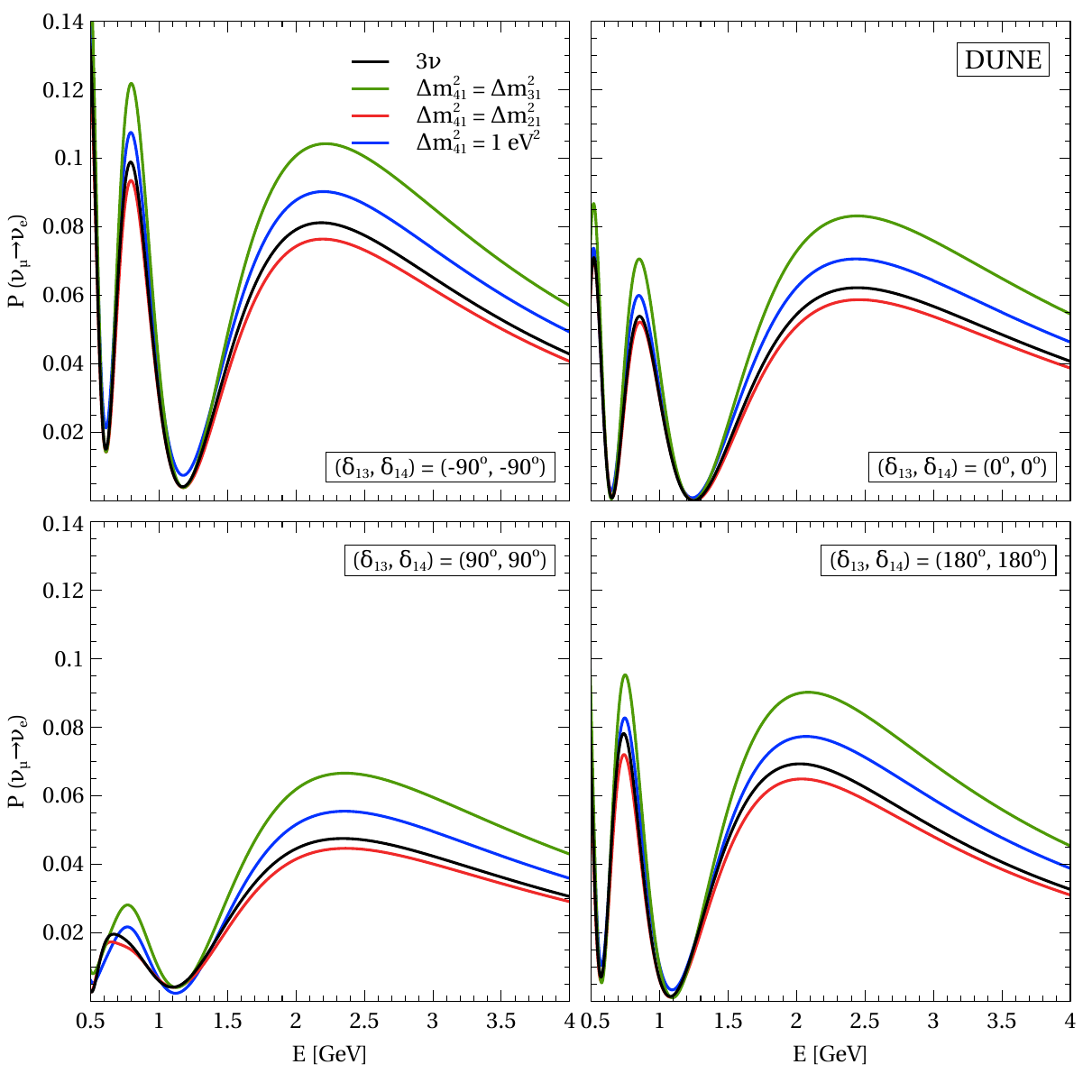}
	\vspace*{-10mm}
	\mycaption {$\nu_{\mu}\rightarrow\nu_{e}$ transition probability as a function of neutrino energy for DUNE with a baseline of 1285 km. The four panels correspond to four different combinations of CP phases ($\delta_{13}$, $\delta_{14}$) and in each panel, we show the probabilities for the standard 3$\nu$-case and 3+1 scheme with three benchmark choices of $\Delta{m}^2_{41}$ as labeled in the figure. The values of the other oscillation parameters are fixed at their benchmark values as given in table~\ref{tab:params_value_str}. We assume normal mass ordering. \label{fig: DUNE_prob} }
\end{figure}

In figs.~\ref{fig: JD_prob} and \ref{fig: DUNE_prob}, we show the $\nu_{\mu}\rightarrow\nu_{e}$ oscillation probability as a function of true neutrino energy for JD and DUNE, respectively, for three different values of $\Delta{m}^2_{41}$. The values of the oscillation parameters used for our calculations are listed in table~\ref{tab:params_value_str}, and we consider normal mass ordering (NMO) for our calculations. The four panels in each of these figures correspond to the four different combinations of the CP phases $\delta_{13}$ and $\delta_{14}$, as labeled in the plots. The black curve in each panel shows the 3$\nu$ oscillation probability whereas the other colored lines are for 3+1 oscillation scheme with $\Delta m^2_{41}$ = $\Delta m^2_{31}$ (green curve), $\Delta m^2_{41}$ = $\Delta m^2_{21}$ (red curve), and $\Delta m^2_{41}$ = 1 eV$^2$ (blue curve). For higher values of the sterile mass-squared differences (when $\Delta{m}^2_{41}\gtrsim 0.1~\rm{eV}^2$), the rapid oscillations will be completely averaged-out by the finite energy resolution of the detectors. Proper care has been taken to carry out the averaging, and we present our results after such an averaging. We can see a significant deviation from the standard $3\nu$ transition probability when the sterile frequency becomes equal to the atmospheric one. The deviation is much larger than the $\Delta m^2_{41}=1~{\rm eV}^2$ case, which is usually studied in the literature. This is also evident from \equ{palazzo_atm}, where the strength of the interference terms get {\it{doubled}} as compared to those in the averaged-out case as in \equ{palazzo_avg}. We can observe this effect while studying the CP sensitivity at different sterile mass scales (see sections~\ref{sec: CPV discovery} and \ref{sec: CP reconstruction}). There is only a mild change in the oscillation probability when $\Delta m^2_{41}$ = $\Delta m^2_{21}$, where the overall probability slightly reduces as compared to the 3$\nu$ probability due to the finite active-sterile mixing angles, by a factor of $(1 - s^2_{14} - s^2_{24})$ as evident from \equ{palazzo_sol}.

%================================================
\section{Experimental and simulation details}
\label{sec:experimental details}
%================================================
\subsection{Experimental details}
\subsubsection{T2HK (Tokai-to-Hyper-Kamiokande):}

T2HK~\cite{Abe:2011ts, Abe:2016ero, Hyper-KamiokandeProto-:2015xww, Hyper-Kamiokande:2018ofw, Hyper-KamiokandeWorkingGroup:2014czz}, which we call JD (Japanese Detector) in our work, is a next-generation underground water Cherenkov detector that will act as a far detector for the long-baseline experiment. This experiment is specifically designed for CP violation measurement in the neutrino sector. Various beyond the  Standard Model physics scenarios~\cite{Choubey:2017ppj, Agarwalla:2018nlx, Kelly:2017kch, Choubey:2017cba, Agarwalla:2021owd} have also been explored with the T2HK setup. T2HK (JD) will receive the neutrino beam from the J-PARC proton synchrotron that uses a proton beam of power 1.3 MW and will deliver $27\times 10^{21}$ P.O.T. per year. The fiducial mass of JD is 187 kt, and it is placed at a distance of 295 km from the J-PARC facility and at an off-axis angle of $2.5^\circ$ to get a narrow-band neutrino beam, which peaks at around the first oscillation maximum of 0.6 GeV. The run time in neutrino mode is 2.5 years, whereas, in antineutrino mode, it is 7.5 years to ensure equal contributions from the neutrino and antineutrino modes.

Another identical detector has been proposed~\cite{Seo:2019dpr} to be installed in Korea, which is named as KD (Korean Detector) in our work. It is also assumed to be placed at the same off-axis angle of $2.5^\circ$ from the beamline and at a distance of 1100 km from J-PARC. The KD will operate at around the second oscillation maximum of $\sim$ 0.7 GeV with an average neutrino energy of 0.6 GeV. The T2HK, along with the second detector placed in Korea, is known as the T2HKK setup, and in this work, we refer to this combination as JD+KD.

The above-discussed detectors act as the far detectors. Besides them, two near detectors have also been proposed. The first one is ND280~\cite{T2K:2019bbb}, which is an off-axis detector located at a distance of 280 m from the neutrino source and $2.5^\circ$ away from the beamline. The ND280, situated very close to the source, will help in measuring the unoscillated neutrino flux and the neutrino cross sections. The second near detector is the Intermediate Water Cherenkov Detector (IWCD)~\cite{Drakopoulou:2017qdu,Wilson:2020trq, Andreopoulos:2016rqc}, which has a fiducial volume of 1 kt and is assumed to be located at a distance of 1 km from J-PARC. The IWCD can be moved vertically to assume any off-axis angle while collecting the data. However, in the present work, we consider the IWCD located at the same off-axis angle as JD and KD. The main features of these experiments are listed in second column of table~\ref{tab:experiments}.

\begin{table}[h!]
	\begin{center}
		\begin{tabular}{|c|c|c|}
			\hline\hline
			& JD/KD (IWCD)& DUNE FD (ND)\\
			\hline
			Detector type & LArTPC & Water Cherenkov\\
			\hline
			Detector mass & 187 kt/187 kt (1 kt)& 40 kt (67.2 t)\\
			\hline
			Proton energy & 80 GeV & 120 GeV\\
			\hline
			P.O.T./year & $27\times10^{21}$ & $1.1\times10^{21}$ \\
			\hline
			Run time ($\nu$\,+\,$\bar{\nu}$) & 2.5 yrs + 7.5 yrs & 3.5 yrs + 3.5 yrs \\
			\hline
			Baseline length & 295 km/1100 km (1 km) & 1285 km (574 m) \\
			\hline
			Beam type & off-axis & on-axis \\
			\hline\hline
		\end{tabular}
	\end{center}
	\mycaption{Essential details of the experimental setups used for our analysis. \label{tab:experiments}}
\end{table}

%-------------------------
\subsubsection{DUNE (Deep Underground Neutrino Experiment):}
%-------------------------
DUNE uses the neutrino beam which is produced at the LBNF's Main Injector in Fermilab. Here, the protons having a beam power of 1.2 MW and energy of 120 GeV, delivering 1.1$\times10^{21}$ protons on target (P.O.T) per year, are made to bombard on a graphite target leading to the production of charged mesons. These charged mesons then decay in flight, giving rise to the neutrinos. The DUNE will use a high-intensity, wide-band neutrino beam whose flux peaks at around 2.5 GeV. These neutrinos from Fermilab travel a distance of 1285 km through the outer mantle of Earth having a nearly constant matter density of 2.85 g/cm$^3$ to South Dakota, where a far detector (FD),  which is a liquid argon time projection chamber (LArTPC) with a fiducial mass of 40 kt, is placed underground in the Homestake mine and will see an on-axis neutrino beam. For our analysis, we consider 7 years of total run-time with 3.5 years each in neutrino and antineutrino modes. We follow the configuration details as given in the recent Technical Design Report (TDR)~\cite{DUNE:2021cuw} by the DUNE collaboration. Along with the 40 kt LArTPC far detector, DUNE has a near detector complex, located at a distance of 574 m from the neutrino source, consisting of three modules of near detectors~\cite{DUNE:2021tad}. The first near detector is a Liquid Argon Time Projection Chamber (LArTPC) with a fiducial volume of 67.2 t, aims to measure the unoscillated neutrino flux and cross section. The second one is a multi-purpose detector (MPD), which will use a magnetic spectrometer with one ton of  high-pressure gaseous argon time projection chamber, basically designed to see any new physics signals. The last one is the system for on-axis neutrino detection (SAND), composed mainly of a KLOE magnet and a calorimeter to track the outgoing particles. The SAND has a fiducial volume of one ton. In our analysis, we use only the first near detector module, having a fiducial volume of 67.2 t, placed at a distance of 574 m from the source (following the approach used in \Refe~\cite{DUNE:2020fgq}). The key features of the DUNE setup are mentioned in the third column of table~\ref{tab:experiments}.

%-------------------------------
\subsection{Simulation details}\label{sm}
%-------------------------------

For all our numerical simulations, we use the publicly available software package, namely the General Long-baseline Experiment Simulator (GLoBES)~\cite{Huber:2004ka, Huber:2007ji}. For the implementation of the sterile neutrino scenario in the 3+1 scheme, we use the sterile probability engine with the extension snu.h~\cite{Kopp:2007ne}. To simulate the DUNE experiment, we use the latest DUNE configuration files from Ref.~\cite{DUNE:2021cuw}, and for JD/KD, we take the configuration details from Ref.~\cite{Hyper-Kamiokande:2016srs}. The DUNE, T2HK, and T2HKK setups will mainly probe two oscillation channels i.e., $\nu_{\mu} (\bar\nu_{\mu})$ disappearance and $\nu_{e} (\bar\nu_{e})$ appearance.

For T2HK/T2HKK (JD/JD+KD), the backgrounds in the appearance channels are the intrinsic $\nu_{e}$ from the beam contamination, misidentified $\nu_{\mu}$, and the NC events. We take 5\% normalization and 5\% calibration errors as the systematic uncertainties for the signal components. In the disappearance channel, the backgrounds to the signal are misidentified $\nu_{e}$ and the NC events. The systematic uncertainties in the disappearance channel are 3.5\% normalization and 5\% calibration errors. The efficiencies and energy resolutions are taken from Ref.~\cite{Hyper-Kamiokande:2016srs}.

For DUNE, the predicted backgrounds to the appearance channel are the intrinsic $\nu_{e}$ beam contamination, the misidentified $\nu_{\mu}$, and $\nu_{\tau}$ and the NC events. For the disappearance channel, the backgrounds to the signal include misidentified $\nu_{\tau}$ and the NC events. For the simulation of DUNE, we incorporate two distinct sets of systematics while simulating the standalone DUNE FD and the combined DUNE FD+ND. For DUNE FD+ND analysis, we adopt the nominal systematics for the FD as provided by the collaboration \ie\,2\% signal error in $\nu_e$ channel, 5\% signal in $\nu_\mu$ channel, and 5\% background errors in both channels. However, for the ND in DUNE FD+ND analysis, we amplify the systematics by a factor of three compared to the FD. This amplification reflects the additional constraints by the ND on the systematics related to neutrino flux. In contrast, for the standalone DUNE FD analysis, where the ND is not available to measure the flux, we use a different set of systematics --- 6\% signal error in $\nu_e$ channel, 15\% signal error in $\nu_\mu$ channel, 15\% background errors in both the appearance and disappearance channels. The choice of consistency between the systematics in standalone DUNE FD and combined DUNE FD+ND analysis ensures a comparable treatment of uncertainties despite of differing experimental setups. Efficiency functions and smearing matrices have been provided by the DUNE collaboration in Ref.~\cite{DUNE:2021cuw}.

The sensitivity of the experiments under consideration are given by the Poissonian $\chi^2$ defined as,
\begin{equation}
\chi^2 (\vec{\lambda}, \xi_{s}, \xi_{b}) = \min_{\vec\rho, \xi_{s}, \xi_{b}}~\Big[{{2\sum_{i=1}^{n}}\big(y_{i}-x_{i}-x_{i} \text{ln}\frac{y_{i}}{x_{i}}\big) + \xi_{s}^2 + \xi_{b}^2}~\Big],
\end{equation} 
which gives the median sensitivity of the experiment where $n$ is the total number of reconstructed energy bins.
\begin{equation}
y_{i} = N^{th}_{i}
(\vec\rho)~[1 + \pi^{s}\xi_{s}] + N^{b}_{i}(\vec\rho)~[1+\pi^b\xi_{b}],
\end{equation}

where $N^{th}_{i}$ is the expected number of signal events in the $i$-th energy bin with the set of oscillation parameters $\vec\lambda$ = \{$\theta_{12}, \theta_{13}, \theta_{23}, \Delta{m}^2_{21}, \delta_{13}, \Delta{m}^2_{31}, \theta_{14}, \theta_{24}, \theta_{34}, \delta_{14}, \delta_{34}, \Delta{m}^2_{41}$\}. $N^{b}_{i}$ is the number of background events in the $i$-th energy bin. $\xi_{s}$ and $\xi_{b}$ denote the systematic pulls on the signal and background, respectively. We marginalize the $\chi^2$ over the set of parameters $\vec\rho$ and also over the systematic pulls ($\xi_{s}$ and $\xi_{b}$) in the theory. The variables $\pi^s$ and $\pi^b$ indicate the signal and background normalization errors, respectively. $x_{i} = N^{s}_{i} + N^{b}_{i}$, represents the prospective data from the experiment, where $N^{s}_{i}$ is the number of charged current signal events and $N^b_{i}$, as defined before, is the number of background events. 

We present our exclusion limits (figs.~\ref{fig:th14_delm41}, \ref{fig:th24_delm41}, and \ref{fig:th_mue_delm41}) in terms of the Poissonian $\Delta{\chi^2}$~\cite{Blennow:2013oma}, defined as,
\begin{equation}
\Delta{\chi^2}_{\rm exclusion} = \min_{\vec\rho,\xi_{s},\xi_{b}}\big[\chi^2_{3+1}({\rm test}) - \chi^2_{3+0}({\rm true})\big]\,,
\end{equation}
%where, $\chi^2_{3+0}$ and $\chi^2_{3+1}$ are calculated by fitting the prospective data with standard $3\nu$ oscillation scenario and $4\nu$ case with one sterile neutrino species.

where, the true data is taken in $3\nu$-scenario, and is compared with the test hypothesis of the 3+1 scheme in presence of one light sterile neutrino.  

$\vec\rho$ are the marginalization parameters. We also marginalize the $\Delta\chi^2$ over the systematic pulls on signal ($\xi_{s}$) and background ($\xi_{b}$). For all the simulated results throughout this paper, we consider the mixing angle $\theta_{34}$ and the corresponding new phase $\delta_{34}$ to be zero. When we present the constraints on the active-sterile mixing angle over a wide range of $\Delta{m}^2_{41}$, we marginalize over the atmospheric mixing angle $\theta_{23}$ and also over the remaining active-sterile mixing angle (other than $\theta_{34}$) and its corresponding CP phase.

%==============================================================
\section{Discussion at event level}
\label{sec:event level discussion}
%==============================================================
\subsection{Event numbers as a function of $\Delta{m}^2_{41}$}
%==============================================================
\begin{figure}[h!]
	\centering
	\includegraphics[width=\textwidth]{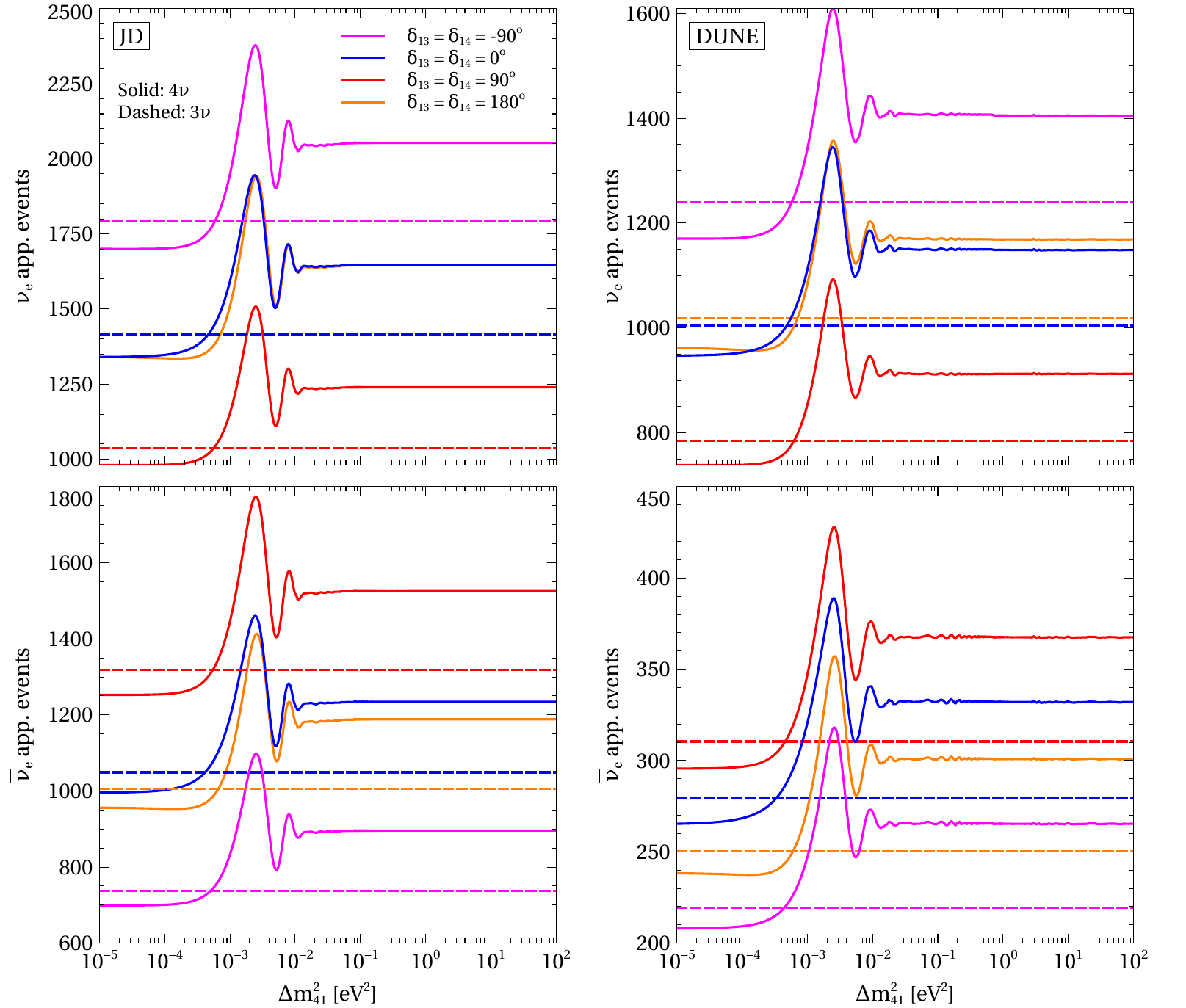}
	\vspace*{-8mm}
	\mycaption {Number of $\nu_e$ (upper panels) and $\bar\nu_e$ (lower panels) appearance events (signal) for JD (left panels) and DUNE (right panels) as a function of $\Delta{m}^2_{41}$. The solid lines correspond to the number of events in presence of active-sterile neutrino oscillation whereas the dashed lines depict the event numbers in absence of active-sterile oscillations. We take four different combinations of the CP phases namely; $\delta_{13}=\delta_{14}=-90^\circ, 0^\circ, 90^\circ, ~\rm {and}~180^\circ$. The values of the other oscillation parameters are fixed at their benchmark values as given in table~\ref{tab:params_value_str}. Here $4\nu$ refers to the 3+1 scheme. \label{fig: event_delm41} }
\end{figure}
%==============================================================
In \figu{ event_delm41}, we show the number of $\nu_e$ (upper panels) and $\bar\nu_e$ (lower panels) appearance events (signal) for JD (left panels) and DUNE (right panels) as a function of $\Delta{m}^2_{41}$. In every panel, we show the 3-neutrino events as dashed lines and the corresponding 4-neutrino events as solid lines. We consider four different combinations of the phases $\delta_{13}$ and $\delta_{14}$ as shown in the legends.

We can have the following observations from the event plots, which are true for all the combinations of the CP phases considered:

\begin{itemize}
	\item In the range $10^{-5}-10^{-4}~\rm{eV}^2$, the curve is almost flat and the event numbers do not change. This is because, in this range, the active-sterile oscillations will not be developed in the long-baseline experiments under consideration. However, we can see a slight decrease in event numbers from that of the corresponding 3-neutrino events.
	
	\item In the range $10^{-4}-10^{-2}~\rm{eV}^2$, the event numbers show oscillatory behavior. This happens due to the interference between the standard atmospheric oscillation frequency $\Delta{m}^2_{31}$ and the sterile frequency $\Delta{m}^2_{41}$.
	
	\item When $\Delta{m}^2_{41}$ is in the range $10^{-2}-10^{2}~\rm{eV}^2$, the active-sterile oscillations get averaged out due to the finite energy resolution of the detectors; hence the event numbers will not change as a function of $\Delta{m}^2_{41}$.
\end{itemize}
\noindent
From the above observations, it is clear that one would expect maximum sensitivity to the active-sterile neutrino oscillation parameters CP phases when the sterile frequency ($\Delta{m}^2_{41}$) becomes nearly the same order as that of the standard atmospheric mass-squared splitting $\Delta{m}^2_{31}$.
%---------------------------------------------------------------------------------
\subsection{Bi-event plots}
%---------------------------------------------------------------------------------

\begin{figure}[h!]
	\centering
	\includegraphics[width=0.99\textwidth]{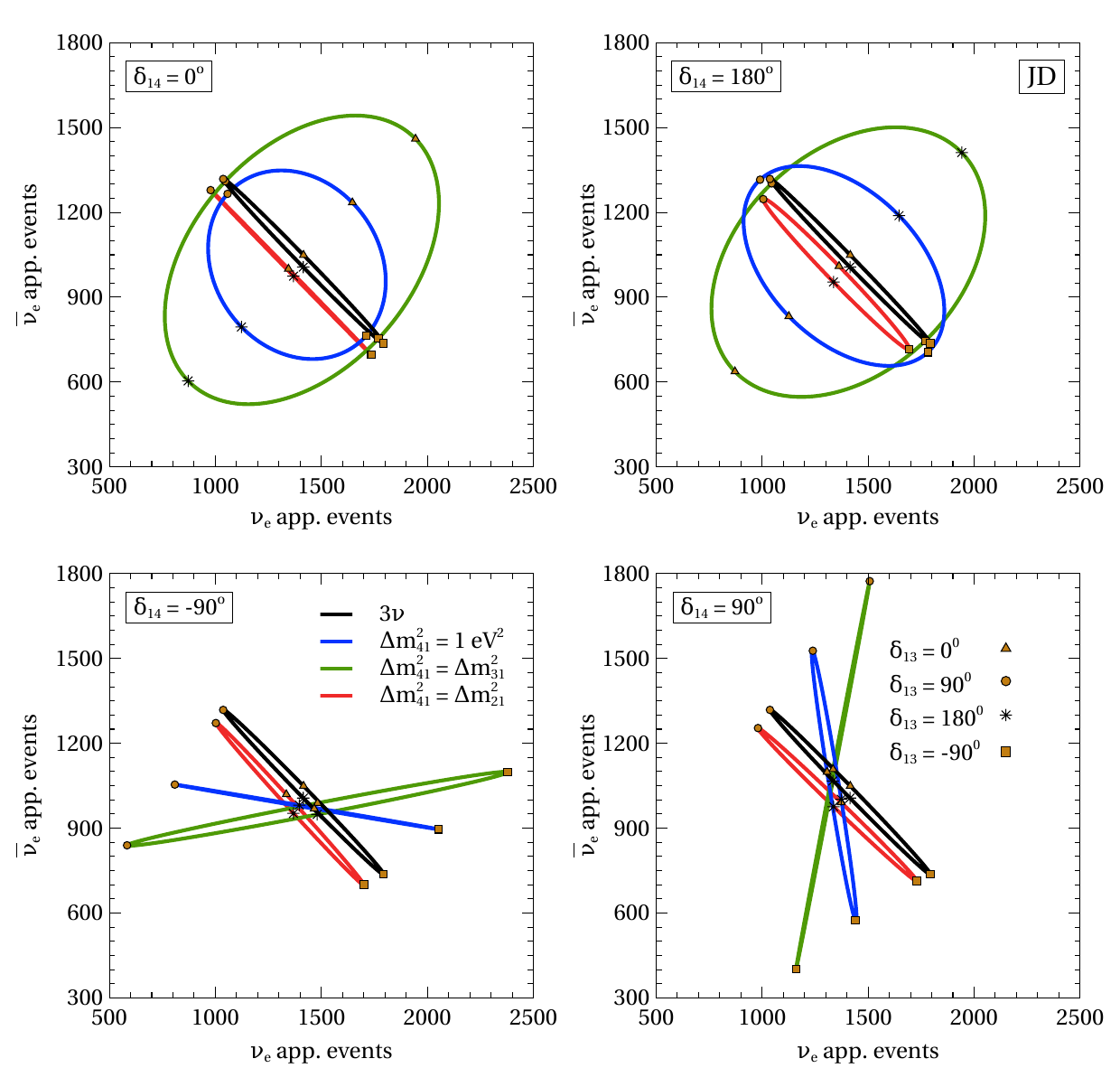}
	\vspace*{-1mm}
	\mycaption {Bi-event plots for JD, considering fixed choice of $\delta_{14}$ for every panel. Here, we consider only the signal events. The black ellipse represents the bi-events in standard 3$\nu$ scenario whereas the blue, green, and red ellipses correspond to the 3+1 scheme with $\Delta{m}^2_{41}$ equal to 1 eV$^2$, $\Delta{m}^2_{31}$, and $\Delta{m}^2_{21}$, respectively. The values of the other oscillation parameters are fixed at their benchmark values given in table~\ref{tab:params_value_str}.\label{fig: bievent_ellipse_JD} }
\end{figure}

\begin{figure}[h!]
	\centering
	\includegraphics[width=\textwidth]{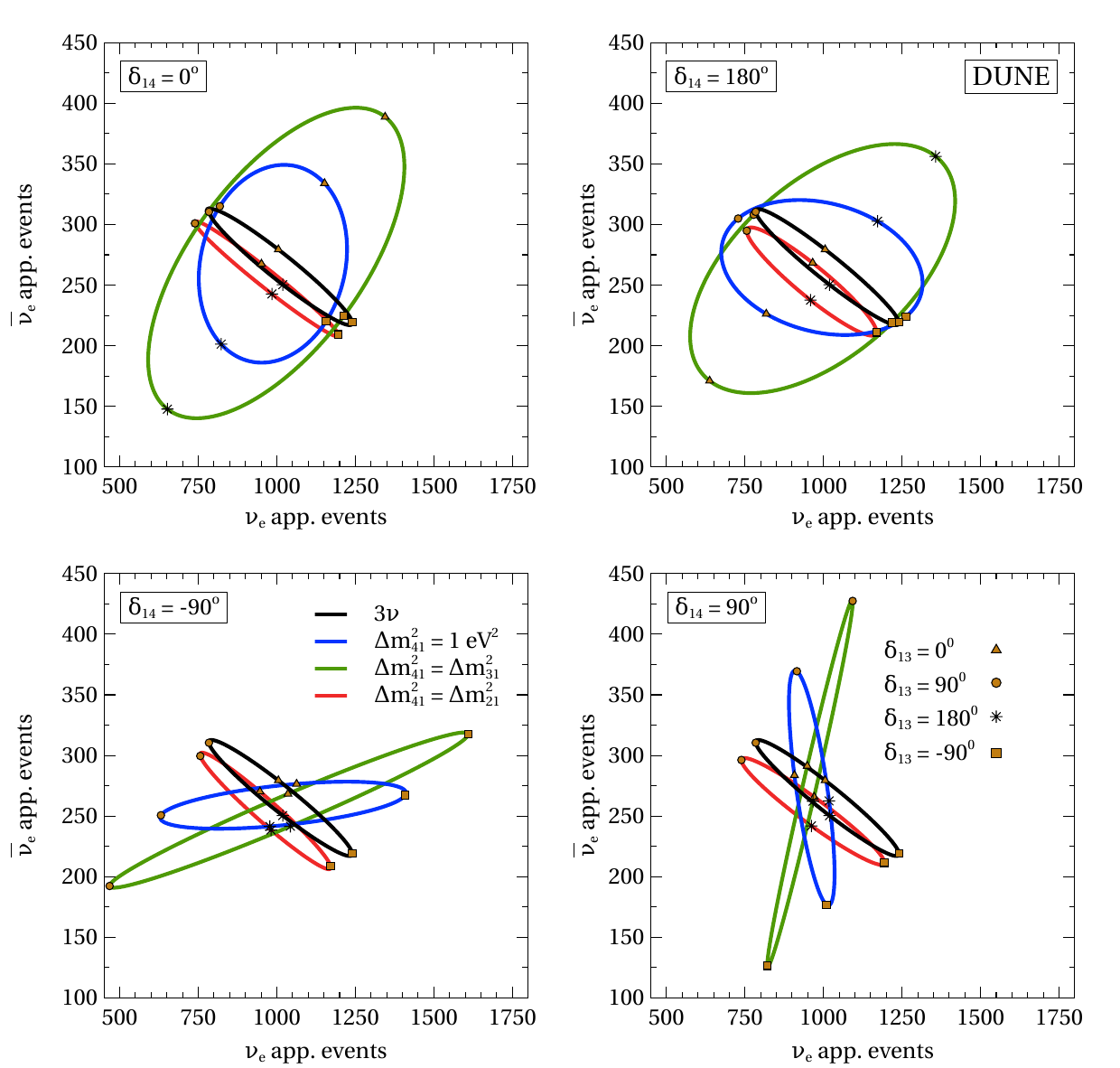}
	\vspace*{-5mm}
	\mycaption {Bi-event plots for DUNE, considering fixed choice of $\delta_{14}$ for every panel. Here, we consider only the signal events. The black ellipse represents the bi-events in standard 3$\nu$ scenario whereas the blue, green, and red ellipses correspond to the 3+1 scheme with $\Delta{m}^2_{41}$ equal to 1 eV$^2$, $\Delta{m}^2_{31}$, and $\Delta{m}^2_{21}$, respectively. The values of the other oscillation parameters are fixed at their benchmark values given in table~\ref{tab:params_value_str}.\label{fig: bievent_ellipse_DUNE} }
\end{figure}

\begin{figure}[h!]
	\centering
	\includegraphics[width=\textwidth]{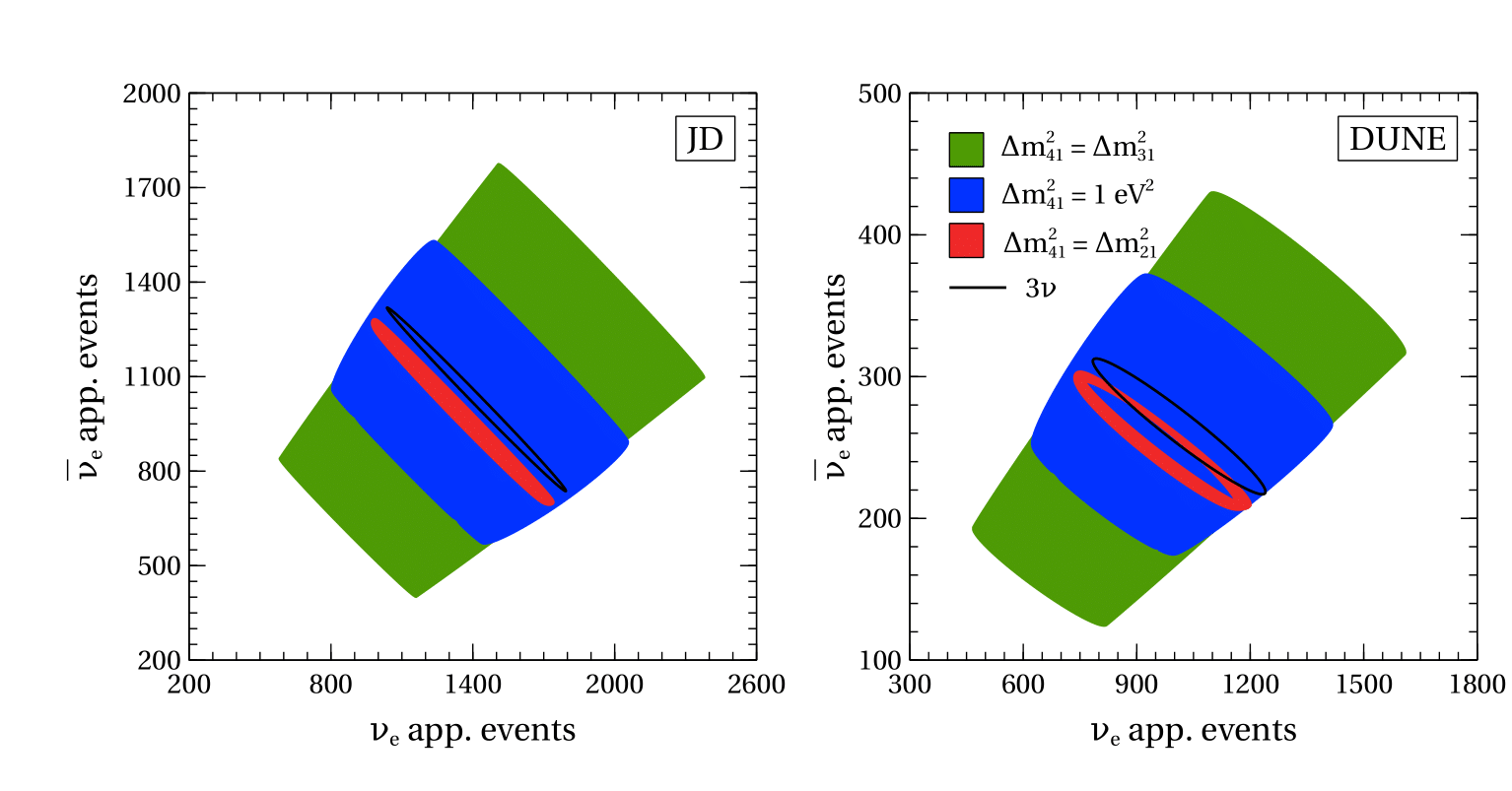}
	\vspace*{-10mm}
	\mycaption {The colored blobs represent the superposition of several bi-event ellipses, each of which is obtained for different $\delta_{14}$ values taken in the range [$-180^\circ:180^\circ$]. Here, we consider only the signal events. The left panel is for JD and right panel is for DUNE. The black ellipse represents the bi-event in standard 3$\nu$ scenario whereas the blue, green, and red blobs correspond to the 3+1 scheme with $\Delta{m}^2_{41}$ equal to 1 eV$^2$, $\Delta{m}^2_{31}$, and $\Delta{m}^2_{21}$, respectively. The values of the other oscillation parameters are fixed at their benchmark values given in table~\ref{tab:params_value_str}. \label{fig: bievent_blob} }
\end{figure}

Figures~\ref{fig: bievent_ellipse_JD} and~\ref{fig: bievent_ellipse_DUNE} show the bi-event plots for JD and DUNE, respectively. These bi-event plots in 3+1 scenario were first introduced in \Refe~\cite{Agarwalla:2016mrc} in the context of T2K and NO$\nu$A. We show the total number of neutrino appearance events (signal) on the x-axis and the same for the antineutrino appearance events (signal) on the y-axis. The four panels in each figure correspond to four different choices of $\delta_{14}$: $\delta_{14}= 0^\circ$ (top left), $\delta_{14}= 180^\circ$ (top right), $\delta_{14}= -90^\circ$ (bottom left), and $\delta_{14}= 90^\circ$ (bottom right) . In each panel, we present the bi-event plots for both $3\nu$ case and 3+1 case with three benchmark choices of $\Delta m^2_{41}$ as shown in the legends. The ellipses are obtained by varying the standard CP phase $\delta_{13}$ in the range [$-180^\circ: 180^\circ$]. We have marked the CP-conserving ($0^\circ~\text{and}~180^\circ$) and CP-violating  ($90^\circ~\text{and}~ -90^\circ$) values of $\delta_{13}$ by various symbols as shown in the plots. The non-zero separation between the CP-conserving and CP-violating values ensures that the event counts are sensitive to CP violation. We can see from the plots that this separation is maximum when the sterile frequency $\Delta m^2_{41}$ becomes equal to the atmospheric frequency $\Delta m^2_{31}$. Hence, one can expect maximum CP-sensitivity when the sterile frequency approaches the atmospheric one. This fact is also clear from \figu{ bievent_blob}, where we display the bi-event blobs for JD (left plot), and DUNE (right plot) in $3\nu$-scenario (black curve) as well as 3+1 framework with $\Delta{m}^2_{41}= 1~\text{eV}^2$ (blue blob), $\Delta{m}^2_{41}=    \Delta{m}^2_{31}$ (green blob), and $\Delta{m}^2_{41}=    \Delta{m}^2_{21}$ (red blob). These blobs are generated by the superposition of various bi-event ellipses (that are obtained by varying $\delta_{13}$ at a fixed $\delta_{14}$), where $\delta_{14}$ is allowed to vary in the range [$-180^\circ: 180^\circ$].

%===================================================================================
\section{Total CP violation discovery potential at JD, JD+KD, and DUNE} 
\label{sec: CPV discovery}
%===================================================================================
In the 3+1 scenario of neutrino oscillation, along with the standard CP phase $\delta_{13}$, the other sources of CP violation (CPV) include the sterile CP phases, $\delta_{14}$ and $\delta_{34}$. For a simplified scenario, we consider $\delta_{34}=0^\circ$ in our analysis. Hence, in this section, we explore the total CP violation, simultaneously induced by $\delta_{13}$ and $\delta_{14}$, at JD, JD+KD, and DUNE. The total CPV discovery potential then corresponds to the significance at which we exclude the test hypothesis of no CP violation, \ie, ($\delta_{13}$, $\delta_{14}$) = ($0^\circ$, $0^\circ$), ($0^\circ$, $180^\circ$), ($180^\circ$, $0^\circ$), and ($180^\circ$, $180^\circ$). 

The corresponding test statistic used to forecast the CP violation discovery potential is given by,
\begin{equation}
	\Delta{\chi^2}_{\!\!\rm CPV} = \min_{\vec\rho,\xi_{s},\xi_{b}}\big[ \chi^2_{3+1}({\delta^{\rm true}_{13},\delta^{\rm true}_{14}}\in[-180^\circ, 180^\circ])-\chi^2_{3+1}({\delta^{\rm test}_{13},\delta^{\rm test}_{14}}=0^\circ,180^\circ)\big]\,.
\end{equation}
Here, we marginalize over the parameter, $\vec\rho=\theta_{23}$, in the range given in table~\ref{tab:params_value_str}.

Figure~\ref{fig:cpv3_3s_5s} depicts the CP violation discovery potential of JD, JD+KD, and DUNE. We study the CP violation discovery potential of these three experimental setups for three different choices of the sterile mass-squared difference $\stfr$, namely, $\stfr$ = $\ldm$ (first row), $\stfr$ = $\sdm$ (second row), and $\stfr$ = 1 eV$^2$ (third row). In each panel, the dark-colored region corresponds to the total CPV discovery potential at more than $5\sigma$ C.L (2 d.o.f.) and the light-shaded colored region shows the same at $\geq 3\sigma$ C.L. (2 d.o.f.). The white regions correspond to the four CP-conserving combinations of $\delta_{13}$ and $\delta_{14}$. We marginalize over the atmospheric mixing angle ($\tz$) in its allowed range. However, all other oscillation parameters are fixed at their benchmark values listed in table~\ref{tab:params_value_str}.

\begin{figure}[h!]
	\centering
	\includegraphics[width=\textwidth]{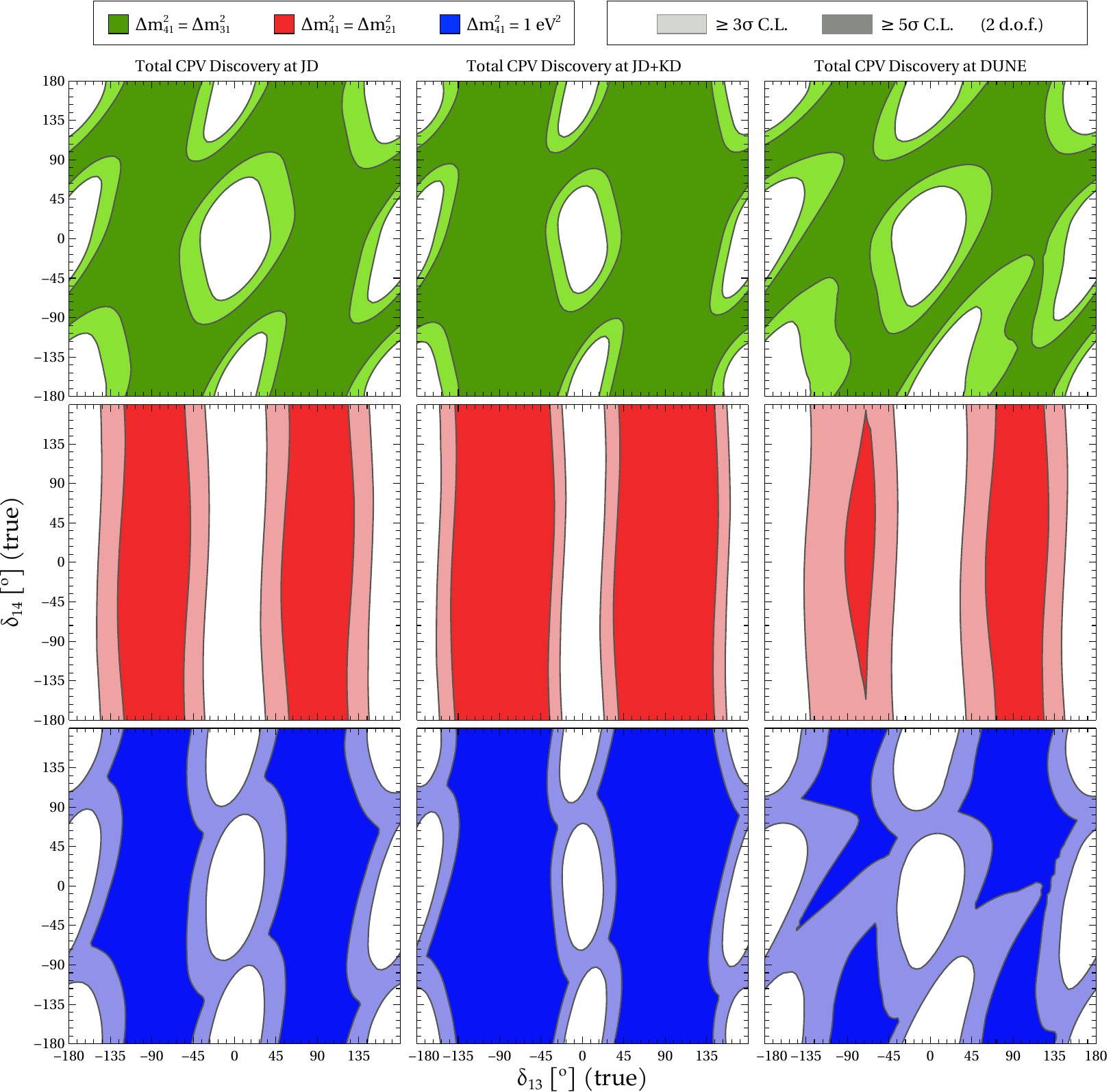}
	\vspace*{-5mm}
	\mycaption {Total CP violation (induced by both $\delta_{13}$ and $\delta_{14}$) discovery potential of JD, JD+KD, and DUNE for $\Delta{m}^{2}_{41}$ = $\Delta{m}^{2}_{31}$ (first row), $\Delta{m}^{2}_{41}$ = $\Delta{m}^{2}_{21}$ (second row), and $\Delta{m}^{2}_{41}$ = 1 eV$^2$ (third row). In the theory, we consider the CP-conserving values of ($\delta_{13}$, $\delta_{14}$) which are ($0^\circ$, $0^\circ$), ($0^\circ$, $180^\circ$), ($180^\circ$, $0^\circ$), and ($180^\circ$, $180^\circ$). The colored regions correspond to the CP violation discovery potential at JD, JD+KD, and DUNE, where the dark-shaded regions in each panel correspond to discovery potential at more than $5\sigma$ C.L. (2 d.o.f) and the light-shaded one depicts the same at more than $3\sigma$ C.L. (2 d.o.f). Marginalization is done over $\theta_{23}$ in its allowed range. The values of the other oscillation parameters are fixed, in both data and theory, at their benchmark values given in table~\ref{tab:params_value_str}.\label{fig:cpv3_3s_5s}}
\end{figure}

\noindent As far as the total CPV discovery potential induced by both $\delta_{13}$ and $\delta_{14}$ is concerned, we make the following observations from \figu{cpv3_3s_5s}:

\begin{itemize}
	\item The combination JD+KD outperforms the sensitivity from the standalone JD and DUNE.
	\item Between JD and DUNE, JD performs comparatively better than DUNE. This is expected because the relatively smaller baseline experiment like JD is mainly designed to measure the CP phases without having much interference from the fake CP asymmetry due to the Earth matter effect.
	\item For any particular experimental setup, among the three choices of $\stfr$, the CPV discovery potential is maximum when $\stfr$ = $\ldm$. The performance further deteriorates for $\stfr$ = 1 eV$^2$, whereas in the case of $\stfr$ = $\sdm$, all experimental setups become almost insensitive to the sterile CP phase $\delta_{14}$.
\end{itemize}

%===================================================================
\section{Reconstructing the CP phases with sterile neutrino}
\label{sec: CP reconstruction}
%===================================================================

\begin{figure}[h!]
	\centering
	\includegraphics[width=0.9\textwidth]{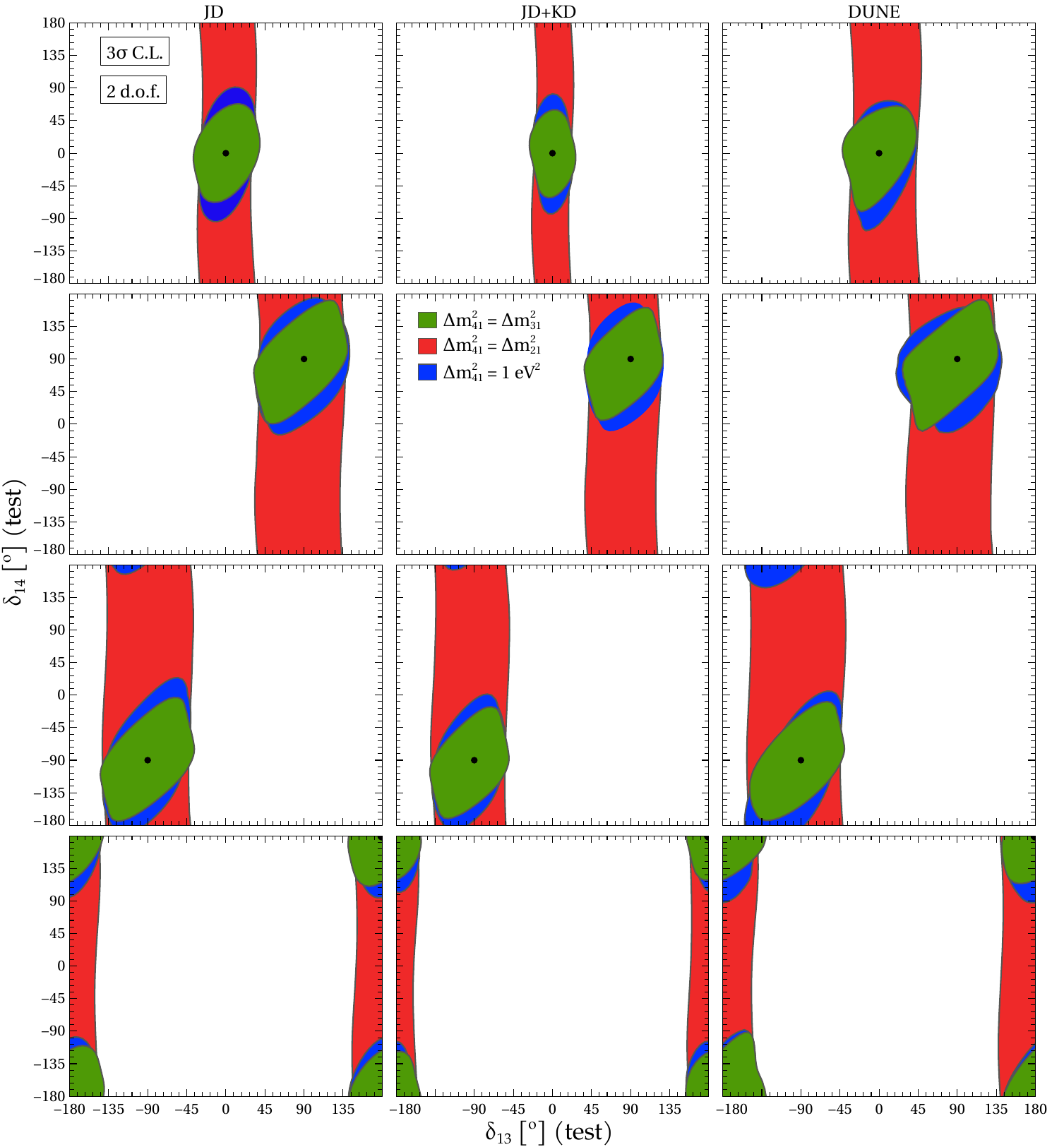}
	\vspace*{-2mm}
	\mycaption {Reconstructed regions at 3$\sigma$ C.L. (2 d.o.f.) for JD (first column), JD+KD (second column), and DUNE (third column) in test ($\delta_{13} -\delta_{14}$) plane for four different set of true values of $\delta_{13}$ and $\delta_{14}$ namely ($0^{\circ}$, $0^{\circ}$) (first row), ($90^{\circ}$, $90^{\circ}$) (second row), ($-90^{\circ}$, $-90^{\circ}$) (third row), and ($180^{\circ}$, $180^{\circ}$) (forth row) indicated by ($\bullet$) mark in each panel. We have shown the results for three benchmark values of $\Delta{m}^{2}_{41}$.  The orange contour is for $\Delta{m}^{2}_{41}$ = $\Delta{m}^{2}_{31}$, red contour is for $\Delta{m}^{2}_{41}$ = $\Delta{m}^{2}_{21}$, and blue contours are for $\Delta{m}^{2}_{41}$ = 1 eV$^2$. Marginalization is done over $\theta_{23}$ in its allowed range. The values of the other oscillation parameters are fixed, in both data and theory, at their benchmark values given in table~\ref{tab:params_value_str}.\label{fig:cp_precision_w_marg}}
\end{figure}			

In the previous section, we mainly focused on the CPV due to the phases $\delta_{13}$ and $\delta_{14}$. This section discusses the CP phase reconstruction capabilities of T2HK (JD)/T2HKK (JD+KD) and DUNE setups, independent of the amount of CPV present in Nature. 
The corresponding test statistic used to forecast the CP reconstruction capabilities is given by,
\begin{equation}
\Delta{\chi^2}_{\!\!\rm CP~recon.} = \min_{\vec\rho,\xi_{s},\xi_{b}}\big[\chi^2_{3+1}({\delta^{\rm test}_{13},\delta^{\rm test}_{14}}\in[-180^0, 180^0]) - \chi^2_{3+1}({\delta^{\rm true}_{13},\delta^{\rm true}_{14}})\big]\,.
\end{equation}
Here, we marginalize over the parameter, $\vec\rho=\theta_{23}$, in the range given in table~\ref{tab:params_value_str}. 
%The values of the unseen oscillation parameters are taken from table~\ref{tab:params_value_str}. 
%Here, we marginalize over $\theta_{23}$ in the range as given in table~\ref{tab:params_value_str}. 

In \figu{cp_precision_w_marg}, the reconstructed regions are given around four different representative points for the true values of the CP phases $(\delta^{\rm true}_{13}, 
\delta^{\rm true}_{14})$: ($0^\circ, 0^\circ$), ($90^\circ, 90^\circ$),
($-90^\circ, -90^\circ$), and ($180^\circ, 180^\circ$). 
The various colored areas represent the reconstructed regions for three benchmark choices of the sterile frequency; $\Delta m^2_{41}$ = $\Delta m^2_{31}$, $\Delta m^2_{41}$ = $\Delta m^2_{21}$, and $\Delta m^2_{41} = 1~\text{eV}^2$ as shown by the legends in the plots. The reconstructed regions are presented at $3\sigma$ C.L. (2 d.o.f.). The left column of the plot is for JD, the middle column is for JD+KD, and the right column depicts the CP phase reconstruction for DUNE, respectively. The important thing to note here is that the reconstruction capability is maximum when the sterile frequency is equal to the atmospheric one. This is due to the enhancement of the interference term in the probability by a factor of two (as evident from \equ{palazzo_atm}), increasing the sensitivity towards the CP phases. The long red strip in each panel corresponds to the reconstructed region when $\Delta m^2_{41}$ equals the solar frequency, which shows no sensitivity towards the new CP phase $\delta_{14}$ in this scale. This can be understood from \equ{palazzo_sol}, where the $\delta_{14}$-dependent interference terms do not appear in the probability expression up to the fourth order in $\epsilon$. Similar to the previous section, we also notice here that the CP phase reconstruction capability of JD is better than DUNE. However, when we combine the data from JD and KD, the performance gets even better.

%=========================================================================
\section{Constraints on sterile neutrino parameter space}
\label{sec: exclusion contours}
%=========================================================================

In order to derive the constraints on the sterile neutrino parameter space, we generate the data in the $3\nu$ scenario, whereas in the fit, we consider 3+1 scenario with three active and one sterile neutrino. We follow the simulation methods described in section~\ref{sm}. In this section, we present the expected constraints on the active-sterile mixing angles $\theta_{14}$, $\theta_{24}$, and the effective mixing angle $\theta_{\mu e}$ with JD, JD+KD, JD+KD+IWCD, DUNE FD, and DUNE (FD+ND), over a wide range of the sterile frequency, $\Delta m^2_{41}$ ($10^{-5}-10^2~\rm {eV}^2$).

The corresponding test statistic is given by (see section~\ref{sm} for more detail),
\begin{equation}
\Delta{\chi^2}_{\!\!\rm exclusion} = \min_{\vec\rho,\xi_{s},\xi_{b}}\big[\chi^2_{3+1}({\rm test}) - \chi^2_{3+0}({\rm true})\big]\,.
\end{equation}
%------------------------------------------------------------------------------------
\subsection{Exclusion plots in test ($\Delta{m}^2_{41}-\sin^{2}\theta_{14}$) plane}
%------------------------------------------------------------------------------------

\begin{figure}[h!]
	\centering
	\includegraphics[width=13cm]{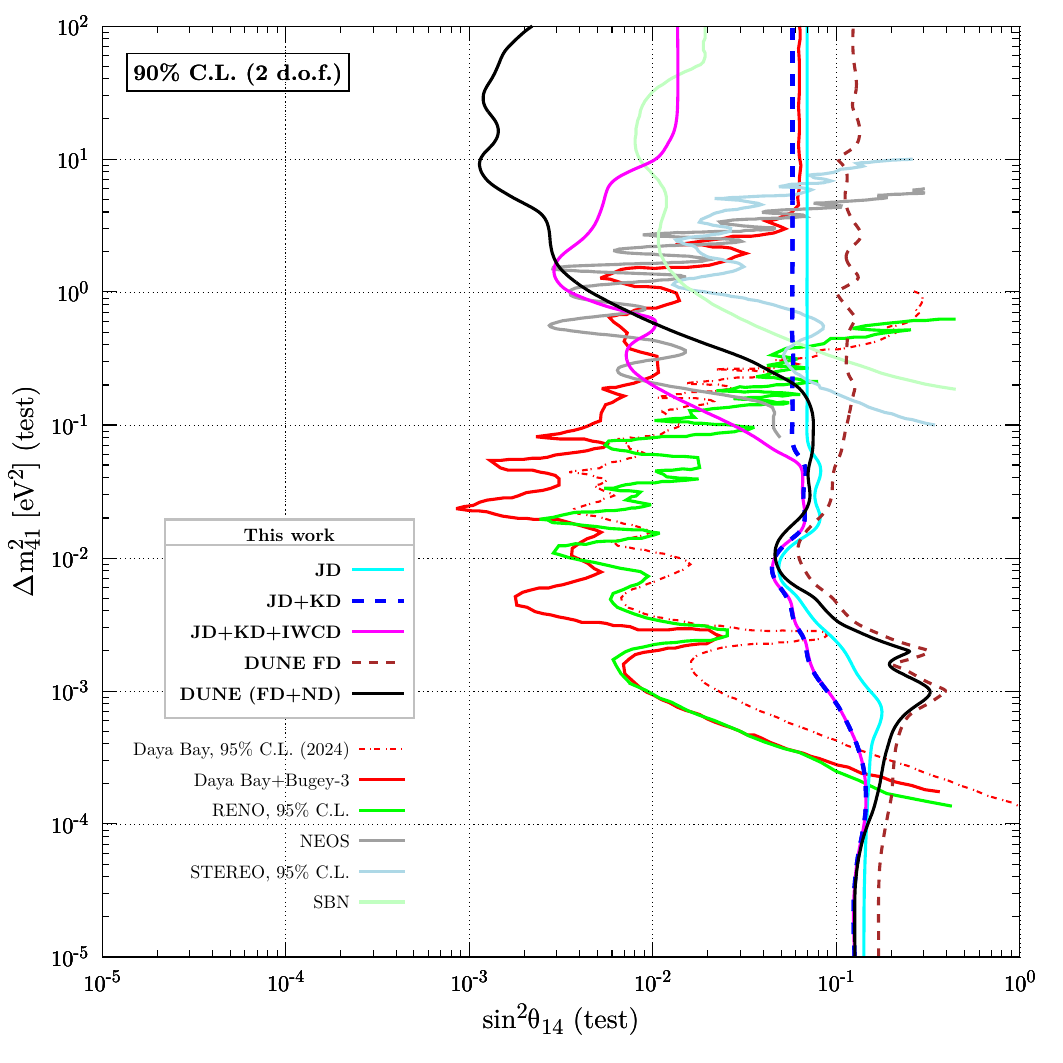}
	\vspace*{-5mm}
	\mycaption { 90$\%$ C.L. (2 d.o.f.) exclusion contours in test ($\Delta{m}^2_{41}-\sin^{2}\theta_{14}$) plane. We generate the data in 3-flavor oscillation scenario with no active-sterile mixing, and the fit is done in the 3+1 scenario. Both appearance and disappearance data are considered in our analysis. We marginalize over $\delta_{13}, \delta_{14}$, $ \theta_{24}$, and $\theta_{23}$ in their allowed ranges. We consider the true value of $\delta_{13}$ to be $-90^\circ$. The values of the other oscillation parameters are fixed, in both data and theory, at their benchmark values given in table~\ref{tab:params_value_str}.\label{fig:th14_delm41}}
\end{figure}

In \figu{th14_delm41}, we present the exclusion limit in $\Delta{m}^2_{41}-\sin^2{\theta_{14}}$ plane. The exclusion contours are given at 90\% C.L. and for 2 d.o.f., where $\Delta \chi^2 = 4.61$. The regions right to the contours are excluded. For the simulation of results, the standard three flavor oscillation parameters are kept constant at their benchmark values in both theory and data as mentioned in table~\ref{tab:params_value_str}, the only exception being $\delta_{13}$, over which $\Delta \chi^2$ has been marginalized in the range [$-180^\circ:180^\circ$] in the theory and in the data its value is fixed at $-90^\circ$. Cyan, blue, and magenta curves are, respectively, the constraints coming from JD, JD+KD, and JD+KD+IWCD. We also place the limits from DUNE with its Far and Near detector, and for comparison, we show the existing constraints from the combination of Daya Bay and Bugey-3 at 95\% C.L.~\cite{DayaBay:2016lkk}, NEOS~\cite{NEOS:2016wee}, RENO~\cite{RENO:2020uip}, and STEREO~\cite{STEREO:2022nzk}. 
The recent result from the analysis of the full dataset of Daya Bay in 2024 is also included~\cite{DayaBay:2024nip}. 
The expected sensitivity from the Short-Baseline Neutrino (SBN) program of Fermilab is also given~\cite{Cianci:2017okw}.  

It can be observed from \figu{th14_delm41} that the sensitivity due to JD suddenly increases at the mass-squared range $2\times 10^{-3}~\rm {eV}^2$  to $2\times 10^{-2}~\rm {eV}^2$. This happens because, in this range, both $\Delta{m}^2_{41}$ and $\Delta{m}^2_{31}$ are of the same order, hence leading to the interference effect. In the higher mass-squared region ($2\times 10^{-2}~\rm {eV}^2$ to $10^{2}~\rm {eV}^2$), the frequency of oscillation becomes too high to give an aliasing effect and hence gets averaged out by the finite energy resolution of the detector. So in this range the bounds on $\theta_{14}$ becomes independent of $\Delta{m}^2_{41}$. There is a slight improvement in the performance when data from both JD and KD are taken together. The improvement in sensitivity is noticeable in the higher mass-squared range (averaged out region) due to combined statistics from both detectors. A dramatic enhancement in the performance happens when we add the Intermediate Water Cherenkov Detector (IWCD) to JD+KD in the range $2\times 10^{-2}~\rm{eV}^2$ to $10^{2}~\rm{eV}^2$. This can be understood from the fact that IWCD being a near detector with a baseline of 1 km, develops active to sterile oscillation in higher mass-squared ranges with a very small active-sterile mixing angle.

It is clear from \figu{th14_delm41} that JD performs better over the entire range of $\Delta{m}^2_{41}$ than the DUNE FD in excluding the sterile parameter space. However, DUNE with both FD and ND is more sensitive than JD+KD+IWCD in higher mass-squared ranges ($1 ~\text{eV}^2\lesssim~\Delta{m}^2_{41}~\lesssim50~\text{eV}^2$). In this range of the sterile mass-squared difference, the sensitivity to the mixing angles mostly comes from the near detectors; hence, the better performance of the DUNE (FD+ND) than JD+KD+IWCD is due to the shorter baseline (nearly half) of DUNE ND than IWCD. The maximum sensitivity reaches of DUNE (FD+ND) and JD+KD+IWCD for $\theta_{14}$ are $\gtrsim1.9^\circ$ and $\gtrsim 3^\circ$ at $\Delta m^2_{41}\simeq10~\rm {eV}^2$ and $\Delta m^2_{41}\simeq1~\rm {eV}^2$, respectively.

%------------------------------------------------------------------------------------
\subsection{Exclusion plots in test ($\Delta{m}^2_{41}-\sin^{2}\theta_{24}$) plane}
%--------------------------------------------------------------------------------------

\begin{figure}[h!]
	\centering
	\includegraphics[width=13cm]{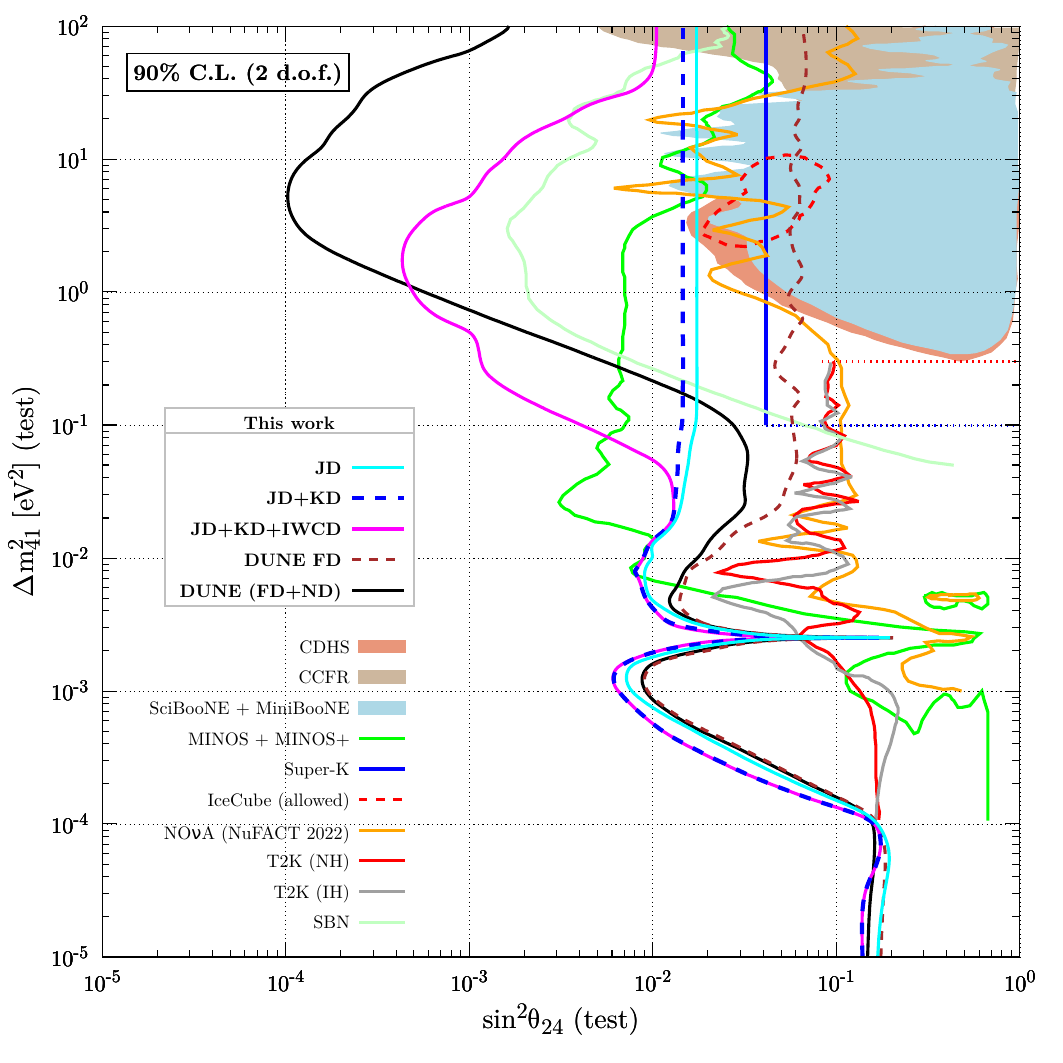}
	\vspace*{-5mm}
	\mycaption { 90$\%$ C.L. (2 d.o.f.) exclusion contours in test ($\Delta{m}^2_{41}-\sin^{2}\theta_{24}$) plane. We generate the data in 3-flavor oscillation scenario with no active-sterile mixing, and the fit is done in the 3+1 scenario. Both appearance and disappearance data are considered in our analysis. We marginalize over $\delta_{13}, \delta_{14}$, $ \theta_{14}$, and $\theta_{23}$ in their allowed ranges. We consider the true value of $\delta_{13}$ to be $-90^\circ$. The values of the other oscillation parameters are fixed, in both data and theory, at their benchmark values given in table~\ref{tab:params_value_str}.\label{fig:th24_delm41}}
\end{figure}

Figure~\ref{fig:th24_delm41} shows 90$\%$ C.L. (2 d.o.f.) exclusion contours in $\Delta{m}^2_{41}-\sin^2{\theta_{24}}$ plane. Here, we consider both the appearance and disappearance data for DUNE and T2HK/\\T2HKK (JD/JD+KD) setups. We keep the standard three flavor oscillation parameters at their benchmark values in both theory and data except $\delta_{13}$ whose value is kept at $-90^\circ$, while generating the data, but in theory, we have marginalized over it in its full range $[-180^\circ:180^\circ]$. The other two CP phases have also been marginalized over the same range in the theory. We also vary the active-sterile mixing angle $\theta_{14}$ in the range $[0^{0}:10^{0}]$ in the theory.
Various colored lines correspond to JD (cyan), JD+KD (blue-dashed), JD+KD+IWCD (magenta), DUNE FD (brown), and DUNE FD+ND (black). The present bounds from CCFR~\cite{Stockdale:1984cg}, CDHS~\cite{Dydak:1983zq}, combination of SciBooNE and MiniBooNE~\cite{MiniBooNE:2012meu}, combination of MINOS and MINOS+~\cite{MINOS:2017cae}, Super-Kamiokande~\cite{Super-Kamiokande:2014ndf}, T2K~\cite{T2K:2019efw}, and NO$\nu$A (refer to the talk by Jeremy Wolcott in NuFACT 2022) are given for comparison. The parameter space right to the contours is excluded. The 90\% C.L. allowed region from IceCube~\cite{IceCubeCollaboration:2024nle} is also shown in the same plot. We also show the sensitivity from Fermilab's SBN program~\cite{MicroBooNE:2015bmn}.

As observed in the previous section, it can also be seen from \figu{th24_delm41} that JD+KD can exclude slightly more parameter space than JD alone and the performance substantially enhances when we add IWCD to JD+KD for higher values of $\Delta{m}^2_{41}$ ($>2\times10^{-2}$ eV$^2$). The lowest value of $\theta_{24}$ that the combination JD+KD+IWCD can probe is $\sim1.2^\circ$ at $\Delta{m}^2_{41}\sim$ 1 eV$^2$.  It is noticeable that for the individual and the combinations of the experiments, the sensitivity decreases when $\Delta{m}^2_{41}$ becomes nearly equal to the atmospheric mass-square splitting. This happens due to the interference between the standard atmospheric frequency and the new frequency responsible for active-sterile oscillation sitting at the same value. DUNE FD+ND outperforms the other experimental setups at higher mass-squared ranges ($1 ~\text{eV}^2\lesssim~\Delta{m}^2_{41}~\lesssim50~\text{eV}^2$) and the lowest value of $\theta_{24}$ that DUNE FD+ND can reach is $\sim0.6^\circ$ at $\Delta m^2_{41}\sim5-10$ eV$^2$.

%--------------------------------------------------------------------------------------
\subsection{Exclusion plots in test ($\Delta{m}^2_{41}-\sin^{2}2\theta_{\mu{e}}$) plane}
%--------------------------------------------------------------------------------------

\begin{figure}[h!]
	\centering
	\includegraphics[width=13cm]{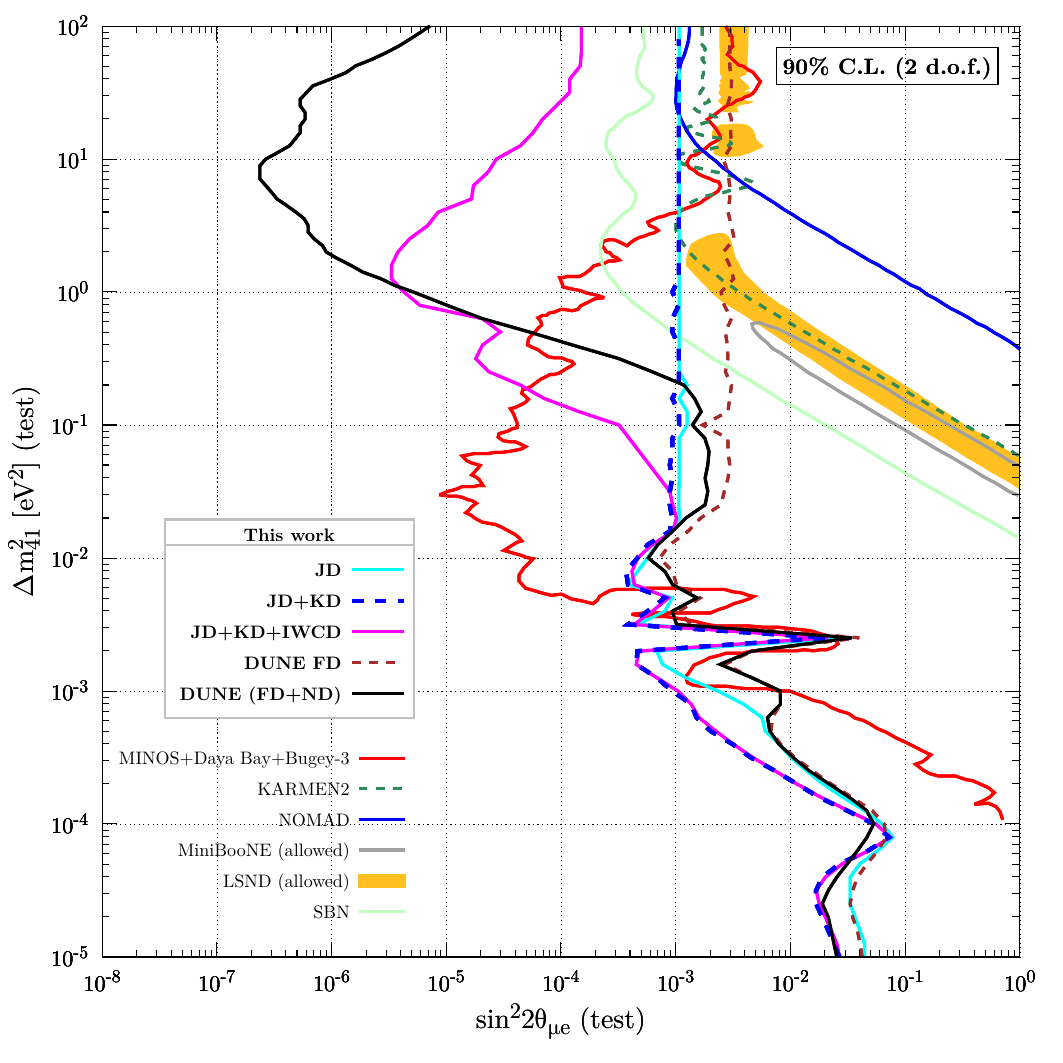}
	\vspace*{-5mm}
	\mycaption{90$\%$ C.L. (2 d.o.f.) exclusion contours in test ($\Delta{m}^2_{41}-\sin^{2}2\theta_{\mu{e}}$) plane. We generate the data in 3-flavor oscillation scenario with no active-sterile mixing, and the fit is done in the 3+1 scenario. Both appearance and disappearance data are considered in our analysis. We marginalize over $\delta_{13}, \delta_{14}$, and $\theta_{23}$ in their allowed ranges. We consider the true value of $\delta_{13}$ to be $-90^\circ$. The values of the other oscillation parameters are fixed, in both data and theory, at their benchmark values given in table~\ref{tab:params_value_str}.\label{fig:th_mue_delm41}}
\end{figure}

In \figu{th_mue_delm41}, we show the exclusion limit at $90\%$ C.L. in $\Delta m^2_{41}-\sin^{2}{2\theta_{\mu e}}$ plane using both the appearance and disappearance channels for T2HK/T2HKK (JD/JD+KD) and DUNE. We present the existing exclusion limits from KARMEN2~\cite{KARMEN:2002zcm}, NOMAD~\cite{NOMAD:2003mqg}, and the combination of MINOS, MINOS+, Daya Bay, and Bugey-3~\cite{MINOS:2020iqj} for comparison with our results. We also show the 90\% C.L. allowed regions from LSND and MiniBooNE. The future sensitivity of the SBN program at Fermilab is also given for comparison~\cite{MicroBooNE:2015bmn}. The effective mixing angle ($\theta_{\mu e}$) depends on two sterile mixing angles ($\theta_{14}$, $\theta_{24}$) and is defined as $\sin^{2}{2\theta_{\mu e}}= 4 \sin^{2}\theta_{14} \cos^{2}\theta_{14} \sin^{2}\theta_{24}$. Various combinations of $\theta_{14}$ and $\theta_{24}$ can give the same values of $\sin^{2}{2\theta_{\mu e}}$ \textit{i.e.} there exits a degeneracy in defining the effective mixing angle in terms of $\theta_{14}$ and $\theta_{24}$. Hence, we get different $\Delta\chi^2$ for a particular value of $\sin^{2}{2\theta_{\mu e}}$ depending on the values of $\theta_{14}$ and $\theta_{24}$. To draw the constraint region at $\Delta\chi^2=4.61$, we choose the $\sin^{2}{2\theta_{\mu e}}$ value such that the region towards the right of the contour always gives $\Delta\chi^2>4.61$ for any choices of the sterile mixing angles while in the region towards the left, we always find a combination of $\theta_{14}$ and $\theta_{24}$ such that $\Delta\chi^2<4.61$. Therefore, the contour represents the most conservative bound. We can observe from \figu{th14_th24_t2hk_dune} that if either of the sterile mixing angles ($\theta_{14}$ and $\theta_{24}$) is zero, then the detector can also provide significant constraints on the non-zero sterile mixing angle. Let us consider, for example the $\Delta{m}^2_{41}$= 1 eV$^2$ contour (solid green for JD+KD+IWCD) of the left panel of \figu{th14_th24_t2hk_dune}. For the point $(1^\circ,3^\circ)$ in the $\theta_{14}-\theta_{24}$ plane, we get the $\Delta \chi^2$ $>$ 4.61, however, for $(3^\circ,1^\circ)$ the $\Delta \chi^2$ is less than 4.61. These two points provide almost same values of $\sin^{2}{2\theta_{\mu e}}$ while giving very different $\Delta \chi^2$. The bounds on \figu{th_mue_delm41} are approximately given by the diagonal points of the contours in the $\theta_{14}-\theta_{24}$ plane of \figu{th14_th24_t2hk_dune} leading to the largest $\sin^{2}{2\theta_{\mu e}}$ for a given confidence level.

%-------------------------------------------------------------------------------
\subsection{Exclusion plots in test ($\theta_{14} -\theta_{24}$) plane}
%-------------------------------------------------------------------------------

\begin{figure}[h!]
	\centering
	\includegraphics[width=\textwidth]{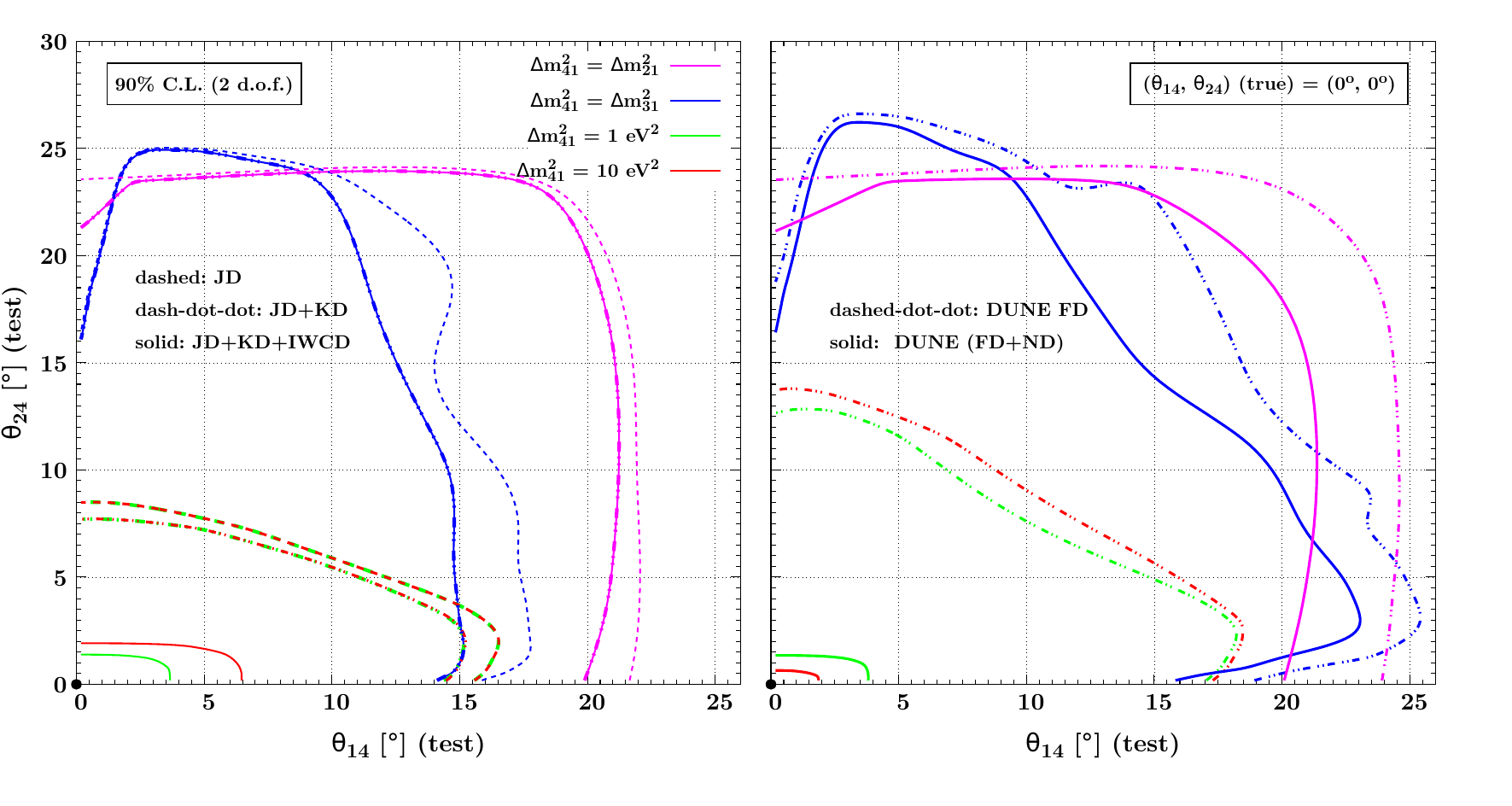}
	\vspace*{-10mm}
	\mycaption{90\% C.L. (2 d.o.f.) contours in test ($\theta_{14} -\theta_{24}$) plane for four different choices of $\Delta{m}^{2}_{41}$ (1 eV$^2$, 10 eV$^2$, $\Delta{m}^{2}_{41}$ = $\Delta{m}^{2}_{31}$, and $\Delta{m}^{2}_{41}$ = $\Delta{m}^{2}_{21}$). We generate the data in 3-flavor oscillation scenario with no active-sterile mixing, and the fit is done in the 3+1 scenario. Both appearance and disappearance data are considered in our analysis. We marginalize over $\delta_{13}, \delta_{14}$, and $\theta_{23}$ in their allowed ranges. We consider the true value of $\delta_{13}$ to be $-90^\circ$. The values of the other oscillation parameters are fixed, in both data and theory, at their benchmark values given in table~\ref{tab:params_value_str}.\label{fig:th14_th24_t2hk_dune}}
\end{figure}

In \figu{th14_th24_t2hk_dune}, we show the exclusion contours in test ($\theta_{14}-\theta_{24}$) plane for four different choices of $\Delta m^2_{41}$. Two of the choices for $\Delta m^2_{41}$ are made to be degenerate with the solar and atmospheric mass-squared splittings, and the other two are 1 eV$^2$ and 10 eV$^2$, respectively. In this figure, we also demonstrate the effect of the combinations of the near and far detectors on the sensitivity to the mixing angles. In the left panel, the dashed, dashed-dot-dot, and solid lines correspond to JD, JD+KD, and JD+KD+IWCD, respectively. In the right panel, the dashed-dot-dot and solid lines correspond to the DUNE ND and DUNE FD+ND, respectively. The exclusion contours are drawn at 90$\%$ C.L for 2 d.o.f. ($\Delta \chi^2 = 4.61$). The regions outside the contours are excluded. Here, we fix the standard CP phase at $-90^\circ$ in the data, whereas in theory, we marginalize over it in the range [$-180^\circ:180^\circ$]. We also marginalize over the new CP phase $\delta_{14}$ in the same range and over the mixing angle  $\theta_{23}$ from $40^\circ$ to $50^\circ$ in the theory. The values of the other oscillation parameters are fixed at their benchmark values given in table~\ref{tab:params_value_str}. Both appearance and disappearance data are taken for all three detectors.

From the left panel of \figu{th14_th24_t2hk_dune}, we see that when we take JD-alone, it gives similar sensitivity at 1 $\rm {eV}^2$ and 10 $\rm {eV}^2$. The sensitivity decreases and becomes even weaker when $\Delta m^2_{41}$ is equal to solar (magenta-dashed) and atmospheric (blue-dashed) mass-squared splittings. We observe that $\theta_{14}$ gets a more stringent bound for $\Delta m^2_{41}$ in the atmospheric scale than in the solar scale. The sensitivity to $\theta_{24}$, for $1.5^\circ\lesssim\theta_{14} \lesssim9^\circ$, is more in the case of solar scale than in the case of atmospheric scale. When the combination of JD and KD is taken, the sensitivity increases from that of JD-alone for all the four choices of $\Delta m^2_{41}$. The addition of IWCD to JD+KD has no effect when $\Delta m^2_{41}$ is in solar and atmospheric scale; however, there is a significant improvement in the performance when $\Delta m^2_{41}= 1~\rm{eV}^2$ and $\Delta m^2_{41}= 10~\rm{eV}^2$. The best sensitivity for $\theta_{14}$ and $\theta_{24}$ are respectively around 3.5$^\circ$ and 1.5$^\circ$ in case of JD+KD+IWCD for $\Delta m^2_{41}= 1~\rm{eV}^2$. From the right panel of \figu{th14_th24_t2hk_dune}, it is evident that in the case of DUNE, the bounds on the mixing angles $\theta_{14}$ and $\theta_{24}$ improve for DUNE FD+ND than DUNE FD for all the four choices of the $\Delta m^2_{41}$. The enhancement in the sensitivity at higher $\Delta m^2_{41}$ ($1~\rm{eV}^2$ and $10~\rm{eV}^2$) is due to the contribution from ND and at lower values of $\Delta m^2_{41}$ ($\Delta m^2_{31}$ and $\Delta m^2_{21}$), it is mainly due to the reduction in the systematics of FD due to presence of ND. In the range $1.5^\circ\lesssim\theta_{14} \lesssim9^\circ$, DUNE FD+ND puts stronger bound on $\theta_{24}$ in the case of $\Delta m^2_{41} = \Delta m^2_{21}$ than that of $\Delta m^2_{41} = \Delta m^2_{31}$. When $1^\circ\lesssim\theta_{24} \lesssim6.5^\circ$, $\theta_{14}$ is more tightly constrained by the setup DUNE FD+ND if the active-sterile oscillations take place with $\Delta{m}^2_{41} = \Delta{m}^2_{21}$ than $\Delta{m}^2_{41} = \Delta{m}^2_{31}$. The best bounds from the DUNE FD+ND setup on the two mixing angles are, respectively, $\theta_{14}\lesssim1.9^{\circ}~\text{and}~\theta_{24}\lesssim0.6^{\circ}$, when $\Delta m^2_{41} = 10~\text{eV}^2$.

%==============================================================================
\section{Summary and conclusions}
\label{sec: conclusion}
%==============================================================================

The three neutrino oscillation parameters have been measured very accurately with great precision (apart from a few remaining issues like measurement of standard CP phase, $\theta_{23}$ octant and neutrino mass ordering) with massive data from various neutrino oscillation experiments. This makes us look for new sub-leading effects like the presence of sterile neutrino, Lorentz Invariance Violation, Non-unitarity of lepton mixing matrix, and Non-standard Interactions in the neutrino oscillation experiments. Though the three-neutrino oscillation scenario fits most of the experimental results, there are various anomalous results from short-baseline experiments which point towards the existence of an extra species of neutrino, which we call the sterile neutrino. The sterile neutrino is a singlet under the Standard Model gauge group that does not take part in weak interactions. However, it gives its signature via mixing with the active neutrinos. So, this makes them worth studying in long-baseline neutrino oscillation experiments. We, in the present work, probe the active-sterile oscillation over a broad range of $\Delta m^2_{41}$ from $10^{-5}-10^{2}~\text{eV}^2$ with the long-baseline experiments DUNE, T2HK/JD, T2HKK/JD+KD. In our work, we include the Near Detectors of these experimental setups, which enable us to probe the parameter space allowed by the short-baseline anomalies. The sensitivity at the lower $\Delta m^2_{41}$ ($10^{-5}-10^{-2}~\text{eV}^2$) region comes due to interference between the standard mass-squared differences, $\Delta m^2_{21}$ and $\Delta m^2_{31}$, and the sterile mass-squared difference $\Delta m^2_{41}$. This way, our setups cover the whole range of $\Delta m^2_{41}$ considered in this work.

Here, we study the sensitivity of the two future long-baseline experimental setups, T2HK/T2HKK (JD/JD+KD) and DUNE, to the parameter space of active-sterile neutrino mixing in 3+1 scheme, where the presence of one sterile neutrino is considered. We discuss the CP violation discovery potential (refer to \figu{cpv3_3s_5s}) of JD, JD+KD, and DUNE at more than 5$\sigma$ significance level and the CP phase reconstruction capability (see \figu{cp_precision_w_marg}) of those experiments at 3$\sigma$ significance level. Mainly three representative values of $\Delta m^2_{41}$ are considered for our study, out of which two are degenerate cases, where the sterile mass-squared difference is equivalent to that of atmospheric ($\Delta m^2_{41}= \Delta m^2_{31}$) and the solar mass-squared splitting ($\Delta m^2_{41}= \Delta m^2_{21}$) and the third choice being $\Delta m^2_{41}= 1~\text{eV}^2$ as indicated by the SBL anomalies. We derive the approximate analytical expressions for the $\nu_{\mu}\to\nu_e$ appearance probabilities at these three scales to understand the essential features of the simulated results presented in this work. The sensitivity to the CP phases is maximum when $\Delta m^2_{41}= \Delta m^2_{31}$, which is also clear from the probability plots (figs.~\ref{fig: JD_prob} and \ref{fig: DUNE_prob}), the events counts as a function of $\Delta m^2_{41}$ (see \figu{ event_delm41}) and the bi-event plots (refer to figs.~\ref{fig: bievent_ellipse_JD},~\ref{fig: bievent_ellipse_DUNE}, and~\ref{fig: bievent_blob}). We find that the CP violation discovery potential of JD is better than DUNE. The combination JD+KD improves the results further.

We investigate the sensitivity of JD, JD+KD, and DUNE in constraining the active-sterile mixing angles; $\theta_{14}$ (\figu{th14_delm41}), $\theta_{24}$ (\figu{th24_delm41}), and the effective appearance angle $\theta_{\mu e}$ (\figu{th_mue_delm41}) for a wide range of $\Delta m^2_{41}$ from $10^{-5}~\text{eV}^2$ to $10^{2}~\text{eV}^2$ when there is no evidence of sterile neutrino in our experiment. Also, we show the exclusion contours in $\theta_{14} (\rm {test})-\theta_{24} (\rm test)$ plane for specific values of $\Delta m^2_{41}$ (refer to \figu{th14_th24_t2hk_dune}). For the first time, we consider the Intermediate water Cherenkov Detector (IWCD) for the T2HK/T2HKK (JD/JD+KD) setup to explore its sensitivity at the higher mass-squared differences. We report from the exclusion plots that the minimum values of $\theta_{14}$ that JD+KD+IWCD (DUNE (FD+ND)) can probe is $\approx3.5^\circ$ ($\approx1.9^\circ$) when $\Delta m^2_{41}$ $\approx$ 1 eV$^2$ ($\approx$ 10 eV$^2$). Similarly the best bound on $\theta_{24}$ for JD+KD+IWCD (DUNE (FD+ND))  is $\theta_{24}$~$\lesssim~$1.5$^\circ$ ($\theta_{24}$~$\lesssim$~0.6$^\circ$) when $\Delta m^2_{41}$ $\approx$ 1 eV$^2$ ($\approx$ 5--10 eV$^2$). These future long-baseline experimental facilities can well probe the parameter spaces allowed by the SBL anomalies with their Near Detectors (see \figu{th_mue_delm41}). We hope our analysis can provide an independent investigation of active-sterile neutrino oscillation with two different experimental setups and will contribute substantially to the saga of sterile neutrino search.

%============================================================================

\blankpage 
%%%%%%%%%%%%%%%%%%%%%%%%%%%% Chapter-6 %%%%%%%%%%%%%%%%%%%%%%%%%%%%%%%
%%%%%%%%%%%%%%%%%%%%%%%%% CHAPTER - 6 %%%%%%%%%%%%%%%%%%%%%%%%%
\chapter{Summary and conclusions}
\label{C6}
%%%%%%%%%%%%%%%%%%%%%%%%%%%%%%%%%%%%%%%%%%%%%%%%%%%%%%%%%%%%%%%

After photon, neutrinos are the second most abundant particles in Nature. Despite being present everywhere, they are the most elusive particles. Their existence was first postulated by Wolfgang Pauli in 1930 to explain the apparent violation of energy conservation in beta decay and was later confirmed experimentally by Clyde Cowan and Frederick Reines in 1956. They are neutral leptons forming the $SU(2)_L$ doublets with their respective charged leptons. Being neutral and colorless particles, they do not have electromagnetic and strong interactions. Neutrinos interact only via weak interactions mediated by the weak gauge bosons, $W^{\pm}$ and $Z^0$. This makes their detection extremely difficult. The challenging nature of neutrino detection has paradoxically fueled intense scientific interest and compelled researchers to develop innovative detection techniques. Today, we have state-of-the-art facilities worldwide to catch these elusive particles. In the Standard Model (SM) of particle physics, neutrinos remain massless due to the absence of their right-handed partners. However, various world-class experiments have now established the phenomenon of neutrino oscillation, which is induced by the non-zero and non-degenerate neutrino masses. This is the first ever experimental evidence of beyond the Standard Model (BSM) physics. The oscillation in three-neutrino framework involves three mixing angles, namely, the atmospheric mixing angle ($\theta_{23}$), the reactor mixing angle ($\theta_{13}$), and the solar mixing angle ($\theta_{12}$), one Dirac CP-violating phase ($\delta_{\rm CP}$), and two independent mass-squared differences -- atmospheric mass-squared difference ($\Delta m^2_{31}\equiv m^2_{3}-m^2_{1}$) and solar mass-squared difference ($\Delta m^2_{21}\equiv m^2_{2}-m^2_{1}$). 

As discussed in chapter~\ref{C2}, the joint efforts of the solar, atmospheric, reactor, and long-baseline experiments have been proven spectacular in measuring the above parameters with reasonable precision. Some of the yet-to-be-solved issues include (i) the octant ambiguity of $\theta_{23}$ ($\theta_{23}>45^\circ$ -- higher octant or $\theta_{23}<45^\circ$ -- lower octant), (ii) whether CP is violated in the neutrino sector and (iii) whether neutrino mass ordering is normal ($\Delta m^2_{31}>0$) or inverted ($\Delta m^2_{31}<0$).
Next-generation neutrino oscillation experiments are expected to resolve these issues with high confidence. Now, in the precision era of the oscillation parameters, neutrino oscillation experiments are becoming sensitive to various sub-leading new physics effects, if any, over the expected SM background. The new physics effects could manifest as new neutrino interactions, new neutrino properties, or any additional particles beyond the SM particle content or violation of any fundamental physics principles at a high-energy scale, showing their impacts in low-energy experiments. In this thesis, we explore three such new physics scenarios, namely, (i) long-range interactions of neutrinos, (ii) Lorentz invariance violation, and (iii) active-sterile neutrino oscillation, in the context of next-generation long-baseline (LBL) neutrino oscillation experiments, Deep Underground Neutrino Experiment (DUNE), Tokai-to-Hyper-Kamiokande (T2HK), and Tokai to Hyper-Kamiokande with a second detector in Korea (T2HKK). Below, we summarize our main results and conclusions from these studies.

\vspace{0.3cm}

\noindent{\bf (i) Long-range interactions of neutrinos:} The robustness of neutrino oscillation framework, by the measurement of oscillation parameters to sub-percent level precision, allows us to go beyond the SM to explain the non-zero neutrino mass and mixing pattern. One possible way to go beyond the SM is by imposing additional symmetries. These extra symmetries may lead to new neutrino interactions. In this context, long-range interactions (LRIs) of neutrinos are mediated by the new light gauge boson $Z^{\prime}$ when the SM gauge group $SU(3)_{C} \times SU(2)_{L} \times U(1)_{Y}$ is extended by the abelian guage group $U(1)^{\prime}$. These interactions are super-feeble. However, if $Z^{\prime}$ is sufficiently light, below $10^{-10}$ eV down to $10^{-35}$ eV, the interaction range goes up to Gpc and the neutrinos at terrestrial experiments can feel the presence of matter inside the Sun, Moon, Milky Way galaxy, and the cosmological distribution of matter. 

In chapter~\ref{C3}, we explore the sensitivity of two next-generation long-baseline experiments, DUNE and T2HK in probing long-range neutrino interactions arising from a plethora of gauged $U(1)^\prime$ symmetries that are combinations of baryon number ($B$) and lepton number ($L$), namely, $B-3L_e$, $B-3L_\mu$, $B-3L_\tau$, $B-L_e-2L_\tau$, $B_y+L_\mu+L_\tau$, $B-\frac{3}{2}(L_\mu+L_\tau)$, $L-3L_e$, $L-3L_\mu$, $L-3L_\tau$, $L_e-\frac{1}{2}(L_\mu+L_\tau)$, $L_e+2L_\mu+2L_\tau$, $L_e-L_\mu$, $L_e-L_\tau$, and $L_\mu-L_\tau$.
Note that, $B_y \equiv B_1-yB_2-(3-y)B_3$, where $B_i$ ($i=1,2,3$) denotes the baryon number of $i\rm th$ generation quark. The parameter $y$ is an arbitrary constant and we assume to $y=0$ because we focus on the interactions of neutrinos with the first-generation quarks only. We do not consider the flavor-universal symmetries, say, $B-L$, as they will not show up in the neutrino oscillation experiments. The maximal abelian group $U(1)^\prime=U(1)_{B-L}\times U(1)_{L_{\mu}-L_{\tau}}\times U(1)_{L_\mu-L_e}$ can be gauged in an anomaly-free way by the addition of three right-handed neutrinos to the SM particle spectrum. Each of the new $U(1)^\prime$ symmetries mentioned above, being a subset of $(B-L)  \otimes (L_\mu-L_\tau)  \otimes  (L_\mu-L_e)$, can thus be gauged in an anomaly-free way with the extended particle spectrum.
Any one of the above candidate symmetry, after being gauged, results in a neutral gauge boson $Z^\prime$, which induces a Yukawa potential sourced by electron, proton, or neutron depending on the specific symmetry. Additionally, the flavor-dependent nature of the symmetries affects only $\nu_e$, $\nu_\mu$, $\nu_\tau$, or a combination of them, resulting in distinct modifications to the flavor transitions. We group the symmetries according to the texture of long-range potential ($V_{\rm LRI}$) that they induce (see table~\ref{tab:charges}). 
We find from our study that the oscillation gets suppressed when $V_{\rm LRI}$ dominates over the standard contributions to the Hamiltonian, due to $V_{\rm LRI}$ being diagonal. However, when $V_{\rm LRI}$ is comparable to the standard contributions, it resonantly enhances the oscillation parameters and significantly affects the oscillation probabilities. The symmetries mainly affecting the $\mu-\tau$ sector get the strongest constraints as they modify significantly, the disappearance channels, $\nu_\mu\to\nu_\mu$ and $\bar\nu_\mu\to\bar\nu_\mu$. Both DUNE and T2HK have significant number of events in their disappearance channels, offering better sensitivity to these symmetries. In contrast, the symmetries that affect mostly the electron sector get the weakest limits. These symmetries get contribution from the $\nu_\mu\to\nu_{e}$ and $\bar\nu_\mu\to\bar\nu_{e}$ appearance channels. The appearance channels in DUNE and T2HK have lower event rates --- hence offer weaker sensitivity. Similarly, the symmetries affecting mostly the disappearance channels need comparatively lower value of $V_{\rm LRI}$ to be discovered than the symmetries affecting the appearance channels, which need larger value of $V_{\rm LRI}$. We find that regardless of which symmetry is responsible for inducing long-range neutrino-matter interaction, DUNE and T2HK may constrain them more strongly than ever before or may discover them when the new matter potential is of the order of the standard ones.

For the first time, in this work, we also explore the possibility of distinguishing between the symmetries with the combined setup, DUNE+T2HK. The separation between different symmetries is more pronounced when their matter potential matrices, $\mathbf{V}_{\rm LRI}$, have different textures. In contrast, symmetries with similar matter potential textures, such as $B - 3 L_e$ and $L_e + 2 L\mu + 2 L_\tau$, are harder to distinguish, and symmetries with identical textures, like $B - 3 L_e$ and $L - 3 L_e$, cannot be separated at all. This challenge persists regardless of the value of $V_{\rm LRI}$ and is a fundamental limitation.  Similar to constraining and discovering the new interaction, symmetries that predominantly affect disappearance probabilities, such as $L_\mu - L_\tau$ and $B - L_e - 2 L_\tau$, create noticeable changes in the event rates, making them easier to distinguish due to higher event counts. Conversely, symmetries that primarily influence appearance probabilities are more challenging to be distinguished against the alternate symmetries.

\noindent{\bf  (ii) Lorentz invariance violation (LIV):} 
The possibility of Lorentz invariance violation arises in the unified theories like loop quantum gravity and string theory where the tensor field having non-trivial Lorentz transformation acquires vacuum expectation value to break Lorentz symmetry spontaneously.  
Using the effective field theory formulation, the Lorentz invariance violation can manifest at low energies with its strength suppressed by the Planck mass ($M_{\rm P}$). The Standard Model Extension (SME) framework is popular in studying the Lorentz violation in low energy experiments like neutrino oscillation. The SME Lagrangian is observer frame-independent. The Lorentz violating parameters in the SME Lagrangian are proportional to the vacuum expectation value of the tensors with non-trivial Lorentz transformation at the high-energy scales and the Planck suppression factor ($1/M_{\rm P}$). These parameters can be CPT-violating or CPT-conserving depending on the nature of the CPT properties of their associated operators.

In chapter~\ref{C4}, we probe the isotropic components of the off-diagonal CPT-violating LIV parameters ($a_{e\mu}$, $a_{e\tau}$, and $a_{\mu\tau}$) and for the first time, CPT-conserving LIV parameters ($c_{e\mu}$, $c_{e\tau}$ and $c_{\mu\tau}$). Through the derivation of approximate analytical expressions for both the $\nu_\mu\to\nu_e$ appearance and $\nu_\mu\to\nu_\mu$ disappearance probabilities, we verify our numerical findings that the LIV in the $e-\mu$ and $e-\tau$ sectors mostly affect the $\nu_\mu\to\nu_e$ appearance channel and the LIV in the $\mu-\tau$ sector alters the $\nu_\mu\to\nu_\mu$ disappearance probability.    
From the approximate analytical probability expressions, we see that the terms containing the CPT-violating LIV parameters are proportional to the baseline of the experiments, whereas the terms with CPT-conserving ones are proportional to the product of the baseline and neutrino energy. While exploiting the correlations of the LIV parameters with the most uncertain oscillation parameters, $\delta_{\rm CP}$ and $\theta_{23}$, we find that there exist some non-trivial solutions in the opposite octant of the true value of $\theta_{23}$ in case of $a_{e\mu}$ and $a_{e\tau}$. This happens due to the degeneracies between $\theta_{23}$, $\delta_{\rm CP}$, and the LIV phases in the $\nu_\mu\to\nu_{e}$ appearance channel. These non-trivial degeneracies cause local minima in the test statistics while placing the constraints on the LIV parameters, particularly in case of $a_{e\mu}$ and $a_{e\tau}$. These local minima can be lifted once we combine the prospective data from DUNE and T2HK. This happens because the baseline lengths and the accessible neutrino energies in these two experiments are different and these two setups complement each other. We find through our analysis that due to having a longer baseline and access to multi-GeV energies, DUNE has better sensitivity to all the CPT-violating as well as the CPT-conserving LIV parameters. However, given that the CPT-conserving LIV parameters comes with explicit energy dependence in the oscillation probabilities, T2HK shows almost no sensitivity to these parameters due to its access to sub-GeV neutrinos.

For comparison purpose, we also derive the expected bound from the combination of the currently running long-baseline experiments, T2K and NO$\nu$A with their full exposures --- 84.4 kt·MW·yrs for T2K and 58.8 kt·MW·yrs for NO$\nu$A. We observe that DUNE will have an upper hand in constraining both the CPT-violating and CPT-conserving LIV parameters than the combined T2K+NO$\nu$A setup, because the former has a larger baseline, higher energy, and less systematic uncertainties.  However, the expected constraints on the CPT-violating LIV parameters from T2K+NO$\nu$A setup are close to that of T2HK. Since NO$\nu$A has a comparatively larger baseline, it has a better sensitivity to the CPT-violating LIV parameters. However, T2HK has less systematic uncertainties to compensate for its small baseline. Similarly, for the CPT-conserving LIV parameters, the limits from the combined T2K+NO$\nu$A setup are of the same order as T2HK, where the former gives slightly better constraints. It is because, apart from the larger baseline, the energy of the neutrino beam is also high for NO$\nu$A, which plays a vital role in constraining the CPT-conserving LIV parameters.

While comparing our results with the existing limits on the LIV parameters from the atmospheric neutrino experiments, IceCube and Super-K, we find that the expected constraints on $a_{e\mu}$ and $a_{e\tau}$ from the standalone DUNE setup are better than that from Super-K. This is because $a_{e\mu}$ and $a_{e\tau}$ mostly get contribution from the $\nu_\mu\to\nu_e$ appearance channel which is the major oscillation channel for long-baseline setups, however for Super-K, the major channel is $\nu_\mu\to\nu_\mu$. For $a_{\mu\tau}$, the sensitivities from both DUNE+T2HK and Super-K are comparable. 
Since the atmospheric experiments, Super-K and IceCube, have access to a wide range of neutrino energies and baselines, they put stringent constraints on the CPT-conserving LIV parameters, with IceCube being the best.

\vspace{0.3cm}

\noindent{\bf (iii) Active-sterile neutrino oscillation:} 
Though the three-flavor framework of neutrino oscillation accommodates almost all the available oscillation data, there have been various motivations from both the theory and experiment sides to believe in the existence of extra neutrino states. Since the number of active neutrino flavors is constrained to be three from the measurement of the invisible decay width of $Z^0$ boson at the LEP experiment, any other neutrino states, if present, should be sterile, in the sense that they do not have any SM interactions. From the theory side, various neutrino mass models require sterile states, and from the experiment side, there are various short-baseline (SBL) anomalies, namely, LSND, MiniBooNE, Gallium, and Reactor antineutrino anomalies, which hint towards the existence of a sterile neutrino. The number of sterile states and their masses can, in principle, take any value. The TeV-scale sterile neutrino is required to explain the small active neutrino masses via the seesaw mechanism. The keV-scale sterile neutrinos can be the viable dark matter candidate. The eV-scale sterile neutrinos are required to explain the above-mentioned SBL anomalies, the super-light sterile neutrinos with masses very close to the active ones can explain the absence of the expected upturn in the solar neutrino energy spectrum below 8 MeV.

In the 3+1 framework of active-sterile neutrino oscillation, where one sterile neutrino is assumed along with the three active neutrinos, the mixing matrix between the flavor ($\nu_{e}\,,\nu_{\mu}\,,\nu_{\tau}\,,\nu_{s}$) and the mass ($\nu_{1}\,,\nu_{2}\,,\nu_{3}\,,\nu_{4}$) eigenstates is a $4\times4$ unitary matrix, parameterized as
$U =   \tilde R({\theta_{34}\,,\delta_{34}})\,R({\theta_{24}})\,\tilde R({\theta_{14}\,,\delta_{14}})\,R({\theta_{23}})\,\tilde R({\theta_{13}\,,\delta_{13}})\,R({\theta_{12}})$, where the $R_{ij}$ and $\tilde R_{ij}$ are, respectively, the $4\times4$ real and complex rotation matrices in the ($i,j$) plane. The 3+1 oscillation framework contains six rotation angles, namely, $\theta_{12}\,,\theta_{13}\,,\theta_{23}\,,\theta_{14}\,,\theta_{24}\,,{\rm and}\,\,\theta_{34}$, three CP-violating phases, namely, $\delta_{13}~(\equiv\delta_{\rm CP})\,,\delta_{14}\,,{\rm and}\,\,\delta_{34}$, and three independent mass-squared differences, namely, $\Delta m^2_{21}$, $\Delta m^2_{31}$, and $\Delta m^2_{41}$. Here, $\Delta m^2_{41}(\equiv m^2_{4}- m^2_{1})$ governs the active-sterile neutrino oscillation.

In chapter~\ref{C5}, we probe the active-sterile neutrino oscillation for a wide range of the sterile mass-squared splitting, $\Delta m^2_{41}$, from $10^{-5}$ eV$^2$ to 100 eV$^2$ using DUNE, the Japanese detector (JD, also called T2HK), and the Japanese detector with another detector in Korea (JD+KD, also called T2HKK). We particularly exploit the phenomenological implications of three benchmark values of $\Delta m^2_{41}$, \eg~(i) $\Delta m^2_{41}$ = $\Delta m^2_{31}$, (ii) $\Delta m^2_{41}$ = $\Delta m^2_{21}$, and (iii) $\Delta m^2_{41}$ = 1 eV$^2$ in probing the CP-violating phases. The effect of the CP phases occur due to the interference terms between any two of the mass-squared splittings, $\Delta m^2_{41}$, $\Delta m^2_{31}$, and $\Delta m^2_{21}$. We find from the approximate $\nu_\mu\to\nu_e$ probability expressions, derived in our work, that the strength of the interference terms in presence of sterile neutrino becomes double in case of $\Delta m^2_{41}=\Delta m^2_{31}$ as compared to $\Delta m^2_{41}$ = 1 eV$^2$. In case of $\Delta m^2_{41}=\Delta m^2_{21}$, no interference terms are present at the leading order in the probability expression. Due to this, the total CP violation discovery potential induced by $\delta_{13}$ and $\delta_{14}$ is maximum when $\Delta m^2_{41}$ = $\Delta m^2_{31}$. Also, while reconstructing the true values of $\delta_{13}$ and $\delta_{14}$, maximum sensitivity comes when $\Delta m^2_{41}$ = $\Delta m^2_{31}$. Comparing the sensitivities from the considered experimental setups, as far as the sensitivity to the CP phases is considered, JD performs better than DUNE. This is because JD with a relatively smaller baseline can measure the CP phases without having much interference from the fake CP asymmetry due to the Earth matter effect. The JD+KD setup gives the best CP sensitivity. We give exclusion limits in the plane of ($\Delta m^2_{41}-{\sin^2 \theta_{14}}$), ($\Delta m^2_{41}-{\sin^2 \theta_{24}}$), and ($\Delta m^2_{41}-{\sin^2 2\theta_{\mu e}}$), where, $\sin^{2}{2\theta_{\mu e}}\equiv 4 \sin^{2}\theta_{14} \cos^{2}\theta_{14} \sin^{2}\theta_{24}$. While doing so, we consider the near detector of DUNE and, for the first time, the Inter-mediate Water Cherenkov Detector (IWCD) in addition to JD+KD. Note that JD+KD setup also has a near detector, named ND280, which constrains the systematic uncertainties. With the far detectors only, the sensitive region is around $\Delta m^2_{41}\approx\Delta m^2_{31}$. At the far detectors for larger $\Delta m^2_{41}\sim O(\rm eV^2)$, the sensitivity to the sterile mixing angles are independent of $\Delta m^2_{41}$. This is because, the rapid oscillations which are induced by large $\Delta m^2_{41}$ cannot be probed due to the finite energy resolution of the experiments under consideration and hence are averaged out. At intermediate values of $\Delta m^2_{41}\sim 10^{-3}-10^{-2}~\rm eV^2$, the sensitivity increases because of the interference between the atmospheric and sterile mass-squared splittings. The maximum exclusion capability happens with the inclusion of the near detectors. We find that DUNE with both its far and near detector (DUNE FD+ND) can exclude the values of $\theta_{14}\geq1.9^\circ$ for $\Delta m^2_{41}\sim10~ {\rm eV}^2$ and JD+KD+IWCD can exclude $\theta_{14}\geq 3^\circ$ for $\Delta m^2_{41}\sim1~ {\rm eV}^2$, at 90\% C.L.. Due to higher event rates in the disappearance channels, $\theta_{24}$ gets stronger limits than $\theta_{14}$. The best sensitivities, at 90\% C.L., for $\theta_{24}$ are $1.2^\circ$ when $\Delta m^2_{41}\sim1~ {\rm eV}^2$ from JD+KD+IWCD and $0.6^\circ$ when $\Delta m^2_{41}$ is around 5--10 eV$^2$ from DUNE FD+ND, respectively. We find that at 90\% C.L., our long-baseline setups can probe the entire allowed regions by the LSND and MiniBooNE anomalies in the plane of $\Delta m^2_{41}-{\sin^2 2\theta_{\mu e}}$.

In summary, we study three possible BSM scenarios, long-range neutrino matter interaction, possible violation of Lorentz symmetry, and active-sterile neutrino oscillation, in the context of the next-generation long-baseline experiments DUNE, T2HK, and T2HKK. Using the publicly available simulation package, General Long-Baseline Experiment Simulator (GLoBES), we perform a detailed simulation study of these long-baseline experiments and estimate their sensitivities towards various new physics parameters that appear in the three above-mentioned BSM scenarios. We demonstrate how these next-generation high-precision LBL experiments hold promise to constrain the parameter space of these new physics scenarios with the possibility of discovering them in some favorable cases. Our study will open up new avenues for the next-generation LBL experiments to search for these new physics effects in their data, which will either lead to discovery or if not, they will put stringent constraints on these new physics parameters.       

\blankpage 
%%%%%%%%%%%%%%%%%%%%%%%%%%%%%%%%%%
%\chapter{Appendices}
%\label{C3-appendix} 
%=================================================
\begin{appendix}
	%=================================================	
\chapter{Appendix to Chapter~\ref{C3}}

	%=================================================
	\section{$U(1)^\prime$ charges of fermions}
	\label{app:U1-charges}
	%=================================================
	
	\renewcommand\thefigure{A\arabic{figure}}
	\renewcommand\theHfigure{A\arabic{figure}}
	\renewcommand\thetable{A\arabic{table}}
	\renewcommand\theequation{A\arabic{equation}}
	\setcounter{figure}{0} 
	\setcounter{table}{0}
	\setcounter{equation}{0}
	
	%-------------------------------------------------------------
	Table~\ref{tab:charges2} shows the $U(1)^\prime$ charges of fermions for each of our candidate symmetries (table~\ref{tab:charges}).  The charges are used to compute the long-range matter potential, $V_{\rm LRI}$, using eqs.~(\ref{equ:pot_total_general}) and (\ref{equ:pot_total_simplified}) in the main text.
	
	%=================================================
	\begin{table}[t!]
		\centering
		\renewcommand{\arraystretch}{1.3}
		\begin{center}
			\begin{tabular}{|c|c|c|c|c|c|c|c|}
				\hline
				\mr{2}{*}{$U(1)^\prime$ symmetry} &  \mc{6}{c|}{$U(1)^\prime$ charge}
				\\
				&\mr{1}{*}{$a_{u}$} & \mr{1}{*}{$a_{d}$} & \mr{1}{*}{$a_{e}$} & \mr{1}{*}{$b_{e}$} & \mr{1}{*}{$b_{\mu}$} & \mr{1}{*}{$b_{\tau}$} 
				\\
				\hline
				\mr{1}{*}{$B-3L_{e}$}& \mr{1}{*}{$\frac{1}{3}$}&\mr{1}{*}{$\frac{1}{3}$} & \mr{1}{*}{$-3$} & \mr{1}{*} {$-3$} & \mr{1}{*}{0} & \mr{1}{*}{0} 
				\\
				\hline
				\mr{1}{*}{$L-3L_{e}$}& \mr{1}{*}{0}&\mr{1}{*}{0} & \mr{1}{*}{$-2$} & \mr{1}{*} {$-2$} & \mr{1}{*}{1} & \mr{1}{*}{1} 
				\\
				\hline
				\mr{1}{*}{$B-\frac{3}{2}(L_{\mu}+L_{\tau})$}& \mr{1}{*}{$\frac{1}{3}$}&\mr{1}{*}{$\frac{1}{3}$} & \mr{1}{*}{0} & \mr{1}{*} {0} & \mr{1}{*}{$-\frac{3}{2}$} & \mr{1}{*}{$-\frac{3}{2}$} 
				\\
				\hline
				\mr{1}{*}{$L_{e}-\frac{1}{2}(L_{\mu}+L_{\tau})$}& \mr{1}{*}{0}&\mr{1}{*}{0} & \mr{1}{*}{1} & \mr{1}{*} {1} & \mr{1}{*}{$-\frac{1}{2}$} & \mr{1}{*}{$-\frac{1}{2}$} 
				\\
				\hline
				\mr{1}{*}{$L_{e}+2L_{\mu}+2L_{\tau}$}& \mr{1}{*}{0}&\mr{1}{*}{0} & \mr{1}{*}{1} & \mr{1}{*} {1} & \mr{1}{*}{2} & \mr{1}{*}{2} 
				\\
				\hline
				\mr{1}{*}{$B_{y}+L_{\mu}+L_{\tau}$}& \mr{1}{*}{$\frac{1}{3}$}&\mr{1}{*}{$\frac{1}{3}$} & \mr{1}{*}{0} & \mr{1}{*} {0} & \mr{1}{*}{1} & \mr{1}{*}{1} 
				\\
				\hline
				\mr{1}{*}{$B-3L_{\mu}$}& \mr{1}{*}{$\frac{1}{3}$}&\mr{1}{*}{$\frac{1}{3}$} & \mr{1}{*}{0} & \mr{1}{*} {0} & \mr{1}{*}{$-3$} & 0 
				\\
				\hline
				\mr{1}{*}{$L-3L_{\mu}$}& \mr{1}{*}{0}&\mr{1}{*}{0} & \mr{1}{*}{1} & \mr{1}{*} {1} & \mr{1}{*}{$-2$} & \mr{1}{*}{1}
				\\
				\hline
				\mr{1}{*}{$B-3L_{\tau}$}& \mr{1}{*}{$\frac{1}{3}$}&\mr{1}{*}{$\frac{1}{3}$} & \mr{1}{*}{0} & \mr{1}{*} {0} & \mr{1}{*}{0} & \mr{1}{*}{$-3$}
				\\
				\hline
				\mr{1}{*}{$L-3L_{\tau}$}& \mr{1}{*}{0}&\mr{1}{*}{0} & \mr{1}{*}{1} & \mr{1}{*} {1} & \mr{1}{*}{1} & \mr{1}{*}{$-2$}
				\\
				\hline
				\mr{1}{*}{$L_{e}-L_{\mu}$}& \mr{1}{*}{0}&\mr{1}{*}{0} & \mr{1}{*}{1} & \mr{1}{*} {1} & \mr{1}{*}{$-1$} & \mr{1}{*}{0} 
				\\
				\hline
				\mr{1}{*}{$L_{e}-L_{\tau}$}& \mr{1}{*}{0}&\mr{1}{*}{0} & \mr{1}{*}{1} & \mr{1}{*} {1} & \mr{1}{*}{0} & \mr{1}{*}{$-1$} 
				\\
				\hline
				\mr{1}{*}{$L_{\mu}-L_{\tau}$}& \mr{1}{*}{0}&\mr{1}{*}{0} & \mr{1}{*}{0} & \mr{1}{*} {0} & \mr{1}{*}{1} & \mr{1}{*}{$-1$} 
				\\
				\hline					
				\mr{1}{*}{$B-L_{e}-2L_{\tau}$}& \mr{1}{*}{$\frac{1}{3}$}&\mr{1}{*}{$\frac{1}{3}$} & \mr{1}{*}{$-1$} & \mr{1}{*} {$-1$} & \mr{1}{*} {0} & \mr{1}{*} {$-2$}
				\\
				\hline
			\end{tabular}
			\caption{\textit{$U(1)^{\prime}$ charges of the fermions for the candidate symmetries.}  Charges $a_u$, $a_d$, and $a_e$ are, respectively, of the up quark, down quark, and the electron; and $b_e$, $b_\mu$, and $b_\tau$ are, respectively, of $\nu_e$, $\nu_\mu$, and $\nu_\tau$.  For protons, the charge is $a_p = 2 a_u + a_d$; for neutrons, it is $a_n = 2 a_d + a_u$.  The charges are used to compute the long-range potential, $V_{\rm LRI}$, using eqs.~(\ref{equ:pot_total_general}) and (\ref{equ:pot_total_simplified}) in the main text.
				\label{tab:charges2}}
		\end{center}
	\end{table}
	%=================================================
	
	%=================================================
	\section{Effect of long-range interactions on neutrino oscillation parameters}
	\label{app:param_run}
	%=================================================
	
%	\renewcommand\thefigure{B\arabic{figure}}
%	\renewcommand\theHfigure{B\arabic{figure}}
%	\renewcommand\thetable{B\arabic{table}}
%	\renewcommand\theequation{B\arabic{equation}}
%	\setcounter{figure}{0} 
%	\setcounter{table}{0}
%	\setcounter{equation}{0}
	
	Figures~\ref{fig:theta_run} and~\ref{fig:delm_run} show the modification with energy of the mixing angles and mass-squared differences modified under the new matter potential introduced by our candidate $U(1)^\prime$ symmetries (table~\ref{tab:charges}).  To produce these figures, we compute the modified oscillation parameters using the approximate expressions from \Refe~\cite{Agarwalla:2021zfr}; to produce all other results, we compute them numerically and implicitly as part of the calculation of the oscillation probabilities.  We show results for DUNE; the results for T2HK, not shown, are analogous.
	
	%=================================================
	\begin{figure}
		\centering
		\includegraphics[width=0.8\textwidth]{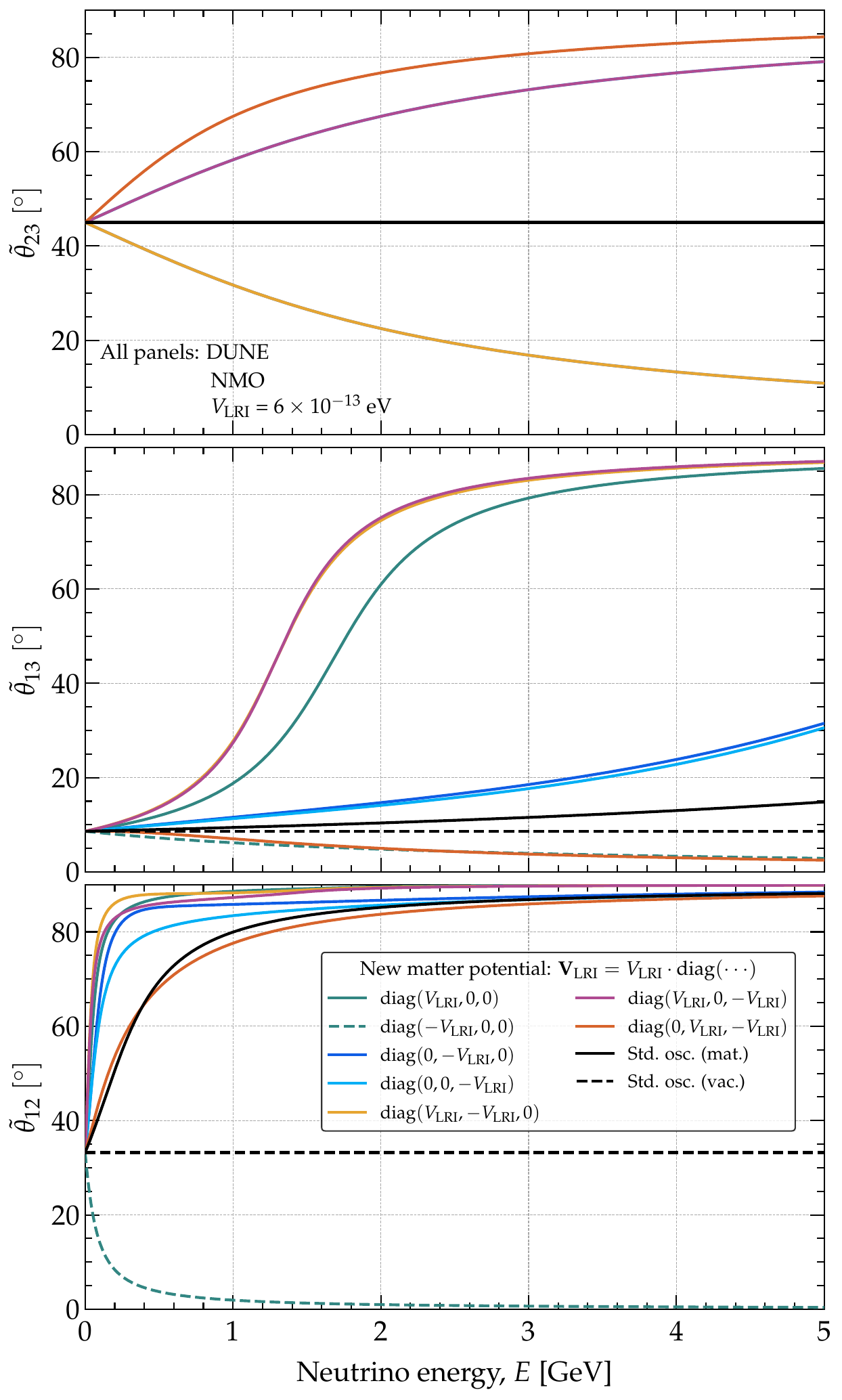}
		\caption{\textit{Modification of the mixing angles with energy.}  We compare their modification in the presence of the new matter potential induced by our candidate $U(1)^\prime$ symmetries {\it vs.}~their standard values in vacuum and modified by matter inside Earth.  We assume the DUNE baseline,  an illustrative value of the new matter potential, of $V_{\rm LRI} = 6 \cdot 10^{-13}$~eV, and the values of the oscillation parameters from table~\ref{tab:params_value1}, except for $\theta_{23}$, which we set to $45^\circ$.  See \figu{delm_run} for the modification of the neutrino mass-squared differences and appendix~\ref{app:param_run} for details.}
		\label{fig:theta_run}
	\end{figure}
	%=================================================
	
	We group symmetries according to the texture of the new matter potential that they induce, $\mathbf{V}_{\rm LRI}$ in table~\ref{tab:charges}.  Symmetries with equal or similar potential texture yield equal or similar modification of the oscillation parameters, which, in turn yields equal or similar oscillation probabilities (figs.~\ref{fig:probability} and \ref{fig:event_spectra}).  Below, we point out the salient features in the modification of the oscillation parameters:
	
	%=================================================
	\begin{description}
		%%%
		\item[Modification of $\tilde\theta_{23}$ (\figu{theta_run}):]
		The value of $\tilde\theta_{23}$ drives both the transition and survival probabilities, eqs.~(\ref{equ:app_mat}) and (\ref{equ:surv_mat}).  Symmetries that induce a new potential in the muon sector, tau sector, or both, of the Hamiltonian, \equ{hamiltonian_tot} in the main text, affect the modification of $\tilde\theta_{23}$.  This includes symmetries that induce a matter potential, $\mathbf{V}_{\rm LRI}$ of the form $\textrm{diag}(0,\bullet,0)$ (\ie, $B-3L_\mu$ and $L-3L_\mu$), $\textrm{diag}(0,0,\bullet)$ (\ie, $B-3L_\tau$, $L-3L_\tau$), $\textrm{diag}(\bullet,\bullet,0)$ (\ie, $L_e-L_\mu$), $\textrm{diag}(\bullet,0,\bullet)$ (\ie, $L_e-L_\tau$), and, especially, $\textrm{diag}(0,\bullet,\bullet)$ (\ie, $L_\mu-L_\tau$ and $B-L_e-2L_\tau$).  Depending on the signs of the nonzero elements of $\mathbf{V}_{\rm LRI}$, the value of $\tilde\theta_{23}$ either increases or decreases {\it vs.}~its value in vacuum, $\theta_{23}$.  Symmetries that induce a new potential only in the electron sector, \ie, with texture $\textrm{diag}(\bullet,0,0)$, do not affect the modification of $\tilde\theta_{23}$.  This encompasses  symmetries  $B-3L_e$, $L-3L_e$, $B-\frac{3}{2}(L_\mu+L_\tau)$, $L_e-\frac{1}{2}(L_\mu+L_\tau)$, $L_e + 2L_\mu + 2L_\tau$, and $B_y + L_\mu + L_\tau$.
		%%%
		\item[Modification of $\tilde\theta_{13}$ (\figu{theta_run}):]
		The value of $\tilde\theta_{13}$ also drives both probabilities, eqs.~(\ref{equ:app_mat}) and (\ref{equ:surv_mat}).  Most of our candidate symmetries induce a new potential in either the electron sector, tau sector, or both, and so directly affect the modification of $\tilde\theta_{13}$.  The exceptions are the symmetries that induce a potential only in the muon sector, of the form $\textrm{diag}(0,\bullet,0)$ (\ie, $B-3L_\mu$ and $L-3L_\mu$), which, however, still affect $\tilde\theta_{13}$ indirectly due to the mixing of the muon sector with the electron and tau sectors via $\mathbf{H}_{\rm vac}$.  Like for $\tilde\theta_{23}$ above, depending on the signs of the nonzero elements of $\mathbf{V}_{\rm LRI}$, the value of $\tilde\theta_{13}$ either increases or decreases {\it vs.}~its value in vacuum, $\theta_{13}$.  The largest deviations from the vacuum value occur under symmetries that induce a potential that affects both sectors: $\textrm{diag}(\bullet,0,\bullet)$ (\ie, $L_e-L_\tau$) and $\textrm{diag}(\bullet,\bullet,0)$ (\ie, $L_e-L_\mu$, which affects the tau sector via the standard mixing between it and the muon sector). In the presence of new matter potentials arising from most of our candidate symmetries, $\tilde\theta_{13}$ reaches $45^\circ$ at the resonance energy and continues to increase as the energy rises. 
		From \equ{app_mat}, it is clear that the probability approaches maximum as soon as $\tilde\theta_{13}$ attains resonance. 
		
		The resonance energy under LRI in the one-mass-scale-dominance (OMSD, $\Delta m^2_{31}L/4E$ $>>\Delta m^2_{21}L/4E$) approximation and assuming $\theta_{23} = 45^\circ$, is given by~\cite{Agarwalla:2021zfr},
		\begin{equation}
			\label{equ:resonance_lri}
			E_{\text{res}}^{\text{LRI}} \simeq  
			\left[E_{\text{res}}^{\text{SI}}\right]_\text{OMSD} \cdot V_{\rm CC} \cdot \bigg[\frac{1-(\alpha s^2_{12} c_{13}^{2} /\cos 2\theta_{13})}{V_{\rm CC} - \frac{1}{2}(V_{{\rm LRI}, \mu}+V_{{\rm LRI}, \tau}-2V_{{\rm LRI}, e})}\bigg] ,
		\end{equation}  
		where 
		\begin{equation}
			\label{equ:resonance_si}
			\left[E_{\text{res}}^{\text{SI}}\right]_\text{OMSD} = \frac{\Delta m^2_{31}\cos2\theta_{13}}{2V_{CC}} , 
		\end{equation}
		is the resonance energy in presence of standard charged-current interactions.
		Figure~\ref{fig:res_energy} shows the resonance energy as a function of $V_{\rm LRI}$ for the symmetry textures that are expected to give $\tilde\theta_{13}$ resonance. The resonance energies decrease from that of the standard oscillation value as $V_{\rm LRI}$ grows. The resonance is attained at lower energies for the symmetries with (i) ${\bf V}_{\rm LRI}={\rm diag}(V_{\rm LRI}, 0, -V_{\rm LRI})$ than (ii) ${\bf V}_{\rm LRI}={\rm diag}(V_{\rm LRI}, 0, 0)$ than (iii) ${\bf V}_{\rm LRI}={\rm diag}(0, 0, -V_{\rm LRI})$. This can be understood from \equ{resonance_lri}, as the denominator of the term inside the bracket increases more for (i) than (ii) than (iii) leading to the decrease in the energy needed to attain the resonance. The resonance energies are equivalent for symmetries with ${\bf V}_{\rm LRI}$ of the form ${\rm diag}(V_{\rm LRI}, -V_{\rm LRI}, 0)$ \& ${\rm diag}(V_{\rm LRI}, 0, -V_{\rm LRI})$ and ${\rm diag}(0, -V_{\rm LRI}, 0)$ \& ${\rm diag}(0, 0, -V_{\rm LRI})$. For all these symmetries, the $\tilde\theta_{13}$ resonance for DUNE baseline happens below 2 GeV when the strength of the new potential reaches around $10^{-12}-10^{-11}$ eV, which is clearly seen in \figu{dune_prob_events}, where we illustrate one particular texture, ${\rm diag}(0, 0, -V_{\rm LRI})$. For the textures, ${\rm diag}(-V_{\rm LRI}, 0, 0)$ and ${\rm diag}(0, V_{\rm LRI}, -V_{\rm LRI})$, $\tilde\theta_{13}$ decreases with energy even from its vacuum and will never achieve resonance. 
		%%%
		\item[Modification of $\tilde\theta_{12}$ (\figu{theta_run}):]
		Most of our candidate symmetries induce a new potential in either the electron sector, muon sector, or both, and so directly affect the modification of $\tilde\theta_{12}$.  The exceptions are the symmetries that induce a potential only in the tau sector, of the form $\textrm{diag}(0,0,\bullet)$ (\ie, $B-3L_\tau$ and $L-3L_\tau$), which, however, still affect $\tilde\theta_{12}$ indirectly due to the mixing of the tau sector with the electron and muon sectors via $\mathbf{H}_{\rm vac}$.  In nearly all cases, the value of $\tilde\theta_{12}$ saturates to $90^\circ$ early in its energy modification, which justifies our use of the approximate expression for the $\nu_\mu \to \nu_e$ probability, \equ{app_mat}, to interpret our results in the main text.  The exceptions are the symmetries that induce a potential of the form $\textrm{diag}(-V_{\rm LRI}, 0, 0)$ (\ie, $L_e+2L_\mu+2L_\tau$ and $B_y+L_\mu+L_\tau$), which instead quickly drives $\tilde\theta_{12}$ to zero; however, in this case, $\tilde\theta_{12}$ instead saturates early to $90^\circ$ for antineutrinos, since they are affected by the potential $-\mathbf{V}_{\rm LRI}$.
		%%%
		\item[Modification of $\Delta \tilde m^2_{31}$ (\figu{delm_run}):] 
		The modification of $\Delta \tilde m^2_{31}$ affects the oscillation phase of the $\nu_\mu \to \nu_\mu$ probability, \equ{surv_mat}.  Figure~\ref{fig:osc_length_run} shows the modification with energy of the oscillation length associated to $\Delta \tilde m^2_{31}$, \ie, $L_{\rm osc}^{31} \equiv 2.47~{\rm km}~(E/{\rm GeV}) / (\Delta \tilde{m}_{31}^2/{\rm eV}^2)$, which helps understand the impact of the modification of $\Delta \tilde m^2_{31}$ on the $\nu_\mu \to \nu_\mu$ probability.  At low energies, below about 0.8~GeV, the value of $\Delta \tilde m^2_{31}$ decreases slightly below its vacuum value, $\Delta m^2_{31}$, for most of our candidate symmetries.  However, this is overcome by the low energies and, as a result, the $\nu_\mu \to \nu_\mu$ oscillation length is shorter than in vacuum (\figu{osc_length_run}) and grows more slowly than in vacuum, and so the first oscillation maximum of the probability is shifted to slightly higher energies to compensate for the slower growth in $L_{\rm osc}^{31}$; see \figu{probability}.  At higher energies, \figu{delm_run} shows that the value of $\Delta \tilde m^2_{31}$ increases quickly, but because the energy is also growing, the net effect is to first stall and then overturn the growth of $L_{\rm osc}^{31}$, which, again, shifts the position of the second maximum of the probability further to higher energies.
		%%%
		\item[Modification of $\Delta \tilde m^2_{21}$ (\figu{delm_run}):]
		The value of $\Delta \tilde m^2_{21}$ grows with energy under all of our candidate symmetries.  However, this has only a mild impact on our results, since the transition and survival probabilities for DUNE and T2HK, eqs.~(\ref{equ:app_mat}) and (\ref{equ:surv_mat}), are driven by $\Delta \tilde m^2_{31}$ and $\Delta \tilde m^2_{32}$.
		%%%
		\item[Modification of $\Delta \tilde m^2_{32}$ (\figu{delm_run}):]
		The modification of $\Delta \tilde m^2_{32} \equiv \Delta \tilde m^2_{31}-\Delta \tilde m^2_{21}$ affects the oscillation phase of the $\nu_\mu \to \nu_e$ probability, \equ{app_mat}.  For most of our candidate symmetries, the value of $\Delta \tilde m^2_{32}$ is smaller than the vacuum value, $\Delta m^2_{32}$, across most of the energy range in \figu{delm_run}, roughly below 3~GeV.  As a result, in this range, the oscillation length associated to $\Delta \tilde m^2_{32}$, \ie, $L_{\rm osc}^{32} \equiv 2.47~{\rm km}~(E/{\rm GeV}) / (\Delta \tilde{m}_{32}^2/{\rm eV}^2)$, grows with energy faster than in vacuum (\figu{osc_length_run}) which, in turn, shifts the first and second oscillation maxima in the $\nu_\mu \to \nu_e$ probability to lower energies; see \figu{probability}.
	\end{description}

	%=================================================
	\begin{figure}
		\centering
		\includegraphics[width=0.82\textwidth]{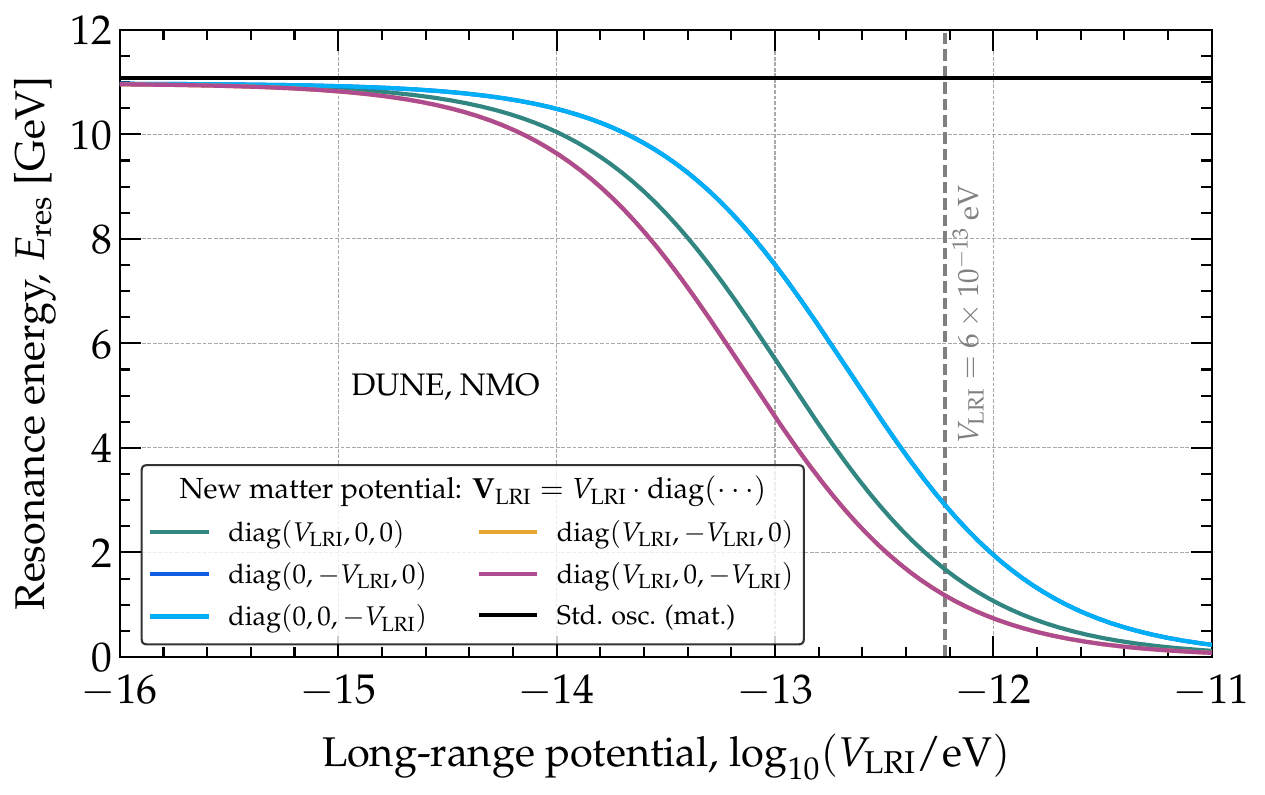}
		\caption{\textit{Resonance neutrino energy as a function of long-range interaction potential.} The energies are calculated for the DUNE baseline assuming normal neutrino mass ordering, and the values of the oscillation parameters are from table~\ref{tab:params_value1}, except for $\theta_{23}$, which we set to $45^\circ$. The grey vertical line corresponds to the illustrative value of long-range potential taken when we showcase its impact on the modification of mixing parameters in figs.~\ref{fig:theta_run}, \ref{fig:delm_run}, and \ref{fig:osc_length_run}, the oscillation probabilities and the event spectra in figs.~\ref{fig:dune_prob_events}, \ref{fig:t2hk_prob_events}, \ref{fig:probability}, and \ref{fig:event_spectra}. It reaffirms that $\tilde\theta_{13}$ attains the resonance value between 1-2 GeV for some symmetries, as shown in the middle panel of \figu{theta_run} and validates our approximate expression, \equ{resonance_lri}.}
		\label{fig:res_energy}
	\end{figure}
	%=================================================
	
	%=================================================
	\begin{figure}
		\centering
		\includegraphics[width=0.8\textwidth]{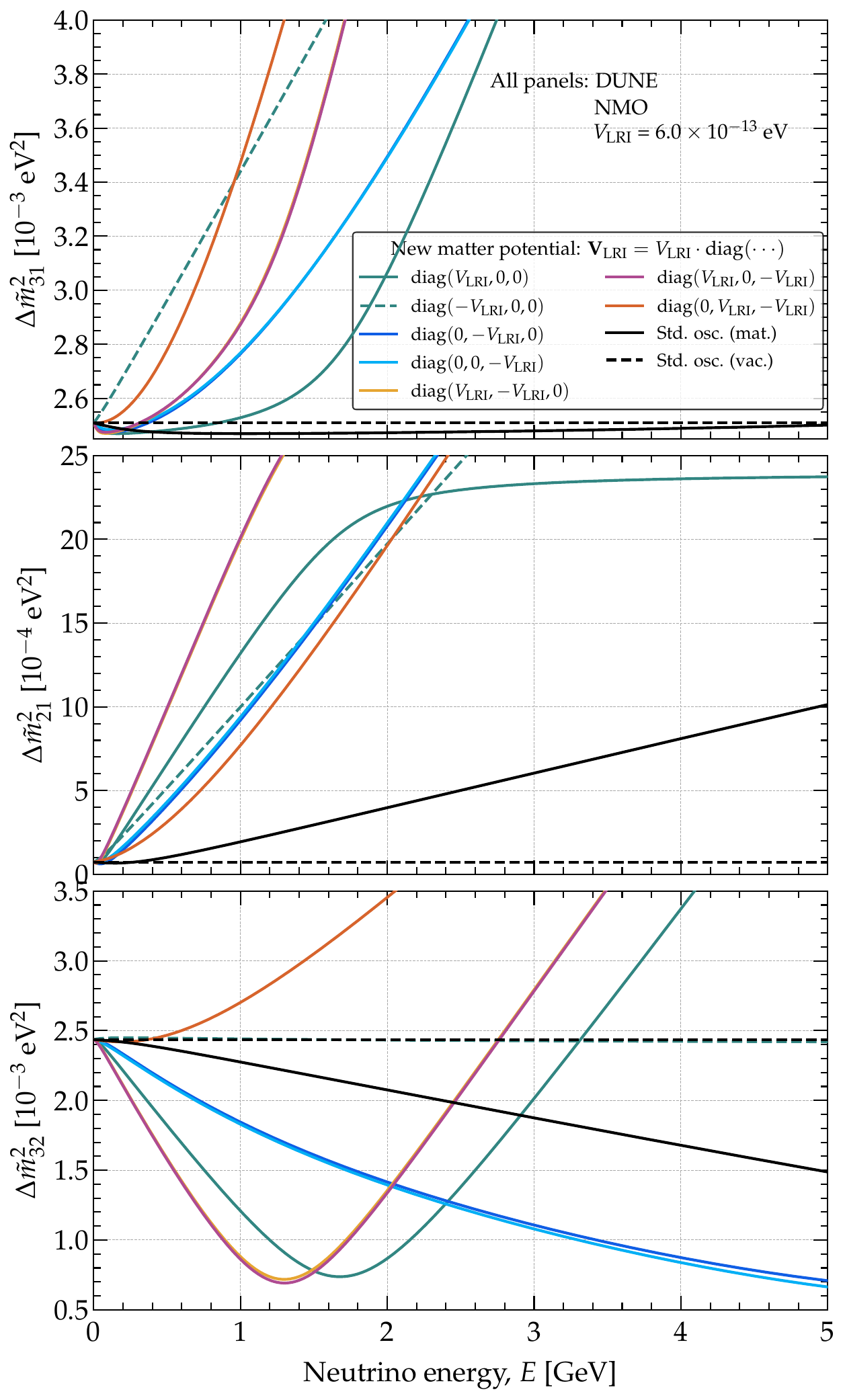}
		\caption{\textit{Modification of the neutrino mass-squared differences with energy.}  We compare their modification in the presence of the new matter potential induced by our candidate $U(1)^\prime$ symmetries {\it vs.}~their standard values in vacuum and modified by matter inside Earth.  We assume the DUNE baseline,  an illustrative value of the new matter potential, of $V_{\rm LRI} = 6 \cdot 10^{-13}$~eV, and the values of the oscillation parameters from table~\ref{tab:params_value1}, except for $\theta_{23}$, which we set to $45^\circ$.  See \figu{theta_run} for the modification of the mixing angles and appendix~\ref{app:param_run} for details.}
		\label{fig:delm_run}
	\end{figure}
	%=================================================
	
	%=================================================
	\begin{figure}
		\centering
		\includegraphics[width=0.82\textwidth]{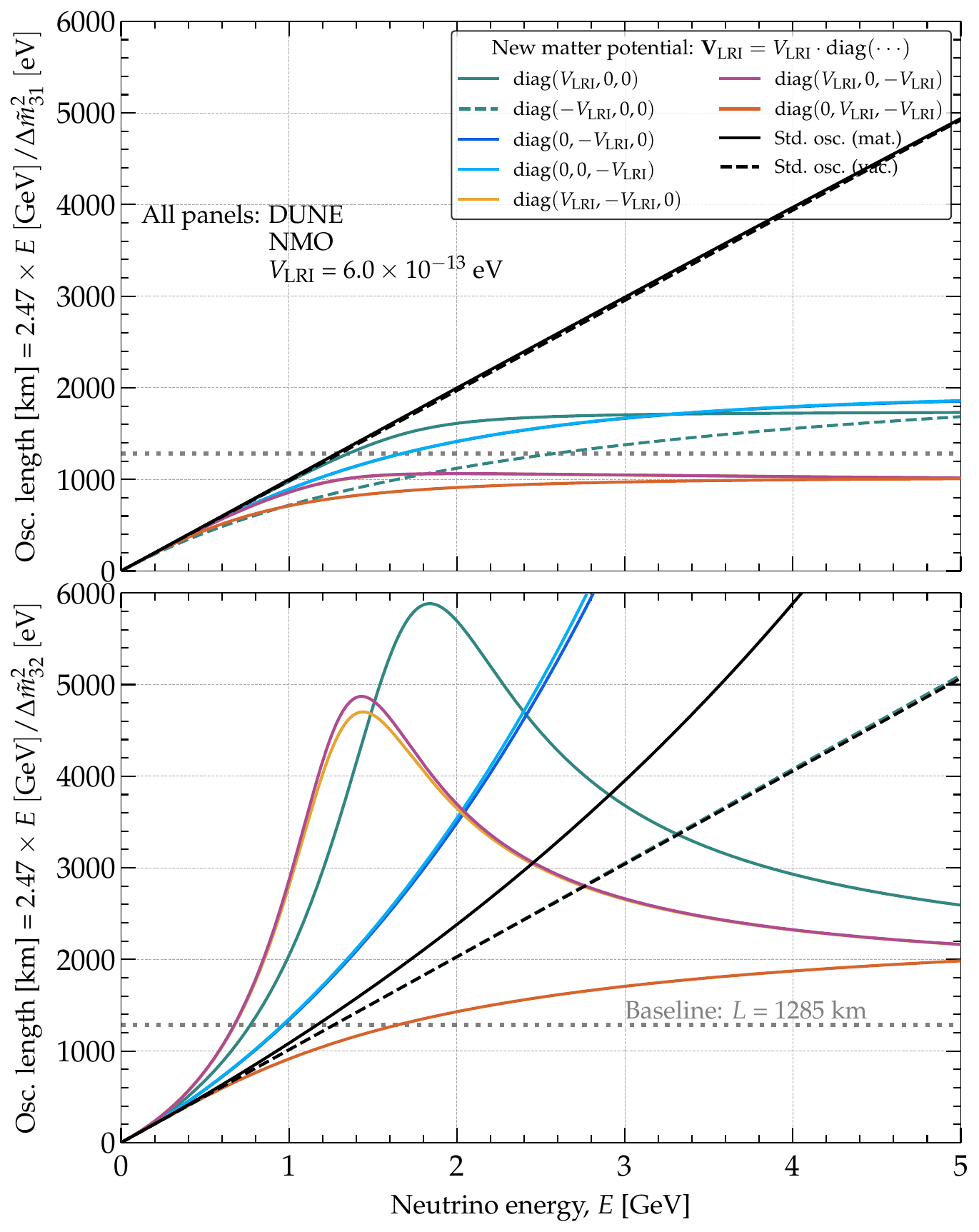}
		\caption{\textit{Modification of the neutrino oscillation length with energy.}  We show the modification of the oscillation length associated with the squared-mass difference modified by the new matter, $\Delta \tilde{m}_{31}^2$ ({\it top}) and $\Delta \tilde{m}_{32}^2$ ({\it bottom}), for our candidate $U(1)^\prime$ symmetries.  We compare them against the standard values in vacuum and modified by matter inside Earth.  We assume the DUNE baseline, an illustrative value of the new matter potential, of $V_{\rm LRI} = 6 \cdot 10^{-13}$~eV, and the values of the oscillation parameters from table~\ref{tab:params_value1}, except for $\theta_{23}$, which we set to $45^\circ$.  See \figu{delm_run} for the modification of the mass-squared differences and appendix~\ref{app:param_run} for details.}
		\label{fig:osc_length_run}
	\end{figure}
	%=================================================
	
%	\renewcommand\thefigure{C\arabic{figure}}
%	\renewcommand\theHfigure{C\arabic{figure}}
%	\renewcommand\thetable{C\arabic{table}}
%	\renewcommand\theequation{C\arabic{equation}}
%	\setcounter{figure}{0} 
%	\setcounter{table}{0}
%	\setcounter{equation}{0}
	
	%=================================================
	\begin{figure}
		\centering
		\includegraphics[width=\textwidth]{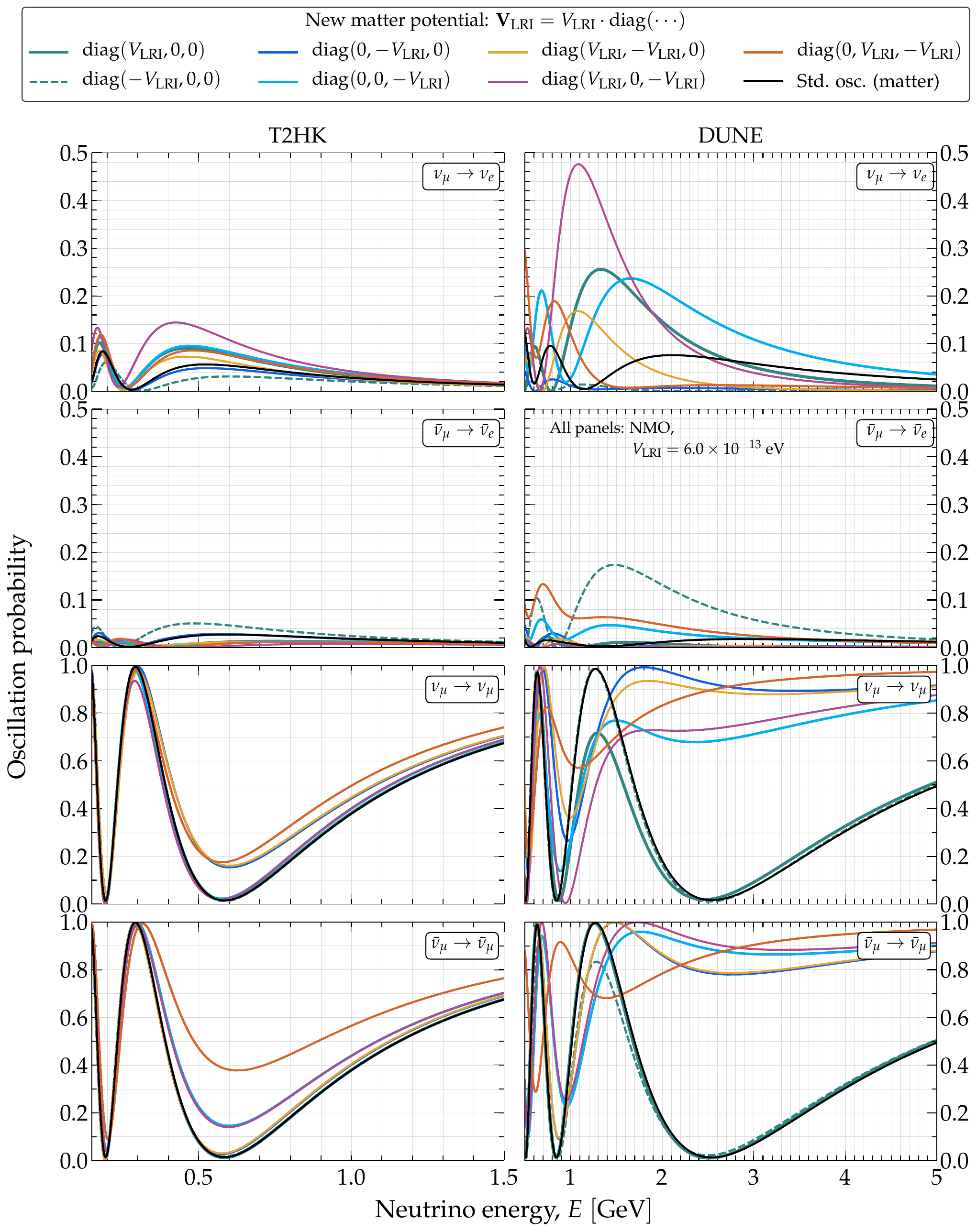}
		\caption{\textit{Neutrino oscillation probabilities in the presence of a new matter potential.}  The new matter potential is induced by each of our candidate $U(1)^\prime$ symmetries (table~\ref{tab:charges}).  In this figure, the neutrino mass ordering is normal, the values of the standard oscillation parameters are the best-fit values from table~\ref{tab:params_value1}, and we pick an illustrative value of the potential, of $V_{\rm LRI} = 6 \cdot 10^{-13}$~eV.  The probabilities are for T2HK ({\it left column}) and DUNE ({\it right column}), and for all the detection channels that we consider in our analysis: $\nu_\mu \to \nu_e$, $\bar{\nu}_\mu \to \bar{\nu}_e$, $\nu_\mu \to \nu_\mu$, and $\bar{\nu}_\mu \to \bar{\nu}_\mu$.  This figure extends the results shown in figs.~\ref{fig:dune_prob_events} and \ref{fig:t2hk_prob_events}.  See section~\ref{sec:prob_var_pot} for details and \figu{event_spectra} for corresponding results for the distribution of detected events.
			\label{fig:probability}} 
	\end{figure}
	%=================================================
	
	%=================================================
	\begin{figure}
		\centering
		\includegraphics[width=1\textwidth]{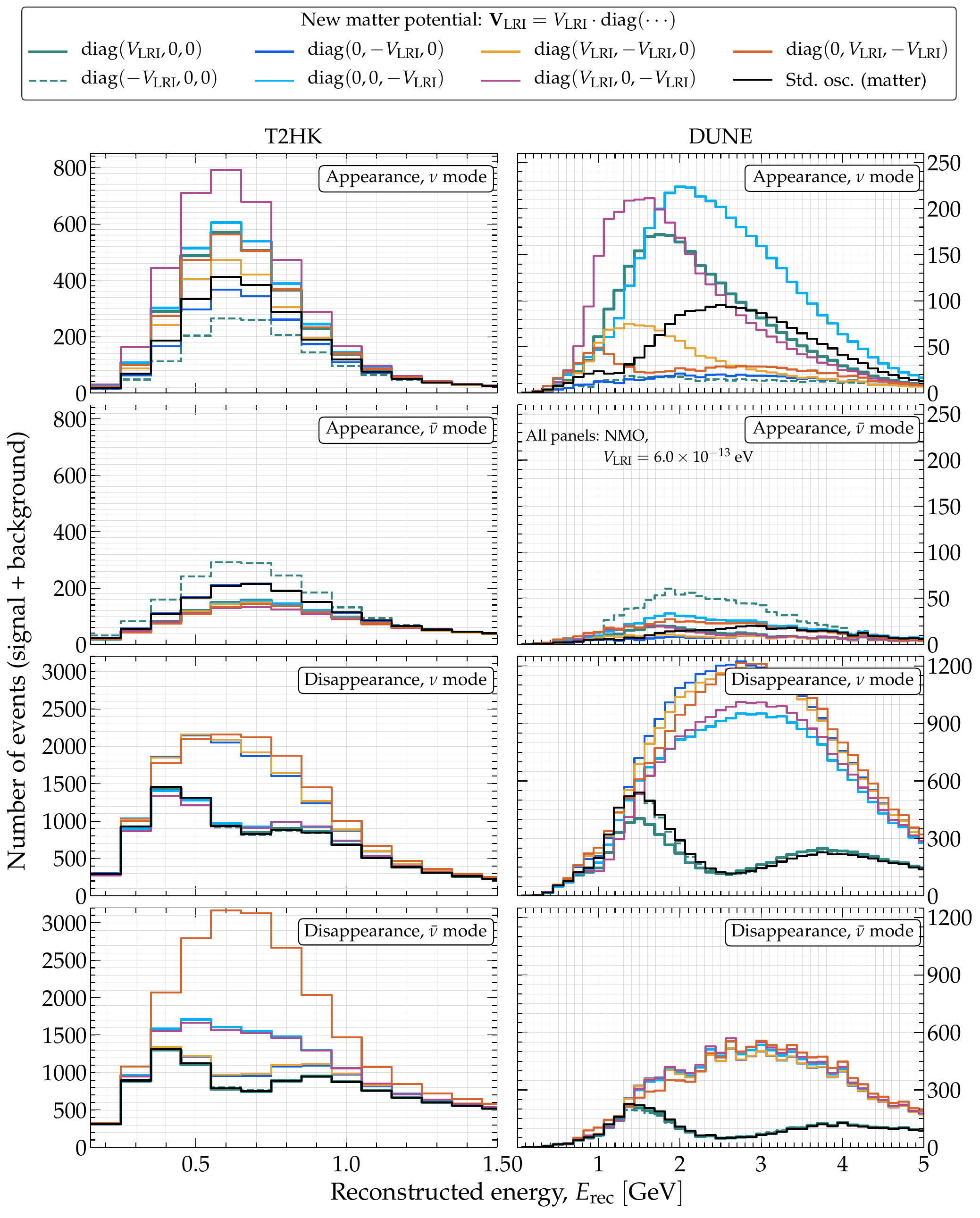}
		\caption{\textit{Spectra of detected neutrino-initiated events in the presence of a new matter potential.}  The new matter potential is induced by each of our candidate $U(1)^\prime$ symmetries (table~\ref{tab:charges}).  In this figure, the neutrino mass ordering is normal, the values of the standard oscillation parameters are the best-fit values from table~\ref{tab:params_value1}, and we pick an illustrative value of the potential, of $V_{\rm LRI} = 6 \cdot 10^{-13}$~eV.  The spectra are for T2HK ({\it left column}) and DUNE ({\it right column}), and for all the detection channels that we consider in our analysis: appearance and disappearance in neutrino and antineutrino models.  This figure extends the results shown in figs.~\ref{fig:dune_prob_events} and \ref{fig:t2hk_prob_events}.  See section~\ref{sec:expt-details} for details and \figu{probability} for corresponding results for the neutrino oscillation probabilities.
			\label{fig:event_spectra}} 
	\end{figure}
	%=================================================
	
	%=================================================
	\section{Effect of a new matter potential on oscillations and event rates}
	\label{app:prob_plots}
	%=================================================
	
	Figures~\ref{fig:probability} and~\ref{fig:event_spectra} show, respectively, the oscillation probabilities and event spectra across all the detection channels of DUNE and T2HK, for all our candidate $U(1)^\prime$ symmetries (table~\ref{tab:charges}).  They extend the illustrative case for a single choice of symmetry shown in figs.~\ref{fig:dune_prob_events} and \ref{fig:t2hk_prob_events} in the main text.  The features of the event spectra reflect the features of the oscillation probabilities.  The behavior of the latter results from the running of the oscillation parameters in the presence of the new matter potential (figs.~\ref{fig:theta_run} and~\ref{fig:delm_run}).

	%=================================================
	\section{Detailed results on constraints}
	\label{app:other}
	%=================================================
	
%	\renewcommand\thefigure{D\arabic{figure}}
%	\renewcommand\theHfigure{D\arabic{figure}}
%	\renewcommand\thetable{D\arabic{table}}
%	\renewcommand\theequation{D\arabic{equation}}
%	\setcounter{figure}{0} 
%	\setcounter{table}{0}
%	\setcounter{equation}{0}
	
	Figure~\ref{fig:lrf_bounds_NH} shows the test statistic that we use to place constraints on the new matter potential, \equ{delta_chi2_dune} for DUNE and analogous expressions for T2HK and DUNE + T2HK, for all our candidate symmetries, assuming the true neutrino mass ordering is normal.  This figure includes and extends \figu{lrf_bounds_NH_selective} in the main text, where we showed a single illustrative case.  In \figu{lrf_bounds_NH}, the symmetries are grouped according to the texture of the matter potential they induce, $\mathbf{V}_{\rm LRI}$ in table~\ref{tab:charges}, since symmetries with equal or similar potential texture yield equal or similar constraints on $V_{\rm LRI}$ (section~\ref{sec:constraints_pot}).  The results in \figu{lrf_bounds_NH} reaffirm and extend those in \figu{lrf_bounds_NH_selective}: while the constraints are driven by DUNE, it is only by combining it with T2HK that the parameter degeneracies that plague each experiment separately  --- the ``dips'' in their individual test statistics --- are lifted (section~\ref{sec:constraints_pot}).  
	
	In line with section~\ref{sec:constraints_pot}, \figu{lrf_bounds_NH} shows that the tightest limits on $V_{\rm LRI}$ are obtained for symmetries that affect primarily $\tilde\theta_{23}$, since this is the mixing angle that drives the amplitude of the oscillation probabilities, eqs.~(\ref{equ:app_mat}) and (\ref{equ:surv_mat}).  Among those symmetries, the ones that induce $\mathbf{V}_{\rm LRI}$ with texture of the form $\textrm{diag}(0,\bullet,\bullet)$ (\ie, $L_\mu-L_\tau$ and $B-L_e-2L_\tau$), and, therefore, affect primarily the muon and tau sector of the Hamiltonian, yield the best limits on $V_{\rm LRI}$ (table~\ref{tab:upper_limits_potential_NMO_organized}); see also appendix~\ref{app:param_run}.  Conversely, symmetries that induce a potential texture of the form $\textrm{diag}(\bullet,0,0)$ and that, therefore, affect predominantly the electron sector, yield the weakest limits, since they do not modify $\tilde\theta_{23}$; see appendix~\ref{app:param_run}.
	
	In \figu{lrf_bounds_NH}, the degeneracies in the test statistic are larger for symmetries whose matter potential contains negative entries; see table~\ref{tab:charges}.  These negative entries partially cancel the standard matter potential $\mathbf{V}_{\rm mat}$ (section~\ref{sec:hamiltonians}), hindering the capability of DUNE to single out the neutrino mass ordering, and resulting in the large dips in some of the test statistics seen in \figu{lrf_bounds_NH}.  Because T2HK has a shorter baseline than DUNE, it is less affected by standard matter effects, and therefore less impacted by the above issue.  This is why combining DUNE and T2HK strengthens the resulting constraints on $V_{\rm LRI}$.
	
	Figure~\ref{fig:lrf_bounds_IH} shows the same test statistic as \figu{lrf_bounds_NH}, but computed assuming that the true neutrino mass ordering is inverted.  Compared to \figu{lrf_bounds_NH}, the impact of the parameter degeneracies on the test statistic is milder (except for symmetries with a potential of the form $\mathbf{V}_{\rm LRI} = (-V_{\rm LRI}, 0, 0)$, which we explain below).  This is related to the degeneracy between $\theta_{23}$ and $\delta_{\rm CP}$, and the need to detect comparable event rates of neutrinos and antineutrinos in order to resolve it~\cite{Agarwalla:2013ju}.  Under normal mass ordering, the antineutrino rates are suppressed by the smaller interaction cross section and flux, which makes the rates of events due to neutrinos and antineutrinos uneven.  Under inverted ordering, the antineutrino rates (not shown) are significantly enhanced due to the matter effects, thereby making them comparable to those of neutrinos.  This helps to break the degeneracy between $\theta_{23}$ and $\delta_{\rm CP}$~\cite{Ballett:2016daj, Bernabeu:2018twl, Kelly:2018kmb, King:2020ydu, Singh:2021kov, Agarwalla:2022xdo}, and to remove the dips in the test statistic in \figu{lrf_bounds_IH}. However, for the texture ${\textbf V_{\rm LRI}}=(-V_{\rm LRI}, 0, 0)$, it is instead the degeneracy between $V_{\rm LRI}$ and the mass ordering that affects the test statistic, which leads its exhibiting a deeper dip under inverted mass ordering, where the determination of the mass ordering is impaired, than under normal ordering ({\it cf.}~\figu{lrf_bounds_NH} and \figu{lrf_bounds_IH}).
	
	Figure~\ref{fig:constraints_on_pot_dune_t2hk-IMO} shows the upper limits on $V_{\rm LRI}$ obtained from \figu{lrf_bounds_IH}.  Because of the above explanation, the limits on symmetries that induce the new matter potential in the electron sector, \ie, those that have $\mathbf{V}_{\rm LRI} = \textrm{diag}(\bullet,0,0)$, improve compared to assuming normal ordering, except for the case ${\textbf V_{\rm LRI}}=(-V_{\rm LRI}, 0, 0)$, {\it cf.}~\figu{constraints_on_pot_dune_t2hk-IMO} and \figu{constraints_on_pot_dune_t2hk-NMO} in the main text.
	
	Table~\ref{tab:upper_limits_potential_NMO_organized} shows the numerical values of the upper limits on $V_{\rm LRI}$ from figs.~\ref{fig:constraints_on_pot_dune_t2hk-NMO} and \ref{fig:constraints_on_pot_dune_t2hk-IMO}, for all our candidate symmetries, for normal and inverted mass ordering, and for DUNE and T2HK, separate and together. 
	
	%=================================================
	\begin{figure}
		\centering
		\includegraphics[width=1\textwidth]{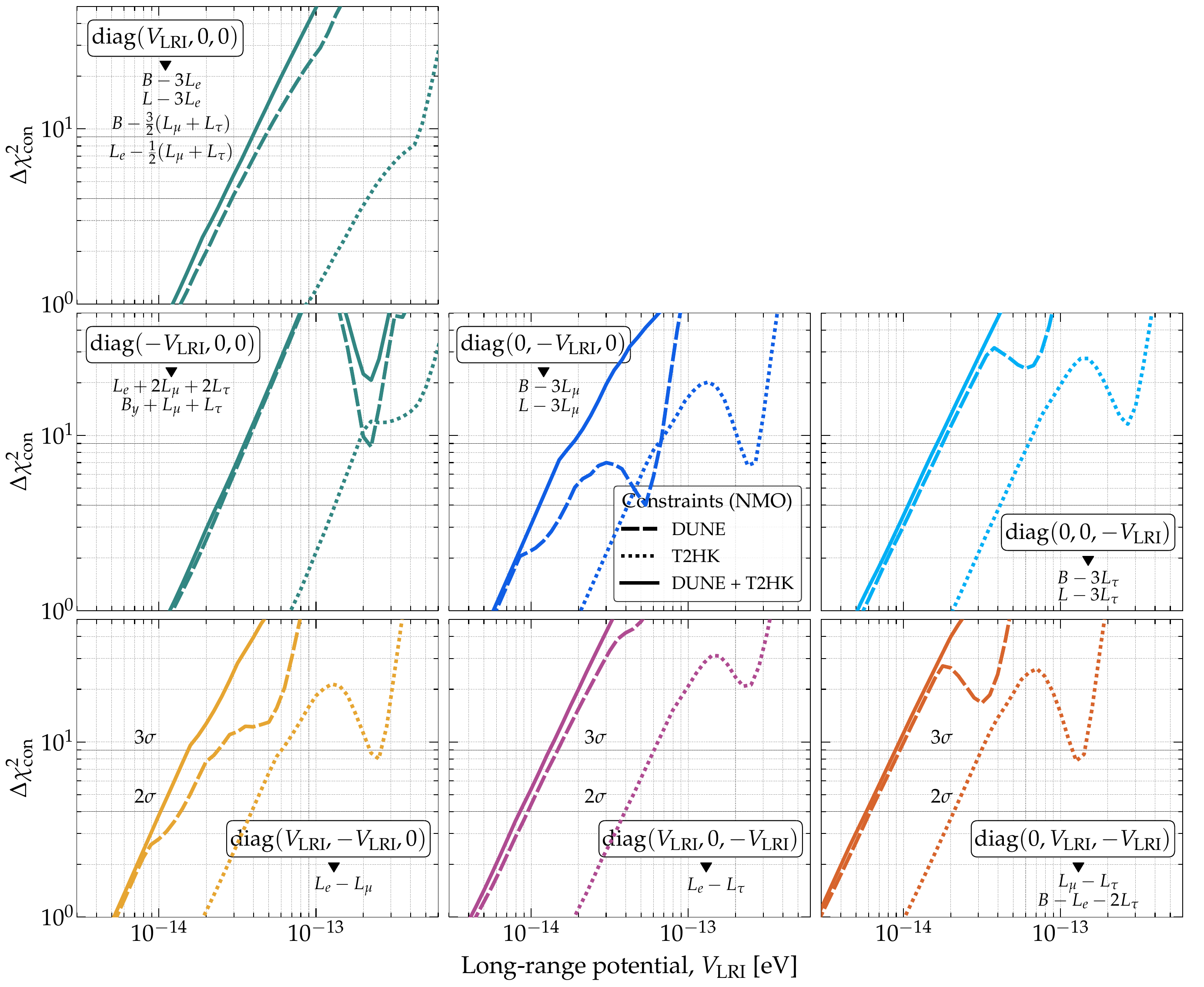}
		\caption{\textit{Projected test statistic used to constrain the new matter potential induced by our candidate $U(1)^\prime$ symmetries, assuming normal mass ordering}  This figure extends the illustrative case shown in \figu{lrf_bounds_NH_selective}.  The test statistic is \equ{delta_chi2_dune}, computed for DUNE and T2HK separately and combined.  Figure~\ref{fig:lrf_bounds_IH} shows results under inverted mass ordering.  See sections~\ref{sec:stat_methods} and \ref{sec:constraints_pot} for details.
			\label{fig:lrf_bounds_NH}} 
	\end{figure}
	%=================================================
	
	%=================================================
	\begin{figure}
		\centering
		\includegraphics[width=1\textwidth]{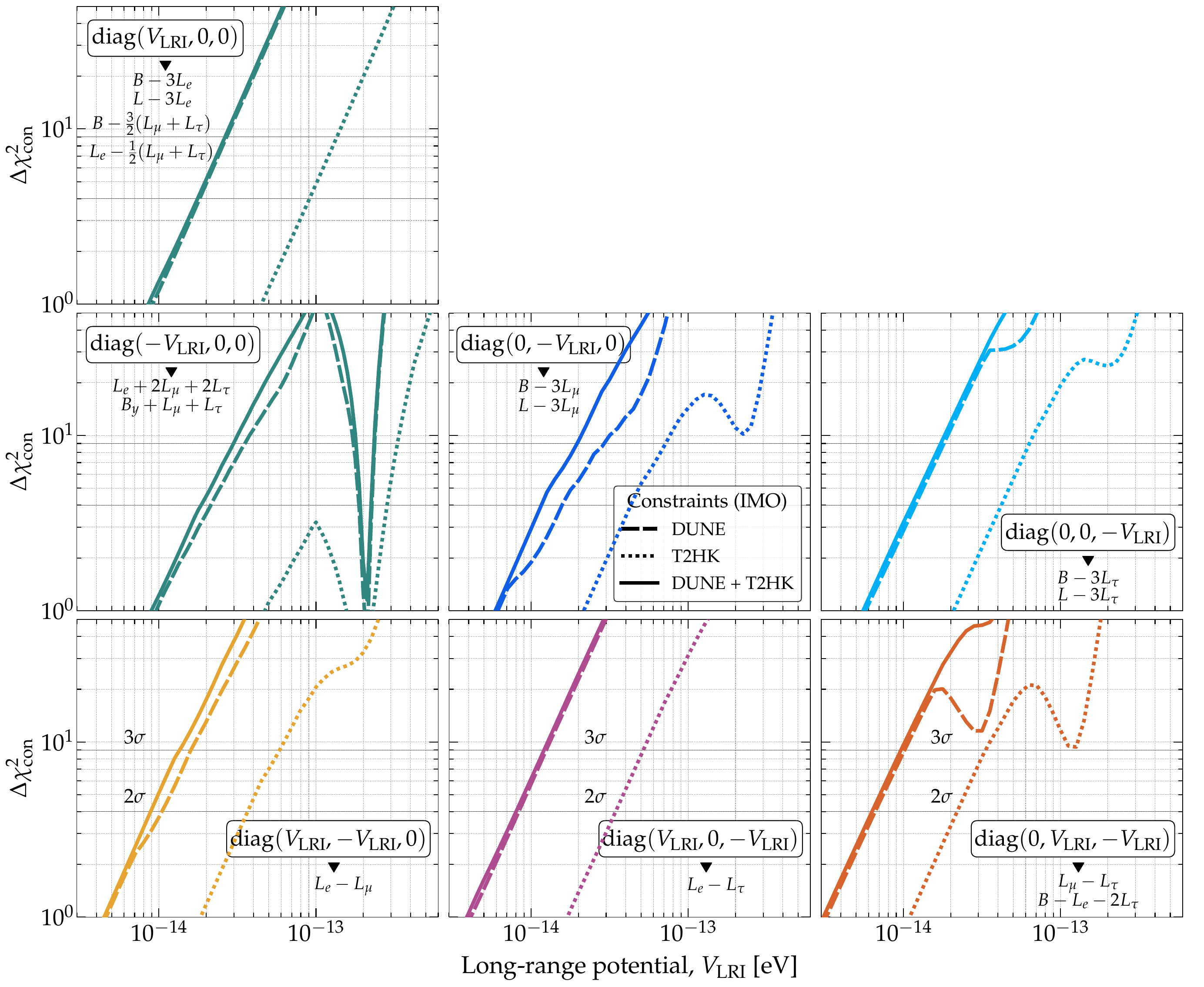}
		\caption{\textit{Projected test statistic used to constrain the new matter potential induced by our candidate $U(1)^\prime$ symmetries, assuming inverted mass ordering.}  Same as \figu{lrf_bounds_NH}, but for inverted mass ordering.  See sections~\ref{sec:stat_methods} and \ref{sec:constraints_pot} for details.
			\label{fig:lrf_bounds_IH}} 
	\end{figure}
	%=================================================
	
	%=================================================
	\begin{figure}
		\centering
		\includegraphics[width=1\textwidth]{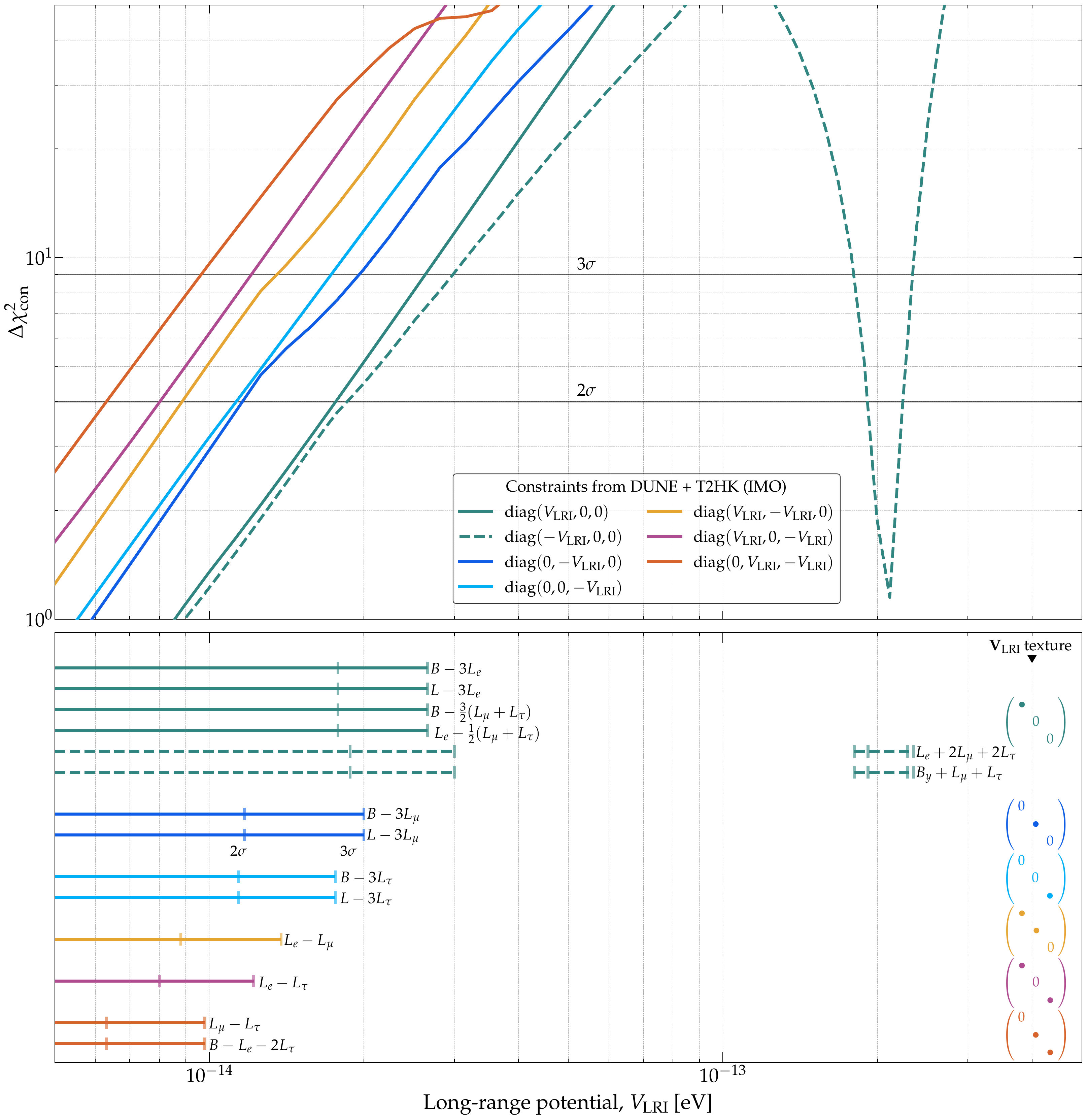}
		\caption{\textit{Projected test statistic (top) used to place upper limits (bottom) on the new matter potential induced by our candidate $U(1)^\prime$ symmetries.}  Same as \figu{constraints_on_pot_dune_t2hk-NMO}, but assuming that the true neutrino mass ordering is inverted.
			\label{fig:constraints_on_pot_dune_t2hk-IMO}} 
	\end{figure}
	%=================================================
	
	%=================================================
	\begin{table}[t!]
		\centering
		\renewcommand{\arraystretch}{1.3}
		\resizebox{\columnwidth}{!}{%
			\begin{tabular}{|c|c|c|c|c|c|c|c|c|c|c|c|c|}
				\hline
				\multirow{4}{*}{$U(1)^{\prime}$ symmetry} &
				\multicolumn{12}{c|}{Upper limit on the new matter potential, $V_{\mathrm{LRI}}$ [$10^{-14}$~eV]} 
				\\ 
				& \mc{6}{c|}{Normal mass ordering (NMO)} & \mc{6}{c|}{Inverted mass ordering (IMO)}
				\\
				& \mc{2}{c|}{DUNE} & \mc{2}{c|}{T2HK} &\mc{2}{c|}{ DUNE+T2HK} & \mc{2}{c|}{DUNE} & \mc{2}{c|}{T2HK} &\mc{2}{c|}{ DUNE+T2HK} 
				\\
				& 2$\sigma$ & 3$\sigma$ & 2$\sigma$ & 3$\sigma$ & 2$\sigma$ & 3$\sigma$ & 2$\sigma$ & 3$\sigma$ & 2$\sigma$ & 3$\sigma$ & 2$\sigma$ & 3$\sigma$
				\\ 
				\hline
				$B-3L_e$  & 3.0 & 4.80 & 21.60 & 45.0 & 2.52 & 3.96 & 1.82 & 2.66 & 9.13 & 13.51 & 1.78 & 2.66
				\\
				$L-3L_e$ & $\vert$ & $\vert$ & $\vert$ & $\vert$ & $\vert$ & $\vert$ & $\vert$ & $\vert$ & $\vert$ & $\vert$ & $\vert$ & $\vert$  
				\\
				$B-\frac{3}{2}(L_{\mu}+L_{\tau})$ & $\vert$ & $\vert$ & $\vert$ & $\vert$ & $\vert$ & $\vert$ & $\vert$ & $\vert$ & $\vert$ & $\vert$ & $\vert$ & $\vert$  
				\\
				$L_e-\frac{1}{2}(L_\mu+L_\tau)$ & $\vert$ & $\vert$ & $\vert$ & $\vert$ & $\vert$ & $\vert$ & $\vert$ & $\vert$ & $\vert$ & $\vert$ & $\vert$ & $\vert$  
				\\ 
				$L_e+2L_\mu+2L_\tau$ & 2.50 & 22.50 & 13.60 & 18.60 & 2.36 & 3.56 & 22.68 & 23.52 & 28.62 & 33.35 & 22.62 & 23.52 
				\\
				$B_y+L_\mu+L_\tau$ & $\vert$ & $\vert$ & $\vert$ & $\vert$ & $\vert$ & $\vert$ & $\vert$ & $\vert$ & $\vert$ & $\vert$ & $\vert$ & $\vert$  
				\\
				\hline
				$B-3L_\mu$ & 5.40 & 6.72 & 4.20 & 28.20 & 1.14 & 1.86 & 1.58 & 2.95 & 4.37 & 7.22 & 1.17 & 2.0 
				\\
				$L-3L_\mu$ & $\vert$ & $\vert$ & $\vert$ & $\vert$ & $\vert$ & $\vert$ & $\vert$ & $\vert$ & $\vert$ & $\vert$ & $\vert$ & $\vert$  
				\\ 
				\hline
				$B-3L_\tau$ & 1.20 & 1.80 & 4.20 & 6.36 & 1.08 & 1.68 & 1.21 & 1.82 & 4.28 & 6.31 & 1.14 & 1.76 
				\\
				$L-3L_\tau$ & $\vert$ & $\vert$ & $\vert$ & $\vert$ & $\vert$ & $\vert$ & $\vert$ & $\vert$ & $\vert$ & $\vert$ & $\vert$ & $\vert$  
				\\ 
				\hline
				$L_e-L_\mu$ &  1.40 & 2.50 & 4.10 & 25.70 & 1.03 & 1.56 & 1.05 & 1.61 & 3.74 & 6.06 & 0.88 & 1.38
				\\
				\hline
				$L_e-L_\tau$ & 0.98 & 1.50 & 4 & 6.10 & 0.84 & 1.32 & 0.84 & 1.26 & 3.49 & 5.21 & 0.8 & 1.22 
				\\
				\hline
				$L_\mu-L_\tau$ & 0.62 & 0.95 & 2.12 & 14.30 & 0.58 & 0.9 & 0.66 & 1.03 & 2.25 & 3.41 & 0.63 & 0.98 
				\\
				$B-L_{e}-2L_{\tau}$ & $\vert$ & $\vert$ & $\vert$ & $\vert$ & $\vert$ & $\vert$ & $\vert$ & $\vert$ & $\vert$ & $\vert$ & $\vert$ & $\vert$
				\\
				\hline
			\end{tabular}
		}
		\caption{\textit{Projected upper limits on the new matter potential, $V_{\rm LRI}$, induced by our candidate $U(1)^\prime$ symmetries.}  As in table~\ref{tab:charges}, the symmetries are grouped according to the texture of the matter potential, $\mathbf{V}_{\rm LRI}$, that they induce.  Symmetries with equal or similar potential texture yield equal or similar upper limits.  See figs.~\ref{fig:constraints_on_pot_dune_t2hk-NMO} and \ref{fig:constraints_on_pot_dune_t2hk-IMO} for a graphical representation of the limits in this table, and sections~\ref{sec:stat_methods} and \ref{sec:constraints_pot} for details.  The contents of this table are available in \Refe~\cite{GitHub_lrf-repo}.}
		\label{tab:upper_limits_potential_NMO_organized}
	\end{table}
	%=======f==========================================
	
	%=================================================
	\section{Detailed results on discovery prospects}
	\label{app:discovery}
	%=================================================
	
%	\renewcommand\thefigure{E\arabic{figure}}
%	\renewcommand\theHfigure{E\arabic{figure}}
%	\renewcommand\thetable{E\arabic{table}}
%	\renewcommand\theequation{E\arabic{equation}}
%	\setcounter{figure}{0} 
%	\setcounter{table}{0}
%	\setcounter{equation}{0}
	
	Figure~\ref{fig:discovery_b-3ltau} shows the test statistic, \equ{delta_chi2_disc}, that we use to forecast discovery prospects of $V_{\rm LRI}$, assuming, for illustration, a matter potential with the texture $\mathbf{V}_{\rm LRI} = {\rm diag}(0,0, -V_{\rm LRI})$, as in figs.~\ref{fig:dune_prob_events}, \ref{fig:t2hk_prob_events}, and \ref{fig:lrf_bounds_NH_selective}.  Unlike in the test statistic that we use to place constraints (figs.~\ref{fig:lrf_bounds_NH_selective}, \ref{fig:lrf_bounds_NH}, and \ref{fig:lrf_bounds_IH}), there are no large degeneracies between $V_{\rm LRI}$ and the standard oscillation parameters since, when computing the test statistic, we fix the test value of $V_{\rm LRI}$ to zero while varying the standard oscillation parameters.
	
	Table~\ref{tab:discovery_strength} gives the numerical values of $V_{\rm LRI}$ that lead to discovery for all our candidate symmetries, as shown in \figu{discovery_3s_5s_dune_t2hk} in the main text.  
	
	%=================================================
	\begin{figure}
		\centering
		\includegraphics[width=0.85\textwidth]{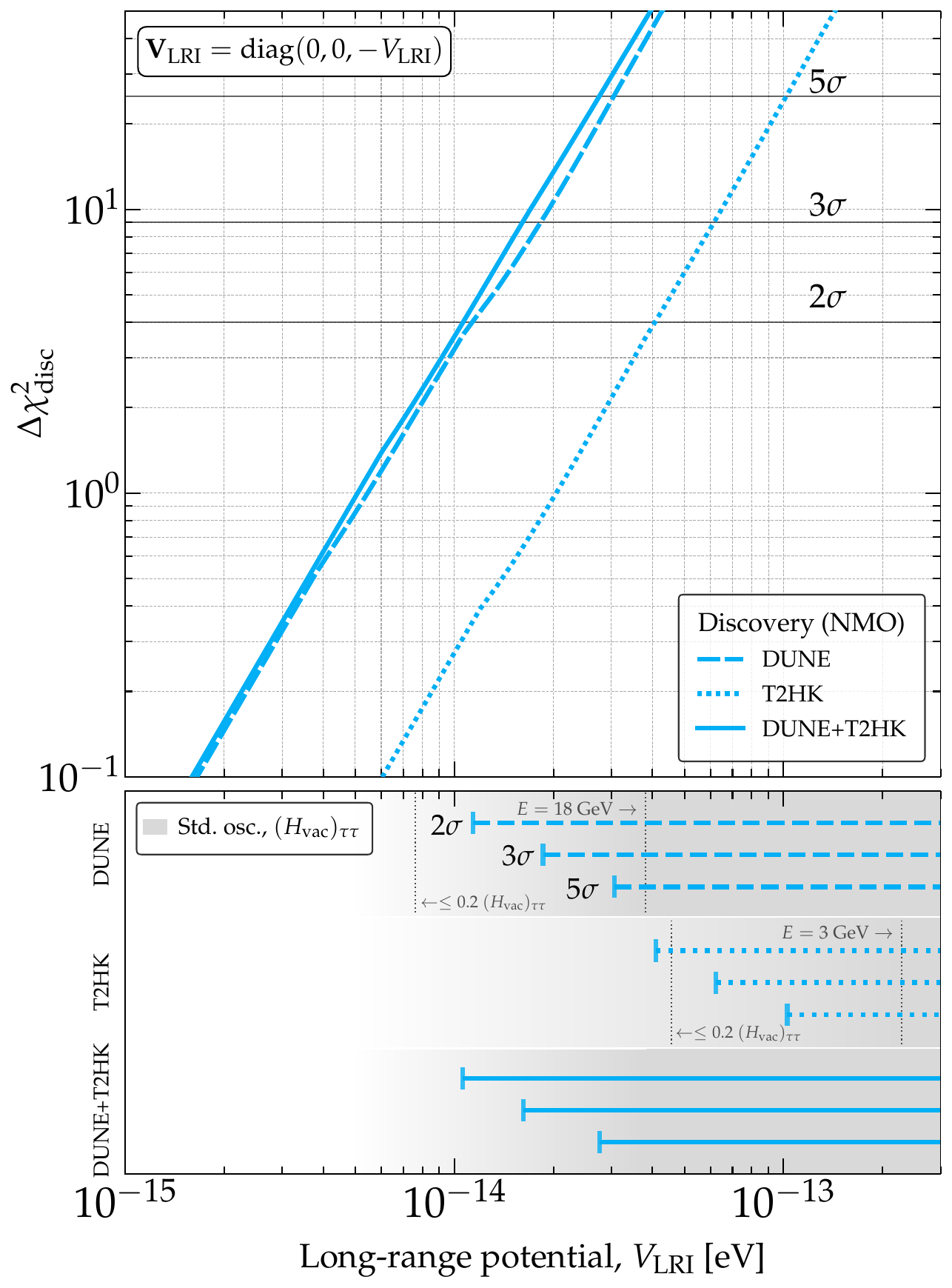}
		\caption{\textit{Projected test statistic used to compute discovery prospects on the new matter potential induced by a $U(1)^\prime$ symmetry.} For this plot, as illustration, we show limits on a potential of the form $\mathbf{V}_{\rm LRI} = \textrm{diag}(0, 0, -V_{\rm LRI})$ for neutrinos and $-\mathbf{V}_{\rm LRI}$ for antineutrinos, as would be introduced by symmetries $L - 3L_\tau$ or $B - 3L_\tau$ (table~\ref{tab:charges}).  The test statistic is \equ{delta_chi2_disc}.  Results are for DUNE and T2HK separately and combined.  The true neutrino mass ordering is assumed to be normal.  See sections~\ref{sec:stat_methods} and \ref{sec:discovery} for details.  Like when placing constraints (\figu{lrf_bounds_NH_selective}), the experiments are sensitive to values of $V_{\rm LRI}$ that are comparable to the standard-oscillation terms in the Hamiltonian; for the choice of $\mathbf{V}_{\rm LRI}$ texture in this figure, this is $(\mathbf{H}_{\rm vac})_{\tau\tau}$.  Figure~\ref{fig:discovery_3s_5s_dune_t2hk} shows the discovery prospects for all of our candidate symmetries.} 
		\label{fig:discovery_b-3ltau}
	\end{figure}
	%=================================================
	
	%=================================================
	\begin{table}
		\centering
		\renewcommand{\arraystretch}{1.3}
		\begin{tabular}{|c|c|c|c|c|c|c|}
			\hline
			\multirow{4}{*}{$U(1)^{\prime}$ symmetry} & 
			\multicolumn{6}{c|}{\makecell{Discovery strength of LRI potential}} 
			\\
			& \mc{6}{c|}{[$10^{-14}$~eV], NMO}
			\\
			\cline{2-7}
			& \mc{2}{c|}{DUNE} & \mc{2}{c|}{T2HK} & \mc{2}{c|}{DUNE+T2HK}
			\\
			& $3\sigma$ & $5\sigma$ & $3\sigma$ & $5\sigma$ & $3\sigma$ & $5\sigma$
			\\
			\hline
			$B-3L_e$ &  3.75 & 6.30 & 18.30 & 29.10 & 3.60 & 6.0 
			\\
			$L-3L_e$ & $\vert$ & $\vert$ & $\vert$ & $\vert$ & $\vert$ & $\vert$ 
			\\
			$B-\frac{3}{2}(L_{\mu}+L_{\tau})$ & $\vert$ & $\vert$ & $\vert$ & $\vert$ & $\vert$ & $\vert$ 
			\\ 
			$L_e-\frac{1}{2}(L_\mu+L_\tau)$ & $\vert$ & $\vert$ & $\vert$ & $\vert$ & $\vert$ & $\vert$ 
			\\
			$L_e+2L_\mu+2L_\tau$ & 22.40 & 29.0 & 50.0 & 62.80 & 4.0 & 7.0 
			\\
			$B_y+L_\mu+L_\tau$ & $\vert$ & $\vert$ & $\vert$ & $\vert$ & $\vert$ & $\vert$ 
			\\ 
			\hline
			$B-3L_\mu$ & 2.16 & 3.72 & 6.24 & 10.50 & 1.68 & 3.0  
			\\
			$L-3L_\mu$ & $\vert$ & $\vert$ & $\vert$ & $\vert$ & $\vert$ & $\vert$ 
			\\
			\hline
			$B-3L_\tau$ & 1.86 & 3.06 & 6.24 & 10.26 & 1.62 & 2.76 
			\\
			$L-3L_\tau$ & $\vert$ & $\vert$ & $\vert$ & $\vert$ & $\vert$ & $\vert$ 
			\\ 
			\hline
			$L_e-L_\mu$ & 1.90 & 3.20 & 5.90 & 9.90 & 1.50 & 2.65  
			\\
			\hline
			$L_e-L_\tau$ & 1.40 & 2.30 & 5.60 & 9.0  & 1.28 & 2.12  
			\\
			\hline
			$L_\mu-L_\tau$ & 1.03 & 1.70 & 3.20 & 5.40 & 0.91 & 1.50 
			\\
			$B-L_{e}-2L_{\tau}$ & $\vert$ & $\vert$ & $\vert$ & $\vert$ & $\vert$ & $\vert$ 
			\\  
			\hline
		\end{tabular}
		\caption{\textit{Projected discovery prospects of the new matter potential induced by our candidate $U(1)^\prime$ symmetries.}  As in table~\ref{tab:charges}, the symmetries are grouped according to the texture of the matter potential, $\mathbf{V}_{\rm LRI}$, that they induce.  Symmetries with equal or similar potential texture yield equal or similar discovery prospects.  See \figu{discovery_3s_5s_dune_t2hk} for a graphical representation of the values in this table, and sections~\ref{sec:stat_methods} and \ref{sec:discovery} for details.  The contents of this table are available in \Refe~\cite{GitHub_lrf-repo}.}
		\label{tab:discovery_strength}
	\end{table}

%%%%%%%%%%%%%%%%%%%%%%%%%%%%%%%%%%%%%%%%%%%%%%%
\section{A note on constraints on long-range interaction other than ours}
\label{sec:other-constraints}
%%%%%%%%%%%%%%%%%%%%%%%%%%%%%%%%%%%%%%%%%%%%%%%
In this section, we discuss in brief the constraints on long-range neutrino interaction, placed by the previous studies.
\begin{description}
	\item [Early universe constraints:] 
	Reference~\cite{Dror:2020fbh} discusses that light gauge bosons ($X$) lead to additional neutrino decay channels $(\nu_i\to\nu_j+X)$, which can impact the evolution of the early Universe. If neutrinos decay too quickly, they disrupt them from free-streaming, altering the Cosmic Microwave Background. This limits the neutrino decay via light gauge boson and, ultimately, the mass and coupling of the new gauge boson.
	
	\item [Constraints from compact binary stars:] 
	Reference~\cite{KumarPoddar:2019ceq} explores the possibility of decay in the orbital period due to radiation of ultra-light gauge boson from the binary systems of Neutron Star-Neutron Star (NS-NS) or Neutron Star-White Drarf (NS-WD) and derive constraints on the new vector boson coupling.    
	
	\item [Constraints from atmospheric, solar, and reactor data:] 
	The studies in \Refes~\cite{Joshipura:2003jh, Bandyopadhyay:2006uh} placed stringent new constraints on long-range forces that are stronger than previous fifth-force experiments, showing for the first time that neutrino oscillations are a sensitive probe of exotic physics beyond the Standard Model. Reference~\cite{Joshipura:2003jh} used the atmospheric data from Super-K to place $90\%$ C.L. upper limits on the gauge boson coupling associated with the gauged $L_e-L_\mu$ and $L_e-L_\tau$: $g_{{Z^\prime}_{L_e-L_\mu}}<8.32 \times 10^{-26}$ and $g_{{Z^\prime}_{L_e-L_\tau}}<8.97 \times 10^{-26}$. Reference~\cite{Bandyopadhyay:2006uh} used the solar and reactor data from KamLAND to put $3\sigma$ upper limits: $g_{{Z^\prime}_{L_e-L_\mu}}<2.06 \times 10^{-26}$ and $g_{{Z^\prime}_{L_e-L_\tau}}<1.77 \times 10^{-26}$, when $\theta_{13}=0$.
	
	\item [Constraints from Black-hole Superradiance:] 
	In \Refe~\cite{Baryakhtar:2017ngi}, the authors used Black-hole superradiance -- a process where a rotating Black-hole (BH) can lose energy and angular momentum -- to constrain the ultra-light vector bosons. If an ultralight vector boson exists, it can form a gravitationally bound cloud around a rapidly rotating BH. The bosonic cloud extracts angular momentum from the BH, causing its spin to decrease. By the measurement of the spin of the rotating BH through X-ray observations, this paper rules out three separate mass windows as shown in Figs.~\ref{fig:constraints_discovery_dune_t2hk-NMO}, \ref{fig:all_symmetry}, and \ref{fig:discovery_all_DUNE+T2HK}.
	
	\item [Constraints from Weak gravity conjucture:] 
	Reference~\cite{Arkani-Hamed:2006emk} assumes gravity to be the weakest force and argues that in any consistent quantum gravity theory with gravity and $U(1)^\prime$ gauge symmetry, the gauge force must be stronger than gravity for at least one charged particle. This leads to a lower bound on the gauge coupling, $g_{Z^\prime}$, in any $U(1)^\prime$ gauge theory.
	
	\item [Constraints from astrophysical neutrinos:] 
	The new interaction induced by the $U(1)^\prime$ gauge symmetries may also modify the flavor ratio of the astrophysical neutrinos. References~\cite{Bustamante:2018mzu} gives the estimated limits on the long-range interaction form the neutrino flavor ratio measurement at IceCube which has been revited in \Refe~\cite{Agarwalla:2023sng} with more robust statistical method.
	
	\item [Constraints from global $U(1)^\prime$ study:] Reference~\cite{Coloma:2020gfv} place constraints on the mass and coupling of the new gauge boson associated with the gauged $U(1)^\prime$ symmetry with the global neutrino oscillation data.
\end{description}
	%=================================================
	
	\chapter{Appendix to Chapter~\ref{C4}}

	\renewcommand\thefigure{B\arabic{figure}}
	\renewcommand\theHfigure{B\arabic{figure}}
	\renewcommand\theequation{B\arabic{equation}}
	\setcounter{figure}{0} 
	\setcounter{equation}{0}
	%=============================%
	\section{Comparison between numerical and analytical probabilities}
	\label{appndx}
	%=============================%
	
	\begin{figure}[h!]
		\centering
		\includegraphics[width=0.86\textwidth]{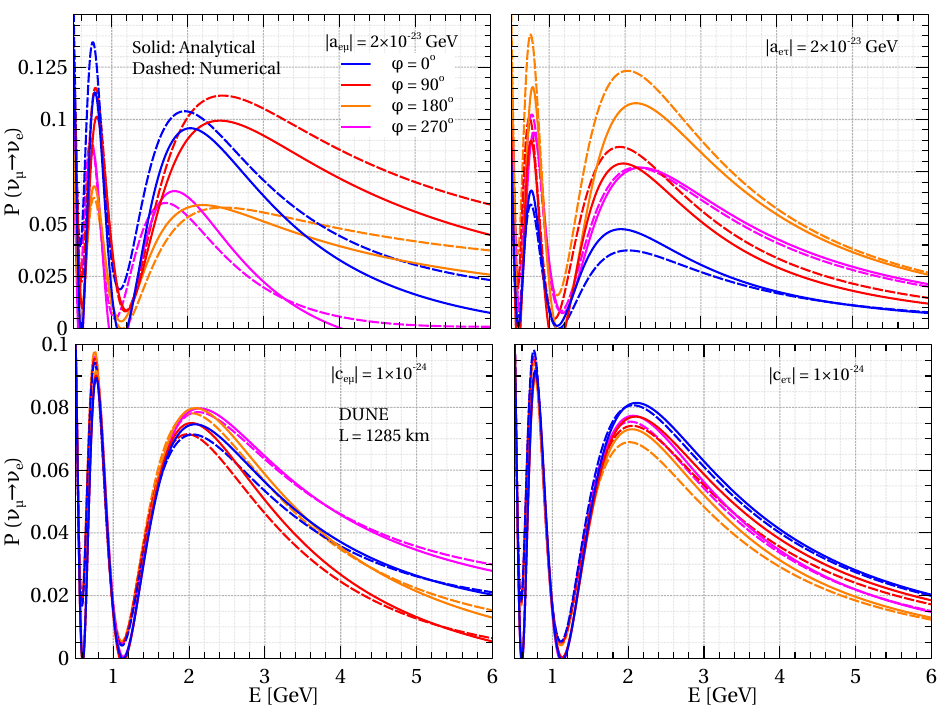}
		\mycaption{Comparison between exact $\nu_{\mu}\to\nu_e$ appearance probability calculated numerically (dashed lines) with the same calculated analytically (solid lines) using \equ{pme_liv} for a baseline $L=1285$ km. The top (bottom) row corresponds to the probability in the presence of CPT-violating (CPT-conserving) LIV parameters $a_{e\beta}$ ($c_{e\beta}$) with $\beta=\mu $ in left column, and $\beta=\tau$ in right column. Four colored curves correspond to four values of the associated phase, as mentioned in the legend. The values of the standard oscillation parameters used in the plot are given in table~\ref{tab:params_value} with NMO.\label{fig:Pme-comparison}}
	\end{figure}
	
	In this section, we check the accuracy of the approximate analytical expressions of the oscillation probabilities derived in section~\ref{sec:LIV}. In \figu{Pme-comparison}, we compare the $\nu_\mu\rightarrow\nu_e$ appearance probability in the presence of $a_{e\mu}/c_{e\mu}$ and $a_{e\tau}/c_{e\tau}$, calculated using the analytical expression with the exact oscillation probability calculated numerically using GLoBES software, for DUNE ($L = 1285$ km). We consider only these LIV parameters since they appear in the first order in the expansion parameters. We show the results for four values of the phase associated with the LIV parameters, namely, $0^\circ$, $90^\circ$, $180^\circ$, and $270^\circ$. We observe that for the CPT-violating LIV parameters, matching with the exact appearance probability is slightly worse compared to the CPT-conserving case, as the assumed strength of LIV parameters in the CPT-violating case is one order larger than the corresponding CPT-conserving case.
	
	However, the features of the appearance probabilities for different values of LIV-phases are preserved by the analytical expression. In \figu{Pmm-comparison}, we show the same for the $\nu_\mu\to\nu_\mu$ disappearance probability. 
	
		\begin{figure}[t]
		\centering
		\includegraphics[width=0.86\textwidth]{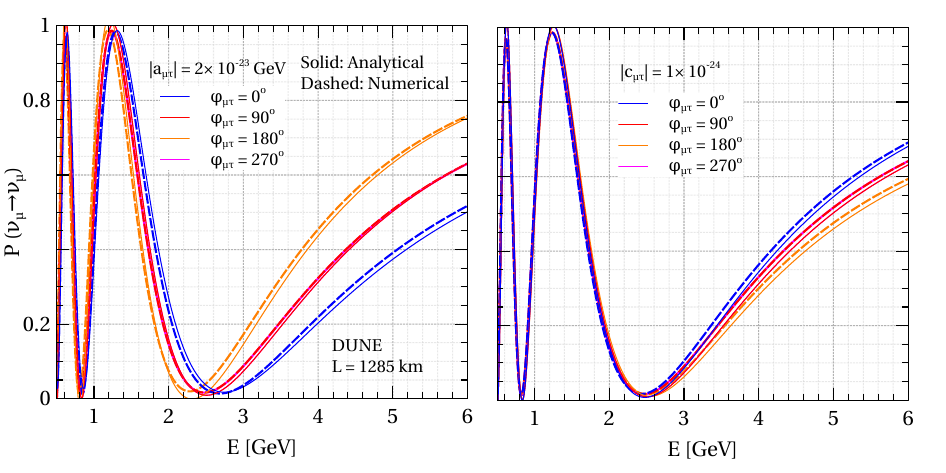}
		\mycaption{Comparison between exact $\nu_\mu\to\nu_{\mu}$ disappearance probability calculated numerically (dashed lines) with the same calculated analytically (solid lines) using \equ{pmm} for a baseline $L=1285$ km. The Left (right) column corresponds to the probability in the presence of CPT-violating (CPT-conserving) LIV parameter $a_{\mu\tau}$ ($c_{\mu\tau}$). Four colored curves correspond to four values of the associated phase, as mentioned in the legend. The values of the standard oscillation parameters used in the plot are given in table~\ref{tab:params_value} with NMO.\label{fig:Pmm-comparison}}
	\end{figure}
    \noindent
    Here, we show the effect from the LIV parameters, $a_{\mu\tau}/c_{\mu\tau}$, since only these terms appear till the first order. 
	Again, we consider four values of associated phases as in \figu{Pme-comparison}. We find the oscillation probabilities calculated analytically match quite well with the numerical ones. It is valid for all assumed values of the associated phases, both in CPT-violating and CPT-conserving scenarios. The same is true for T2HK.
	\newpage
	%==============================================
%	\renewcommand\thefigure{B\arabic{figure}}
%	\renewcommand\theHfigure{B\arabic{figure}}
%	\renewcommand\theequation{B\arabic{equation}}
%	\setcounter{figure}{0} 
%	\setcounter{equation}{0}
	%==============================================
	\section{$\nu_{e}$ appearance and $\nu_{\mu}$ disappearance event spectra in DUNE and T2HK in presence of the CPT-violating LIV parameters}
	\label{appndx-B}
	%==============================================
	
	\begin{figure}[h!]
		\centering
		\includegraphics[width=\textwidth]{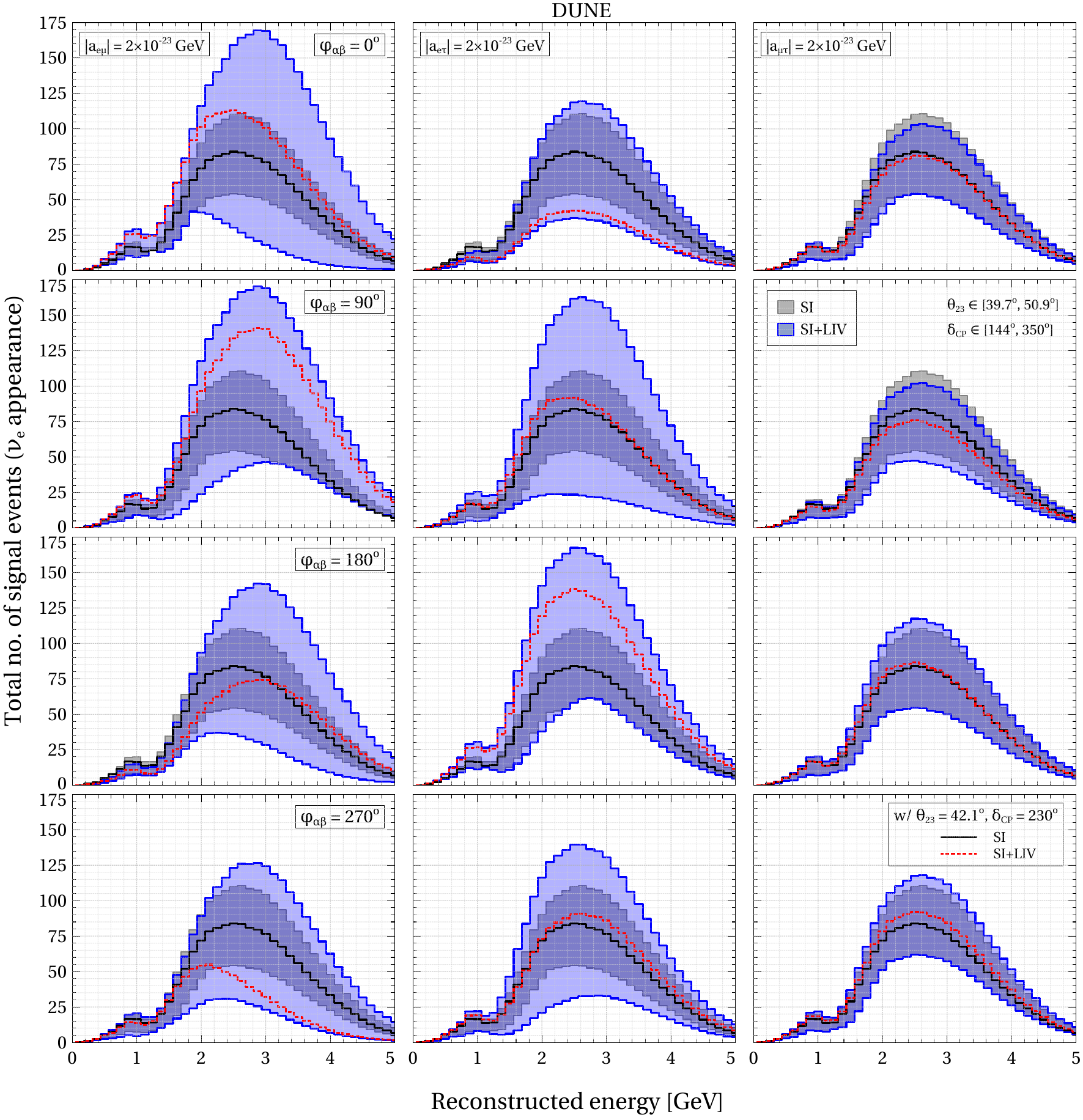}
		\vspace*{-5mm}
		\mycaption{The distribution band of reconstructed $\nu_e$ events (signal) obtained via $\nu_\mu \to \nu_e$ channel for both SI and SI+LIV scenarios at DUNE. The bands in each panel appear due to the uncertainties in the standard oscillation parameters, $\theta_{23}$ and $\delta_{\rm CP}$ in their $3\sigma$ range. The gray bands correspond to SI scenarios, whereas the blue ones represent CPT-violating LIV scenarios. The solid black and the dashed red lines in each panel represent the event spectra in case of SI and SI+LIV, respectively, with the oscillation parameters fixed at their benchmark values given in table~\ref{tab:params_value}. The strength of these LIV parameters: $|a_{e\mu}|$ (first column), $|a_{e\tau}|$ (second column), and $|a_{\mu\tau}|$ (third column) are considered to be $2\times 10^{-23}$ GeV for four choices of the corresponding LIV phases, namely, $\phi_{\alpha\beta}$ = $0^\circ$ (first row), $90^\circ$ (second row), $180^\circ$ (third row), and $270^\circ$ (fourth row).\label{fig:event_rate_nu_app_dune}}
	\end{figure}
	
	\begin{figure}[h!]
		\centering
		\includegraphics[width=\textwidth]{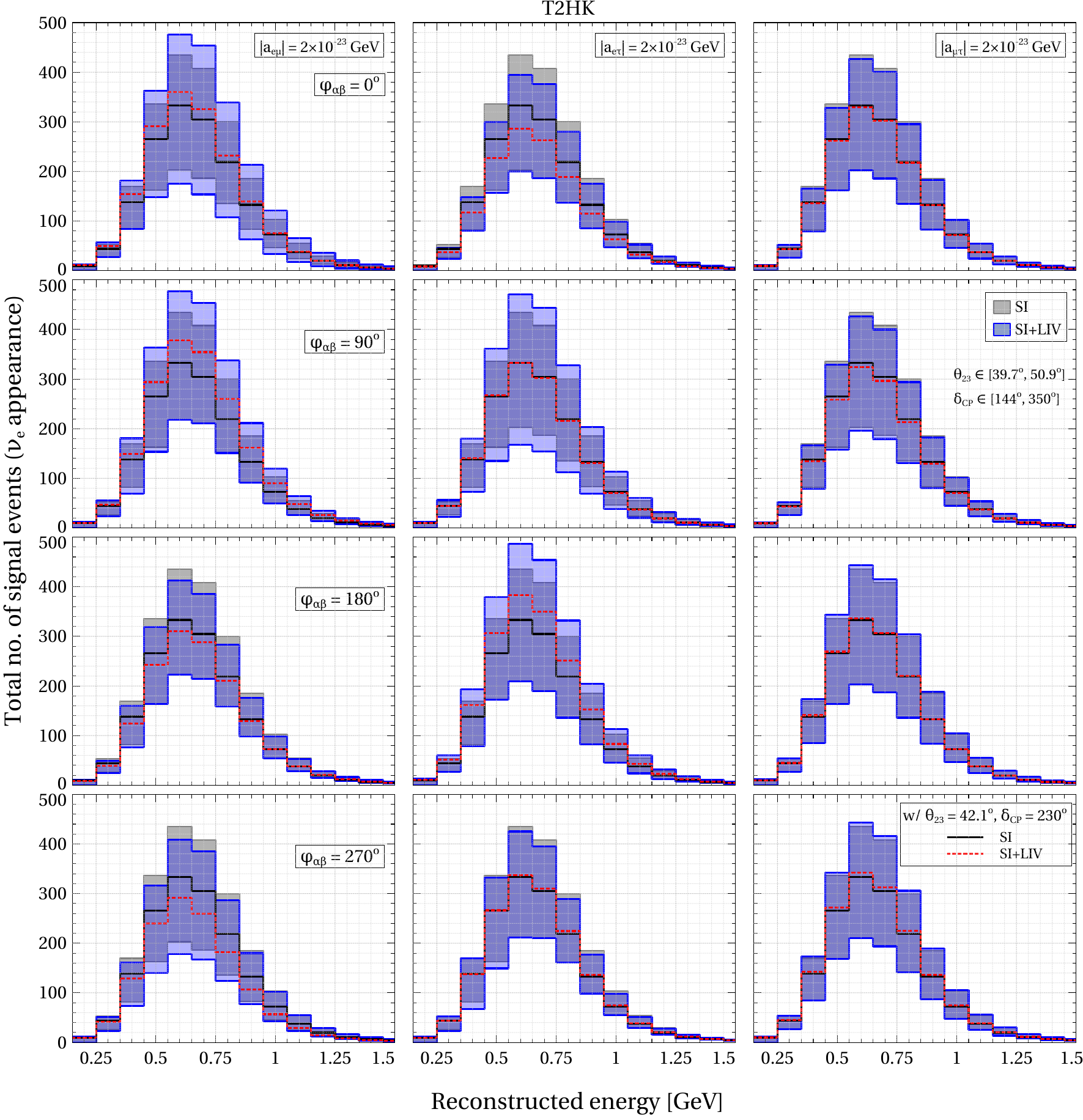}
		\vspace*{-5mm}
		\mycaption{The distribution band of reconstructed $\nu_e$ events (signal) obtained via $\nu_\mu \to \nu_e$ channel for both SI and SI+LIV scenarios at T2HK. The bands in each panel appear due to the uncertainties in the standard oscillation parameters, $\theta_{23}$ and $\delta_{\rm CP}$ in their $3\sigma$ range. The gray bands correspond to SI scenarios, whereas the blue ones represent CPT-violating LIV scenarios. The solid black and the dashed red lines in each panel represent the event spectra in case of SI and SI+LIV, respectively, with the oscillation parameters fixed at their benchmark values given in table~\ref{tab:params_value}. The strength of these LIV parameters: $|a_{e\mu}|$ (first column), $|a_{e\tau}|$ (second column), and $|a_{\mu\tau}|$ (third column) are considered to be $2\times 10^{-23}$ GeV for four choices of the corresponding LIV phases, namely, $\phi_{\alpha\beta}$ = $0^\circ$ (first row), $90^\circ$ (second row), $180^\circ$ (third row), and $270^\circ$ (fourth row). \label{fig:event_rate_nu_app_hk}}	
	\end{figure}
	
	\begin{figure}[h!]
		\centering
		\includegraphics[width=0.9\textwidth]{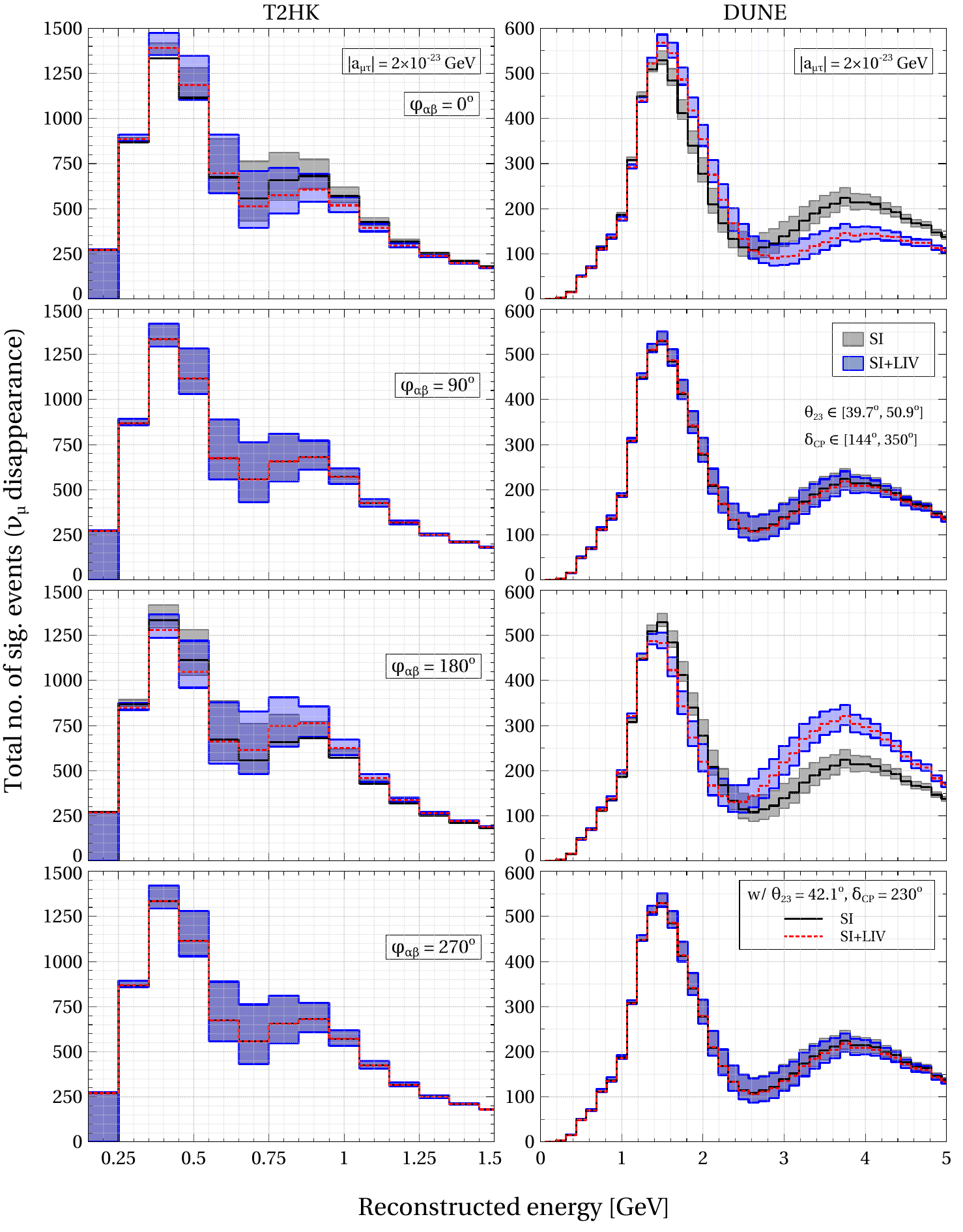}
		\vspace*{-2.5mm}
		\mycaption{The distribution band of reconstructed $\nu_\mu$ events (signal) obtained via $\nu_\mu \to \nu_\mu$ channel for both SI and SI+LIV scenarios at T2HK (left column) and DUNE (right column). The bands in each panel appear due to the uncertainties in the standard oscillation parameters, $\theta_{23}$ and $\delta_{\rm CP}$ in their $3\sigma$ range. The gray bands correspond to SI scenarios, whereas the blue ones represent CPT-violating LIV scenarios. The solid black and the dashed red lines in each panel represent the event spectra in case of SI and SI+LIV, respectively, with the oscillation parameters fixed at their benchmark values given in table~\ref{tab:params_value}. Here, we consider only the effect of non-zero $|a_{\mu\tau}|$ whose value is considered to be $2\times 10^{-23}$ GeV for four choices of the corresponding LIV phase, namely, $\phi_{\mu\tau}$ = $0^\circ$ (first row), $90^\circ$ (second row), $180^\circ$ (third row), and $270^\circ$ (fourth row).\label{fig:event_rate_nu_disapp_hk_dune}}	
	\end{figure}
	
	In figs.~\ref{fig:event_rate_nu_app_dune} and \ref{fig:event_rate_nu_app_hk}, we show the $\nu_e$ appearance signal event spectra in DUNE and T2HK, respectively, in presence of the CPT-violating LIV parameters, $a_{e\mu}$, $a_{e\tau}$, and $a_{\mu\tau}$. Since $a_{\mu\tau}$ mainly affects the $\nu_\mu\to\nu_{\mu}$ disappearance channel, we show the disappearance event spectra in \figu{event_rate_nu_disapp_hk_dune}, for T2HK and DUNE for non-zero $a_{\mu\tau}$. The gray band in each panel shows the neutrino signal rate considering $3\sigma$ uncertainties in the present best-fit values of the standard oscillation parameters~\cite{Esteban:2020cvm}, $\theta_{23}$ and $\delta_{\rm CP}$ in the SI case. Blue bands show the same in the presence of the off-diagonal CPT-violating LIV parameters one-at-a-time for four choices of the associated phase, namely, $0^\circ$, $90^\circ$, $180^\circ$, and $270^\circ$. The benchmark values of the LIV parameters are mentioned in the top panels. Features in the signal rate bands reflect the features in the oscillation probabilities shown in figs.~\ref{fig:app_prob} and \ref{fig:disapp_prob}. 
	
	In \figu{event_rate_nu_app_dune}, we observe that in the presence of $|a_{e\mu}|$ and $|a_{e\tau}|$,  the $\nu_e$ appearance signal rates exceed the SI band with present $3\sigma$ uncertainties in the oscillation parameters. This is valid for all four values of the associated phases. For $|a_{\mu\tau}|$, at the mentioned strength, deviation is small, as this parameter appears in the higher order terms in the appearance probability expression. For T2HK (see \figu{event_rate_nu_app_hk}), all deviations from expected signal rates in the SI case are comparatively smaller for all choices of the associated phase. The deviation is almost negligible for $|a_{\mu\tau}|$. It is expected from the fact that T2HK operates at lower energy and baseline as compared to DUNE. 
	
	Similarly, from \figu{event_rate_nu_disapp_hk_dune}, it is evident that the deviation between the $\nu_\mu$ disappearance events rates, in SI case and in the presence of $a_{\mu\tau}$, is significant for DUNE. 
	Hence, it is clear from these three figures that DUNE will have comparatively better potential to probe off-diagonal LIV parameters due to its access to higher neutrino energy and longer baseline.
	
	%==============================================
	\chapter{Appendix to Chapter~\ref{C5}}

	\renewcommand\thefigure{C\arabic{figure}}
	\renewcommand\theequation{C\arabic{equation}}
	\setcounter{figure}{0} 
	\setcounter{equation}{0}
	
	%===============================================================================
	\section{Validating the approximate probability expressions}
	\label{validation_prob}
	%===============================================================================
	
	\begin{figure}[h!]
		\centering
		\includegraphics[width=15cm]{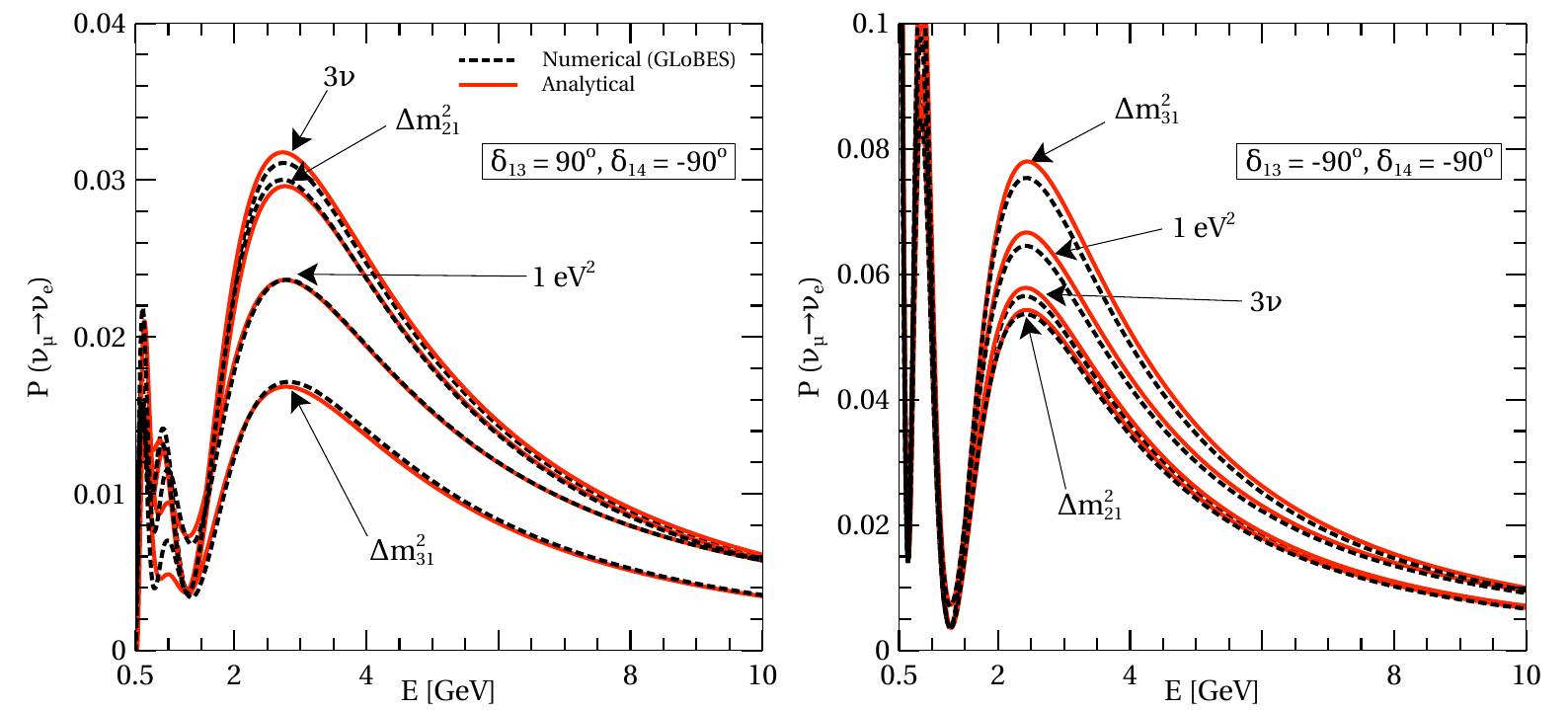}
		\mycaption{Comparison between the $\nu_{\mu}\rightarrow\nu_{e}$ transition probabilities calculated numerically using GLoBES and those calculated analytically using eqs.~(\ref{equ:palazzo_3n})--(\ref{equ:palazzo_sol}), as a function of neutrino energy for DUNE. The two panels correspond to two different combinations of CP phases ($\delta_{13}$, $\delta_{14}$) and in each panel  we show the probability for standard 3$\nu$-scenario and 3+1 scenario with three benchmark choices of $\Delta{m}^2_{41}$, namely, $\Delta{m}^2_{41}~=~\Delta{m}^2_{31}$, $\Delta{m}^2_{41}~=~\Delta{m}^2_{21}$, and $\Delta{m}^2_{41}~=~1~\rm{eV}^2$. The values of the other oscillation parameters are fixed at their benchmark values given in table~\ref{tab:params_value_str}.\label{fig:validation}}
	\end{figure}
	
	In \figu{validation}, we have shown the transition probability as a function of energy for DUNE. The two panels correspond to two different combinations of CP phases $\delta_{13}$ and $\delta_{14}$ as labeled in the plots. The solid red curves are obtained using the approximate expressions of the transition probabilities in $3\nu$ scenario as well as in 3+1 scenario with $\Delta m^2_{41} = \Delta m^2_{31}$, $\Delta m^2_{41} = \Delta m^2_{21}$, and $\Delta m^2_{41} = 1~\text{eV}^2$. The dotted black curves are the same generated numerically using GLoBES software. The figure shows that the analytical results with approximate expressions~(refer to section~\ref{sec:basics}, eqs.~(\ref{equ:palazzo_3n})--(\ref{equ:palazzo_sol})) and the numerical results obtained using GLoBES are in very good agreement.
	
	%=============================================================================	
	\section{Full probability expressions in vacuum in case of atmospheric, solar, and averaged out scales of the sterile frequency\label{Probability_limit}}
	\label{appendix:full_prob}
	%=============================================================================		
For $3\nu$ case, $\theta_{14}=\theta_{24}=\theta_{34}=0$, $\delta_{14}=\delta_{24}=\delta_{34}=0$, and $\Delta m^2_{41}=\Delta m^2_{42}=\Delta m^2_{43}=0$. Using these in \equ{pme_general}, we get the $3\nu$ oscillation probability as,
	\begin{align} \label{eq:pme_3n}
		\nonumber P_{\mu e}^{3\nu} &=
		\sin^22\tmet
		\big[\cos^2\tx\sin^2\Delta_{31}+\sin^2\tx\sin^2\Delta_{32} \big]\\
		\nonumber&+\big[b^2- \frac{1}{4} \sin^22\tmet \sin^22\tx] \sin^2\Delta_{21}\\
		\nonumber &+ b \cos\da \sin2\tmet 
		\big[\cos2\tx\sin^2\Delta_{21}+\sin^2\Delta_{31}-\sin^2\Delta_{32} \big]\\
		\nonumber &-  \frac{1}{2} b \sin\da \sin2\tmet \big[\sin2\Delta_{21}-\sin2\Delta_{31}+\sin2\Delta_{32}\big]\\
	\end{align}

For $4\nu$ oscillation scenario, when $\Delta m^2_{41}\sim~1$ eV$^2$, the oscillations due to $\Delta m^2_{41}$, $\Delta m^2_{42}$, and $\Delta m^2_{43}$ are averaged out. Using these in \equ{pme_general}, we get the $4\nu$ oscillation probability at the averaged out scale as,	
	\begin{eqnarray} \label{eq:pme_1eV2}
		\nonumber P_{\mu e}^{4\nu}\big|_{\text{high}~\Delta m^2_{41}} &=& \frac{1}{2}\sin^22\tmef \\
		\nonumber &+&  (a^2\sin^22\tmet - \frac{1}{4}\sin^22\ty\sin^22\tmef)
		\big[\cos^2\tx\sin^2\Delta_{31}+\sin^2\tx\sin^2\Delta_{32} \big]\\
		\nonumber &+& a^2b \cos\da \sin2\tmet 
		\big[\cos2\tx\sin^2\Delta_{21}+\sin^2\Delta_{31}-\sin^2\Delta_{32} \big]\\
		\nonumber &+& ab \cos\db \sin2\tmef 
		\big[\cos2\tx\cos^2\ty\sin^2\Delta_{21}-\sin^2\ty(\sin^2\Delta_{31}
		-\sin^2\Delta_{32}) \big]\\
		\nonumber &+&  a \cos(\da - \db) \sin2\tmet \sin2\tmef
		\Big[-\frac{1}{2}\sin^22\tx\cos^2\ty\sin^2\Delta_{21} \\ 
		\nonumber &+& \cos2\ty(\cos^2\tx\sin^2\Delta_{31}+\sin^2\tx\sin^2\Delta_{32})
		\Big]\\
		\nonumber &+&
		(a^2b^2-\frac{1}{4}a^2\sin^22\tx\sin^22\tmet-\frac{1}{4}
		\cos^4\ty\sin^22\tx\sin^22\tmef) \sin^2\Delta_{21}\\
		\nonumber &-& \frac{1}{2} a^2b \sin\da \sin2\tmet 
		\big[\sin2\Delta_{21}-\sin2\Delta_{31}+\sin2\Delta_{32} \big]\\
		\nonumber &-& \frac{1}{2} ab \sin\db \sin2\tmef 
		\big[\cos^2\ty\sin2\Delta_{21}+\sin^2\ty(\sin2\Delta_{31}-\sin2\Delta_{32})
		\big]\\
		\nonumber &+& \frac{1}{2} a \sin(\da - \db) \sin2\tmet \sin2\tmef 
		\big[\cos^2\tx\sin2\Delta_{31}+\sin^2\tx\sin2\Delta_{32} \big]\\
	\end{eqnarray}
For $4\nu$ oscillation scenario, when $\Delta m^2_{41}$ is at the atmospheric scale, \ie~ $\Delta m^2_{41}\sim~\Delta m^2_{31}$, then we have $\Delta_{41}\sim\Delta_{31}$, and hence, $\Delta_{43}\sim0$ and $\Delta_{42}\sim\Delta_{32}$. Using these in \equ{pme_general}, we get the $4\nu$ oscillation probability at the atmospheric scale as,		
	\begin{align} \label{eq:pme_atm}
		\nonumber P_{\mu e}^{4\nu}\big|_{\Delta m^2_{41}\,\simeq\,\Delta m^2_{31}} &= 
		(a^2\sin^22\tmet - \frac{1}{4}\sin^22\ty\sin^22\tmef\\ \nonumber&+\sin^22\tmef\cos^2{\theta_{13}})
		\big[\cos^2\tx\sin^2\Delta_{31}+\sin^2\tx\sin^2\Delta_{32} \big]\\
		\nonumber &+ a^2b \cos\da  \sin2\tmet 
		\big[\cos2\tx\sin^2\Delta_{21}+\sin^2\Delta_{31}-\sin^2\Delta_{32} \big]\\
		\nonumber &+ ab \cos\db \sin2\tmef 
		\big[\cos2\tx\cos^2\ty\sin^2\Delta_{21}+\cos^2\ty(\sin^2\Delta_{31}
		-\sin^2\Delta_{32}) \big]\\
		\nonumber &+ a \cos(\da - \db) \sin2\tmet \sin2\tmef
		\Big[-\frac{1}{2}\sin^22\tx\cos^2\ty\sin^2\Delta_{21} \\ 
		\nonumber &+ 2\cos^{2}\ty(\cos^2\tx\sin^2\Delta_{31}+\sin^2\tx\sin^2\Delta_{32})
		\Big]\\
		\nonumber &+
		(a^2b^2-\frac{1}{4}a^2\sin^22\tx\sin^22\tmet-\frac{1}{4}
		\cos^4\ty\sin^22\tx\sin^22\tmef) \sin^2\Delta_{21}\\
		\nonumber &- \frac{1}{2} a^2b \sin\da \sin2\tmet 
		\big[\sin2\Delta_{21}-\sin2\Delta_{31}+\sin2\Delta_{32} \big]\\
		\nonumber &- \frac{1}{2} ab \sin\db \sin2\tmef 
		\cos^2\ty\big[\sin2\Delta_{21}-\sin2\Delta_{31}+\sin2\Delta_{32}
		\big]\\
	\end{align}
For $4\nu$ oscillation scenario, when the sterile frequency is very high, $\Delta m^2_{41}$ is at the solar scale, \ie~ $\Delta m^2_{41}\sim~\Delta m^2_{21}$, then we have $\Delta_{41}\sim\Delta_{21}$, and hence, $\Delta_{42}\sim0$ and $\Delta_{43}\sim(-\Delta_{32})$. Using these in \equ{pme_general}, we get the $4\nu$ oscillation probability at the solar scale as,	
	\begin{align} \label{eq:pme_sol}
		\nonumber P_{\mu e}^{4\nu}\big|_{\Delta m^2_{41}\,\simeq\,\Delta m^2_{21}} 	 
		&= \sin^22\tmef\big[ \cos^2\theta_{13}\cos^2\theta_{12}\sin^2\Delta_{21}+\sin^2\theta_{13}\sin^2\Delta_{32}\big]\\\nonumber 	 
		&+(a^2\sin^22\tmet - \frac{1}{4}\sin^22\ty\sin^22\tmef)
		\big[\cos^2\tx\sin^2\Delta_{31}+\sin^2\tx\sin^2\Delta_{32} \big]\\
		\nonumber &+ a^2b \cos\da  \sin2\tmet 
		\big[\cos2\tx\sin^2\Delta_{21}+\sin^2\Delta_{31}-\sin^2\Delta_{32} \big]\\
		\nonumber &+ ab \cos\db \sin2\tmef 
		\big[\cos2\tx\cos^2\ty\sin^2\Delta_{21}-\sin^2\ty(\sin^2\Delta_{31}
		-\sin^2\Delta_{32})+\sin^2\Delta_{21} \big]\\
		\nonumber &+ a \cos(\da - \db)  \sin2\tmet \sin2\tmef
		\Big[-\frac{1}{2}\sin^22\tx\cos^2\ty\sin^2\Delta_{21} \\ 
		\nonumber &+ \cos 2\ty(\cos^2\tx\sin^2\Delta_{31}+\sin^2\tx\sin^2\Delta_{32}) + \cos^2\tx \sin^2 \Delta_{21}-\sin^2\Delta_{32}
		\Big]\\
		\nonumber &+
		(a^2b^2-\frac{1}{4}a^2\sin^22\tx\sin^22\tmet-\frac{1}{4}
		\cos^4\ty\sin^22\tx\sin^22\tmef) \sin^2\Delta_{21}\\
		\nonumber &- \frac{1}{2} a^2b \sin\da \sin2\tmet 
		\big[\sin2\Delta_{21}-\sin2\Delta_{31}+\sin2\Delta_{32} \big]\\
		\nonumber &- \frac{1}{2} ab \sin\db \sin2\tmef 
		\sin^2\ty\big[-\sin2\Delta_{21}+\sin2\Delta_{31}-\sin2\Delta_{32}
		\big]\\
		\nonumber &+ \frac{1}{2} a \sin(\da - \db)  \sin2\tmet \sin2\tmef 
		\cos^2\tx\big[-\sin2\Delta_{21}+\sin2\Delta_{31}-\sin2\Delta_{32} \big]\\
	\end{align}
	
\end{appendix}

%=================================================
\blankpage  
%%%%%%%%%%%%%%%%%%%%%%%%%%%%%%%%%%%%%%%%%%%%%%%%%%%%%%%%%%%%%%%%%%%%

\addcontentsline{toc}{chapter}{REFERENCES}
\bibliographystyle{hcdas}
\bibliography{pragyanprasu-thesis}
\end{document}